\documentclass[11pt,letterpaper]{article}
\ifdefined\pdfoutput\pdfoutput=1\fi

\usepackage[margin=1in]{geometry}
\usepackage[utf8]{inputenc}
\usepackage[T1]{fontenc}
\usepackage[nopatch=footnote]{microtype}
\usepackage{amsfonts}
\usepackage{amssymb}
\usepackage{amsmath}
\usepackage{amsthm}
\usepackage{booktabs}
\usepackage{longtable}
\usepackage[shortlabels,inline]{enumitem}
\usepackage{setspace}
\usepackage[hidelinks,bookmarks,bookmarksopen,bookmarksnumbered]{hyperref}
\usepackage[noend,noline,ruled,linesnumbered]{algorithm2e}
\usepackage[capitalize,noabbrev]{cleveref}
\usepackage{mathtools}
\usepackage{bm}
\usepackage{dsfont}
\usepackage{calc}
\usepackage{tikz}
\usepackage[font=small,labelfont=bf]{caption}
\usepackage{titlesec}
\usepackage{thmtools}
\usepackage{wrapfig}
\usepackage{tablefootnote}
\usepackage{thm-restate}
\usepackage{soul}
\usepackage[draft]{fixme}
\usepackage{framed}
\usepackage{lineno}
\usepackage{subcaption}
\usepackage[norefs,nocites,nomsgs]{refcheck}
\usepackage{xspace}

\usetikzlibrary{arrows}
\usetikzlibrary{arrows.meta}
\usetikzlibrary{decorations.pathreplacing}
\usetikzlibrary{shapes.geometric}

\input{settings}

\SetKwComment{Comment}{$\triangleright$\ \rm}{}

\usetikzlibrary{arrows,arrows.meta,shapes.misc, decorations.pathreplacing,decorations.pathmorphing,calligraphy, patterns.meta}
\tikzset{dotmark/.style={circle,fill,inner sep=1.5pt}}
\tikzset{emptymark/.style={circle,draw,fill=white,inner sep=1.5pt}}
\tikzset{crossmark/.style={thick,inner sep=1.5pt}}

\newcommand{\A}{\mathcal{A}}
\newcommand{\B}{\mathcal{B}}

\newcommand{\canD}{\mathfrak{D}}
\newcommand{\prD}{\mathfrak{D}^{\circ}}
\newcommand{\E}{E}
\newcommand{\Es}{\mathsf{E}}

\DeclareRobustCommand{\twoheadleadsto}{\tikz[baseline=(a.base)]{%
        \node at (.2,0) {\(\leadsto\)};%
        \fill[white] (-.1,-.1+.012) -- (.25,-.1+.012) -- (.345,0.011) -- (.25,.113) --
        (-.1,.113) --cycle;
        \node at (.125,0) {\(\leadsto\)};%
        \node (a) at (.4/2,-.0) {\phantom{\(\leadsto\)}};%
}}

\def\onto{\twoheadleadsto}
\def\ponto#1{\twoheadleadsto\kern-.3em{}_{#1}\kern.3em}

\SetKwFunction{densecover}{DenseCover}

\def\fragmentco#1#2{\bm{[}\,#1\,\bm{.\,.}\,#2\,\bm{)}}
\def\fragmentoc#1#2{\bm{(}\,#1\,\bm{.\,.}\,#2\,\bm{]}}
\def\fragmentoo#1#2{\bm{(}\,#1\,\bm{.\,.}\,#2\,\bm{)}}
\def\fragment#1#2{\bm{[}\,#1\,\bm{.\,.}\,#2\,\bm{]}}
\newcommand{\fragmentcc}{\fragment}
\def\position#1{\bm{[}\,#1\,\bm{]}}

\def\emptyset{\varnothing}

\newcommand{\Oh}{\mathcal{O}}
\newcommand{\tOh}{{\rlap{\raisebox{-0.2ex}{$\widetilde{\phantom{\Oh}}$}}\Oh}}

\newcommand{\Ohtilde}{\tOh}

\newcommand{\OccE}{\mathrm{Occ}}
\newcommand{\ed}{\mathsf{ed}}
\newcommand{\cost}{\operatorname{cost}}

\newcommand{\edl}[2]{{\ed}(#1,{}^*\!#2^*)}

\def\modelname{{\tt PILLAR}\xspace}

\def\ipmOpName{{\tt IPM}\xspace}
\def\accOpName{{\tt Access}\xspace}
\def\extractOpName{{\tt Extract}\xspace}
\def\lenOpName{{\tt Length}\xspace}

\def\lceOpName{{\tt LCP}\xspace}
\def\lcbOpName{{\tt LCP$^R$}\xspace}

\newcommand{\ceil}[1]{\left\lceil #1 \right\rceil}
\newcommand{\floor}[1]{\left\lfloor #1 \right\rfloor}

\def\Q{\ensuremath Q^{\infty}}

\newcommand{\brp}[1]{\mathsf{bp}(#1)}

\SetKwFunction{swaps}{TrimmedPM}

\newcommand{\SM}{\textsc{\ref{prob:sm}}\xspace}

\long\def\SMproblem{
    \begin{problem}[AlignedPeriodicMatches]{Aligned\-Periodic\-Matches{\tt($P$, $T$, $k$, $d$, $Q$, $\brp{\A_P}$, $\brp{\A_T}$)}}
    \label{prob:sm}

    \PInput{A pattern $P$ of length $m$,\\
    an integer threshold $k\in\fragment{0}{m}$,\\
    a text~$T$ of length
    $n \in \fragmentco{m-k}{\ceil{\threehalfs m} + k}$,\\
    a positive integer structural budget $d\ge 2k$,\\
    a primitive string $Q$ of length
    $q \coloneqq |Q| \le {m}/{8d}$,\\
    the breakpoint representation $\brp{\A_P}$ of an edit-distance
    alignment $\A_P: P \onto Q^\infty\fragmentco{0}{y_P}$
        of cost $d_P\coloneqq \edl{P}{Q}\le d$,\\
        the breakpoint representation $\brp{\A_T}$ of an edit-distance
        alignment $\A_T : T \onto Q^\infty{\fragmentco{x_T}{y_T}}$ of cost
        $d_T \coloneqq \edl{T}{Q}\le 3d$, where
        $x_T \in \fragmentco{0}{q}$.}

    \POutput{$\OccE_k(P,T)$, represented as
    $\Oh(d^3)$ disjoint arithmetic progressions of difference~$q$.}
    \end{problem}}

\usepackage{mathtools} %

\usepackage{pgf}

\def\betav{8}

\pgfmathsetmacro{\betavh}{int(\betav/2)}
\def\threehalfs{{}^3{\mskip -4mu/\mskip -3.5mu}_2\,}

\newcommand{\AG}{\mathsf{AG}}
\newcommand{\oAG}{\overline{\mathsf{AG}}}

\newcommand{\dist}{\mathsf{dist}}
\newcommand{\dtob}{\mathsf{d}_{\downarrow}}
\newcommand{\tdtob}{\widetilde{\mathsf{d}}_{\downarrow}}
\newcommand{\cancost}{\mathsf{cc}}

\newcommand{\meq}[1]{\mathop{\smash{\stackrel{#1}{=}}\vphantom{{}^{|^|}}}}
\newcommand{\nmeq}[1]{\mathop{\smash{\stackrel{#1}{\neq}}\vphantom{{}^{|^|}}}}

\def\udmat#1#2#3{\smash{\ensuremath D_{#1,#2}^{\mathstrut}
\smash{\raisebox{-.025ex}{\big|}}_{#3}^{\mathstrut}}}

\newcommand{\hG}{\hat{G}}

\newcommand{\dens}[1]{#1^{\square}}

\makeatletter
\renewenvironment{cases}{%
    \matrix@check\cases\env@cases
}{%
    \endarray\right.%
}
\def\env@cases{%
    \let\@ifnextchar\new@ifnextchar
    \left\lbrace
        \def\arraystretch{1.1}%
        \array{@{\;}c@{\quad}l@{}}%
}
\makeatother

\SetKwFunction{locked}{Locked}

\newcommand{\light}{\mathbb{L}}

\def\agn#1{\sim_{#1}}
\def\agneq#1{\simeq_{#1}}

\newcommand{\hA}{\hat{\A}}

\newcommand{\hT}{\hat{T}}
\newcommand{\hP}{\hat{P}}
\newcommand{\hF}{\hat{F}}

\newcommand{\Z}{\mathbb{Z}}

\newcommand{\htp}{\hat{p}}
\newcommand{\htt}{\hat{t}}

\newcommand{\ep}{\tilde{p}}
\newcommand{\et}{\tilde{t}}
\newcommand{\eA}{\tilde{\mathcal{A}}}
\newcommand{\eAG}{\widetilde{\AG}}
\newcommand{\eD}[1]{\tilde{\mathcal{D}}_{#1}}
\newcommand{\hD}[1]{\hat{\mathcal{D}}_{#1}}

\newcommand{\pP}{\pi_P}
\newcommand{\phP}{\pi_{\hat{P}}}
\newcommand{\pT}{\pi_T}
\newcommand{\phT}{\pi_{\hat{T}}}
\newcommand{\pPT}{\pi_{P,T}}
\newcommand{\phPT}{\pi_{\hat{P},\hat{T}}}

\newcommand{\qp}{\htp}

\newcommand{\qt}{\htt}

\newcommand{\tpena}{\widetilde{\mathsf{pen}}}

\newcommand{\tClose}{\widetilde C}
\newtheorem{observation}[theorem]{Observation}
\addtotheorempostheadhook[theoremqq]{\let\qed\relax}
\crefname{observation}{Observation}{Observations}
\Crefname{observation}{Observation}{Observations}

\SetCommentSty{textnormal}

\hypersetup{pageanchor=false}
\begin{document}

\title{\vspace*{-2em}Pattern Matching with Edits in \texorpdfstring{$\Ohtilde(n+k^3 \cdot n/m)$}{Õ(n+k³·n/m)} Time}

\author{
  \large Tomasz Kociumaka\\[-0.3ex]
  \normalsize Max Planck Institute for Informatics,\\[-0.3ex]
  \normalsize Saarland Informatics Campus,\\[-0.3ex]
  \normalsize Saarbrücken, Germany\\[-0.3ex]
  \normalsize \texttt{tomasz.kociumaka@mpi-inf.mpg.de}
}

\date{\vspace{-1.0cm}}
\maketitle

\begin{abstract}
In the Pattern Matching with Edits problem, given a pattern $P$
of length $m$, a text $T$ of length $n$, and a threshold~$k$, the
task is to decide whether $T$ contains a substring that can be
transformed into $P$ using at most $k$ character insertions, deletions,
and substitutions or, in the reporting version, to find all starting
positions where such substrings occur.  This textbook problem is arguably the most natural
formalization of approximate pattern matching.  Its formulation goes back
to Sellers [J.~Algorithms'80], whose dynamic-programming procedure runs in $\Oh(nm)$
time; the classic algorithm of Landau and Vishkin [STOC'86,
J.~Algorithms'89] takes $\Oh(nk)$ time.  Both running times are near-optimal for $k=\Theta(m)$,
conditioned on the Orthogonal Vectors Hypothesis, and subsequent work
focused on small~$k$.  Cole and Hariharan [SODA'98, SICOMP'02] achieved
$\Oh(n+k^4\cdot n/m)$ time, and more than two decades later,
Charalampopoulos, Kociumaka, and Wellnitz (CKW) [FOCS'22] improved the running
time to $\Oh(n+k^{3.5}\log k\cdot n/m)$.  This is essentially
halfway to the conjecture of Cole and Hariharan, which
postulates that ${\Oh(n+k^3\cdot n/m)}$ time is possible.

We confirm this long-standing conjecture up to a polylogarithmic factor and present an
$\Oh(n+k^3\log^2 k\cdot n/m)$-time algorithm.  Our algorithm works in the
\modelname model [CKW; FOCS'20],
which accesses strings only through a few basic operations.  Hence, our algorithm extends to
\mbox{grammar-compressed} and dynamic strings and to the quantum and small-space
\mbox{read-only} settings, where these operations admit efficient
implementations.

After almost three decades, our algorithm closes, up to the $\Oh(\log^2 k)$ factor,
the gap between the worst case and certain aperiodic patterns.
Cole and Hariharan achieved $\Oh(n+k^3\cdot n/m)$ time for such patterns, and
CKW [FOCS'20] provided an alternative approach with a clear
remaining bottleneck: instances in which the pattern and the text are close to
periodic strings with a common period, to which the general case reduces.

For such instances, the task amounts
to evaluating many min-plus products of distance matrices
between short, highly similar substrings of the pattern and the text.  We
replace every long periodic stretch in these products by one fixed
matrix and show that this replacement preserves the distances that determine the
occurrences unless the parts of the pattern and the text that break periodicity
interact strongly near the corresponding period shift, that is, the relative shift
between the periodic structures of the pattern and the text.  Our main structural result, a planar
variation theorem, bounds, in terms of such interactions, how much optimal
alignment costs vary across neighboring period shifts beyond a simple baseline.
Algorithmically, the theorem confines the remaining hard cases to
regions so narrow that we can \mbox{afford exact distance computation}.
\end{abstract}
\thispagestyle{empty}
\clearpage
\pagestyle{empty}
\tableofcontents
\clearpage
\pagestyle{plain}
\hypersetup{pageanchor=true}
\setcounter{page}{1}

\section{Introduction}\label{sec:introduction}

Finding the occurrences of a pattern in a text, that is, the positions
where the pattern appears as a contiguous substring of the text, is among
the most basic tasks on strings, and efficient algorithms for this exact
pattern matching problem
date back to the 1970s~\cite{KMP77,BoyerM77}; see
also~\cite{GalilS83,KarpR87,CrochemoreP91}.  In many applications, however, exact
occurrences are too restrictive: a single typo in a document or a single
sequencing error in a DNA read hides an occurrence.  The most common way
to tolerate such errors is the \emph{edit distance} $\ed(X,Y)$ of two
strings $X$ and $Y$~\cite{Levenshtein66}, the minimum number of character
insertions, deletions, and substitutions (collectively called
\emph{edits}) that transform $X$ into~$Y$.  Its pattern-matching variant, also known as the $k$-differences problem,
asks whether a text $T$ has a substring at edit distance at most $k$
from a pattern $P$.  This problem has been studied since
1980~\cite{Sellers80}, it is covered by textbooks on string
algorithms~\cite{Gusfield97,CrochemoreR02,NavarroR02,AlgorithmsOnStrings,Kingsford26},
and it is a core primitive in
computational biology~\cite{MBCT23}, for example, in read
mapping~\cite{AlserEtAl21,SahlinBCM23}.

\newcommand{\PMWE}{\hyperref[prob:pmwe]{\textsc{\mbox{Pattern}\allowbreak\mbox{Matching}\allowbreak\mbox{With}\allowbreak\mbox{Edits}}}\xspace}
\begin{problem}[PatternMatchingWithEdits]{Pattern\-Matching\-With\-Edits{\tt($P$, $T$, $k$)}}
\label{prob:pmwe}
\PInput{A pattern $P$ of length $m\ge1$, a text $T$ of length $n\ge1$,
and an integer threshold $k\ge0$.}
\POutput{The set of starting positions of \emph{$k$-error occurrences}
of $P$ in $T$, that is, of fragments\footnotemark{} $T\fragmentco{s}{e}$
with $\ed\bigl(P,T\fragmentco{s}{e}\bigr)\le k$:
\[
    \OccE_k(P,T)
    \coloneqq
    \left\{
        s\in\fragment{0}{|T|}:
        \min_{e\in\fragment{s}{|T|}}
            \ed\bigl(P,T\fragmentco{s}{e}\bigr)\le k
    \right\}.
\]
In the decision version, the task is to decide whether
$\OccE_k(P,T)\ne\emptyset$.}
\end{problem}
\footnotetext{Strings are indexed from zero, and a \emph{fragment} $T\fragmentco{a}{b}$ of $T$
is an occurrence of ${T[a]T[a+1]\cdots T[b-1]}$ for $0\le a\le b\le|T|$;
the notations $\fragment{a}{b}$ and $\fragmentco{a}{b}$ denote
inclusive and half-open integer intervals.}
Most algorithms, including ours, solve the reporting version: they compute
the full set $\OccE_k(P,T)$.

The dynamic-programming (DP) algorithm of Sellers~\cite{Sellers80} solves Pattern Matching
with Edits in $\Oh(nm)$ time, and Masek and Paterson~\cite{MasekP80}
saved a polylogarithmic factor by precomputing the results for all small
blocks of the DP table.
Landau and Vishkin reduced the running time to
$\Oh(nk^2)$~\cite{LandauV88} and then to $\Oh(nk)$~\cite{LandauV89}, and
Galil and Park~\cite{GalilP90} obtained the same bound while
preprocessing only the pattern.
Many further algorithms are fast in practice or on random inputs,
using filtering, bit parallelism, or
automata~\cite{Ukkonen85,WuM92,TarhioU93,ChangL94,Baeza-YatesN99,Myers99},
and tools such as Edlib~\cite{SosicS17} and Sassy~\cite{BeelooGK26} build
on them; see the surveys~\cite{GalilG88,Navarro01}.

Subsequent theoretical work focused on small $k$.  Sahinalp and Vishkin~\cite{SV96} were the
first to achieve linear time for $k\lesssim m^{0.07}$.  Cole
and Hariharan~\cite{ColeH98} gave a simpler and faster algorithm, which
took $\Oh(n+k^4\cdot n/m)$ time, linear for $k=\Oh(m^{1/4})$.  They conjectured that $\Oh(n+k^3\cdot n/m)$ time could
be achieved.  The structural approach of Charalampopoulos, Kociumaka, and
Wellnitz~\cite{CKW20} yielded another $\Oh(n+k^4\cdot n/m)$-time
algorithm, which also extended to compressed and dynamic settings.  More
than two decades after Cole and Hariharan,
the same authors~\cite{CKW22} broke the quartic barrier with an
$\Oh(n+k^{3.5}\lg k\cdot n/m)$-time algorithm,\footnote{Throughout, $\lg z\coloneqq\log_2(\max(2,z))$ for real
$z$, and $\Ohtilde(\cdot)$ hides factors polylogarithmic in the lengths of the strings.} halfway to the conjecture.

In this work, we confirm the conjecture of Cole and Hariharan up to a polylogarithmic
factor.

\begin{theoremq}[\PMWE in $\Ohtilde(n+k^3\cdot\ceil{n/m})$ time]\label{thm:standard_setting}
The \PMWE problem can be solved in
${\Oh(n+k^3\lg^2 k\cdot\ceil{n/m})}$ time.
\end{theoremq}

For every constant $\varepsilon>0$, the running time of \cref{thm:standard_setting} is linear for
$k \le m^{1/3-\varepsilon}$, and it is polynomially faster than all previous algorithms when
$m^{2/7+\varepsilon}\le k\le m^{1/2-\varepsilon}$.  The natural question
is what happens for $m^{1/3+\varepsilon}\le k\le m^{1-\varepsilon}$ and,
in particular, how much further the exponent of~$k$ can be reduced.  As a
benchmark, the edit distance~$k$ of two
strings of length at most~$n$ can be computed in $\Oh(n+k^2)$
time~\cite{LandauV88,Myers86}, and this is optimal up to subpolynomial
factors conditioned on the Orthogonal Vectors Hypothesis
(OVH)~\cite{Williams05}: an $\Oh(n+k^{2-\varepsilon})$-time algorithm for a
constant $\varepsilon>0$ would refute OVH~\cite{bi18}.  This lower bound
holds pointwise, that is, also for instances with ${k=\Theta(n^{\kappa})}$
for any fixed $\kappa\in(1/2,1]$.  Padding
arguments~\cite[Section~1]{CKW22} extend this lower bound to \mbox{\PMWE}: an
$\Oh(n+k^{2-\varepsilon}\cdot n/m)$-time algorithm would refute OVH even
if restricted to instances with ${m=\Theta(n)}$ and
${k=\Theta(m^{\kappa})}$ for a fixed $\kappa\in(1/2,1]$.  Whether \PMWE
can be solved in $\Ohtilde(n+k^{3-\Omega(1)}\cdot n/m)$ time remains open;
Cole and Hariharan~\cite{ColeH98} believed $k^3$ to be the right exponent and beating it to be hard.

The exponent~$3$ in the conjecture of Cole and Hariharan stems from
patterns without periodic structure: their algorithm takes
$\Oh(n+k^4\cdot n/m)$ time only if the pattern is \emph{$k$-break
periodic}, that is, it consists of periodic pieces separated by $\Oh(k)$
short breaks, and $\Oh(n+k^3\cdot n/m)$ time otherwise.  The approach
of~\cite{CKW20} isolated the bottleneck more precisely: instances in which
$P$ and $T$ are close to periodic strings with a common period, to which
the general case reduces.  For these instances, the dependence on~$k$
remained $k^4$ in~\cite{CKW20} and dropped to $k^{3.5}\lg k$
in~\cite{CKW22}; we reduce it to $k^3\lg^2 k$.  Thus, after almost three
decades, \cref{thm:standard_setting} closes the gap between the worst
case and aperiodic patterns, up to the $\lg^2 k$ factor.  Yet, no
${\tOh(n+k^{3-\Omega(1)}\cdot n/m)}$-time algorithm is known, not even for
aperiodic patterns or any other broad class of patterns that might be considered typical.

\paragraph*{Our \modelname result and its consequences.}
Like the algorithms of~\cite{CKW20,CKW22}, ours can be formulated
in the \emph{\modelname model}~\cite[Section~2.1]{CKW20}, in which
strings are accessed only through a few primitive operations, such as
retrieving a character, computing the longest common prefix of two
fragments, and finding the exact occurrences of a fragment in a fragment
at most twice as long (see \cref{def:pillar_model}).  The running time counts
these operations together with ordinary computation.

\begin{restatable*}[\PMWE in $\Ohtilde(k^3\cdot\ceil{n/m})$ time in the \modelname model]{mtheorem}{edalgII}
    \dglabel{thm:edalgII}[fact:reduction,lem:sthold]
    For ${k>0}$, the \PMWE problem can be solved in
    $\Oh(k^3\lg^2 m\cdot\ceil{n/m})$ time in the \modelname model, with
    $\OccE_k(P,T)$ given as $\Oh(k^3\cdot\ceil{n/m})$ pairwise disjoint arithmetic progressions with a common difference.
\end{restatable*}

The output representation in \cref{thm:edalgII} follows~\cite{CKW20}, whose main structural insight is
that $|\OccE_k(P,T)|=\Oh(k^2\cdot\ceil{n/m})$ unless $P$ is approximately
periodic, that is, within $\Oh(k)$ edits from a string with a period of
length $\Oh(m/k)$.  If $P$ is approximately periodic, $\OccE_k(P,T)$ is the union of
$\Oh(k^3\cdot\ceil{n/m})$ pairwise disjoint arithmetic progressions whose common
difference is the approximate period of~$P$.

The \modelname model admits efficient implementations in several
settings.  Combining \cref{thm:edalgII} with these implementations as
in~\cite{CKW22,gk24} improves, for $k>0$, the dependence on~$k$ from $k^{3.5}$ to $k^3$
in each of these settings, up to factors polylogarithmic in the lengths
of the strings.
\begin{itemize}
    \item \emph{Standard setting.} For ordinary strings, every \modelname
    operation can be implemented in $\Oh(1)$ time after $\Oh(n+m)$-time
    preprocessing; see~\cite[Theorem~7.2]{CKW20}.
    With \cref{thm:edalgII}, this yields $\Oh(n+k^3\lg^2 m\cdot\ceil{n/m})$
    time for $k>0$ (the remaining $k=0$ is exact pattern matching), which implies \cref{thm:standard_setting}: we may assume
    $m\le n+k$, as $\OccE_k(P,T)=\emptyset$ otherwise; if $k<m^{1/4}$, then
    ${k^3\lg^2 m\cdot\ceil{n/m}=\Oh(m\cdot\ceil{n/m})=\Oh(n+k)}$, and
    otherwise $\lg m\le 4\lg k$.
    \item \emph{Small-space read-only setting.} If $P$ and $T$ are
    in read-only memory, then, for every ${\tau\in\fragment{1}{m}}$, \modelname
    operations on strings of length $\Oh(m)$ take $\tOh(\tau)$ time and
    $\tOh(m/\tau)$ extra space~\cite{KosolobovS24,BathieCS24}.  Thus,
    computing $\OccE_k(P,T)$ takes $\tOh(n+k^3\tau\cdot n/m)$ time and
    $\tOh(m/\tau+k^3)$ extra space, and $\tau=\ceil{m/k^3}$ yields
    $\tOh(n+k^3\cdot\ceil{n/m})$ time and $\tOh(k^3)$ space.
    \item \emph{Fully compressed setting.} If $T$ and $P$ are given as
    straight-line programs (context-free grammars generating single
    strings) of sizes $g_T$ and $g_P$, respectively, then $|\OccE_k(P,T)|$
    can be computed in $\tOh(g_P+g_T\cdot k^3)$ time, and the set itself can be
    reported in $\Oh(|\OccE_k(P,T)|)$ extra time~\cite{I17,DuysterK24}.
    \item \emph{Quantum setting.} The quantum algorithm of~\cite{KNW25}
    runs a \modelname algorithm on compressed strings derived from the
    input.  With \cref{thm:edalgII}, it decides whether
    $\OccE_k(P,T)\ne\emptyset$ for $n\ge m$ in
    $\widehat\Oh(\sqrt{nk}+\sqrt{n/m}\cdot k^3)$ time, where
    $\widehat\Oh(\cdot)$ hides factors $n^{o(1)}$.
    \item \emph{Dynamic setting.} For strings in a collection maintained
    under concatenations and splits \cite{ods,DuysterK24}, $\OccE_k(P,T)$
    can be computed in $\tOh(k^3\cdot\ceil{n/m})$ time with high probability.
\end{itemize}

\paragraph*{Our techniques in a nutshell.}
We use the tools of~\cite{CKW20}, as stated in~\cite[Fact 1.1]{CKW22}, to reduce arbitrary instances to the
\emph{periodic case}, in which $P$ and $T$ are
within $\Oh(k)$ edits from periodic strings with a common short period $Q$ of length $q$.
As in~\cite{CKW22}, the candidate starting positions then fall into
groups of nearby \emph{period shifts}, that is, numbers of copies of $Q$
by which the periodic structure of~$T$ is shifted relative to that of~$P$.  For
each group, the required distances are the row minima of a min-plus
product of $\Oh(k)\times \Oh(k)$ Monge distance matrices, one for each factor of a fixed
factorization of~$P$.  Most factors are \emph{plain}: they follow the
periodic structure exactly, so the product consists of long runs of one
fixed matrix separated by $\Oh(k)$ other matrices.

The algorithm of~\cite{CKW22} trims every long run to length~$\alpha$,
so the products change $\Oh(k^2\alpha)$ times, at $\tOh(k)$ time each.
This preserves the required distances unless the periodicity-breaking
parts of $P$ and $T$ interact near the corresponding period shift.
Such interactions can cause false positives in $\Oh(k^2/\alpha)$ groups, each verified naively in
$\Oh(k^2)$ time, and $\alpha=\Theta(\sqrt k)$ balances the costs at
$\Ohtilde(k^{3.5})$.  Instead, we replace every run longer than a constant by one
fixed matrix and apply the Monge toolbox
of~\cite{GK25}; the products then change only $\Oh(k^2)$ times.
The replacement never underestimates distances, so its errors are
false negatives, which a minimum with local upper bounds repairs,
unlike false positives.  After a thorough redesign of the tools
of~\cite{CKW20,CKW22}, false negatives remain possible in only
$\Oh(k^2/q)$ groups, and in none if the period is much smaller
than~$\sqrt k$ or much larger than~$k$.

Our main contribution is to localize the repairs for the remaining groups
more precisely.  We assign a \emph{mark load} to each individual period shift,
rather than to entire groups as in~\cite{CKW22}, measuring how strongly the periodicity-breaking parts
interact near it; the loads still sum up to $\Oh(k^2)$.  The alignment that
matches the periodic structures of $P$ and $T$ with a given period shift
traces a path, called a \emph{rail}, in the alignment graph.  Our main structural result, the
\emph{planar variation theorem} (\cref{thm:combi}), states that, along any row of this graph, the optimal costs of aligning the rest
of~$P$ from the points where the rails cross this row vary in total, between
consecutive rails, by at most the variation of a
simple baseline plus a constant times the mark load near these rails.  Hence,
optimal alignments switch rails only near the few neighboring rails with
very different distances, confining the exact computations to
narrow regions whose processing cost the mark loads pay for.

\paragraph*{Related work.}
\PMWE has been studied in numerous settings.  For compressed texts, algorithms that avoid
decompression~\cite{KarkkainenNU03,DBLP:conf/isaac/GawrychowskiS13,t14,BilleLRSSW15}
preceded the \modelname-based ones~\cite{CKW20,CKW22}.  Text indexes
preprocess~$T$ so that the $k$-error occurrences of a pattern given at
query time can be reported quickly~\cite{ColeGL04,ChanLSTW10}.  In the
online setting, $T$ arrives one character at a time, and after each
character, the algorithm reports whether a \mbox{$k$-error} occurrence ends
there: building on~\cite{LandauMS98}, Clifford and Sach~\cite{CliffordS10}
spend $\Oh(k)$ time per character, the algorithms
of~\cite{KPS21,BathieKS23} use small space, and approximate answers take
amortized sublinear time per character~\cite{ChakrabortyDK19}.  In the
streaming model, where the space bound covers all information stored
about~$P$ and the processed prefix of~$T$, the most recent algorithm~\cite{BhattacharyaK23},
improving upon~\cite{Starikovskaya17,KPS21}, uses $\tOh(k^2)$ space and
$\tOh(k^2)$ time per character; related edit distance sketches appear
in~\cite{KPS21,BhattacharyaK23stoc,KouckyS24}.  In terms
of communication complexity, the output can be encoded in
$\Oh(k\log m\cdot\ceil{n/m})$ bits~\cite{KNW24,KNW26}, and such encodings
underlie quantum algorithms with near-optimal query
complexity~\cite{KNW24,KNW25}.  The structural results
of~\cite{CKW20} also apply to approximate
circular pattern matching~\cite{CharalampopoulosKPRRWZ22,CharalampopoulosPRRWZ24,CharalampopoulosEEGS26}.  Closer to our periodic case, the
longest common subsequence of periodic strings has been
studied~\cite{abs-0707-3619,Tiskin09,GZT25}.

For weighted edit distance, where edit costs are at least~$1$ and depend
on the characters involved, the best running times known are $\tOh(nk)$ and ${\tOh(n+k^4\cdot n/m)}$, and
${\tOh(n+k^{3.5}W^4\cdot n/m)}$ for integer metric weights bounded
by~$W$~\cite{gk24}, building on~\cite{DasGHKS23,CassisKW23,GK25}.

Pattern Matching with Mismatches, the analogous problem for
Hamming distance, can be solved in ${\tOh(n+k\sqrt m\cdot n/m)}$
time~\cite{GawrychowskiU18,cgkkp20,CKW20,ChanJVX23}, improving
upon \cite{Abrahamson,Kosaraju,LandauV86,GG86,AmirLP04,CliffordFPSS16}.
The $k$-mismatch occurrences form $\Oh(k^2\cdot\ceil{n/m})$ arithmetic
progressions, and this bound is tight~\cite{bkw19,CKW20}.  Designing a polynomially faster
algorithm, even for ${k=\Theta(m)}$, remains a major open problem and likely requires fast matrix
multiplication~\cite{GawrychowskiU18}.

\section{Technical Overview}\label{sec:technical_overview}

In this section, we provide a high-level description of the algorithm
behind \cref{thm:edalgII}, with the emphasis on
where it departs from the earlier solution~\cite{CKW22} and its weighted
counterpart~\cite{gk24}.  The outer structure is the same: a reduction
to a structured periodic problem, a grouping of the candidate starting
positions by period shifts, and min-plus products of distance matrices
that serve many groups at once.  Some departures clean up language and
notation, others redesign familiar notions into stronger and more robust
ones.
The most substantial novel contributions include the planar variation theorem and
the algorithm for high-load block indices (\cref{sec:overview_variation}).

\paragraph*{Notation.}
For a nonempty string $Q$ of length $q$, let $Q^\infty$ denote the
bi-infinite string with $Q^\infty[i]=Q[i\bmod q]$ for all
$i\in\mathbb Z$; we call positions in $Q^\infty$ \emph{reference
coordinates} and address its fragments by them, as in
$Q^\infty\fragmentco{x}{y}$.  If $Q$ is \emph{primitive}, that is, not
of the form $U^e$ for a string $U$ and an integer $e\ge2$, then the
\emph{synchronization property} holds: $Q^\infty\fragmentco{i}{i+q}=Q$
if and only if $q$ divides $i$. We use $\edl{X}{Q}$ to denote the minimum edit distance between $X$ and a
fragment of $Q^\infty$.

An \emph{alignment} $\A:X\onto Y$ is a
sequence of matches, substitutions, insertions, and deletions
transforming $X$ into $Y$, and $\cost(\A)$ is its number of edits.  Its
\emph{breakpoint representation} $\brp{\A}$ lists the edits in
$\Oh(\cost(\A)+1)$ space.
The \emph{alignment graph} of strings $X$ and $Y$ has vertices $(x,y)$
for $x\in\fragment{0}{|X|}$ and $y\in\fragment{0}{|Y|}$, edges
$(x,y)\to(x+1,y)$ and $(x,y)\to(x,y+1)$ of cost $1$, and edges
$(x,y)\to(x+1,y+1)$ of cost $0$ if $X[x]=Y[y]$ and $1$ otherwise.
Paths from $(0,y)$ to $(|X|,y')$ correspond to alignments of $X$ with
$Y\fragmentco{y}{y'}$.  In the alignment graph of $P$ and $T$, let
$\dtob(u)$ be the distance from a vertex $u=(x,y)$ to the bottom row,
that is, ${\min_{y'}\ed(P\fragmentco{x}{m},T\fragmentco{y}{y'})}$.

\subsection{Reduction to the Periodic Case}\label{sec:overview_reduction}

The characterization of~\cite{CKW20} distinguishes three cases.  In the
first case, $P$ contains many \emph{breaks}: short fragments with no
short period.  Then, $P$ has few
occurrences, and they can be verified one cluster at a time.  In the
second case, $P$ contains \emph{repetitive regions}: long fragments $H$,
each at a calibrated distance $\edl{H}{Q_H}=\ceil{8k|H|/m}$ from
periodicity, for its own short primitive string $Q_H$.  Then, $P$ again
has few occurrences, and they can be
found from the occurrences of the regions.  In the remaining case, $P$
itself is close to a substring of $Q^\infty$ for a short primitive
string $Q$.  In the last two cases, these results reduce the
computation to finding the occurrences of a new pattern $P'$: either
$P$ itself, which is then approximately periodic, or one of its
repetitive regions.  For each such $P'$, the reduction
of~\cite{CKW22} considers overlapping windows of $T$ of length about
$\threehalfs|P'|$, so that every occurrence of $P'$ lies in one of
them, and makes their periodic structure explicit by
aligning $P'$ and each window with fragments of the same
$Q^\infty$.  This yields instances of the following problem:

\SMproblem

An instance is \emph{valid} if it satisfies all promises of the input
specification.  The \emph{structural budget} $d$ bounds the edit
distances of $P$ and $T$ from the \emph{periodic references} $Q^\infty\fragmentco{0}{y_P}$ and $Q^\infty\fragmentco{x_T}{y_T}$.

\begin{fact}[Reduction to \texttt{AlignedPeriodicMatches},~{\cite[Fact 1.1 and its proof]{CKW22}}]
    \dglabel{fact:reduction}
    Let \(P\) be a pattern of length \({m\ge1}\), let \(T\) be a
    text of length \({n\ge1}\), and let \(k\in\fragment{1}{m}\).
    One can compute $\OccE_k(P,T)$, represented as
    $\Oh(k^3\cdot\ceil{n/m})$ pairwise disjoint arithmetic progressions
    with one common difference, in $\Oh(k^3\cdot\ceil{n/m})$ time in the
    \modelname model plus the time needed to solve valid instances
    $\SM(P^{(i)},T^{(i)},k^{(i)},d^{(i)},Q^{(i)},\brp{\A_{P^{(i)}}},\brp{\A_{T^{(i)}}})$
    with $|P^{(i)}|\le m$ for every $i$ and
    ${\sum_i(d^{(i)})^3=\Oh(k^3\cdot\ceil{n/m})}$.
\end{fact}

The bound on $\sum_i(d^{(i)})^3$ follows from the structural budgets
chosen in~\cite{CKW22} (see \cref{sec:main_theorem_proof}).  Hence,
\cref{thm:edalgII} reduces to the next theorem, proved in
\cref{sec:central_algorithm}.

\begin{restatable}[\texttt{AlignedPeriodicMatches($P$, $T$, $k$, $d$, $Q$, $\brp{\A_P}$,
    $\brp{\A_T}$)}]{theoremq}{newalg}
    \dglabel{lem:sthold}(\texttt{AlignedPeriodicMatches($P$, $T$, $k$, $d$, $Q$, $\brp{\A_P}$, $\brp{\A_T}$)} is in $\Oh(d^3\lg^2 m)$ time in the \modelname model)
    The \SM problem can be solved in $\Oh(d^3\lg^2 m)$ time in the
    \modelname model.
\end{restatable}

\begin{example}\label{ex:overview_periodic}
Consider the following instance with $k=1$ and $d=2$, which ignores
the length constraints of \SM:
${P=(\texttt{abc})^2\,\texttt{ab}\,(\texttt{abc})^3}$,
${T=(\texttt{abc})^4\,\texttt{ab}\,(\texttt{abc})^3\,\texttt{ab}\,\texttt{d}\,\texttt{c}\,\texttt{abc}}$,
and $Q=\texttt{abc}$.  Here, $\A_P:P\onto Q^\infty\fragmentco{0}{18}$
inserts the missing $\texttt{c}$ at reference coordinate $8$, so
$d_P=1$, and $\A_T:T\onto Q^\infty\fragmentco{0}{30}$ inserts the
missing $\texttt{c}$ at reference coordinate $14$ and deletes the extra
$\texttt{d}$ at reference coordinate $26$, so $d_T=2$.  In the drawing
below, the top row shows $Q^\infty$ with reference coordinates, and
$\A_T$ places every character of $T$ below the reference character it
is matched with; dashes mark the gaps of the alignments.  The lower rows
show the pattern with $\A_P$ applied after shifting its reference by $6$
and by $3$, together with the cost of aligning it with $T$ there.
\begin{center}
\begin{tikzpicture}[x=0.36cm,y=0.5cm,
    ch/.style={font=\ttfamily\small,inner sep=0pt,minimum width=0.3cm,anchor=mid},
    lab/.style={font=\small,anchor=mid east}]
  \fill[color3!12] (14,0.45) rectangle (15,-1.5);
  \fill[color3!12] (26,0.45) rectangle (27,-1.5);
  \fill[color1!14] (14,-1.55) rectangle (15,-2.45);
  \fill[color1!14] (11,-2.55) rectangle (12,-3.45);
  \foreach \i in {0,3,...,24} {\draw[black!25] (\i,0.55) -- (\i,-3.5);}
  \foreach \i in {28,31} {\draw[black!25] (\i,0.55) -- (\i,-1.5);}
  \foreach \i in {0,3,...,24} {\node[font=\scriptsize] at (\i+0.5,1.05) {\i};}
  \node[font=\scriptsize] at (28.5,1.05) {27};
  \node[lab] at (-0.4,0) {$Q^\infty$};
  \node[lab] at (-0.4,-1) {$T$};
  \node[lab] at (-0.4,-2) {$P$};
  \node[lab] at (-0.4,-3) {$P$};
  \foreach \i in {0,...,29} {
    \pgfmathsetmacro{\ch}{{"a","b","c"}[mod(\i,3)]}
    \pgfmathtruncatemacro{\xx}{\i<26 ? \i : \i+1}
    \ifnum\i=14 \node[ch,color3] at (\xx+0.5,0) {\ch};
      \node[ch,color3] at (\xx+0.5,-1) {--};
    \else \node[ch] at (\xx+0.5,0) {\ch};
      \node[ch] at (\xx+0.5,-1) {\ch};\fi
  }
  \node[ch,color3] at (26.5,0) {--};
  \node[ch,color3] at (26.5,-1) {d};
  \foreach \i in {6,...,23} {
    \pgfmathsetmacro{\ch}{{"a","b","c"}[mod(\i,3)]}
    \ifnum\i=14 \node[ch,color1] at (\i+0.5,-2) {--};
    \else \node[ch] at (\i+0.5,-2) {\ch};\fi
  }
  \foreach \i in {3,...,20} {
    \pgfmathsetmacro{\ch}{{"a","b","c"}[mod(\i,3)]}
    \ifnum\i=11 \node[ch,color1] at (\i+0.5,-3) {--};
    \else \node[ch] at (\i+0.5,-3) {\ch};\fi
  }
  \node[font=\scriptsize,anchor=mid west] at (24.3,-2) {cost $0$: $P=T\fragmentco{6}{23}$};
  \node[font=\scriptsize,anchor=mid west] at (24.3,-3) {cost $2$};
\end{tikzpicture}
\end{center}
The two missing characters cancel at position~$6$, where $P$ occurs
exactly, and $\OccE_1(P,T)=\{5,6,7\}$.
\end{example}

\subsection{Block Indices and the Common Factorization}\label{sec:overview_blocks}

To prove \cref{lem:sthold}, we fix a valid instance for the rest of the
overview and write
$\hP\coloneqq Q^\infty\fragmentco{0}{y_P}$ and
$\hT\coloneqq Q^\infty\fragmentco{x_T}{y_T}$ for the \emph{pattern
reference} and the \emph{text reference}.

Our first goal is to group the possible starting positions of
occurrences according to the periodic structure, so that each group can
be handled at once.  To this end, note that, via $\A_T$, every
occurrence corresponds to $\hP$ shifted by a whole number of periods,
up to $\kappa\coloneqq d+d_P+d_T\le 5d$ positions at either end
(\cref{lem:widened_endpoint_localization}); we call such a whole number
$b$ a \emph{period shift} of the occurrence.

There are about $(n-m)/q$ period shifts, and we group them into blocks
of $\tau/q$ consecutive ones, where $\tau$ is the least
multiple of $q$ exceeding $\kappa$.  The set $J$ of \emph{block
indices} consists of the $j$ for which $\hP$ shifted by $j\tau$ fits
into $\hT$, up to $\kappa$ positions at either end.  Block index $j$
has \emph{origin} $b_j\coloneqq j\tau/q$, and its \emph{start interval}
$\fragmentco{r_j}{r_{j+1}}$ is obtained by pulling the reference
coordinate $j\tau-\kappa$ back to $T$ through $\A_T$.  The start
intervals are disjoint and contain all occurrences, and every
occurrence of cost at most $d$ that starts in $\fragmentco{r_j}{r_{j+1}}$
starts within $\fragment{r_j}{r_j+4\kappa}$
(\cref{lem:working_budget_ckw_slicing}); we say that it is
\emph{assigned} to block index $j$.  These notions adapt~\cite[Definition~3.6]{CKW22}
to larger block length $\tau$.

Verifying the occurrences assigned to a single block index takes
$\Oh(d^2)$ time, which is too slow, since $|J|$ can be
$\Theta(m/\tau)$.  Instead, we exploit that most block indices look
alike.  The occurrences assigned to $j$ are the starts
${v\in\fragmentco{r_j}{r_{j+1}}\cap\fragment{r_j}{r_j+4\kappa}}$ with
$\dtob(0,v)\le k$, so it suffices to know the $\Oh(d)$ values
$\dtob(0,v)$ at these starts up to \emph{$k$-equivalence}, where
$a\meq{k}b$ means that $a=b$ or both $a$ and $b$ exceed $k$.  Our goal is to group
most block indices into $\Oh(d^2)$ intervals such that, within each
interval, the points $r_j$ are spaced exactly $\tau$ apart and the
values do not change when $j$ grows by one and $v$ by $\tau$.  Then,
the occurrences assigned to an interval form $\Oh(d)$ progressions with
difference $\tau$.

\begin{figure}[t]
\centering
\begin{tikzpicture}[x=.9cm,y=-0.9cm]
  \foreach \i/\fil in {0/white,1/white,2/white,3/white} {
    \pgfmathsetmacro{\xl}{0.3+1.15*\i}
    \pgfmathsetmacro{\xr}{2.05+1.15*\i}
    \filldraw[fill=\fil,draw=black!60] (\xl,\i) rectangle (\xr,\i+1);
  }
  \filldraw[fill=white,draw=black!60] (4.9,4) rectangle (6.08,4.5);
  \foreach \i in {1,2,3,4} {
    \pgfmathsetmacro{\sl}{0.3+1.15*\i}
    \pgfmathsetmacro{\sr}{2.05+1.15*(\i-1)}
    \draw[line width=2.2pt,color4] (\sl,\i) -- (\sr,\i);
  }
  \draw[line width=2.2pt,color4] (0.3,0) -- (0.9,0);
  \draw[line width=2.2pt,color4] (5.48,4.5) -- (6.08,4.5);
  \foreach \i in {0,...,3} {
    \pgfmathsetmacro{\xr}{2.05+1.15*\i}
    \node[anchor=mid west,font=\small] at (\xr+0.05,\i+0.5) {$I_{5,\i}$};
  }
  \node[anchor=mid west,font=\small] at (6.13,4.25) {$I_{5,4}$};
  \draw[very thick,color2,rounded corners=2pt]
     (0.6,0) -- (1.0,0.45) -- (1.2,0.45) -- (1.72,1) -- (2.35,1.5) -- (2.88,2)
     -- (3.35,2.6) -- (3.55,2.6) -- (3.92,3) -- (4.6,3.8) -- (4.8,3.8) -- (5.1,4)
     -- (5.35,4.25) -- (5.5,4.25) -- (5.75,4.5);
  \foreach \i/\fil in {0/white,1/white,2/white,3/white}
     {\filldraw[fill=\fil,draw=black!60] (-0.45,\i) rectangle (-0.15,\i+1);}
  \filldraw[fill=white,draw=black!60] (-0.45,4) rectangle (-0.15,4.5);
  \foreach \y/\l in {0.5/P_0,1.5/P_1,2.5/P_2,3.5/P_3,4.25/P_R}
     {\node[anchor=mid east,font=\small] at (-0.55,\y) {$\l$};}
  \node[anchor=mid,font=\small] at (-0.3,4.9) {$P$};
  \foreach \i/\l/\fil in {0/T_5/white,1/T_6/white,2/T_7/white,3/T_8/white} {
    \pgfmathsetmacro{\xl}{0.3+1.15*\i}
    \pgfmathsetmacro{\xr}{2.05+1.15*\i}
    \pgfmathsetmacro{\yt}{mod(\i,2)==0 ? -0.75 : -1.1}
    \filldraw[fill=\fil,draw=black!60] (\xl,\yt) rectangle (\xr,\yt+0.3);
    \node[font=\scriptsize,anchor=mid] at ({(\xl+\xr)/2},\yt+0.15) {$\l$};
  }
  \filldraw[fill=white,draw=black!60] (4.9,-0.75) rectangle (6.08,-0.45);
  \node[font=\scriptsize,anchor=mid] at (5.49,-0.6) {$T_{5,R}$};
  \node[anchor=mid east,font=\small] at (-0.55,-0.775) {$T$};
  \draw[decorate,decoration={brace,amplitude=2pt}] (1.45,-1.15) -- (2.05,-1.15)
     node[midway,above=1pt,font=\scriptsize] {$\Delta$};
  \node[font=\scriptsize,anchor=north] at (5.78,4.6) {$\Delta+1$ states};
\end{tikzpicture}
\caption{The common factorization for the block index $j=5$.  The pattern $P$ (left) is cut at the positions $a_i$ aligned with
the reference coordinates $i\tau$ into $P_0,\ldots,P_3$ and $P_R$, and
the text factors $T_5,\ldots,T_8$ and $T_{5,R}$ (top) overlap in
$\Delta$ positions.  Every alignment of cost at most $d$ (purple) crosses
each internal seam within the $\Delta+1$ shared states (orange), so
$I_5=I_{5,0}\oplus\cdots\oplus I_{5,4}$ records all such alignments.}
\label{fig:overview_factorization}
\end{figure}
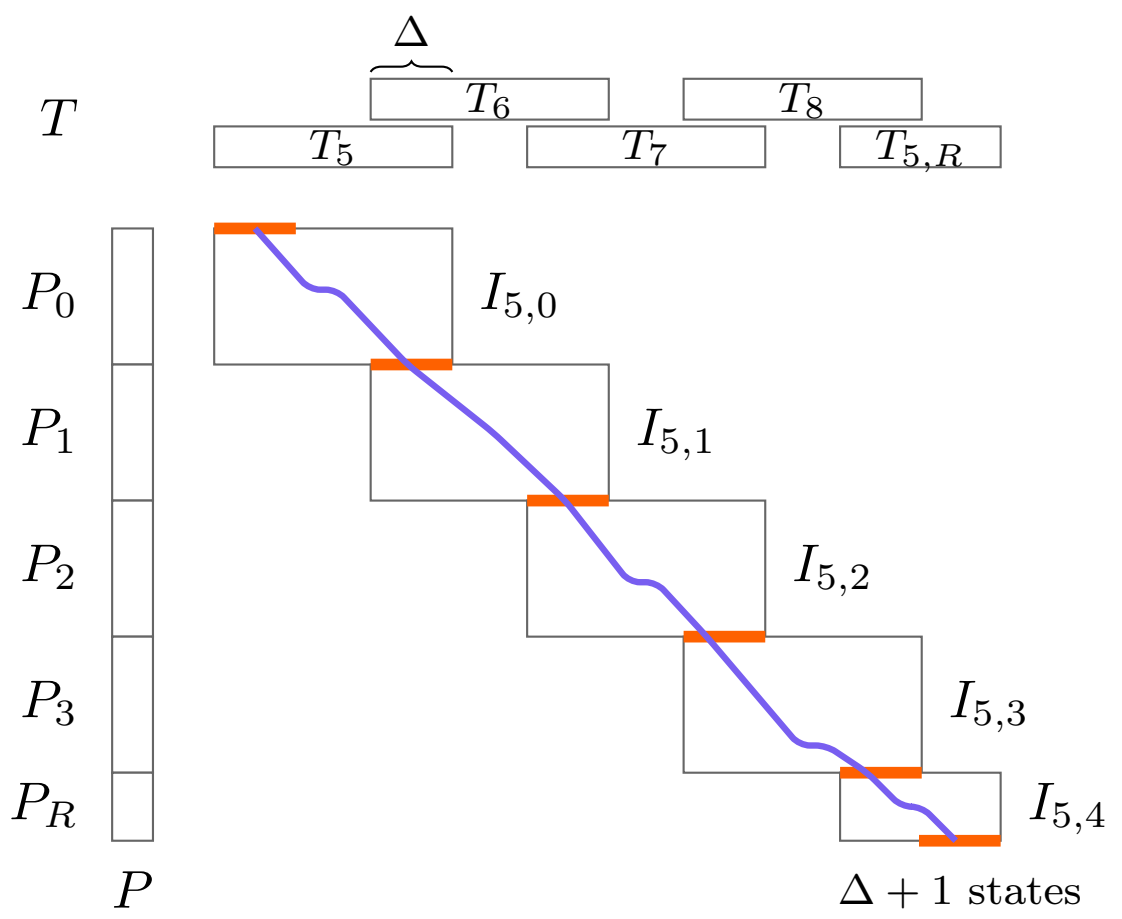

To derive these intervals, we partition $P$ and $T$ into \emph{factors} as follows.
We cut $\hP$ at the
multiples $i\tau$ for $i\in\fragment{0}{t}$, the last of which lies
at least $\kappa$ positions before the end of $\hP$, and $\hT$ at all
multiples $u\tau$; let $a_i$ and $c_u$ be the corresponding positions of
$P$ and $T$.  This yields the \emph{internal factors}
\[
    P_i\coloneqq P\fragmentco{a_i}{a_{i+1}},
    \qquad
    T_u\coloneqq T\fragmentco{c_u-f}{c_{u+1}+f},
\]
where the margins of length $f=\Theta(d)$ make consecutive text factors overlap in
$\Delta\coloneqq2f$ positions.  Block index $j$ pairs $P_i$ with
$T_{j+i}$; the margins guarantee that, in the occurrences assigned to
$j$, the factor $P_i$ is aligned with a fragment contained in
$T_{j+i}$.\footnote{In~\cite{CKW22}, the pattern factors also have
margins, for symmetry, but this unnecessarily complicates the
arguments.}  Together with an endpoint pair, which matches the pattern
suffix $P_R\coloneqq P\fragmentco{a_t}{m}$ with a text factor $T_{j,R}$
ending $f$ positions after the shifted end of $\hP$, these \emph{factor
pairs of $j$}
cover the region of the alignment graph around the diagonal corresponding to $j$
(\cref{fig:overview_factorization}).  The factors are well-defined for block indices sufficiently far from the endpoints of $T$; the remaining $\Oh(1)$ \emph{extreme} block indices are processed naively in $\Oh(d^2)$ time.

For the $i$th factor pair $(X,Y)$ of $j$, let
$I_{j,i}\coloneqq\udmat{X}{Y}{\Delta}$ be its \emph{restricted distance
matrix}, which stores the distances in the alignment graph of $X$ and
$Y$ from the leftmost $\Delta+1$ vertices of the top row to the
rightmost $\Delta+1$ vertices of the bottom row.
Moreover, let $I_j\coloneqq I_{j,0}\oplus\cdots\oplus I_{j,t}$, where
$\oplus$ is the \emph{min-plus product}
$(A\oplus B)[i,i']=\min_h(A[i,h]+B[h,i'])$.  Since an optimal alignment
of $P$ with a fragment of $T$ that starts in the start interval of $j$
crosses every \emph{seam} between consecutive factor pairs within the
$\Delta+1$ vertices that the two pairs share, called \emph{states}, the
row minima of $I_j$ give the values $\dtob(0,v)$ at the starts $v$
assigned to $j$, up to $d$-equivalence (see the proof of
\cref{lem:few_mark_ckw_uniform_instantiation}).

Most internal factors are \emph{plain}.  If $\A_P$ aligns $P_i$ with
$Q^\infty\fragmentco{i\tau}{(i+1)\tau}$ without edits, then
${P_i=Q^\infty\fragmentco{0}{\tau}}$, and similarly
${T_u=Q^\infty\fragmentco{-f}{\tau+f}}$ if no edit of $\A_T$ lies close to
$\fragmentco{u\tau}{(u+1)\tau}$.
The other factors are all designated as \emph{exceptional}; their $\Oh(d)$ indices are
collected in the \emph{exceptional lists} $\mathcal X_P$ and~$\mathcal X_T$.
  If $i\notin\mathcal X_P$
and $j+i\notin\mathcal X_T$, then $I_{j,i}$ equals the single matrix\vspace{-.18cm}
\[
    \smash{A\coloneqq\udmat{Q^\infty\fragmentco{0}{\tau}}{Q^\infty\fragmentco{-f}{\tau+f}}{\Delta}.}
\]
Thus, $I_j$ consists of runs of factors $A$, with products
$A^{\oplus h}$, separated by $\Oh(d)$ other factors.
\pagebreak[2]

Instead of the Dynamic Puzzle Matching data structure of~\cite{CKW22},
which internally maintains products of unit-Monge matrices in Tiskin's seaweed
monoid~\cite{Tiskin15}, we reason directly about distance matrices with
the tools of~\cite{GK25,gk24}.  Restricted distance matrices are
\emph{Monge} and bounded-difference.  As noted in~\cite{GGK25,GK25}, such a matrix admits an $\Oh(\Delta)$-size representation, a \emph{core-based matrix oracle} (CMO), and the
min-plus product of two such matrices takes $\Oh(\Delta\lg\Delta)$
time.  Since the later sections compare distances only up to
$W=\Oh(d)$, it suffices to store a \emph{fern} \cite{gk24} of
each matrix, that is, a $W$-equivalent matrix.  $W$-equivalence is
preserved under min-plus products, so ferns can be multiplied without
accumulating errors.  The \emph{fern catalogue} stores ferns for all
$\Oh(d^2)$ distinct pairs of factors; we build it in $\Oh(d^3\lg^2m)$
time with a data structure of \cite{GK25,gk24} that maintains
strings under character edits and returns ferns of the current pair.

\subsection{Sweeping with the Long-Run Matrix}\label{sec:overview_sweep}

Since all block indices share one factorization, every factor matrix
$I_{j,i}$ corresponds to a catalogue entry, and moving from $j$ to $j+1$ only shifts
the text factors: among the factor pairs of $j$, those with $i\in\mathcal X_P$
stay in place, while the pair $u-j$ of each $u\in\mathcal X_T$ moves one
step to the left.  The maximal runs of consecutive factors $A$ in $I_j$
between them, called \emph{plain runs}, grow or shrink, but an
exceptional pattern factor $P_i$ and an exceptional text factor $T_u$
meet only at $j=u-i$, so there are only $\Oh(d^2)$ \emph{crossing events}; the
difficulty lies in the changing run lengths, as large as
$\Theta(m/\tau)$.

Earlier algorithms \emph{trim} long runs: to length $d+1$, which is
exact because an alignment of cost at most $d$ crosses one of any $d+1$
consecutive plain factor pairs along a zero-cost diagonal, in the
$\Ohtilde(d^4)$-time warm-up of~\cite{CKW22}, also used in~\cite{gk24};
or to length $\alpha=\widetilde\Theta(\sqrt d)$, at the price of verifying
false positives, in the $\Ohtilde(d^{3.5})$-time algorithm of~\cite{CKW22}.

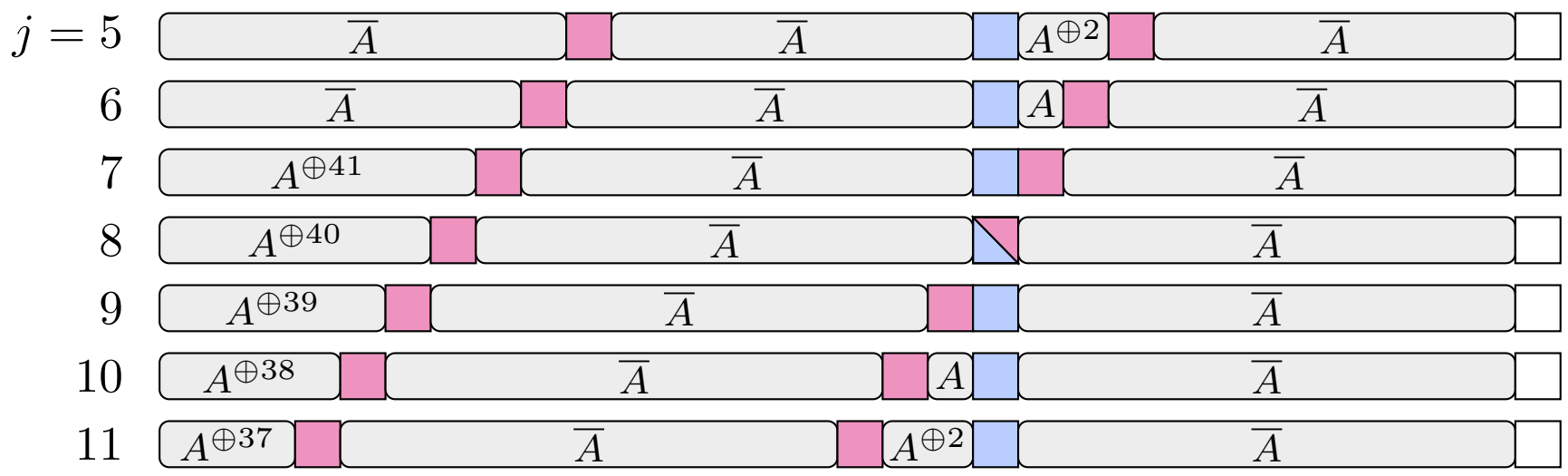
\begin{figure}[t]
\centering
\begin{tikzpicture}[x=0.66cm,y=0.8cm]
  \newcommand{\run}[4]{\draw[rounded corners=2pt,fill=black!7] (#1,#3-0.21) rectangle (#2,#3+0.21);
     \node[font=\scriptsize] at ({(#1+#2)/2},#3) {$#4$};}
  \newcommand{\pr}[2]{\filldraw[draw=black,fill=color1!45] (#1,#2-0.21) rectangle (#1+0.5,#2+0.21);}
  \newcommand{\trd}[2]{\filldraw[draw=black,fill=color3!50] (#1,#2-0.21) rectangle (#1+0.5,#2+0.21);}
  \newcommand{\both}[2]{\filldraw[draw=black,fill=color1!45] (#1,#2-0.21) -- (#1+0.5,#2-0.21) -- (#1,#2+0.21) -- cycle;
     \filldraw[draw=black,fill=color3!50] (#1+0.5,#2-0.21) -- (#1+0.5,#2+0.21) -- (#1,#2+0.21) -- cycle;
     \draw (#1,#2-0.21) rectangle (#1+0.5,#2+0.21);}
  \newcommand{\en}[2]{\filldraw[draw=black,fill=white] (#1,#2-0.21) rectangle (#1+0.5,#2+0.21);}
  \node[anchor=east,font=\small] at (-0.2,0) {$j=5$};
  \run{0}{4.5}{0}{\overline A}\trd{4.5}{0}\run{5}{9}{0}{\overline A}\pr{9}{0}\run{9.5}{10.5}{0}{A^{\oplus 2}}\trd{10.5}{0}\run{11}{15}{0}{\overline A}\en{15}{0}
  \node[anchor=east,font=\small] at (-0.2,-0.62) {$6$};
  \run{0}{4}{-0.62}{\overline A}\trd{4}{-0.62}\run{4.5}{9}{-0.62}{\overline A}\pr{9}{-0.62}\run{9.5}{10}{-0.62}{A}\trd{10}{-0.62}\run{10.5}{15}{-0.62}{\overline A}\en{15}{-0.62}
  \node[anchor=east,font=\small] at (-0.2,-1.24) {$7$};
  \run{0}{3.5}{-1.24}{A^{\oplus 41}}\trd{3.5}{-1.24}\run{4}{9}{-1.24}{\overline A}\pr{9}{-1.24}\trd{9.5}{-1.24}\run{10}{15}{-1.24}{\overline A}\en{15}{-1.24}
  \node[anchor=east,font=\small] at (-0.2,-1.86) {$8$};
  \run{0}{3}{-1.86}{A^{\oplus 40}}\trd{3}{-1.86}\run{3.5}{9}{-1.86}{\overline A}\both{9}{-1.86}\run{9.5}{15}{-1.86}{\overline A}\en{15}{-1.86}
  \node[anchor=east,font=\small] at (-0.2,-2.48) {$9$};
  \run{0}{2.5}{-2.48}{A^{\oplus 39}}\trd{2.5}{-2.48}\run{3}{8.5}{-2.48}{\overline A}\trd{8.5}{-2.48}\pr{9}{-2.48}\run{9.5}{15}{-2.48}{\overline A}\en{15}{-2.48}
  \node[anchor=east,font=\small] at (-0.2,-3.1) {$10$};
  \run{0}{2}{-3.1}{A^{\oplus 38}}\trd{2}{-3.1}\run{2.5}{8}{-3.1}{\overline A}\trd{8}{-3.1}\run{8.5}{9}{-3.1}{A}\pr{9}{-3.1}\run{9.5}{15}{-3.1}{\overline A}\en{15}{-3.1}
  \node[anchor=east,font=\small] at (-0.2,-3.72) {$11$};
  \run{0}{1.5}{-3.72}{A^{\oplus 37}}\trd{1.5}{-3.72}\run{2}{7.5}{-3.72}{\overline A}\trd{7.5}{-3.72}\run{8}{9}{-3.72}{A^{\oplus 2}}\pr{9}{-3.72}\run{9.5}{15}{-3.72}{\overline A}\en{15}{-3.72}
\end{tikzpicture}
\caption{Powered words for $j=5,\ldots,11$.  Exceptional pattern
positions (blue) stay fixed; exceptional text positions (magenta)
move left as $j$ grows.  A plain run of length $h<42$ is drawn as
$A^{\oplus h}$, a longer one as $\overline A$; the white box is
$I_{j,t}$.  The word changes only when a run shorter than $42$ before
or after the step changes its length (the first run, from $j=7$ on) or
when two exceptional positions meet (as for $j=8$).}
\label{fig:overview_sweep}
\end{figure}

We go to the other extreme and replace every maximal plain run of
length at least $\chi\coloneqq\Delta/d=42$ by one matrix, the
\emph{long-run matrix} $\overline A$, keeping the factors $A$ of shorter
runs.  For states $g$ and $g'$, the entry $\overline A[g,g']$ is the
cost of the following path: move from $g$ to a state $c$ where the
copies of $Q$ in $P$ and $T$ are synchronized, follow the zero-cost
diagonal from $c$, and then move to $g'$; that is,
$\overline A[g,g']=\min_c(|g-c|+|g'-c|)$ over all such states $c$
(\cref{def:canonical_run_powers}).  Such a path deletes up to
$\Delta=\chi d$ pattern characters, and a run of $h\ge\chi$ factors has
$h\tau>\chi d$ of them, so $\overline A$ bounds $A^{\oplus h}$ from above.  Let $K_j$ be the
resulting \emph{powered product}, and call its sequence of factors the
\emph{powered word} of $j$.  A run now changes
its matrices only when its length crosses one of $0,1,\ldots,\chi$, so
the powered words of all block indices change only $\Oh(d^2)$ times in
total (\cref{fig:overview_sweep}).  The intervals between consecutive
changes, further cut where the spacing of the points $r_j$ changes,
form the desired partition of block indices, provided that $K_j$ can replace~$I_j$.

Since $\overline A$ bounds the powers of $A$ that it replaces from
above, we have $K_j\ge I_j$, and the powered product never reports false
occurrences.  For the converse, fix a \emph{witness}, an optimal
alignment of $P$ with a fragment of $T$ that starts at an occurrence start assigned
to $j$.  We say that it \emph{survives} if the corresponding entry of
$K_j$ does not exceed its cost, so that $K_j$ still finds the
occurrence.  Both strings are fragments of
$Q^\infty$ inside a plain run, so each state of a seam there pairs a
reference coordinate of the pattern with one of the text.  We call the
state \emph{in phase} if these two coordinates are congruent modulo $q$,
that is, if the copies of $Q$ in $P$ and in $T$ are synchronized there.
By the synchronization property, the diagonals through the in-phase
states, one in every $q$ consecutive states, are exactly the zero-cost
diagonals.  A witness survives if it passes an
in-phase state in every long plain run, since it then pays at least the
corresponding entry of $\overline A$.  A witness that crosses a long run
\emph{off phase} pays for mismatches, and its cost may depend on the
run length.  A witness may reach a plain run off phase after leaving the
zero-cost diagonals near edits of $\A_P$ and $\A_T$, as studied in
\cref{sec:overview_marks}.

One simple but useful novel insight is that, for $q<\sqrt d$, every
witness passes an in-phase state in every run of $h\ge\chi$ factors.
Otherwise, it would enter and leave the run between the same two
consecutive in-phase states, where the entry of $\overline A$ is at
most $q$, and pay about one edit per period, that is, about
$h\tau/q>\chi d/q>\chi q$ edits; the path of $\overline A$ would be
cheaper (\cref{lem:ckw_clean_powered_exactness}).  For larger $q$, this fails, and we must study how witnesses move in long runs.

\begin{example}\label{ex:overview_drift}
Let $Q=\texttt{a}^{q-1}\texttt{b}$ for some $\sqrt{d} \le q\le d$, and consider two
copies of a long fragment of $Q^\infty$ shifted against each other by
$s$ positions, where $0<|s|\le q/2$.  Restoring the phase by $|s|$
insertions at one end and $|s|$ deletions at the other end costs $2|s|$,
while the substitution-only alignment costs $2$ per period, as it
mismatches the two $\texttt{b}$'s of every period.  Hence, an optimal
alignment may stay off phase over about $|s|q$ positions, and the
corresponding entries of $A^{\oplus h}$ keep growing until $h\tau$ is of
order $|s|q$, that is, up to order $q^2$.  Thus, runs of length $h$ with $\chi\le h\lesssim q^2/\tau$, which exist
once $q^2$ is large compared to $\chi\tau=\Theta(d)$, cannot simply be
replaced by $\overline A$.
\end{example}

\subsection{Rails, Mark Loads, and Eligible Block Indices}\label{sec:overview_marks}

Our next goal is to understand which witnesses may fail to survive, that is, when a
witness may stay off phase for at least $\chi$ consecutive plain factor
pairs.  As in~\cite{CKW20,CKW22,gk24}, we do so with marks, which
record where edits of $\A_P$ and $\A_T$ may interact.  However, we
place marks more often, also for edits that lie farther apart, and in
individual reference strips, defined below, rather than at whole block
indices, for a finer characterization of the witnesses that do not
survive.

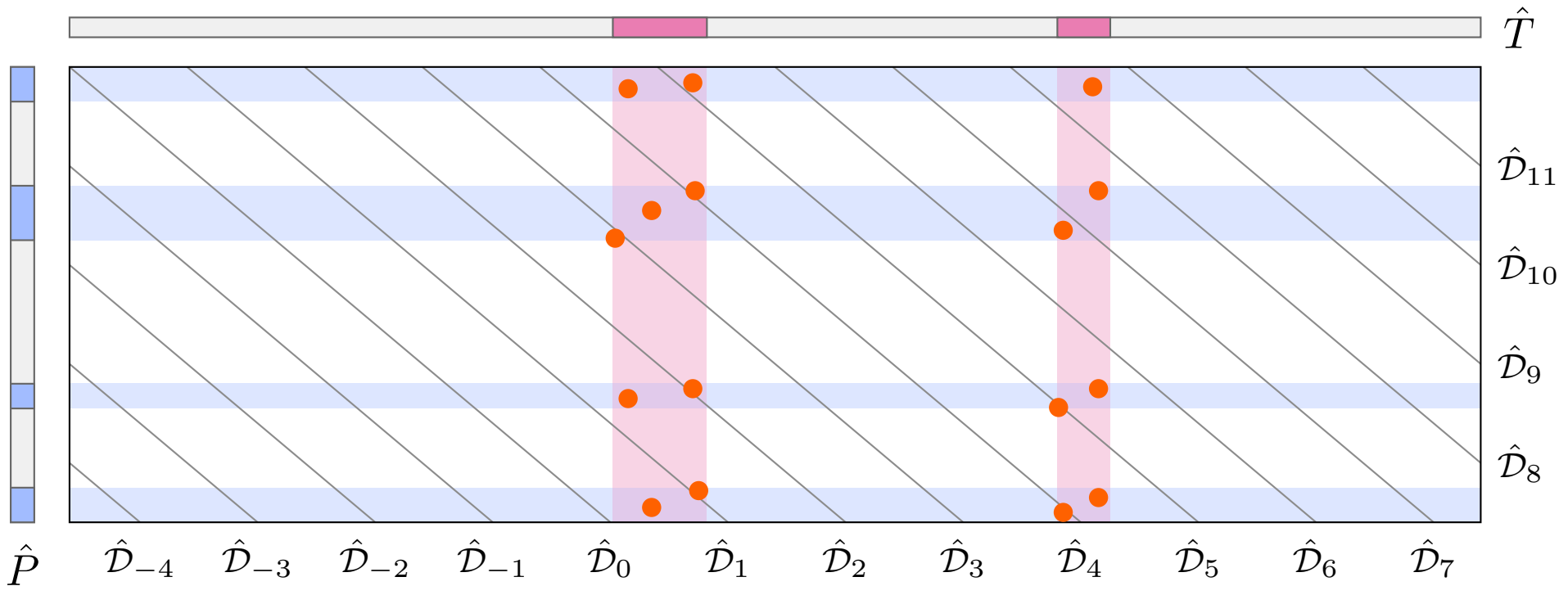
\begin{figure}[t]
\centering
\begin{tikzpicture}[x=0.95cm,y=-0.8cm]
  \begin{scope}
  \clip (0,0) rectangle (12,4.6);
  \fill[color1!22] (0,1.2) rectangle (12,1.75);
  \fill[color1!22] (0,3.2) rectangle (12,3.45);
  \fill[color1!22] (0,0) rectangle (12,0.35);
  \fill[color1!22] (0,4.25) rectangle (12,4.6);
  \fill[color3!20] (4.62,0) rectangle (5.42,4.6);
  \fill[color3!20] (8.4,0) rectangle (8.85,4.6);
  \foreach \ya/\yb in {0/0.35, 1.2/1.75, 3.2/3.45, 4.25/4.6} {\foreach \xa/\xb in {4.62/5.42, 8.4/8.85} {\fill[color1!50!color3!32] (\xa,\ya) rectangle (\xb,\yb);}}
  \foreach \b in {-4,...,11} {\draw[black!45] (\b,0) -- (\b+4.6,4.6);}
  \end{scope}
  \draw (0,0) rectangle (12,4.6);
  \filldraw[draw=black!60,fill=black!6] (0,-0.5) rectangle (12,-0.3);
  \filldraw[draw=black!60,fill=color3!60] (4.62,-0.5) rectangle (5.42,-0.3);
  \filldraw[draw=black!60,fill=color3!60] (8.4,-0.5) rectangle (8.85,-0.3);
  \filldraw[draw=black!60,fill=black!6] (-0.5,0) rectangle (-0.3,4.6);
  \filldraw[draw=black!60,fill=color1!60] (-0.5,1.2) rectangle (-0.3,1.75);
  \filldraw[draw=black!60,fill=color1!60] (-0.5,3.2) rectangle (-0.3,3.45);
  \filldraw[draw=black!60,fill=color1!60] (-0.5,0) rectangle (-0.3,0.35);
  \filldraw[draw=black!60,fill=color1!60] (-0.5,4.25) rectangle (-0.3,4.6);
  \node[font=\small,anchor=west] at (12.05,-0.4) {$\hT$};
  \node[font=\small,anchor=north] at (-0.4,4.65) {$\hP$};
  \foreach \x/\y in {4.64/1.73, 4.95/1.45, 5.32/1.25, 8.45/1.65, 8.75/1.25, 4.75/3.35, 5.3/3.25, 8.41/3.44, 8.75/3.25,
      4.75/0.22, 5.3/0.16, 8.7/0.2, 4.95/4.45, 5.35/4.28, 8.45/4.5, 8.75/4.35} {\fill[color4] (\x,\y) circle (2.2pt);}
  \foreach \b in {-4,...,7} {\node[font=\scriptsize,anchor=north] at (\b+4.6,4.6) {$\hD{\b}$};}
  \foreach \b in {8,...,11} {\node[font=\scriptsize,anchor=west] at (12,12-\b) {$\hD{\b}$};}
\end{tikzpicture}
\caption{The reference plane with the reference rails $\hD{b}$
(gray), the diagonals $\hat t-\hat p=bq$.  The strings $\hP$ and
$\hT$ are drawn with the fragments of their dense covers $\Es_P$
(blue) and $\Es_T$ (magenta), which shade rows and columns.  A pair of fragments places a mark
(orange) in every reference strip that meets their intersection.}
\label{fig:overview_reference}
\end{figure}

We work in the \emph{reference plane} of pairs $(\hat p,\hat t)$ of
reference coordinates of $\hP$ and $\hT$, onto which $\A_P$ and
$\A_T$ map the alignment graph of $P$ and $T$.  \Cref{fig:overview_reference}
shows the objects introduced below.  For every integer $b$, the
diagonal $\hat t-\hat p=bq$ is the \emph{reference rail} $\hD{b}$ (gray);
it aligns $\hP$ with $\hT$ in phase and without edits.  The region
between the reference rails $\hD{b}$ and $\hD{b+1}$ is the
\emph{reference strip}~$b$.

A fragment $F$ of $\hP$ is \emph{dense} if $|F|<5q\cdot c_F$, where
$c_F$ counts the edits that $\A_P$ makes within $F$.\footnote{The formal
definitions treat the ends of $\hP$ specially: the edit counts of
fragments there include extra edits, so that
$\Es_P$ always contains a fragment at each end of $\hP$, and these
fragments get infinite weight in the marks below.}  Merging overlapping
or touching dense fragments yields the \emph{dense cover} $\Es_P$; the
cover $\Es_T$ of $\hT$ is analogous.  In the reference plane, fragments
of $\Es_P$ are ranges of rows (blue), and fragments of $\Es_T$ are
ranges of columns (magenta).  Both covers consist of $\Oh(d)$
disjoint fragments of total length $\Oh(qd)$
(\cref{lem:dense_decomposition}).  Dense covers redesign the locked
fragments of~\cite{CKW20}, defined through a characterization roughly corresponding to $|F|<q\cdot c_F$; dense covers allow controlling the edit density.
We also add to the exceptional lists $\mathcal X_P$ and $\mathcal X_T$
the indices of all factors that overlap or touch a fragment of the
respective cover; the lists still have $\Oh(d)$ elements
(\cref{lem:exceptional_list_construction}).

Fragments $E_P\in\Es_P$ and $E_T\in\Es_T$ \emph{meet} the reference
strip $b$ if the rectangle $E_P\times E_T$ meets its closure, that is,
the region $bq\le\hat t-\hat p\le(b+1)q$.  Each such meeting places a
\emph{pair mark} (orange) of weight $\min(w_P(E_P),w_T(E_T))$, where the
weights count the edits inside the fragments.  The \emph{mark load} $\mu(b)$ is the total weight of
the pair marks in reference strip $b$, and
$\sum_b\mu(b)=\Oh(d^2)$ (\cref{sec:marks}).

We now explain how these objects capture the witnesses that may fail to
survive.  For this, we move from the reference plane to the
\emph{expanded alignment graph}, whose rows and columns are indexed by
the vertices of $\A_P$ and $\A_T$, respectively
(\cref{sec:expanded_canonical_geometry}).  This is more convenient than
the alignment graph of $P$ and $T$: every vertex in the expanded graph corresponds to unique positions in
$P$ and $T$ and reference coordinates, whereas a position of $P$ or $T$
may correspond to several reference coordinates.  The expanded graph is
defined so that its distances agree with those in the alignment graph
of $P$ and $T$ (\cref{obs:expanded_projection}).  In the expanded alignment
graph, the reference rail $\hD{b}$ lifts to the \emph{rail} $\eD{b}$,
which follows $\A_P$ and $\A_T$ simultaneously; it corresponds to the
period shift~$b$.

For vertices $u$ and $u'$, the \emph{canonical cost}
$\mathsf{cc}(u,u')$ is the number of edits of $\A_P$ and $\A_T$ between
the reference coordinates of $u$ and $u'$; it depends only on $u$ and
$u'$, not on a path between them.  A rail segment costs at most its
canonical cost, and possibly less: in \cref{ex:overview_periodic}, the
rail $\eD{2}$ passes both missing characters at once and costs $0$.
Inside a plain run, the rails are exactly the diagonals through the
in-phase states, so a witness survives if it meets a rail in every long
plain run.

Leaving the rails is expensive: since $Q$ is primitive, a path avoiding
rail edges pays about one edit per period it traverses, minus the
edits of $\A_P$ and $\A_T$ there (\cref{lem:noblue}), and switching
between rails costs about $q$ per rail (\cref{lem:blue_jump_limit}).
Consequently, the witnesses for block index $j$ stay within the
\emph{band} of $j$, which consists of the rails $\eD{b}$ with
$|b-b_j|\le w=\Oh(1+d/q)$ and the strips between them
(\cref{lem:relevant_path,lem:ckw_start_relevance}).  The parts of a
path between the edges that it shares with rails are its
\emph{detours}.  A detour is \emph{clean} if it stays inside the dense
covers, that is, if each of its edges has its pattern projection inside
a fragment of $\Es_P$ or its text projection inside a fragment of
$\Es_T$, and a path is clean if all its detours are clean
(\cref{def:detours,def:clean_paths}).  Outside the dense covers, $\A_P$ and
$\A_T$ have fewer than one edit per $5q$ positions, whereas a detour
pays about one edit per period.  Hence, a detour that leaves a rail and
returns to it at no more than the canonical cost of the skipped segment
is clean (\cref{cor:nobluesame}).  Since plain runs now avoid the covers, a
clean witness has no detour inside a plain run: it crosses every plain
run along a single rail, that is, along the zero-cost diagonal through
an in-phase state, and thus survives.

It remains to understand witnesses with detours that are not clean.
By the above, such a detour of an optimal path switches rails.  It pays about $q$ per rail, whereas outside
the covers its canonical cost is below a fifth of its length in periods,
so it costs at least $q/5$ more than its canonical cost
(\cref{lem:blocks_with_few_marks}).  Such a witness competes with the
path that, from the start of the detour, follows the rail to the bottom
row.  Extended along the bottom row to a common endpoint, this path
costs at most its canonical cost, which depends only on the endpoints.
Hence, the witness must save on its canonical cost elsewhere, which
happens only where edits of $\A_P$ and $\A_T$ interact, as in
\cref{ex:overview_periodic}.  The marks capture such
interactions: the total savings of a path below its canonical cost are
at most twice the total mark load of the strips it passes (\cref{lem:strip_path_savings}).

\paragraph*{Eligible block indices.}
The \emph{load} $M_j$ of a block index $j$ is the total mark load
of the strips in the band of $j$ widened by $\Oh(1+d/q)$ rails.  Each strip
belongs to the widened bands of $\Oh(1)$ block indices, so
$\sum_jM_j=\Oh(d^2)$.  We call $j$
\emph{high-load} if $\sqrt d\le q\le\kappa$ and $M_j\ge q/10$, and
\emph{eligible} otherwise; there are only $\Oh(d^2/q)$ high-load block
indices.\footnote{In~\cite{CKW22}, block indices with at least
$\alpha=\widetilde\Theta(\sqrt d)$ marks are verified separately; our
threshold $q/10$ grows with~$q$.}
For eligible block indices, the powered product is exact:
\begin{itemize}
\item If $q<\sqrt d$, every witness survives, as shown in
\cref{sec:overview_sweep}.
\item If $q>\kappa$, switching rails costs more than $d$, so every witness
is clean and thus survives.
\item If $M_j<q/10$, the savings of the witness are at most $2M_j<q/5$,
too little to pay for a detour that is not clean, so the witness is
again clean (\cref{lem:blocks_with_few_marks}).
\end{itemize}
In the last two cases, the witness crosses every plain run along one
rail, in phase, and survives (\cref{lem:ckw_clean_powered_exactness}).
Thus, for eligible block indices, no witness stays off phase across a
long plain run, and, with $K_j\ge I_j$, the occurrences assigned to $j$
can be read off the row minima of $K_j$
(\cref{lem:few_mark_ckw_uniform_instantiation}).

Generalizing how Dynamic Puzzle Matching is implemented in \cite{CKW22,gk24},
a \emph{persistent} balanced tree over the powered words, with CMOs of subtree
products, yields a \emph{range oracle} that applies the powered product
of any range of factors to a vector in $\Oh(d\lg^2m)$ time
(\cref{lem:range_oracle}).  One range query per
interval of eligible indices with a common powered word
(\cref{lem:profile_partition}) yields $\Oh(d)$ progressions: $\Oh(d^3)$
in total, in $\Oh(d^3\lg^2m)$ time.  This
settles the cases of $q<\sqrt d$ and $q>\kappa$.

\subsection{The Planar Variation Theorem and High-Load Block Indices}\label{sec:overview_variation}

For a high-load block index, the powered product may err, and we must
find where optimal paths switch rails.  Our main tool shows that few
pairs of neighboring rails have very different distances to the
bottom row.  It bounds the \emph{variation} of these distances along a
row, that is, the sum of their absolute differences between neighboring rails,
by the canonical cost between the extreme rails plus a charge to the
mark loads near them.  Applying the charging statement of
\cref{sec:overview_marks} to each rail separately would charge the mark
load of the whole band once per rail, up to $\Theta(1+d/q)$ times; the
theorem charges each strip's load once per row.  It has no counterpart
in~\cite{CKW22}.

\begin{theorem}[Informal version of \cref{thm:combi}]
\label{thm:overview_variation}
For every row, every range $C=\fragment{c_1}{c_2}$ of rails, and every
choice of a vertex $u_b$ of $\eD{b}$ in that row for each $b\in C$,
\[
    \sum_{b=c_1+1}^{c_2}\bigl|\dtob(u_b)-\dtob(u_{b-1})\bigr|
    \le
    \mathsf{cc}(u_{c_1},u_{c_2})+4\sum_{b=c_1-\eta}^{c_2+\eta}\mu(b),
\]
where $\eta\coloneqq\floor{8d/q}+1=\Oh(1+d/q)$, provided that the rails
$\eD{b}$ for all $b\in\fragment{c_1-\eta}{c_2+\eta}$ run from the top row
to the bottom row of the alignment graph.
\end{theorem}

The proof (\cref{sec:mark_distance_variation}) works in the \emph{band
graph}: the part of the expanded alignment graph below the fixed row and
between the rails $\eD{c_1-\eta}$ and $\eD{c_2+\eta}$, in which moves
along the bottom row are initially free.  Hence, $\dtob(u_b)$ is the
distance from $u_b$ to the \emph{target} $z$, the bottom endpoint of
$\eD{c_2+\eta}$.  Shortest paths from the vertices $u_b$
never reach the extreme rails: following their own
rail costs at most $d_P+d_T$, whereas moving $\eta$ rails away costs
more.  Hence, the band graph preserves the distances $\dtob(u_b)$
(\cref{lem:comparison_graph_faithfulness}).  It is a planar DAG
with the vertices $u_b$ in order on its outer face.

In the proof, rail edges, which may be cheaper than canonical, and the
free bottom-row edges are raised only up to their canonical costs, the
numbers of edits that they pass.  The charging statement of
\cref{sec:overview_marks} rests on at most $\mu(b)$ edits
\emph{selected} for each strip $b$: a detour saves at most
two units per selected edit that it passes (\cref{lem:strip_path_savings}).
For a selected edit of strip $b$, a \emph{unit increase} raises by one
the costs of the edges of the strip that pass this edit, subject to
these caps.  The raised edges cross a cut through the strip, so they
form a path in the dual graph, and every path pays for at most one of
them.  By planarity, a unit increase raises the distances from a
contiguous range of the vertices $u_b$ by one and leaves the others
unchanged: if $u_a$ and $u_c$ were raised but not $u_b$ for some
$a<b<c$, then the shortest paths from $u_a$ and $u_c$, joined through
the raised edges, would meet a shortest path from $u_b$ that avoids
them.  Hence, the variation changes by at most $2$
(\cref{lem:planar_unit_increase}).  After two unit increases for every
selected edit, at most $2\sum_{b=c_1-\eta}^{c_2+\eta}\mu(b)$ in total,
no path is cheaper than its canonical cost, and following a rail and
then the bottom row costs exactly its canonical cost.  Then,
$\dtob(u_b)=\mathsf{cc}(u_b,z)$ does not increase with $b$, so the
variation is $\mathsf{cc}(u_{c_1},u_{c_2})$, and the theorem follows.

\paragraph*{High-load block indices.}\label{sec:overview_high}
In~\cite{CKW22}, the groups at which trimming may fail are verified one
by one, in $\Oh(d^2)$ time each, regardless of where their marks lie.
For each of our $\Oh(d^2/q)$ high-load block indices, where
${\sqrt d\le q\le\kappa\le5d}$, the planar variation theorem confines the
exact computations to regions whose total size is bounded in terms of
$M_j$, and we spend $\Oh(d(q+M_j)\lg^2m)$ time; with
$\sum_jM_j=\Oh(d^2)$, this gives $\Oh(d^3\lg^2m)$ in total.  Proceeding
from the bottom row upward, we compute estimates of $\dtob$ at the
seams that bound the slabs defined below.  These estimates are
exact on a recognizable set containing all witness states, and those
below $W$ are upper bounds.

We cut the band along seams into $\Oh(d/q)$ \emph{slabs}.  A slab lies
between two seams, its \emph{entrance} at the top and its \emph{exit}
at the bottom; we compute the estimates at its entrance from those at
its exit.  Each slab is
\emph{short}, spanning $\Oh(q^2)$ reference positions of $\hP$, or
\emph{light}, containing fewer than $q/6$ edits of $\A_P$
and $\A_T$ (\cref{lem:band_block_partition}).  We form the slabs
greedily as maximal stretches of factor pairs that are short or light.
Extended by the next factor pair, a slab is neither short nor light, so
it contains at least $q/6$ edits; each of the
$\Oh(d)$ edits is counted for $\Oh(1)$ slabs, so there are $\Oh(d/q)$
slabs.
For a short slab, we
apply its $\Oh(q^2/d)$ catalogue matrices one by one in $\Oh(q^2\lg m)$
time.

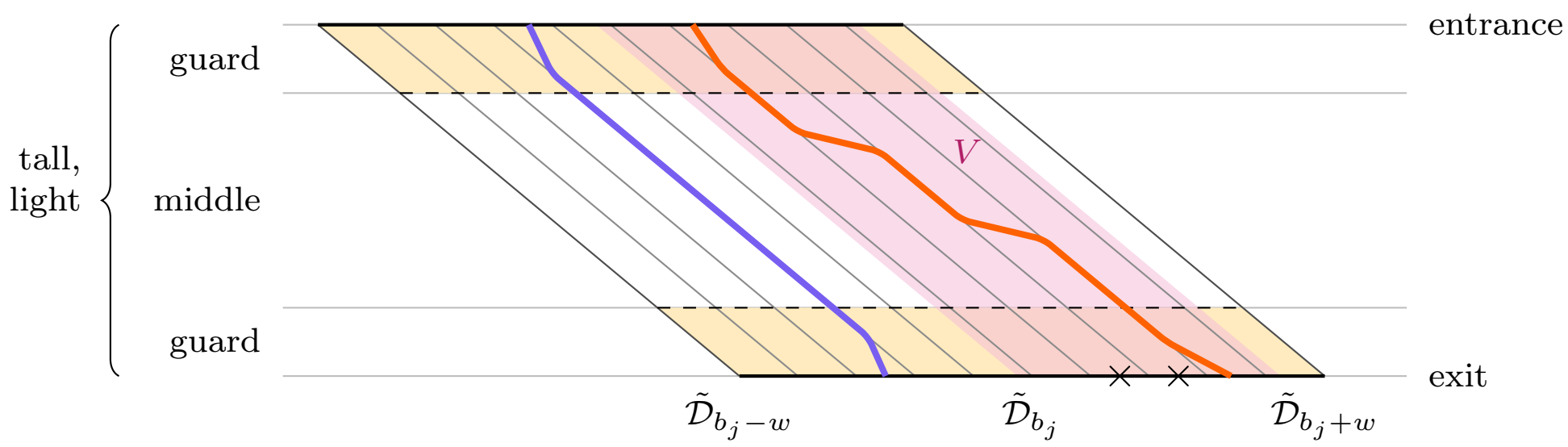
\begin{figure}[t]
\centering
\begin{tikzpicture}[x=0.48cm,y=-0.4cm]
  \draw[black!25] (-0.6,0) -- (18.6,0);
  \draw[black!25] (-0.6,1.4) -- (18.6,1.4);
  \draw[black!25] (-0.6,5.8) -- (18.6,5.8);
  \draw[black!25] (-0.6,7.2) -- (18.6,7.2);
  \fill[color5!25] (0,0) -- (1.4,1.4) -- (11.4,1.4) -- (10,0) -- cycle;
  \fill[color5!25] (5.8,5.8) -- (7.2,7.2) -- (17.2,7.2) -- (15.8,5.8) -- cycle;
  \fill[color3!30,fill opacity=0.55] (4.75,0) -- (11.95,7.2) -- (16.45,7.2) -- (9.25,0) -- cycle;
  \draw[black!70] (0,0) -- (7.2,7.2);
  \foreach \k in {1,...,9} {\draw[black!45] (\k,0) -- (\k+7.2,7.2);}
  \draw[black!70] (10,0) -- (17.2,7.2);
  \draw[thick] (0,0) -- (10,0);
  \draw[thick] (7.2,7.2) -- (17.2,7.2);
  \draw[dashed] (1.4,1.4) -- (11.4,1.4);
  \draw[dashed] (5.8,5.8) -- (15.8,5.8);
  \node[font=\small] at (13.7,7.2) {$\times$};
  \node[font=\small] at (14.7,7.2) {$\times$};
  \draw[line width=1.5pt,color2,rounded corners=2pt] (3.6,0) -- (4,1) -- (9.4,6.4) -- (9.7,7.2);
  \draw[line width=1.5pt,color4,rounded corners=2pt] (6.4,0) -- (6.9,0.9) -- (8.2,2.2) -- (9.6,2.6) -- (11,4) -- (12.4,4.4) -- (14.5,6.5) -- (15.6,7.2);
  \node[font=\scriptsize,anchor=north] at (7.2,7.2) {$\eD{b_j-w}$};
  \node[font=\scriptsize,anchor=north] at (12.2,7.2) {$\eD{b_j}$};
  \node[font=\scriptsize,anchor=north] at (17.2,7.2) {$\eD{b_j+w}$};
  \node[font=\scriptsize,color3!80!black] at (11.1,2.6) {$V$};
  \node[font=\scriptsize,anchor=mid east] at (-0.7,0.7) {guard};
  \node[font=\scriptsize,anchor=mid east] at (-0.7,3.6) {middle};
  \node[font=\scriptsize,anchor=mid east] at (-0.7,6.5) {guard};
  \draw[decorate,decoration={brace,amplitude=4pt}] (-3.4,7.2) -- (-3.4,0) node[midway,anchor=mid east,xshift=-5pt,font=\scriptsize,align=right] {tall,\\light};
  \node[font=\scriptsize,anchor=mid west] at (18.7,0) {entrance};
  \node[font=\scriptsize,anchor=mid west] at (18.7,7.2) {exit};
\end{tikzpicture}
\caption{A tall light slab of the band of a high-load block index $j$ in
expanded coordinates (schematic), with yellow guards.
The pairs of rails marked $\times$ are imbalanced.  A witness meeting one
rail (purple) follows it through the middle except for clean detours, so
the powered product is exact; one meeting several rails (orange) crosses
only imbalanced pairs and stays within the state interval~$V$
(magenta).}
\label{fig:overview_band}
\end{figure}

A light slab that is not short, called \emph{tall}, needs the theorem
(\cref{fig:overview_band}).  The \emph{exit contact} of a rail is its
latest vertex on the exit seam, and two neighboring rails are
\emph{imbalanced} if both exit contacts lie in the recognizable set
above and the
distances $\dtob$ there differ by more than $q/2$.  Since the slab is light, the canonical cost between
the extreme exit contacts is below $q/6$.  Each imbalanced pair adds
more than $q/2$ to the variation, so \cref{thm:overview_variation},
applied to the exit row, yields only $\Oh(M_j/q)$ of them.  An
optimal path that meets several rails inside the slab crosses only
imbalanced pairs: moving from one rail to the next costs about $q$,
while distances along a rail change little between the contacts of the
path and the exit contacts, because the slab contains fewer than $q/6$ edits.
Moreover, an optimal path in a light slab meets a rail at least once
per $\Oh(q^2)$ reference positions: after $\Theta(q^2)$ positions
between rails, it would pay $\Theta(q)$, more than following a rail
through the slab.  With a top and a bottom \emph{guard} of
$\Theta(q^2)$ positions, every witness thus either follows one rail
through the middle part, except for clean detours, or stays within a
\emph{state interval} $V$
(\cref{lem:tall_block_clean_middle,lem:tall_block_witness_localization}).
Here, $V$ is one interval of states, used at every seam of the slab,
that contains the states at which the rails of a maximal run of
consecutive imbalanced pairs cross the seams; one interval suffices
because, in a light slab, the state at which a rail crosses a seam
varies by less than $q/6$ from seam to seam.  Each imbalanced pair contributes $\Oh(q)$ states, so the
intervals have total size $\Oh(M_j)$.  The guards absorb the parts of a
witness near the entrance and the exit of the slab, where it may still
be off the rails; meeting a rail every $\Oh(q^2)$ positions, it meets
one inside each guard.

For witnesses of the first kind, the powered product is exact, as for
eligible block indices.  Accordingly, we compute one candidate vector
with exact products for the guards and a range-oracle query for the
middle part, and, for each interval $V$, one candidate vector with
products restricted to the states in $V$.  The restricted products are
cheap because the slab is light: all but $\Oh(q)$ factor pairs have the
matrix $A$, and a run of such pairs is applied as a power of $A$
restricted to $V$, obtained by repeated squaring.  Each state of the
entrance on a witness receives its exact distance from one of these
candidates (\cref{lem:tall_light_block_profile}), so the estimate at the
entrance is the entrywise minimum of the candidates; this works because
every candidate is an upper bound.  For a tall slab,
the guards and the range query take $\Oh(q^2\lg^2m)$ time, as $d\le q^2$.
Each interval $V$ takes $\Oh(q|V|\lg^2m)$ time:
one restricted product of size $|V|$ for each of the $\Oh(q)$ other
pairs and runs.  With $\sum_V|V|=\Oh(M_j)$, a tall slab takes
$\Oh((q^2+qM_j)\lg^2m)$ time, and all $\Oh(d/q)$ slabs take
$\Oh(d(q+M_j)\lg^2m)$ time.

\paragraph*{Putting everything together.}
The start intervals of the extreme, eligible, and high-load block indices
are disjoint and contain all occurrences, so we obtain $\OccE_k(P,T)$
as $\Oh(d^3)$ progressions of difference $\tau$ in $\Oh(d^3\lg^2m)$ time.
Interval operations convert them in
$\Oh(d^3\lg m)$ time into the $\Oh(d^3)$ progressions of
difference $q$~\cite{CKW20} (\cref{lem:tau_to_q_postprocessing}), which
gives \cref{lem:sthold} and
\cref{thm:edalgII}.

\clearpage
\subsection{Roadmap of the Technical Sections}\label{sec:roadmap}

\begin{figure}[t]
\centering
\begin{tikzpicture}[x=1.05cm,y=0.95cm,>=stealth,
  sec/.style={draw,rounded corners=2pt,align=center,font=\scriptsize,text width=2.5cm,minimum height=0.95cm,inner sep=2pt},
  st/.style={sec,fill=color2!10},
  al/.style={sec,fill=color4!10}]
  \node[st] (s4) at (0,0) {\textsection\ref{sec:ckw_windows} block indices,\\ start intervals};
  \node[st] (s5) at (3,0) {\textsection\ref{sec:expanded_canonical_geometry} expanded graph,\\ rails, band};
  \node[st] (s6) at (6,0) {\textsection\ref{sec:dense_covers_marks} dense covers,\\ detours};
  \node[st] (s7) at (9,0) {\textsection\ref{sec:mark_savings} mark loads,\\ charging, detours};
  \node[st] (s8) at (12,0) {\textsection\ref{sec:mark_distance_variation} planar\\ variation theorem};
  \node[al] (s9) at (0,-2) {\textsection\ref{sec:fern_toolbox} Monge matrices,\\ ferns};
  \node[al] (s10) at (3,-2) {\textsection\ref{sec:ckw_common} factorization,\\ exceptional lists, catalogue};
  \node[al] (s11) at (6,-2) {\textsection\ref{sec:powered_ckw_sweep} eligible indices\\ (Alg.~\ref{alg:eligible})};
  \node[al] (s12) at (9,-2) {\textsection\ref{sec:block_geometry} high-load\\ indices (Alg.~\ref{alg:high_load})};
  \node[al] (s13) at (12,-2) {\textsection\ref{sec:central_algorithm} assembly\\ (Alg.~\ref{alg:central})};
  \draw[->] (s4) -- (s5); \draw[->] (s5) -- (s6); \draw[->] (s6) -- (s7);
  \draw[->] (s7) -- (s8);
  \draw[->] (s9) -- (s10); \draw[->] (s6) -- (s10); \draw[->] (s10) -- (s11);
  \draw[->] (s7) -- (s11); \draw[->] (s7) -- (s10); \draw[->] (s11) -- (s12); \draw[->] (s8) -- (s12);
  \draw[->] (s12) -- (s13);
\end{tikzpicture}
\caption{Dependencies between the sections of the formal part.
Structural sections are purple and algorithmic sections are orange.
\Cref{sec:ckw_windows} is used throughout.}
\label{fig:roadmap}
\end{figure}
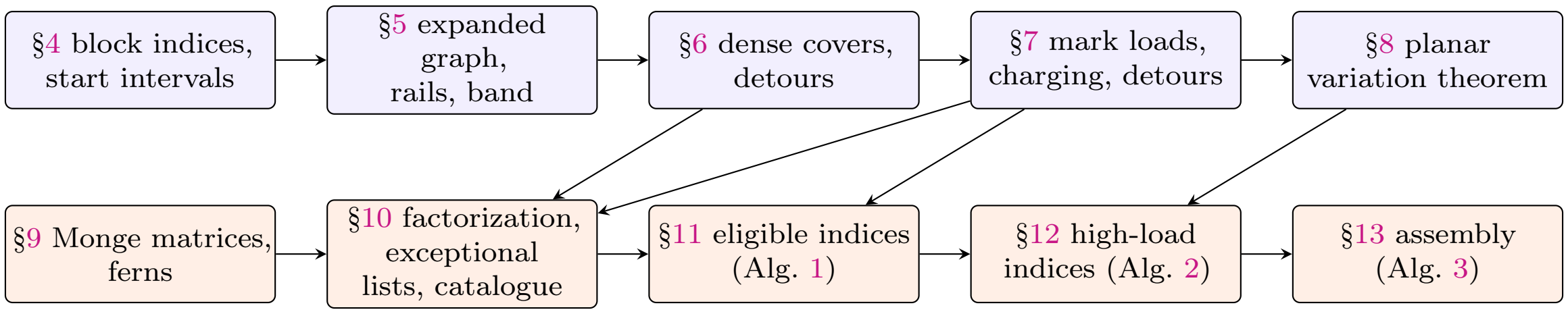

The technical part of this paper consists of structural and algorithmic
sections; \cref{fig:roadmap} shows how the sections depend on each other.
\Cref{sec:ckw_windows} assigns every occurrence start to a
\emph{block index}, the unit of work of all later algorithms.
\Cref{sec:expanded_canonical_geometry,sec:dense_covers_marks,sec:mark_savings,sec:mark_distance_variation}
develop the geometry of alignments relative to the periodic reference:
rails, dense covers, mark loads, a bound on savings inside a strip
(\cref{lem:strip_path_savings}), and the planar
variation theorem (\cref{thm:combi}), which is the main structural
result.
\Cref{sec:fern_toolbox,sec:ckw_common} provide the matrix toolbox and the
common factorization of all block indices.
\Cref{sec:powered_ckw_sweep} handles the eligible block indices,
\cref{sec:block_geometry} handles the high-load ones, and \cref{sec:central_algorithm} assembles the algorithm.
\Cref{sec:powered_ckw_sweep} needs only the clean-detour lemma of
\cref{sec:mark_savings} (\cref{lem:blocks_with_few_marks}); the planar variation theorem
(\cref{sec:mark_distance_variation}) is used only in
\cref{sec:block_geometry}.
\Cref{sec:notation} gathers the notation.

\section{Preliminaries}\label{sec:preliminaries}

This section collects general notation and fixes the conventions for
strings and alignments used throughout the paper.  Most of it
is standard.  As in~\cite{CKW22,gk24},
alignments of fragments use positions in their underlying strings,
without reindexing them from zero.  We also recall the \emph{breakpoint
representation}, which stores an alignment in space proportional to one
plus its cost.  Finally, we specify the \modelname model, in which our
algorithms access strings (\cref{sec:prelims_model}).
\Cref{sec:notation} lists the notation that is used beyond a single
proof, with pointers to its definitions.

\subsection{General Notation}\label{sec:prelims_notation}

We first collect numerical conventions used throughout the paper.

For integers $a,b$, we use the discrete interval notation
\[
\begin{aligned}
    \fragment{a}{b}&=\{z\in\mathbb Z:a\le z\le b\},&
    \fragmentco{a}{b}&=\{z\in\mathbb Z:a\le z<b\},\\
    \fragmentoc{a}{b}&=\{z\in\mathbb Z:a<z\le b\},&
    \fragmentoo{a}{b}&=\{z\in\mathbb Z:a<z<b\}.
\end{aligned}
\]

For $z>0$, $\log z$ denotes the binary logarithm.  For every real $z$,
we write $\lg z\coloneqq\log(\max(2,z))$.
The function $\lg$ satisfies $\lg z=\Theta(1+\log z)$ for $z\ge1$ and
$\lg z\ge1$ for every real $z$, which keeps all logarithmic factors in
our running times positive.

\subsection{Strings and Fragments}\label{sec:prelims_strings}

A \emph{string} $X$ of length $|X|$ is a sequence
$X[0]X[1]\cdots X[|X|-1]$ of characters from a fixed alphabet $\Sigma$;
that is, strings are indexed from zero.  More generally, we index
positions, vertices, and other boundaries from $0$, and the piece
between the boundaries $i$ and $i+1$ has index $i$; for example, the
character $X[i]$ lies between the positions $i$ and $i+1$.  Plain lists
of objects, such as the edits of an alignment, are indexed from $1$.
We write $\varepsilon$ for the
empty string, $\Sigma^+$ for the set of all nonempty strings, $XY$ for
the concatenation of
strings $X$ and $Y$, and $X^e$ for the concatenation of $e\ge0$ copies
of~$X$.  The periodic structure of our instances is described by
infinite repetitions of a short string.

\begin{definition}[Bi-infinite repetitions and primitive strings]
\label{def:bi_infinite_repetition}
For a nonempty string $Q$ of length $q$, the bi-infinite string
$Q^\infty$ satisfies $Q^\infty[i]=Q[i\bmod q]$ for every
$i\in\mathbb Z$.  The string $Q$ is \emph{primitive} if $Q\ne U^e$ for
every string $U$ and every integer $e\ge2$.
\end{definition}

For example, $\texttt{abc}$ is primitive, whereas $\texttt{abab}$ is
not.  Next, we recall fragments, which refer to ranges of positions in
an underlying string.

\begin{definition}[Fragments]
\label{def:fragments}
Let $S$ be a string or a bi-infinite string $Q^\infty$.  For integers
$a\le b$, where additionally $0\le a$ and $b\le|S|$ if $S$ is a string,
the \emph{fragment} $S\fragmentco{a}{b}$ is a reference to $S$ and the
position range $\fragmentco{a}{b}$.  It \emph{starts} at $a$, \emph{ends}
at $b$, has \emph{length} $b-a$, and \emph{spells} the string
$S[a]S[a+1]\cdots S[b-1]$.  The \emph{fragments} of
$S\fragmentco{a}{b}$, also called its \emph{subfragments}, are the
fragments $S\fragmentco{a'}{b'}$ with $a\le a'\le b'\le b$; such a
fragment is a
\emph{prefix} of $S\fragmentco{a}{b}$ if $a'=a$ and a \emph{suffix} if
$b'=b$.
\end{definition}

For instance, if $Q=\texttt{abc}$, then $Q^\infty\fragmentco{5}{9}$
refers to positions $5,6,7,8$ of $Q^\infty$ and spells
$\texttt{cabc}$; $Q^\infty\fragmentco{6}{8}$ is one of its
subfragments.  Fragments at different locations are different
fragments even if they spell the same string.  An empty fragment
$S\fragmentco{a}{a}$ records the boundary position $a$ in $S$.

We implicitly cast a fragment to the string it spells when a string is
expected, for example in string equalities, concatenations, and edit
distances.  When a fragment is expected, a string $X$ stands for its
whole fragment $X\fragmentco{0}{|X|}$.  For example,
$Q^\infty\fragmentco{i}{i+3}=\texttt{abc}$ holds exactly when $3$
divides $i$.  This is the case $j=0$ of the following standard fact,
which is a consequence of the commutation lemma of Lyndon and
Sch\"utzenberger~\cite{LyndonS62}.

\begin{fact}[Synchronization property]
\label{fact:synchronization}
Let $Q$ be a primitive string of length $q$.  For all
$i,j\in\mathbb Z$, we have
$Q^\infty\fragmentco{i}{i+q}=Q^\infty\fragmentco{j}{j+q}$ if and only
if $q$ divides $j-i$.
\end{fact}

\subsection{Alignments and Alignment Graphs}

The \emph{edit distance} $\ed(X,Y)$ of strings $X$ and $Y$ is the
minimum number of insertions, deletions, and substitutions of single
characters that transform $X$ into $Y$.  Alignments describe such
transformations while keeping track of positions.  They are paths in
the following graph.  As in \cite[Definition~2.1]{CKW22} and
\cite[Definition~3.3]{gk24}, the vertices use positions in
the underlying strings.

\begin{definition}[Alignment graph and alignments]
\label{def:alignment_graph}
Let $F=X\fragmentco{x}{x'}$ and $G=Y\fragmentco{y}{y'}$ be fragments
of strings or bi-infinite strings $X$ and $Y$.  The \emph{alignment
graph} $\AG(F,G)$ is the directed graph on the vertex set
\[
    \fragment{x}{x'}\times\fragment{y}{y'}.
\]
For all $i,j$ for which both endpoints are vertices, it contains
\begin{itemize}
    \item the \emph{deletion edge} $(i,j)\to(i+1,j)$ of cost $1$;
    \item the \emph{insertion edge} $(i,j)\to(i,j+1)$ of cost $1$; and
    \item the \emph{diagonal edge} $(i,j)\to(i+1,j+1)$ of cost $0$ if
          $X[i]=Y[j]$, and of cost $1$ otherwise.
\end{itemize}
An \emph{alignment} of $F$ onto $G$, denoted by $\A:F\onto G$ or
$\A:X\fragmentco{x}{x'}\onto Y\fragmentco{y}{y'}$, is a directed path
in $\AG(F,G)$ from $(x,y)$ to $(x',y')$.
The \emph{inverse alignment} $\A^{-1}:G\onto F$ is obtained by swapping
the two coordinates of every vertex of $\A$.
\end{definition}

For fragments $F'$ of $F$ and $G'$ of $G$, the graph $\AG(F',G')$ is
the subgraph of $\AG(F,G)$ induced by its vertex set.

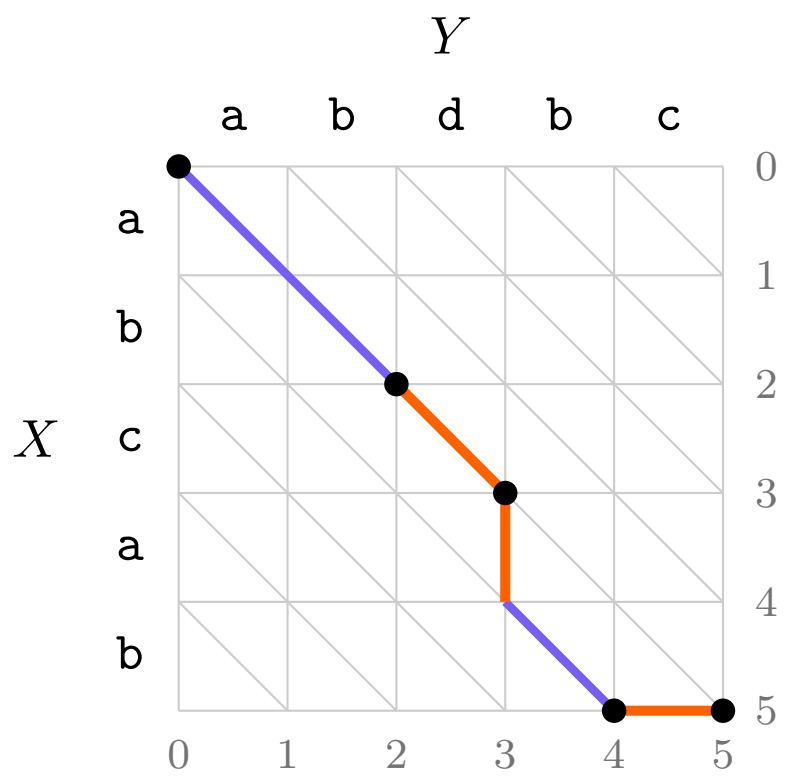
\begin{figure}[t]
\centering
\begin{tikzpicture}[x=0.72cm,y=-0.72cm]
  \foreach \i in {0,...,5} {
    \draw[black!20] (0,\i) -- (5,\i);
    \draw[black!20] (\i,0) -- (\i,5);
  }
  \foreach \i in {0,...,4} {
    \foreach \j in {0,...,4} {
      \draw[black!20] (\j,\i) -- (\j+1,\i+1);
    }
  }
  \foreach \c [count=\i from 0] in {a,b,c,a,b} {
    \node[font=\ttfamily\small,anchor=mid] at (-0.45,\i+0.5) {\c};
  }
  \foreach \c [count=\i from 0] in {a,b,d,b,c} {
    \node[font=\ttfamily\small,anchor=mid] at (\i+0.5,-0.45) {\c};
  }
  \foreach \i in {0,...,5} {
    \node[font=\scriptsize,black!55] at (\i,5.4) {\i};
    \node[font=\scriptsize,black!55] at (5.4,\i) {\i};
  }
  \draw[line width=1.6pt,color2] (0,0) -- (2,2);
  \draw[line width=1.6pt,color2] (3,4) -- (4,5);
  \draw[line width=1.9pt,color4] (2,2) -- (3,3) -- (3,4);
  \draw[line width=1.9pt,color4] (4,5) -- (5,5);
  \foreach \p in {(0,0),(2,2),(3,3),(4,5),(5,5)} {\fill[black] \p circle (2.3pt);}
  \node[font=\small,anchor=east] at (-0.9,2.5) {$X$};
  \node[font=\small,anchor=south] at (2.5,-0.85) {$Y$};
\end{tikzpicture}
\caption{The alignment graph $\AG(X,Y)$ for $X=\texttt{abcab}$
(rows) and $Y=\texttt{abdbc}$ (columns).  The highlighted path is an
optimal alignment $\A$ of cost $3$: after matching $\texttt{ab}$, it
substitutes $\texttt{c}$ by $\texttt{d}$, deletes $\texttt{a}$, matches
$\texttt{b}$, and inserts $\texttt{c}$; the three edits are orange.  The
dots mark the earliest and latest vertices and the breakpoints (the tails
of the edits), whose entries form the breakpoint representation of
\cref{def:breakpoint_representation}.}
\label{fig:prelims_alignment_graph}
\end{figure}

Every path in $\AG(F,G)$ is coordinatewise nondecreasing, and every
subpath of an alignment aligns the fragments determined by its endpoint
coordinates; see \cref{fig:prelims_alignment_graph}.  Inverting an
alignment exchanges insertions and deletions.

For a path $\A$ in an edge-weighted graph, $\cost(\A)$ denotes the
total cost of its edges, and if vertices $u$ and $v$ occur on $\A$ in
this order, then $\A\fragment{u}{v}$ denotes the subpath from $u$ to
$v$, including both endpoints.  We call each positive-cost edge of an
alignment $\A$ an \emph{edit}; thus, $\cost(\A)$ is the number of edits
of $\A$.  For a finite directed graph $H$ with
nonnegative edge weights and vertices $u,v$, we write $\dist_H(u,v)$
for the minimum cost of a directed path from $u$ to $v$, with
$\dist_H(u,v)=+\infty$ if no such path exists.  The edit distance
$\ed(F,G)$ equals the minimum of $\cost(\A)$ over all alignments
$\A:F\onto G$, that is,
$\ed(F,G)=\dist_{\AG(F,G)}((x,y),(x',y'))$.

Alignments can be chained.  For unit costs, the following fact is the
alignment version of the triangle inequality for edit distance.

\begin{fact}[Composition of alignments; {\cite[Lemma~1]{WagnerFischer74}}]
\label{fact:alignment_composition}
Let $X$, $Y$, and $Z$ be strings or bi-infinite strings.  For alignments
$\A:X\fragmentco{x}{x'}\onto Y\fragmentco{y}{y'}$ and
$\B:Y\fragmentco{y}{y'}\onto Z\fragmentco{z}{z'}$, there is an
alignment $\B\circ\A:X\fragmentco{x}{x'}\onto Z\fragmentco{z}{z'}$, the
\emph{composition} of $\A$ and $\B$, such that
$\cost(\B\circ\A)\le\cost(\A)+\cost(\B)$.
\end{fact}

We often restrict an alignment to a pair of fragments.  We first name
the pairs whose endpoints lie on the alignment and record the cost of
the intervening subpath.

\begin{definition}[Aligned fragments and alignment-restriction cost]
    \dglabel{def:alignment_restriction_cost}
    Let $X$ and $Y$ be strings or bi-infinite strings, and let
    $\A:X\fragmentco{x}{x'}\onto Y\fragmentco{y}{y'}$ be an alignment.
    Fragments $F=X\fragmentco{s}{s'}$ and $G=Y\fragmentco{t}{t'}$ are
    \emph{aligned}, denoted by $F\agn{\A}G$, if
    $(s,t)\in\A$ and $(s',t')\in\A$.  For aligned $F,G$, define
    \[
        \ed^{\A}(F,G)
        \coloneqq
        \cost\bigl(\A\fragment{(s,t)}{(s',t')}\bigr).
    \]
    Aligned fragments $F,G$ are \emph{perfectly aligned}, denoted by
    $F\agneq{\A}G$, if $\ed^{\A}(F,G)=0$.
\end{definition}

In \cref{fig:prelims_alignment_graph}, for instance,
$X\fragmentco{0}{2}\agneq{\A}Y\fragmentco{0}{2}$, whereas
$X\fragmentco{2}{4}=\texttt{ca}$ and $Y\fragmentco{2}{3}=\texttt{d}$ are
aligned with $\ed^{\A}(X\fragmentco{2}{4},Y\fragmentco{2}{3})=2$.  The
restriction cost is always at least the edit distance of the two
fragments, and it can be larger if $\A$ is not optimal between them.

An alignment with few edits between long fragments of $Q^\infty$
matches a whole period exactly, and the synchronization property puts
the matched period in phase.  \Cref{sec:ckw_windows,sec:expanded_canonical_geometry,sec:powered_ckw_sweep}
use this argument.

\begin{fact}[Matched periods]
\label{fact:matched_period}
Let $Q$ be a primitive string of length $q$, let $F$ and $G$ be
fragments of $Q^\infty$, and let $\A:F\onto G$ be an alignment with
$\max(|F|,|G|)\ge(\cost(\A)+1)q$.  Then, $\A$ contains $q$ consecutive edges of cost $0$.  Moreover, if $q$
consecutive edges of $\A$ of cost $0$ start at a vertex $(i,j)$,
then $q$ divides $j-i$.
\end{fact}
\begin{proof}
The $E\coloneqq\cost(\A)$ edits of $\A$ split its remaining edges into
at most $E+1$ runs of consecutive edges of cost $0$.  If every run had
at most $q-1$ edges, then, since every edge advances each coordinate by
at most one, we would have $\max(|F|,|G|)\le E+(E+1)(q-1)=(E+1)q-1$,
a contradiction.  For the second claim, edges of cost $0$ are diagonal
edges between equal characters, so
$Q^\infty\fragmentco{i}{i+q}=Q^\infty\fragmentco{j}{j+q}$, and
\cref{fact:synchronization} shows that $q$ divides $j-i$.
\end{proof}

\subsection{Sparse Representations of Alignments}

The alignments supplied with an instance of \SM are long, but
their costs are small.  We therefore store them by listing only their
edits.  We follow the convention of \cite[Section~3]{gk24}: each edit
is given by the vertex at which it starts and the characters involved.

\begin{definition}[Breakpoint representation]
\label{def:breakpoint_representation}
\let\qed\relax %
Let $\A:X\fragmentco{x}{x'}\onto Y\fragmentco{y}{y'}$ be an alignment,
where $X$ and $Y$ are strings or bi-infinite strings.  The
\emph{breakpoints} of
$\A$ are the tails of its edits, that is, the vertices
$(\bar x,\bar y)$ of $\A$ such that $\A$ deletes $X[\bar x]$, inserts
$Y[\bar y]$, or substitutes $X[\bar x]$ by $Y[\bar y]$.  The
\emph{breakpoint representation} $\brp{\A}$ is the sequence that
consists of the entry $((x,\varepsilon),(y,\varepsilon))$, the entries
of all breakpoints in path order, and the entry
$((x',\varepsilon),(y',\varepsilon))$, where the entry of a
breakpoint $(\bar x,\bar y)$ is
\[
    \begin{cases}
        ((\bar x,X[\bar x]),(\bar y,\varepsilon)) & \text{if $\A$ deletes $X[\bar x]$},\\
        ((\bar x,\varepsilon),(\bar y,Y[\bar y])) & \text{if $\A$ inserts $Y[\bar y]$},\\
        ((\bar x,X[\bar x]),(\bar y,Y[\bar y])) & \text{if $\A$ substitutes $X[\bar x]$ by $Y[\bar y]$}.
    \end{cases}
    \tag*{\lipicsEnd}
\]
\end{definition}

Each breakpoint entry determines its edit.  The remaining parts of
$\A$ are runs of zero-cost diagonal edges, and each run is determined
by its endpoints, which are endpoints of edits or the earliest or latest
vertex of $\A$.  Hence, $\brp{\A}$ determines
$\A$ and has $c+2$ entries, where $c=\cost(\A)$.  Swapping the two
components of every entry yields $\brp{\A^{-1}}$.  In
\cref{fig:prelims_alignment_graph}, for example, the breakpoints are
$(2,2)$, $(3,3)$, and $(5,4)$.

\begin{example}\label{ex:breakpoint_representation}
In \cref{ex:overview_periodic}, we have $Q=\texttt{abc}$ and
$P=(\texttt{abc})^2\texttt{ab}(\texttt{abc})^3$.  The supplied
alignment $\A_P:P\onto Q^\infty\fragmentco{0}{18}$ matches
$P\fragmentco{0}{8}$ with $Q^\infty\fragmentco{0}{8}$, uses the
insertion edge $(8,8)\to(8,9)$ to produce the missing $\texttt{c}$ at
position $8$ in $Q^\infty$, and matches $P\fragmentco{8}{17}$ with
$Q^\infty\fragmentco{9}{18}$.  Thus, $\cost(\A_P)=1$, its only
breakpoint is $(8,8)$, and
\[
    \brp{\A_P}=
    ((0,\varepsilon),(0,\varepsilon)),\;
    ((8,\varepsilon),(8,\texttt{c})),\;
    ((17,\varepsilon),(18,\varepsilon)),
\]
even though $\A_P$ has $18$ edges.
\end{example}

\begin{remark}[Comparison with {\cite{CKW22}}]
\label{rem:breakpoints_ckw22}
The representation of \cite[Definition~2.3]{CKW22} lists the same
breakpoints in path order, with edit types instead of characters and
dummy entries $(x-1,y-1,\bot)$ and $(x',y',\bot)$.
For a cost-$c$ alignment, conversion in either direction takes
$\Oh(c+1)$ time in the \modelname model: the positions of
$\varepsilon$ determine the edit type, and each character requires one
access.  Hence, \cref{fact:reduction,fct:alignments} apply to
$\brp{\A}$.
\end{remark}

We often need to translate a position of one fragment into the
corresponding positions of the other.  For a fixed position $\bar y$ of
the target, the positions of the source paired with $\bar y$ form an
interval; it has more than one element exactly when the alignment
deletes characters while staying at $\bar y$.  In
\cref{fig:prelims_alignment_graph}, the position $\bar y=3$ is paired
with the interval $\{3,4\}$.  The breakpoint representation provides
the endpoints of these intervals for all $\bar y$ simultaneously.

\begin{fact}[Alignment recovery from breakpoint representation;
{\cite[Lemma~2.4]{CKW22}}]
    \dglabel{fct:alignments}[def:breakpoint_representation]
    Let
    $\A:X\fragmentco{x}{x'}\onto Y\fragmentco{y}{y'}$ be a cost-$c$
    alignment.  For every $\bar y\in\fragment{y}{y'}$, the positions
    paired with $\bar y$ form the nonempty interval
    \[
        \{\bar x\in\fragment{x}{x'}:(\bar x,\bar y)\in\A\}.
    \]
    Given $\brp{\A}$, in $\Oh(c+1)$ time we can represent both
    endpoint sequences
    \[
        \left(
          \min\{\bar x:(\bar x,\bar y)\in\A\}
        \right)_{\bar y=y}^{y'},
        \qquad
        \left(
          \max\{\bar x:(\bar x,\bar y)\in\A\}
        \right)_{\bar y=y}^{y'}
    \]
    as concatenations of at most $c+1$ arithmetic progressions
    with difference $1$.
\end{fact}

\subsection{Model of Computation}\label{sec:prelims_model}

Our algorithms access strings only through the following interface,
so their running times apply to every representation of strings that
implements it.

\begin{definition}[\modelname model; {\cite[Section~2.1]{CKW20}}]
\label{def:pillar_model}
\let\qed\relax %
Given a family $\mathcal X$ of strings, a \emph{handle} refers to a
fragment of a string in $\mathcal X$.  Initially, a handle to
$X\fragmentco{0}{|X|}$ is available for each $X\in\mathcal X$.  For
fragments $U=S\fragmentco{s_U}{t_U}$ and
$V=R\fragmentco{s_V}{t_V}$ specified by handles, the
\emph{\modelname model} supports the following operations, with coordinates
relative to the beginning of each fragment.
\begin{itemize}
    \item \lenOpName$(U)$ returns $|U|$.
    \item \accOpName$(U,i)$ returns $S[s_U+i]$ for $0\le i<|U|$.
    \item \extractOpName$(U,a,b)$ returns a handle to
          $S\fragmentco{s_U+a}{s_U+b}$ for $0\le a\le b\le|U|$.
    \item \lceOpName$(U,V)$ returns the length of the longest common
          prefix of the strings spelled by $U$ and $V$.
    \item \lcbOpName$(U,V)$ returns the length of the longest common
          suffix of the strings spelled by $U$ and $V$.
    \item \ipmOpName$(U,V)$ (internal pattern matching), for
          $1\le|U|$ and $|V|\le2|U|$, returns
          \[
              \{i\in\fragment{0}{|V|-|U|}:
                    U=R\fragmentco{s_V+i}{s_V+i+|U|}\},
          \]
          represented by an arithmetic progression, or the empty set.\lipicsEnd
\end{itemize}
\end{definition}

Our running times count each primitive call as one
unit, together with ordinary computation.  Preprocessing the underlying
strings is charged separately in each implementation;
\cref{sec:introduction} gives the bound for explicit strings.

The input of \SM contains the string $Q$, and our algorithms also
work with fragments of $Q^\infty$, which we represent by their
endpoints.  For such a fragment $Q^\infty\fragmentco{\ell}{r}$, the
length and the subfragments are given by endpoint arithmetic, and the
characters satisfy $Q^\infty[i]=Q[i\bmod q]$.  Moreover, for every
fragment $U$ specified by a handle, \lceOpName$(U,Q^\infty\fragmentco{\ell}{r})$ and
\lcbOpName$(U,Q^\infty\fragmentco{\ell}{r})$ can be computed with
$\Oh(1)$ \modelname operations
\cite[Lemma~2.8 and Corollaries~2.9 and~2.10]{CKW20}.

\section{Decomposing Periodic Occurrence Starts}\label{sec:ckw_windows}

This section groups the possible occurrence starts of a valid
instance of \SM.  The key tool is
\cref{lem:widened_endpoint_localization}: via $\A_P$ and $\A_T$, every
$d$-error occurrence of $P$ in $T$ corresponds to the pattern reference
shifted by a whole number of periods, up to $\Oh(d)$ positions at each
end.  We cut the periodic reference $Q^\infty$ into blocks
of equal length $\tau$ and assign every occurrence start to exactly one
\emph{block index} $j$.  The starts assigned to $j$ lie in an interval
of length $\Oh(d)$, so they can be verified directly, which is fast enough whenever there are
few block indices (\cref{fct:verifyRj}).
Later sections exploit the fact that the computations for different block indices are
very similar:
\cref{sec:expanded_canonical_geometry,sec:dense_covers_marks,sec:mark_savings,sec:mark_distance_variation}
analyze optimal paths from the starts assigned to a block index, and
\cref{sec:ckw_common,sec:powered_ckw_sweep,sec:block_geometry}
process the block indices.
The construction adapts the periodic-block localization of \cite{CKW22}; an informal account,
with an example, appears in
\cref{sec:overview_reduction,sec:overview_blocks}.

\subsection{Parameters}

For the remainder of this section, fix a valid instance of \SM.
The blocks are defined on the periodic reference using two parameters:
the combined edit budget $\kappa$ and the block length $\tau$.

\begin{definition}[Reference coordinates and working parameters]
\label{def:structural_ckw_parameters}
A position in $Q^\infty$ is a \emph{reference coordinate}, and its
\emph{phase} is its residue modulo $q$.  Define
\[
    \kappa\coloneqq d+d_P+d_T,
    \qquad
    \tau\coloneqq q\left(\floor{\kappa/q}+1\right).
\]
We call $\kappa$ the \emph{combined edit budget} and $\tau$ the
\emph{block length}.
\end{definition}

\begin{remark}[Parameter bounds]
\label{rem:parameter_bounds}
Since $d_P\le d$ and $d_T\le3d$, we have $d\le\kappa\le5d$.  By
definition, $\tau$ is the least multiple of $q$ that exceeds $\kappa$,
so
\[
    \kappa<\tau\le\kappa+q
    \qquad\text{and}\qquad
    \tau/q=\floor{\kappa/q}+1=\Oh(1+d/q).
\]
In particular, $\tau=q$ exactly when $q>\kappa$.
\end{remark}

The supplied alignments map $P$ and $T$ to the following reference
fragments.

\begin{definition}[Periodic reference fragments]
\label{def:periodic_reference_fragments}
Set $x_P\coloneqq0$.  The \emph{pattern reference fragment} and
\emph{text reference fragment} are, respectively,
\[
    \hP\coloneqq Q^\infty\fragmentco{x_P}{y_P},
    \qquad
    \hT\coloneqq Q^\infty\fragmentco{x_T}{y_T}.
\]
We call $\AG(\hP,\hT)$ the \emph{reference alignment graph}.
\end{definition}

Both fragments refer to the same underlying string $Q^\infty$, so
their reference coordinates lie on the same integer axis.

\subsection{Period Shifts of Occurrences}

Let $F$ be a $d$-error occurrence of $P$ in $T$, and let $\widehat F$
be a fragment of $\hT$ that $\A_T$ aligns with $F$.  Composing the
inverse of $\A_P$, an optimal alignment of $P$ with $F$, and $\A_T$
aligns the pattern reference $\hP$ with $\widehat F$ at cost at most
$\kappa$.  Since $\hP$ is longer than $(\kappa+1)q$, this alignment
matches some $q$ consecutive characters exactly, and by synchronization
they are in phase (\cref{fact:matched_period}).  Hence, up to $\kappa$ positions at each
end, $\widehat F$ is $\hP$ shifted by a whole number $b$ of periods;
\cref{ex:overview_periodic} shows the shifts $b=2$ and $b=1$.

\begin{lemma}[Periodic localization of an occurrence]
    \label{lem:widened_endpoint_localization}
    Let $F$ be a $d$-error occurrence of $P$ in $T$, that is, a
    fragment of $T$ satisfying $\ed(P,F)\le d$, and
    let $\widehat F=\Q\fragmentco{\hat v}{\hat e}$ be any fragment such that
    $F\agn{\A_T}\widehat F$.  Then, there exists $b\in\mathbb Z$ such
    that
    \[
        \max\bigl(|\hat v-bq|,\ |\hat e-(bq+y_P)|\bigr)
        \le d_P+\ed(P,F)+d_T
        \le\kappa.
    \]
\end{lemma}
\begin{proof}
    This is the endpoint argument of
    \cite[Lemma~3.4]{CKW22}, with the cost bound enlarged to the
    combined edit budget.

    \emph{Step 1: an alignment between two periodic fragments.}
    Let $\A:P\onto F$ be an optimal alignment.  Composing the inverse of
    $\A_P$, the alignment $\A$, and the restriction of $\A_T$ to
    $F\onto\widehat F$ yields (\cref{fact:alignment_composition}) an
    alignment $\B:\Q\fragmentco{0}{y_P}\onto\widehat F$ of cost
    \[
        \cost(\B)
        \le d_P+\ed(P,F)+d_T
        \le d+d_P+d_T
        =\kappa.
    \]

    \emph{Step 2: one period is matched exactly.}  The promises
    of \SM give
    \[
        y_P\ge m-d_P\ge8dq-d\ge(\kappa+2)q>(\cost(\B)+1)q,
    \]
    where the third inequality uses $\kappa\le5d$ and $d,q\ge1$.
    Hence, by \cref{fact:matched_period}, $\B$ contains $q$ consecutive
    edges of cost $0$, starting at a vertex $(i,j)$ with
    $j-i=bq$ for an integer $b$.

    \emph{Step 3: the endpoints.}  The \emph{offset} $y-x$ of a vertex
    $(x,y)$ of $\B$ equals $bq$ at $(i,j)$.  Each insertion or
    deletion changes it by one and costs one; diagonal edges leave it
    unchanged.  The endpoint offsets $\hat v$ and $\hat e-y_P$ therefore
    differ from $bq$ by at most $\cost(\B)\le d_P+\ed(P,F)+d_T$.
    This proves the lemma.
\end{proof}

\begin{definition}[Period shift]
\label{def:phase_shift}
For $F$ and $\widehat F$ as in
\cref{lem:widened_endpoint_localization}, we call any integer $b$
satisfying the bound in that lemma a \emph{period shift} of $F$ with
respect to $\widehat F$.
\end{definition}

A period shift need not be unique, but
all period shifts of $F$ with respect to the same $\widehat F$ differ by
at most $2\kappa/q$.  In \cref{ex:overview_periodic}, for instance, the
exact occurrence $F=T\fragmentco{6}{23}$ is aligned with
$\widehat F=\Q\fragmentco{6}{24}$, which is the pattern reference
$\Q\fragmentco{0}{18}$ shifted by $6=2q$.  Hence, $b=2$ is a period
shift of $F$ with respect to $\widehat F$, and both differences bounded
in \cref{lem:widened_endpoint_localization} are $0$.

\subsection{Block Indices and Start Intervals}

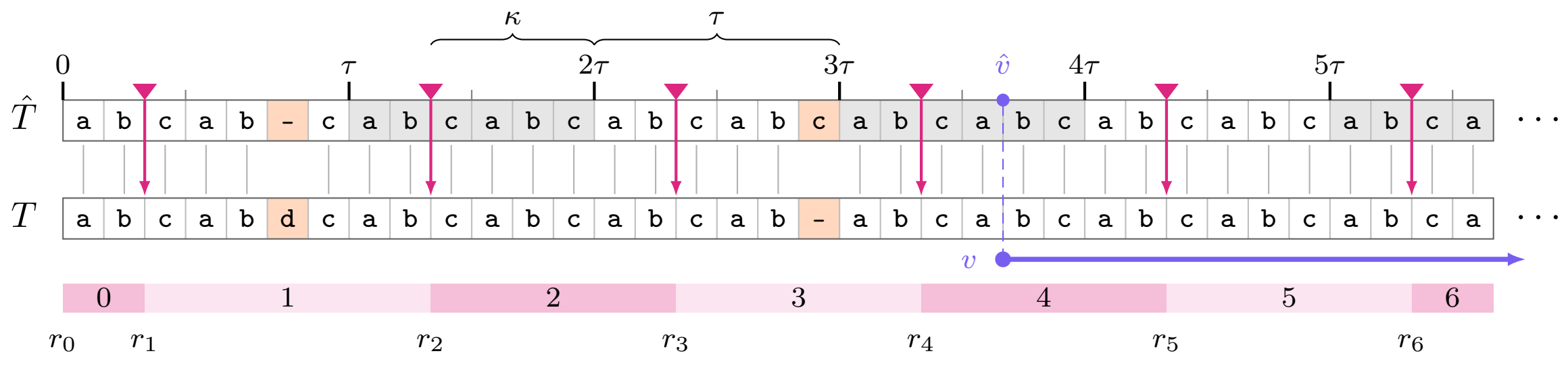
\begin{figure}[t]
\centering
\begin{tikzpicture}[x=0.4cm,y=0.4cm]
  \fill[black!10] (7,2.4) rectangle (13,3.4);
  \fill[black!10] (19,2.4) rectangle (25,3.4);
  \fill[black!10] (31,2.4) rectangle (35,3.4);
  \foreach \c in {5,18} {
    \fill[color4!25] (\c,2.4) rectangle (\c+1,3.4);
    \fill[color4!25] (\c,0) rectangle (\c+1,1);
  }
  \foreach \c in {1,...,34} {
    \draw[black!25] (\c,2.4) -- (\c,3.4);
    \draw[black!25] (\c,0) -- (\c,1);
  }
  \draw[black!60] (0,2.4) rectangle (35,3.4);
  \draw[black!60] (0,0) rectangle (35,1);
  \foreach \ch [count=\c from 0] in {a,b,c,a,b,-,c,a,b,c,a,b,c,a,b,c,a,b,c,a,b,c,a,b,c,a,b,c,a,b,c,a,b,c,a} {
    \node[font=\ttfamily\scriptsize,anchor=mid] at (\c+0.5,2.9) {\ch};
  }
  \foreach \ch [count=\c from 0] in {a,b,c,a,b,d,c,a,b,c,a,b,c,a,b,c,a,b,-,a,b,c,a,b,c,a,b,c,a,b,c,a,b,c,a} {
    \node[font=\ttfamily\scriptsize,anchor=mid] at (\c+0.5,0.5) {\ch};
  }
  \node[font=\small,anchor=mid east] at (-0.3,2.9) {$\hT$};
  \node[font=\small,anchor=mid east] at (-0.3,0.5) {$T$};
  \node[anchor=mid west] at (35.2,2.9) {$\cdots$};
  \node[anchor=mid west] at (35.2,0.5) {$\cdots$};
  \foreach \c in {0,...,4,6,7,...,17,19,20,...,34} {
    \draw[black!30] (\c+0.5,1.1) -- (\c+0.5,2.3);
  }
  \foreach \c in {3,10,16,22,28,34} {\draw[black!50] (\c,3.4) -- (\c,3.65);}
  \foreach \c/\l in {0/0,7/\tau,13/2\tau,19/3\tau,25/4\tau,31/5\tau} {
    \draw[thick] (\c,3.4) -- (\c,3.85);
    \node[font=\scriptsize,anchor=base] at (\c,4.05) {$\l$};
  }
  \draw[decorate,decoration={brace,amplitude=3pt}] (9,4.75) -- (13,4.75)
     node[midway,above=2pt,font=\scriptsize] {$\kappa$};
  \draw[decorate,decoration={brace,amplitude=3pt}] (13,4.75) -- (19,4.75)
     node[midway,above=2pt,font=\scriptsize] {$\tau$};
  \foreach \c in {2,9,15,21,27,33} {
    \fill[color3] (\c-0.3,3.8) -- (\c+0.3,3.8) -- (\c,3.4) -- cycle;
    \draw[color3,thick,-{Latex[length=1.5mm]}] (\c,3.4) -- (\c,1.04);
  }
  \foreach \a/\b/\j/\f in {0/2/0/30,2/9/1/12,9/15/2/30,15/21/3/12,21/27/4/30,27/33/5/12,33/35/6/30} {
    \fill[color3!\f] (\a,-1.8) rectangle (\b,-1.1);
    \node[font=\scriptsize,anchor=mid] at ({(\a+\b)/2},-1.45) {$\j$};
  }
  \foreach \c/\j in {0/0,2/1,9/2,15/3,21/4,27/5,33/6} {
    \node[font=\scriptsize,anchor=base] at (\c,-2.65) {$r_{\j}$};
  }
  \draw[color2,densely dashed] (23,-0.5) -- (23,3.4);
  \draw[color2,line width=1.4pt,-{Latex[length=1.8mm]}] (23,-0.5) -- (35.8,-0.5);
  \fill[color2] (23,-0.5) circle (2.2pt);
  \fill[color2] (23,3.4) circle (1.8pt);
  \node[font=\scriptsize,color2,anchor=mid east] at (22.7,-0.5) {$v$};
  \node[font=\scriptsize,color2,anchor=base] at (23,4.05) {$\hat v$};
\end{tikzpicture}
\caption{Block indices and start intervals for $q=3$, $\kappa=4$, and
$\tau=6$.  Top: the text reference $\hT$ with block boundaries
$j\tau$ (heavy ticks) and the points $j\tau-\kappa$ (magenta triangles).
Bottom: the text $T$ placed by $\A_T$, whose two edits are orange.  The
arrows pull each $j\tau-\kappa$ back to $r_j$, and the bars are the start
intervals $\fragmentco{r_j}{r_{j+1}}$.  The occurrence at $v=22$
(purple) is assigned to $j=4$, and $\hat v\in\fragment{4\tau-\kappa}{4\tau+2\kappa}$
(\cref{lem:working_budget_ckw_slicing}).}
\label{fig:ckw_windows_blocks}
\end{figure}

In total, occurrences can have roughly $(n-m)/q$ different period shifts.  We
group them into blocks of $\tau/q$ consecutive integers, so that the
shifts $bq$ form blocks of reference coordinates of length $\tau$.
Block index $j$ collects the period shifts $b$ with $bq$ near $j\tau$,
and its start interval is
obtained by pulling the reference coordinate $j\tau-\kappa$ back to the
text through $\A_T$; see \cref{fig:ckw_windows_blocks}.  The following
definition adapts \cite[Definition~3.6]{CKW22} to the block length
$\tau$.

\begin{definition}[Block indices and start intervals]
    \dglabel{def:jdef}
    Define the block-index set
    \[
        J\coloneqq
        \fragment{
            \ceil{(x_T-\kappa)/\tau}
        }{
            \floor{(y_T-y_P+\kappa)/\tau}
        }.
    \]
    For every $j\in J$, define
    \[
        r_j\coloneqq
          \min\left\{v:(v,\hat v)\in\A_T
               \text{ for some }\hat v\ge j\tau-\kappa\right\}.
    \]
    When $J$ is nonempty, set $r_{\max J+1}\coloneqq n$.
    We call $\fragmentco{r_j}{r_{j+1}}$ the
    \emph{start interval associated with $j$}, and
    $b_j\coloneqq j\tau/q$ the \emph{origin} of $j$.  An occurrence of
    $P$ in $T$ that starts in $\fragmentco{r_j}{r_{j+1}}$ is
    \emph{assigned} to $j$.
\end{definition}

The minimum defining $r_j$ is over a nonempty set: for $j\in J$,
$j\tau-\kappa\le y_T-y_P\le y_T$, so the final vertex $(n,y_T)$ of
$\A_T$ meets the threshold.

In words, $j\in J$ means that the pattern reference shifted by
$j\tau$, namely $\Q\fragmentco{j\tau}{j\tau+y_P}$, fits into the text
reference $\hT=\Q\fragmentco{x_T}{y_T}$ up to $\kappa$ positions at
either end.  The position $r_j$ is the leftmost text position that $\A_T$
aligns with a reference coordinate of at least $j\tau-\kappa$.  The
origin $b_j$ counts the shift $j\tau$ in periods; it
is an integer because $q$ divides $\tau$.

The next lemma shows that the start intervals partition the
occurrences and that each occurrence starts close to the left end of its
start interval.  For coverage, the proof distinguishes two cases.  If
$q\le\kappa$, then the start intervals cover the whole text.  If
$q>\kappa$, an initial prefix may be uncovered;
\cref{lem:widened_endpoint_localization} rules out occurrence starts
there.  For the distance bound, the period shift places the reference
coordinate of an occurrence start assigned to $j$ near $j\tau$, and,
since $\A_T$ makes at most $d_T$ edits, the start itself lies near
$r_j$.

\begin{lemma}[Occurrence decomposition over start intervals]
    \label{lem:working_budget_ckw_slicing}
    The associated start intervals are pairwise disjoint and cover
    $\OccE_d(P,T)$.  Moreover, let $j\in J$ and
    $v\in\OccE_d(P,T)\cap\fragmentco{r_j}{r_{j+1}}$, and let $\hat v$ be
    maximal such that $(v,\hat v)\in\A_T$.  Then,
    \[
        0\le v-r_j\le4\kappa
        \qquad\text{and}\qquad
        j\tau-\kappa\le\hat v\le j\tau+2\kappa.
    \]
\end{lemma}
\begin{proof}
    \emph{Disjointness.}  As $j$ grows, so does the threshold
    $j\tau-\kappa$ in the definition of $r_j$, so $r_j$ is a minimum over
    a shrinking set.  Hence, $(r_j)_{j\in J}$ is nondecreasing, and its
    start intervals are pairwise disjoint.

    \emph{Setup.}  Fix $v\in\OccE_d(P,T)$ and choose
    $F=T\fragmentco{v}{e}$ with $\ed(P,F)\le d$.
    Among the fragments of $\hT$ aligned with $F$ under $\A_T$, let
    $\widehat F=\Q\fragmentco{\hat v}{\hat e}$ be the one with the
    largest left endpoint $\hat v$ and the smallest right endpoint
    $\hat e$; thus, $\hat v$ is as in the statement.  Let $b$ be a period shift supplied by
    \cref{lem:widened_endpoint_localization}.  Since $\hat v\ge x_T$
    and $\hat e\le y_T$, the lemma gives
    \begin{equation}\label{eq:structural_phase_range}
        x_T-\kappa\le bq
        \le y_T-y_P+\kappa.
    \end{equation}

    \emph{Coverage.}  Note that $v<n$, because the empty fragment
    $T\fragmentco{n}{n}$ is at edit distance $m>d$ from $P$.  It remains
    to show that $J$ is nonempty and $v\ge r_{\min J}$; then, $v$ lies in
    $\fragmentco{r_{\min J}}{n}$, the union of the start intervals.

    If $q\le\kappa$, then $0\le x_T<q\le\kappa<\tau$ by the promises
    of \SM, so $\ceil{(x_T-\kappa)/\tau}=0$.  The length bounds of the
    supplied alignments, $y_T-x_T\ge n-d_T$ and $y_P\le m+d_P$, together
    with the input promise $n\ge m-k$ and $d\ge k$, give
    \[
       y_T-y_P+\kappa
       \ge x_T+n-m+d
       \ge x_T+d-k\ge0,
    \]
    so $\max J\ge0$.  Hence, $\min J=0$, and $r_0=0$ because the
    earliest vertex $(0,x_T)$ of $\A_T$ has reference coordinate
    $x_T\ge-\kappa$.

    If $q>\kappa$, then $\tau=q$ and \cref{eq:structural_phase_range}
    gives $b\in J$.  Hence, $\hat v\ge bq-\kappa\ge(\min J)\cdot\tau-\kappa$,
    which implies $v\ge r_{\min J}$ by the definition of $r_{\min J}$.

    \emph{Position of $\hat v$.}  Let $j$ be the index with
    $v\in\fragmentco{r_j}{r_{j+1}}$.  By monotonicity of $\A_T$ and the
    choice of $\hat v$ as the largest reference coordinate aligned with
    $v$, the definition of $r_j$ gives
    \[
        \hat v\ge j\tau-\kappa
        \qquad\text{and}\qquad
        \hat v<(j+1)\tau-\kappa\quad\text{if }j<\max J.
    \]
    We claim that $bq<(j+1)\tau$.  If $j<\max J$, this follows from
    the second inequality and $bq\le\hat v+\kappa$.  If $j=\max J$, it
    follows from \cref{eq:structural_phase_range}, because
    $\max J=\floor{(y_T-y_P+\kappa)/\tau}$.  Since both $bq$ and
    $(j+1)\tau$ are multiples of $q$, the claim implies
    \[
        bq\le(j+1)\tau-q
          =j\tau+(\tau-q)
          \le j\tau+\kappa.
    \]
    Combined with $\hat v\le bq+\kappa$, this yields the second claim,
    $j\tau-\kappa\le \hat v\le j\tau+2\kappa$.

    \emph{Distance from the left end.}  By the definition of $r_j$, the
    earliest vertex of $\A_T$ with reference coordinate at least
    $j\tau-\kappa$ is $(r_j,\hat r)$ for some $\hat r\ge j\tau-\kappa$.
    Consider the subpath of $\A_T$ from $(r_j,\hat r)$ to $(v,\hat v)$.
    Along any path, the text coordinate can outgrow the reference
    coordinate only through deletions, each of which costs one, and this
    subpath costs at most $d_T$.  Hence,
    \[
        0\le v-r_j
          \le \hat v-\hat r+d_T
          \le \hat v-(j\tau-\kappa)+d_T
          \le3\kappa+d_T
          \le4\kappa.\qedhere
    \]
\end{proof}

The second bound says that, in reference coordinates, every occurrence
assigned to $j$ starts within $2\kappa$ of $j\tau=b_jq$; the proof
also gives $|b-b_j|\le2\kappa/q$ for its period shift~$b$.

\begin{example}[Running example, continued]\label{ex:running_blocks}
We return to \cref{ex:overview_periodic}, where $Q=\texttt{abc}$,
$q=3$, $|P|=17$, $|T|=30$, $y_P=18$, $x_T=0$, $y_T=30$, $d_P=1$, and
$d_T=2$.  As there, we ignore the length constraints of \SM.  We
take $k=1$ and $d=2$, which satisfies $d\ge2k$.  Then, $\kappa=5$ and
$\tau=6$, so
\[
    J=\fragment{\ceil{-5/6}}{\floor{17/6}}=\fragment{0}{2},
    \qquad
    (b_0,b_1,b_2)=(0,2,4).
\]
The alignment $\A_T$ pairs every text position $v\le14$ with the
reference coordinate $v$, every $v\in\fragment{14}{25}$ with $v+1$, and
every $v\ge26$ with $v$.  Thus, the position $14$ is paired with both
$14$ and $15$, and the positions $25$ and $26$ are both paired with
$26$.  Hence, $r_0=0$, $r_1=1$, $r_2=7$, and $r_3=n=30$.  The start
intervals $\fragmentco{0}{1}$, $\fragmentco{1}{7}$, and
$\fragmentco{7}{30}$ split $\OccE_2(P,T)=\{0,3,4,5,6,7,8\}$ into
$\{0\}$, $\{3,\ldots,6\}$, and $\{7,8\}$, and every start lies at most
$5\le4\kappa$ positions after the left end of its interval.  The
occurrences of cost at most $k=1$ start at $5$, $6$, and $7$, and $b=2$
is a period shift of each of them.  The starts $5$ and $6$ are assigned
to $j=1$, whose origin is $b_1=2$, whereas $7$ is assigned to $j=2$,
whose origin $b_2=4$ differs from this period shift by
$2\le2\kappa/q$.  With only $|J|=3$ block indices, this instance is
solved by direct verification (\cref{fct:verifyRj} below).
The later continuations of this
example only illustrate the definitions; the assumptions of
the later sections fail for such a small instance.
\end{example}

\subsection{Computing Start Intervals and Verifying Few Block Indices}

To handle large $J$, consider $j,j+1\in J$ with
$j\tau-\kappa\ge x_T$.  Then, $r_{j+1}=r_j+\tau$ whenever $\A_T$
makes no edit while its reference coordinate advances from
$j\tau-\kappa$ to $(j+1)\tau-\kappa$.  Hence, $(r_j)_{j\in J}$ splits
into arithmetic progressions whose number depends only on $d_T$.

\begin{lemma}[Breakpoint construction of start intervals]
    \dglabel{fct:rj}
    Given the breakpoint representation of $\A_T$, in
    $\Oh(d_T+1)$ time we can represent $(r_j)_{j\in J}$ as a
    concatenation of $\Oh(d_T+1)$ arithmetic progressions with
    difference $\tau$.
\end{lemma}
\begin{proof}
    By \cref{fct:alignments}, applied to $\A_T$, the
    function $\hat v\mapsto\min\{v:(v,\hat v)\in\A_T\}$, which maps a
    reference coordinate to the leftmost text position aligned with it,
    is a concatenation of $\Oh(d_T+1)$ arithmetic progressions with
    difference $1$, which can be constructed in $\Oh(d_T+1)$ time.  If
    $j\tau-\kappa\ge x_T$, then $r_j$ is the value of this function at
    $\hat v=j\tau-\kappa$.  Restricting each progression to the arguments
    of the form $j\tau-\kappa$ yields a progression of values with
    difference $\tau$, and its smallest and largest arguments can be computed
    in constant time.  The remaining indices satisfy
    $x_T-\kappa\le j\tau<x_T+\kappa$.  Since $\tau>\kappa$, there are
    at most two of them, and each satisfies $r_j=0$ because the earliest
    vertex $(0,x_T)$ of $\A_T$ has reference coordinate
    $x_T>j\tau-\kappa$; we record them as singletons.
\end{proof}

For a single block index, the assigned occurrences can be found by
a standard verification procedure, since they start within an
interval of length $\Oh(d)$.

\begin{lemma}[Direct verification for selected block indices]
    \dglabel{fct:verifyRj}
    Let $g\in\fragment{0}{d}$ be an integer and let
    $J_{\mathrm{sel}}\subseteq J$ be given as a list of its elements.
    One can list the elements of
    \[
        \OccE_g(P,T)\cap\bigcup_{j\in J_{\mathrm{sel}}}\fragmentco{r_j}{r_{j+1}}
    \]
    in $\Oh(1+d^2|J_{\mathrm{sel}}|)$ time in the \modelname model;
    there are $\Oh(d|J_{\mathrm{sel}}|)$ of them.
\end{lemma}
\begin{proof}
    If $J_{\mathrm{sel}}$ is nonempty, we first build the representation
    of \cref{fct:rj} in $\Oh(d_T+1)=\Oh(d)$ time.  Fix
    $j\in J_{\mathrm{sel}}$.  Recover $r_j$ and $r_{j+1}$ by queries to the
    representation of \cref{fct:rj}, using the value $n$ when
    $j=\max J$.  Since $g\le d$, every $g$-error occurrence is also a
    $d$-error occurrence.  Hence, by
    \cref{lem:working_budget_ckw_slicing}, every $g$-error occurrence
    assigned to $j$ starts in the integer interval
    \[
        I\coloneqq
          \fragmentco{r_j}{r_{j+1}}
             \cap\fragmentcc{r_j}{r_j+4\kappa},
    \]
    which has $\Oh(d)$ elements; we skip $j$ if $I$ is empty.  For
    $g\ge1$, the procedure $\texttt{Verify}(P,T,g,I)$ of
    \cite[Fact~2.9]{CKW22} returns exactly the starts in $I$ of
    occurrences with at most $g$ edits, using $\Oh(g(g+|I|))=\Oh(d^2)$
    \modelname operations.  For $g=0$, we test each of the $\Oh(d)$
    positions in $I$ with one longest-common-prefix query between $P$
    and the corresponding suffix of $T$.  The queries to the
    representation of $(r_j)$ take $\Oh(\lg d_T)$ time per index,
    which fits the same bound.  The start intervals are disjoint and
    contain all assigned occurrences, so the output lists have total
    size $\Oh(d|J_{\mathrm{sel}}|)$, and the total running time is $\Oh(1+d^2|J_{\mathrm{sel}}|)$.
\end{proof}

By \cref{lem:working_budget_ckw_slicing}, the start intervals cover
$\OccE_g(P,T)\subseteq\OccE_d(P,T)$.  Hence, \cref{fct:verifyRj} with
$J_{\mathrm{sel}}=J$ computes $\OccE_g(P,T)$ in $\Oh(1+d^2|J|)$ time,
which is fast when there are few block indices.

\section{Expanded Alignment Geometry and Rails}\label{sec:expanded_canonical_geometry}

This section sets up the geometric language of the paper for a fixed valid instance of \SM.
The supplied
alignments $\A_P$ and $\A_T$ translate positions of $P$ and $T$ into
reference coordinates in $Q^\infty$.  We study alignments of $P$ with fragments of $T$
in both coordinate systems at once: in input coordinates, their minimum
costs are the edit distances we care about, whereas in reference coordinates,
periodicity makes them easy to analyze.  The main objects are the
\emph{rails}: the paths that
follow $\A_P$ and $\A_T$ simultaneously, with the two references in the
same phase.  The informal counterpart of this section is
\cref{sec:overview_marks}.  \Cref{fig:overview_reference} in the
technical overview shows the rails in reference coordinates, and \cref{ex:overview_periodic}
shows a rail whose cost is smaller than its canonical cost, the number
of edits of $\A_P$ and $\A_T$ that it passes;
\cref{ex:running_rails} revisits this rail in the expanded graph.

The section is organized as follows.
\Cref{sec:expanded_graph} defines the expanded alignment graph, which
records both coordinate systems.  \Cref{sec:suffix_distances} defines
the suffix distances $\tdtob$, whose values at the top row decide where the
occurrences start.  \Cref{sec:canonical_costs} compares the costs of a path in
the two coordinate systems.  \Cref{sec:canonical_diagonals} introduces
the rails and proves that leaving them is expensive.  Finally,
\cref{sec:tight_path_localization} shows that shortest paths from the
occurrence starts assigned to a block index $j$ to the bottom row stay
close to the rail of index~$b_j$.  \Cref{sec:dense_covers_marks,sec:mark_savings,sec:mark_distance_variation}
build on the rails and on
\cref{lem:offset_change,lem:blue_jump_limit,lem:noblue}, and
\cref{sec:powered_ckw_sweep,sec:block_geometry}
use the localization of \cref{sec:tight_path_localization}.

\subsection{Alignment Expansions and the Expanded Graph}
\label{sec:expanded_graph}

\begin{figure}[t]
\centering
\begin{tikzpicture}[x=0.44cm,y=-0.44cm]
  \fill[color1!14] (0,1) rectangle (16,2);
  \fill[color3!14] (2,0) rectangle (3,8);
  \fill[color3!14] (14,0) rectangle (15,8);
  \foreach \i in {0,...,7} {\foreach \j in {0,...,15} {\draw[black!13] (\j,\i) -- (\j+1,\i+1);}}
  \foreach \i in {0,...,8} {\draw[black!22] (0,\i) -- (16,\i);}
  \foreach \j in {0,...,16} {\draw[black!22] (\j,0) -- (\j,8);}
  \foreach \c/\h [count=\i from 0] in {b/b,-/c,a/a,b/b,c/c,a/a,b/b,c/c} {
    \node[font=\ttfamily\scriptsize,anchor=mid] at (-1.1,\i+0.5) {\c};
    \node[font=\ttfamily\scriptsize,anchor=mid] at (-0.45,\i+0.5) {\h};
  }
  \foreach \c/\h [count=\j from 0] in {a/a,b/b,-/c,a/a,b/b,c/c,a/a,b/b,c/c,a/a,b/b,c/c,a/a,b/b,d/-,c/c} {
    \node[font=\ttfamily\scriptsize,anchor=mid] at (\j+0.5,-1.1) {\c};
    \node[font=\ttfamily\scriptsize,anchor=mid] at (\j+0.5,-0.45) {\h};
  }
  \node[font=\scriptsize,color1,anchor=mid] at (-1.1,9.45) {$P$};
  \node[font=\scriptsize,color1,anchor=mid] at (-0.45,9.45) {$\hP$};
  \node[font=\scriptsize,color3,anchor=mid west] at (16.8,-1.1) {$T$};
  \node[font=\scriptsize,color3,anchor=mid west] at (16.8,-0.45) {$\hT$};
  \foreach \j in {0,...,16} {
    \pgfmathtruncatemacro{\s}{\j+12}
    \node[font=\tiny,black!60,anchor=north] at (\j,8.05) {$\s$};
  }
  \foreach \i in {0,...,8} {
    \pgfmathtruncatemacro{\s}{\i+7}
    \node[font=\tiny,black!60,anchor=west] at (16.05,\i) {$\s$};
  }
  \draw[line width=1.5pt,color2] (1,0) -- (9,8);
  \draw[line width=1.5pt,color2] (7,0) -- (8,1);
  \draw[line width=1.8pt,color4] (8,1) -- (9,2);
  \draw[line width=1.5pt,color2] (9,2) -- (14,7);
  \draw[line width=1.8pt,color4] (14,7) -- (15,7);
  \draw[line width=1.5pt,color2] (15,7) -- (16,8);
  \foreach \x in {1,...,9} {\fill[color2] (\x,\x-1) circle (1.5pt);}
  \foreach \p in {(7,0),(8,1),(9,2),(10,3),(11,4),(12,5),(13,6),(14,7),(15,7),(16,8)} {\fill[color2] \p circle (1.5pt);}
  \node[font=\scriptsize,anchor=east,color1] at (-1.6,1.5) {$\A_P$ inserts};
  \node[font=\scriptsize,anchor=south,color3] at (2.5,-1.6) {$\A_T$ inserts};
  \node[font=\scriptsize,anchor=south,color3] at (14.5,-1.6) {$\A_T$ deletes};
  \node[font=\scriptsize,anchor=north] at (8,8.6) {vertices of $\A_T$};
  \node[font=\scriptsize,anchor=north,rotate=90] at (17.1,4) {vertices of $\A_P$};
\end{tikzpicture}
\caption{A window of the expanded alignment graph of
\cref{ex:overview_periodic} (whole in \cref{fig:running_reference}), with
the rows $7,\ldots,15$ and the columns $12,\ldots,28$.  Rows are the
vertices of $\A_P$ and columns those of $\A_T$, with edits shaded blue
and magenta.  The highlighted paths have the
reference offsets $6$ and $12$.  Purple edges cost $0$ and orange ones
$1$, so the left path costs $0$ and the right one $2$, the number of
edits it passes.}
\label{fig:expanded_graph}
\end{figure}
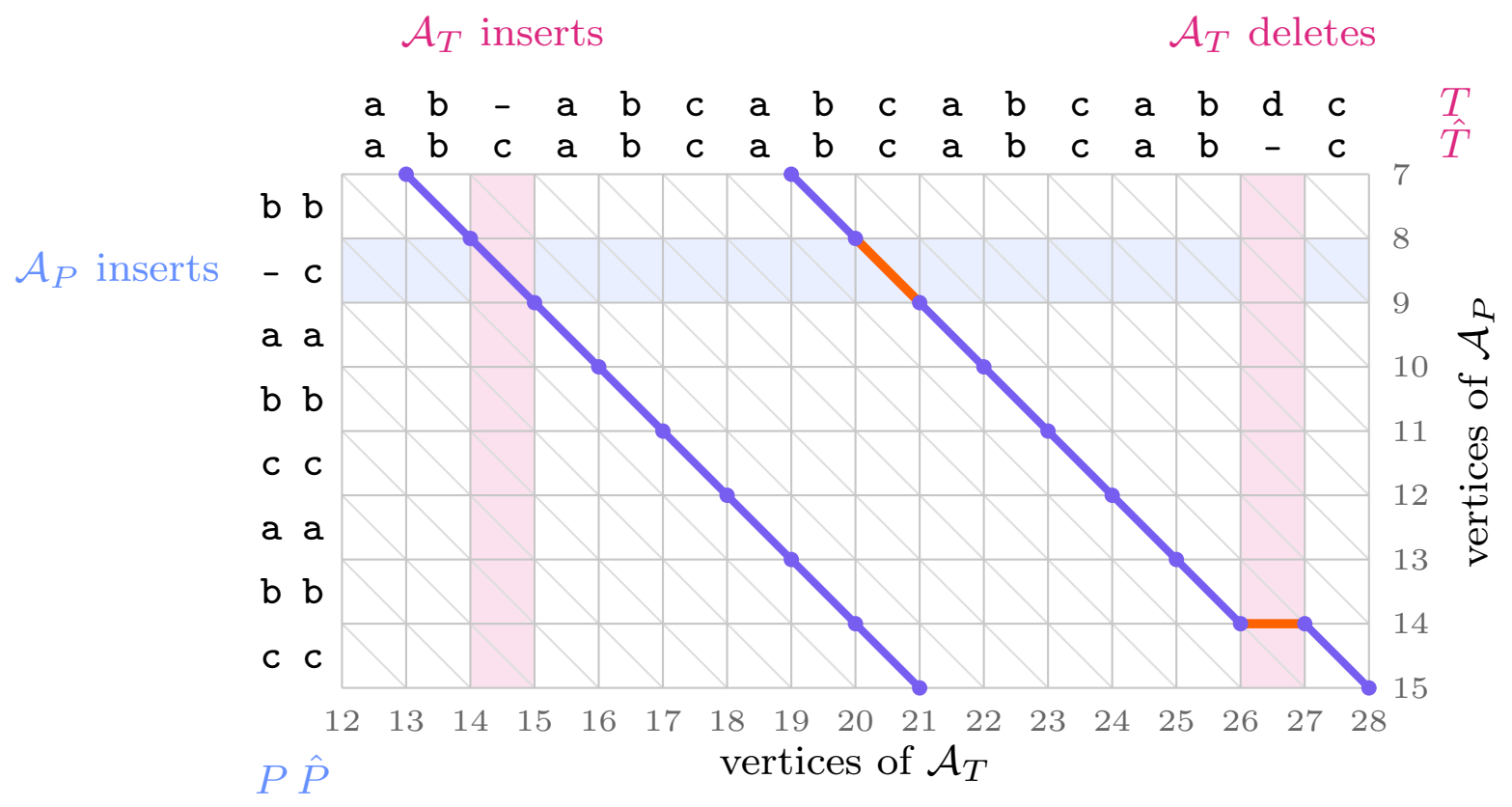

A position of $P$ does not always correspond to a single reference
coordinate: if $\A_P$ inserts reference characters at some position of
$P$, then this position is paired with several consecutive reference
coordinates, and symmetrically for deletions.  To keep both coordinate
systems available, we index positions by the vertices of the supplied
alignments, numbered in path order, rather than by string positions;
the edges of an alignment are its \emph{steps}.  The expanded
graph is a grid whose rows are indexed by the vertices of $\A_P$ and
whose columns are indexed by the vertices of $\A_T$; see
\cref{fig:expanded_graph}.  Its edges correspond to advancing in one or
both alignments, and an edge costs as much as aligning the two input
fragments that it covers.

\begin{definition}[Alignment expansions and the expanded alignment graph]
\label{def:expanded_alignment_graph}
Write $\ell_P$ and $\ell_T$ for the numbers of edges of the supplied
alignments $\A_P$ and $\A_T$, respectively.  Their \emph{expansions} are
the maps $(\pP,\phP):\fragment{0}{\ell_P}\to V(\AG(P,\hP))$ and
$(\pT,\phT):\fragment{0}{\ell_T}\to V(\AG(T,\hT))$ that list the
vertices of the two alignments in path order:
\[
    \A_P=\bigl((\pP(i),\phP(i))\bigr)_{i=0}^{\ell_P}
    \qquad\text{and}\qquad
    \A_T=\bigl((\pT(i),\phT(i))\bigr)_{i=0}^{\ell_T}.
\]

For a pair $u=(\ep,\et)\in \fragment{0}{\ell_P}\times\fragment{0}{\ell_T}$, write $\pP(u)=\pP(\ep)$, $\phP(u)=\phP(\ep)$, $\pT(u)=\pT(\et)$, and $\phT(u)=\phT(\et)$.
For two such pairs $u$ and $v=(\ep',\et')$ with
$\ep\le\ep'$ and $\et\le\et'$, denote
$\pP(u,v)=P\fragmentco{\pP(u)}{\pP(v)}$, $\phP(u,v)=\Q\fragmentco{\phP(u)}{\phP(v)}$,
$\pT(u,v)=T\fragmentco{\pT(u)}{\pT(v)}$,
and $\phT(u,v)=\Q\fragmentco{\phT(u)}{\phT(v)}$.

The \emph{expanded alignment graph} $\eAG(P,T)$ is the directed graph with vertex set $\fragment{0}{\ell_P}\times\fragment{0}{\ell_T}$ and an edge from $u=(\ep,\et)$ to $v=(\ep',\et')$ whenever
\[
    \ep'-\ep\in\{0,1\},\qquad
    \et'-\et\in\{0,1\},\quad\text{and}\quad
    u\ne v.
\]
For an edge $u\to v$, its cost is $\ed(\pP(u,v),\pT(u,v))$.
We call $\{0\}\times\fragment{0}{\ell_T}$ the \emph{top
row} and $\{\ell_P\}\times\fragment{0}{\ell_T}$ the
\emph{bottom row} of $\eAG(P,T)$.
\end{definition}

Since each step of an alignment advances each coordinate by at most
one, the four maps $\pP,\phP,\pT,\phT$ are nondecreasing and increase by
at most one per step; \cref{ex:running_rails} lists them for the
running example.  In particular, both fragments $\pP(u,v)$ and
$\pT(u,v)$ of an edge $u\to v$ have length at most one, and the cost of
the edge is $0$ or $1$.  An
expanded vertex determines four positions: one in each input string and
one in each reference fragment.  Pairing the two input positions or the
two reference positions gives the two projections used throughout the
paper.

\begin{definition}[Projections and lifts]
\label{def:expanded_projections}
The \emph{input projection} and \emph{reference projection} map a
vertex $u$ of $\eAG(P,T)$ to
\[
    \pPT(u)\coloneqq(\pP(u),\pT(u))\in V(\AG(P,T)),
    \qquad
    \phPT(u)\coloneqq(\phP(u),\phT(u))\in V(\AG(\hP,\hT)),
\]
respectively.  The \emph{reference offset} of $u$ is
$\mathrm{off}(u)\coloneqq\phT(u)-\phP(u)$.  For a path $\eA$ in $\eAG(P,T)$, its projections
$\pPT(\eA)$ and $\phPT(\eA)$ are obtained by applying the
corresponding vertex map and removing consecutive repetitions.
A \emph{lift} of a path under either projection is a path in
$\eAG(P,T)$ whose image under that projection is the given path.
\end{definition}

Both projections of a path are paths in the corresponding alignment
graphs.  The input projection preserves costs, as shown next, whereas
the reference projection need not do so; \cref{sec:canonical_costs}
bounds the difference.

\begin{observation}[Cost preservation, lifting, and distances]\label{obs:expanded_projection}
Every path $\eA$ in $\eAG(P,T)$ has the same cost as $\pPT(\eA)$.
For vertices $u,v$ of $\eAG(P,T)$ with $v$ reachable from $u$,
every path in $\AG(P,T)$ from $\pPT(u)$ to $\pPT(v)$ has a lift
from $u$ to $v$ under $\pPT$.  Hence, the distance from $u$ to $v$ in
$\eAG(P,T)$ is $\ed(\pP(u,v),\pT(u,v))$.
\end{observation}
\begin{proof}
An expanded edge projects either to an edge of $\AG(P,T)$ of the same
cost or to a single vertex; in the latter case, its cost is zero by
definition.  Thus, projection preserves path costs.

For lifting, write $u=(\ep,\et)$ and $v=(\ep',\et')$, and restrict
$\pP$ to $\fragment{\ep}{\ep'}$ and $\pT$ to $\fragment{\et}{\et'}$.
Each restricted map is
nondecreasing, advances by at most one per step, and has interval
\emph{fibres}, that is, the sets of indices with a fixed image are
intervals.  Start at $u$ and process the given path in $\AG(P,T)$ edge
by edge.  For each coordinate that the next edge advances, first move
to the largest index of its current fibre, keeping the images of both
coordinates fixed; these moves cost zero.  Then, advance exactly the
coordinates advanced by the edge; the resulting expanded edge has the
prescribed projection and cost.  After the last edge, move to $v$ within
the final fibres, again at zero cost.  The restricted ranges guarantee
that every move is possible and that the lift ends at $v$.

The paths in $\AG(P,T)$ from $\pPT(u)$ to $\pPT(v)$ are the alignments
of $\pP(u,v)$ with $\pT(u,v)$, so projecting and lifting shortest
paths gives the distance formula.
\end{proof}

The diagonal edges of the reference alignment graph play a special
role, because they correspond to synchronized progress along both
references.  Each of them has exactly one preimage in the expanded
graph, which lets us transfer statements about such edges between the
two graphs.

\begin{definition}[Reference-diagonal edges]\label{def:reference_diagonal_edge}
\let\qed\relax %
An edge $u\to v$ of $\eAG(P,T)$ is \emph{reference-diagonal} if $\phPT(u)\to\phPT(v)$ is a diagonal edge of $\AG(\hP,\hT)$; equivalently,
\[
    \phP(v)=\phP(u)+1
    \qquad\text{and}\qquad
    \phT(v)=\phT(u)+1.
    \tag*{\lipicsEnd}
\]
\end{definition}

\begin{observation}[Unique lifts of reference-diagonal edges]
\label{obs:expansion_lifts}
\let\qed\relax %
Let
\[
    (\hat p,\hat t)\longrightarrow(\hat p+1,\hat t+1)
\]
be a diagonal edge of $\AG(\hP,\hT)$.  There is a unique edge $u\to v$
of $\eAG(P,T)$ such that
\[
    \phPT(u)=(\hat p,\hat t)
    \qquad\text{and}\qquad
    \phPT(v)=(\hat p+1,\hat t+1).
    \tag*{\lipicsEnd}
\]
\end{observation}
\begin{proof}
There is a unique step of $\A_P$ that advances its pattern-reference
coordinate from $\hat p$ to $\hat p+1$, say from its vertex $\ep$ to
its vertex $\ep+1$, and a unique step of $\A_T$ that advances its
text-reference coordinate from $\hat t$ to $\hat t+1$, say from its
vertex $\et$ to its vertex $\et+1$.  The desired edge is therefore
\[
    (\ep,\et)\longrightarrow(\ep+1,\et+1),
\]
and no other edge has the prescribed reference projection.
\end{proof}

\subsection{Distances to the Bottom Row}
\label{sec:suffix_distances}

Distances to the bottom row decide where the occurrences start: a text
position $v$ is the starting position of a $k$-error occurrence of $P$
in $T$ if and only if some alignment of $P$ with a fragment
$T\fragmentco{v}{e}$ costs at most $k$, that is, if and only if the
distance from $(0,v)$ to the bottom row of $\AG(P,T)$ is at most~$k$.
We now define these \emph{suffix distances}.  They are
\emph{semiglobal}: all of the remaining pattern is aligned, but only
with some prefix of the remaining text.

\begin{definition}[Semiglobal suffix distances]
\label{def:semiglobal_suffix_distances}
We call the vertices of $\AG(P,T)$ \emph{ordinary vertices}.  For
every ordinary vertex $x$, let $\dtob(x)$ denote its shortest-path
distance in $\AG(P,T)$ to the bottom row
$\{|P|\}\times\fragmentcc{0}{|T|}$; for $x=(p,t)$, we also write
$\dtob(p,t)$.  For every expanded vertex
$u\in\eAG(P,T)$, define
\(\tdtob(u)\coloneqq\dtob(\pPT(u))\).
We call $\dtob$ and $\tdtob$ the \emph{ordinary} and \emph{expanded
semiglobal suffix distances}, respectively.
\end{definition}

The next observation identifies the expanded suffix distance with the
distance to the bottom row of the expanded graph and relates it to
occurrences.

\begin{observation}[Suffix distances and occurrences]
\label{lem:semiglobal_expanded_transfer}
\let\qed\relax %
Let $u\in\eAG(P,T)$ and $x=(p,v)=\pPT(u)$.
The shortest-path distance from $u$ to the bottom row of
$\eAG(P,T)$ is
\[
    \tdtob(u)=\dtob(x)
      =\min_{e\in\fragmentcc{v}{|T|}}
        \ed\bigl(
          P\fragmentco{p}{|P|},
          T\fragmentco{v}{e}
        \bigr).
\]
If $p=0$, then
\[
    v\in\OccE_k(P,T)
      \qquad\text{if and only if}\qquad
    \tdtob(u)\le k.
    \tag*{\lipicsEnd}
\]
\end{observation}
\begin{proof}
The directed paths in $\AG(P,T)$ from $x=(p,v)$ to the bottom row
are exactly the alignments of $P\fragmentco{p}{|P|}$ with fragments
$T\fragmentco{v}{e}$, where $e\in\fragmentcc{v}{|T|}$.  Minimizing their
costs proves the formula for $\dtob(x)$, and $\tdtob(u)=\dtob(x)$ holds
by definition.  The vertices $w$ of the bottom row reachable from $u$
satisfy $\pP(u,w)=P\fragmentco{p}{|P|}$, and $\pT(w)$ takes every value
in $\fragmentcc{v}{|T|}$ on them, because $\pT$ advances by at most one
per step and ends at $|T|$.  Hence, by \cref{obs:expanded_projection},
the expanded distance from $u$ to the bottom row is the same minimum.
When $p=0$, the formula is
exactly the condition defining $\OccE_k(P,T)$.
\end{proof}

Recall that $\dtob(x)$ is the cost of a shortest path from $x$ to the
bottom row.  A path $\mathcal A$ from $x$ to $y$, followed by a shortest
path from $y$ to the bottom row, costs $\cost(\mathcal A)+\dtob(y)$, which
is at least $\dtob(x)$.  We call $\mathcal A$ tight if equality holds,
that is, if $\mathcal A$ is the beginning of a shortest path from $x$ to
the bottom row.  Tight paths will be our witnesses.

\begin{definition}[Tight paths]
\label{def:tight_paths}
A directed path $\mathcal A$ from $x$ to $y$ in $\AG(P,T)$ is
\emph{tight} if $\cost(\mathcal A)=\dtob(x)-\dtob(y)$.  A directed path
in $\eAG(P,T)$ is \emph{tight} if the same holds with $\tdtob$ in place
of $\dtob$.
\end{definition}

By \cref{lem:semiglobal_expanded_transfer}, the inequality
$\tdtob(x)\le\cost(\mathcal A)+\tdtob(y)$ holds in the expanded graph as
well.  Hence, tightness is preserved under taking subpaths: every
subpath of a tight path is tight.

Moving along a directed path changes the suffix distance by at most
the cost of the path.

\begin{lemma}[Path stability of semiglobal distances]
\label{lem:semiglobal_path_stability}
For any vertices $x,y$ of $\AG(P,T)$ and any directed path $\mathcal C$ in
$\AG(P,T)$ from $x$ to $y$,
\[
    |\dtob(x)-\dtob(y)|\le\cost(\mathcal C).
\]
\end{lemma}
\begin{proof}
For strings $X,Y$, let
\[
    \mathrm{ped}(X,Y)
      \coloneqq
      \min\{\ed(X,Z):Z\text{ is a prefix of }Y\}
\]
be the distance from $X$ to the set of prefixes of $Y$, so that
$\dtob(p,t)=\mathrm{ped}(P\fragmentco{p}{|P|},T\fragmentco{t}{|T|})$ by
\cref{lem:semiglobal_expanded_transfer}.

\begin{claim}[A common prefix does not matter]\label{clm:semiglobal_common_prefix}
For all strings $G,S,U$,
\[
    \mathrm{ped}(GS,GU)=\mathrm{ped}(S,U).
\]
\end{claim}
\begin{claimproof}
A common prefix does not change the edit distance:
$\ed(GS,GV)=\ed(S,V)$ for all strings $G,S,V$.  After reversing the
strings, this follows from the monotonicity of the dynamic-programming
table of the edit distance along its diagonals; see, e.g.,
\cite{LandauV89}.  Hence, every prefix $V$ of $U$ yields the prefix
$GV$ of $GU$ with $\ed(GS,GV)=\ed(S,V)$, so
$\mathrm{ped}(GS,GU)\le\mathrm{ped}(S,U)$.
Conversely, a prefix of $GU$ either has the form $GV$ for a prefix $V$
of $U$, or it is a proper prefix $G_0$ of $G$.  In the former case,
$\ed(GS,GV)=\ed(S,V)\ge\mathrm{ped}(S,U)$.  In the latter case,
\[
    \ed(GS,G_0)\ge |GS|-|G_0|\ge |S|\ge\mathrm{ped}(S,U),
\]
because $\mathrm{ped}(S,U)\le\ed(S,\varepsilon)=|S|$.  Hence,
$\mathrm{ped}(GS,GU)\ge\mathrm{ped}(S,U)$.
\end{claimproof}

Now, write $x=(i,j)$ and $y=(i',j')$, and set
\[
    F\coloneqq P\fragmentco{i}{i'},\quad
    G\coloneqq T\fragmentco{j}{j'},\quad
    S\coloneqq P\fragmentco{i'}{|P|},\quad
    U\coloneqq T\fragmentco{j'}{|T|}.
\]
Then, $\dtob(x)=\mathrm{ped}(FS,GU)$ and, by
\cref{clm:semiglobal_common_prefix}, $\dtob(y)=\mathrm{ped}(S,U)=\mathrm{ped}(GS,GU)$.
The distance to a fixed set of strings is $1$-Lipschitz with respect to
the edit distance, and $\mathcal C$ is an alignment of $F$ with $G$, so
\[
    |\dtob(x)-\dtob(y)|
       =|\mathrm{ped}(FS,GU)-\mathrm{ped}(GS,GU)|
       \le\ed(FS,GS)
       \le\ed(F,G)
       \le\cost(\mathcal C).\qedhere
\]
\end{proof}

\subsection{Canonical Costs and Reference Projections}
\label{sec:canonical_costs}

The cost of a path can change when we switch from input coordinates
to reference coordinates, but only because of the edits of $\A_P$ and
$\A_T$ that the path passes.  We count these edits as follows.

\begin{definition}[Canonical cost]\label{def:canonical_cost}
For vertices $u,v\in\eAG(P,T)$ such that $v$ is reachable from $u$,
the \emph{canonical cost} is
\[
    \cancost(u,v)\coloneqq
       \ed^{\A_P}(\pP(u,v),\phP(u,v))
       +\ed^{\A_T}(\pT(u,v),\phT(u,v)).
\]
If the second argument is omitted, it is the bottom-right vertex
$(\ell_P,\ell_T)$ of $\eAG(P,T)$.
\end{definition}

In other words, $\cancost(u,v)$ is the number of edits of $\A_P$
between its vertices $\ep$ and $\ep'$ plus the number of edits of
$\A_T$ between its vertices $\et$ and $\et'$, where $u=(\ep,\et)$ and
$v=(\ep',\et')$; we say that a path from $u$ to $v$ \emph{passes} these
edits.

\begin{observation}[Additivity of canonical costs]\label{obs:cancost_factorizes}
\let\qed\relax %
Let $u,v,w\in\eAG(P,T)$, with $v$ reachable from $u$ and $w$ from $v$.
Then,
\[
    \cancost(u,w)=\cancost(u,v)+\cancost(v,w)
    \qquad\text{and}\qquad
    \cancost(u,v)=\cancost(u)-\cancost(v).
    \tag*{\lipicsEnd}
\]
\end{observation}
\begin{proof}
The two restrictions of each supplied alignment are consecutive and
edge-disjoint, so their costs add.  This proves the first identity;
taking $w=(\ell_P,\ell_T)$ gives the second.
\end{proof}

By the triangle inequality, the costs of a single edge and of its
reference projection differ by at most the number of edits of $\A_P$
and $\A_T$ that the edge passes.  Summing over a path gives the following
bound.

\begin{lemma}[Reference projections and canonical cost]\label{lem:canonical_cost}
Let $\eA$ be a path from $u$ to $v$ in $\eAG(P,T)$
and let $\hA : \phP(u,v)\onto \phT(u,v)$ be its projection under $\phPT$.
Then, the costs of $\eA$ and $\hA$ differ by at most $\cancost(u,v)$.
\end{lemma}
\begin{proof}
    For an edge $x\to y$ of $\eA$, set
    \[
        F\coloneqq\pP(x,y),\quad G\coloneqq\pT(x,y),\quad
        \hF\coloneqq\phP(x,y),\quad \hG\coloneqq\phT(x,y).
    \]
    The cost of $x\to y$ is $\ed(F,G)$, and the cost of its reference
    projection is $\ed(\hF,\hG)$; the latter is zero if the projection
    is a single vertex.  The triangle inequality gives
    \[
        |\ed(F,G)-\ed(\hF,\hG)|
        \le\ed(F,\hF)+\ed(G,\hG)
        \le\cancost(x,y).
    \]
    Summing over the edges of $\eA$ and using
    \cref{obs:cancost_factorizes} bounds $|\cost(\eA)-\cost(\hA)|$ by
    $\sum_{x\to y\in\eA}\cancost(x,y)=\cancost(u,v)$.
\end{proof}

The reference offset changes only along insertions and deletions of the
reference projection.  Hence, a path changes it by at most its cost plus
its canonical cost, so a cheap path changes its reference offset only
a little.

\begin{lemma}[Offset changes]
\label{lem:offset_change}
Let $\eA$ be a path from $u$ to $v$ in $\eAG(P,T)$.  Then,
\[
    |\mathrm{off}(v)-\mathrm{off}(u)|\le\cost(\eA)+\cancost(u,v).
\]
\end{lemma}
\begin{proof}
Let $\hA$ be the reference projection of $\eA$.  A vertex
$(\hat p,\hat t)$ of $\hA$ has reference offset $\hat t-\hat p$, which
equals the reference offset of its preimages.  Each insertion or
deletion edge of $\hA$ changes this offset by one and costs one, and
each diagonal edge leaves it unchanged.  Hence,
$\cost(\hA)\ge|\mathrm{off}(v)-\mathrm{off}(u)|$, and
\cref{lem:canonical_cost} gives $\cost(\eA)\ge\cost(\hA)-\cancost(u,v)$.
\end{proof}

\subsection{Rails and Paths That Leave Them}
\label{sec:canonical_diagonals}

At reference offsets that are multiples of $q$, the two references
are in the same phase and can be aligned without edits.  We lift these
zero-cost alignments into the expanded graph.  In input coordinates,
the lifted paths may have positive cost, but their reference projections
cost nothing, which makes them convenient baselines against which we
compare other paths.

\begin{definition}[Rails]
    \dglabel{def:blue}
    Set $\canD \coloneqq \{b \in \Z : x_T-y_P \le bq \le y_T-x_P\}$.
    For each $b\in \canD$, the \emph{reference rail} $\hD{b}$ of
    index $b$ is the path in $\AG(\hP,\hT)$ formed by the vertices
    $(\hat p,\hat t)$ with $\hat t-\hat p=bq$ and the diagonal edges
    between them.  These edges cost $0$, because
    $Q^\infty[\hat p]=Q^\infty[\hat p+bq]$.

    For a vertex $(\hat p,\hat t)$ of $\hD{b}$, the expanded vertices
    $u$ with $\phPT(u)=(\hat p,\hat t)$ form a rectangle
    $\fragment{\ep_0}{\ep_1}\times\fragment{\et_0}{\et_1}$; its vertical
    edges correspond to the deletions of $\A_P$ at reference coordinate
    $\hat p$, and its horizontal edges to the deletions of $\A_T$ at
    reference coordinate $\hat t$.  Let $\eD{b}$ be the path in $\eAG(P,T)$ that, for the
    vertices of $\hD{b}$ in order, traverses each such rectangle from
    $(\ep_0,\et_0)$ down to $(\ep_1,\et_0)$ and then right to
    $(\ep_1,\et_1)$, and that passes from one rectangle to the next by
    the reference-diagonal edge $(\ep_1,\et_1)\to(\ep_1+1,\et_1+1)$,
    whose head is the vertex $(\ep_0,\et_0)$ of the next rectangle.
    Its projection under $\phPT$ is $\hD{b}$.
    We call the paths $\eD{b}$ \emph{rails}.
    A \emph{rail contact} of a path is a vertex that the
    path shares with a rail.
\end{definition}

\begin{figure}[t]
\centering
\begin{tikzpicture}[x=0.62cm,y=-0.62cm,baseline=(cap.base)]
  \fill[color1!14] (0,1) rectangle (4,2);
  \fill[color3!14] (1,0) rectangle (3,3);
  \foreach \i in {0,...,2} {\foreach \j in {0,...,3} {\draw[black!13] (\j,\i) -- (\j+1,\i+1);}}
  \foreach \i in {0,...,3} {\draw[black!22] (0,\i) -- (4,\i);}
  \foreach \j in {0,...,4} {\draw[black!22] (\j,0) -- (\j,3);}
  \draw[black!60,dashed,rounded corners=2pt] (0.78,0.78) rectangle (3.22,2.22);
  \draw[line width=1.5pt,color2] (0,0) -- (1,1);
  \draw[line width=1.8pt,color4] (1,1) -- (1,2) -- (3,2);
  \draw[line width=1.5pt,color2] (3,2) -- (4,3);
  \foreach \p in {(0,0),(1,1),(1,2),(2,2),(3,2),(4,3)} {\fill[color2] \p circle (1.6pt);}
  \foreach \p in {(2,1),(3,1)} {\draw[black,fill=white] \p circle (1.6pt);}
  \node[font=\scriptsize,anchor=south east,color2] at (0,0) {$\eD{b}$};
  \node[font=\scriptsize,anchor=east,color1] at (-0.1,1.5) {$\A_P$ deletes};
  \node[font=\scriptsize,anchor=south,color3] at (2,-0.05) {$\A_T$ deletes};
  \node[font=\small] (cap) at (2,4.1) {(a)};
\end{tikzpicture}
\hspace{1cm}
\begin{tikzpicture}[x=0.44cm,y=-0.44cm,baseline=(cap.base)]
  \fill[black!8] (2.5,0) .. controls (4.5,2.5) and (5.3,4) .. (7.5,7)
     -- (10.5,7) .. controls (8.3,4) and (7.5,2.5) .. (5.5,0) -- cycle;
  \foreach \s in {-0.5,2.5,5.5,8.5} {
    \draw[line width=1.2pt,color2] (\s,0) .. controls (\s+2,2.5) and (\s+2.8,4) .. (\s+5,7);
  }
  \draw[thick] (-0.8,0) -- (12.2,0);
  \draw[thick] (-0.8,7) -- (14.2,7);
  \node[font=\scriptsize,anchor=south] at (-0.5,0) {$\eD{b-1}$};
  \node[font=\scriptsize,anchor=south] at (2.5,0) {$\eD{b}$};
  \node[font=\scriptsize,anchor=south] at (5.5,0) {$\eD{b+1}$};
  \node[font=\scriptsize,anchor=south] at (8.5,0) {$\eD{b+2}$};
  \draw[line width=1.2pt,color4] (3.077,5.0) -- (6.495,5.6) -- (10.064,6.4);
  \foreach \p in {(3.077,5.0),(6.495,5.6),(10.064,6.4)} {\fill \p circle (1.4pt);}
  \node[font=\scriptsize,anchor=west] at (12.3,0) {top row};
  \node[font=\scriptsize,anchor=north] at (11,7.05) {bottom row};
  \node[font=\small] (cap) at (6.5,8.6) {(b)};
\end{tikzpicture}
\caption{(a)~At a vertex of $\hD{b}$, $\A_P$ makes one deletion
(blue) and $\A_T$ two (magenta), giving six expanded vertices of offset
$bq$ (dashed box).  The rail $\eD{b}$ (purple) takes the deletion of
$\A_P$ first; its three edges there cost $1$ (orange), and the hollow
vertices are not rail contacts.  (b)~Proper rails are disjoint
top-to-bottom paths ordered by index; a path from $\eD{b-1}$ to
$\eD{b+1}$ (orange) has a rail contact on $\eD{b}$.}
\label{fig:rail_separators}
\end{figure}
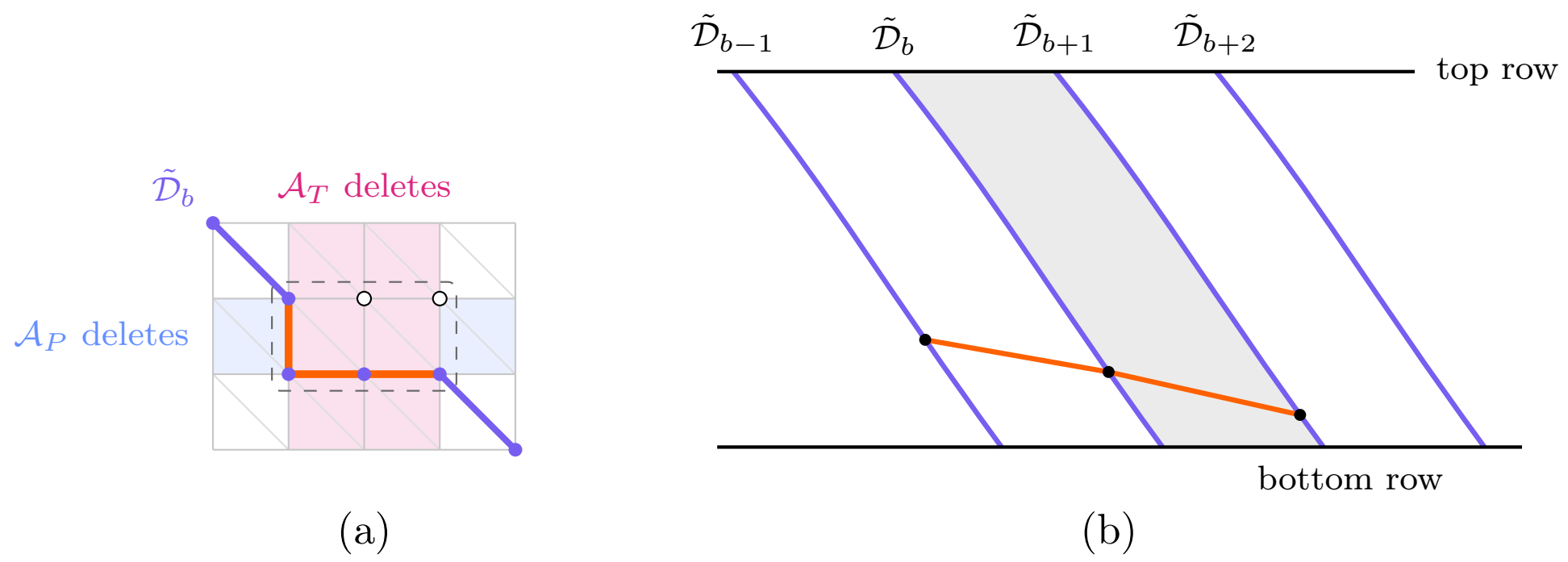

The reference rail $\hD{b}$ is the overlap of $\hP$ and $\hT$ when
$\hP$ is shifted by $bq$, and the rail $\eD{b}$ follows both supplied
alignments in parallel along it; in \cref{fig:expanded_graph}, the
purple paths are segments of the rails $\eD{2}$ and $\eD{4}$.  The reference projection of a rail
costs nothing, so, by \cref{lem:canonical_cost}, the cost of a rail
segment is at most its canonical cost.  At a deletion of $\A_P$ or
$\A_T$, the corresponding reference coordinate does not advance, so the
rail takes this step on its own, at cost $1$.  When both alignments
make deletions at the same vertex of $\hD{b}$, their order does not
affect the cost of the rail, but it determines which expanded vertices
the rail visits; the definition fixes one order.
The expanded vertices that the rail then skips also have reference
offset $bq$, but they are not rail contacts; see
\cref{fig:rail_separators}(a).

\begin{observation}[Rails are disjoint]\label{obs:diagonals_disjoint}
The rails $\eD{b}$, for $b\in \canD$, are vertex-disjoint.
\end{observation}
\begin{proof}
    Every vertex $u$ of $\eD{b}$ has reference offset
    $\mathrm{off}(u)=bq$, because $\phPT(u)$ is a vertex of $\hD{b}$.  A vertex of both
    $\eD{b}$ and $\eD{b'}$ would satisfy $bq=b'q$, and hence $b=b'$.
\end{proof}

The rails that span the whole pattern reference run from the top row
to the bottom row, so each of them splits the expanded graph into a
left and a right part.  The next definition calls these rails proper
and names the two sides.

\begin{definition}[Proper rails and their sides]
\label{def:proper_rails}
A rail $\eD{b}$ is \emph{proper} if $\hD{b}$ has a vertex
$(\hat p,\hat t)$ for every $\hat p\in\fragment{x_P}{y_P}$;
equivalently, $b\in \prD\coloneqq \{b \in \Z : x_T-x_P \le bq \le y_T-y_P\}$.
For $c\in\prD$, a vertex $u$ of $\eAG(P,T)$ lies \emph{left of}
$\eD{c}$ if $\mathrm{off}(u)<cq$, and \emph{right of} $\eD{c}$ if
$\mathrm{off}(u)>cq$, or if $\mathrm{off}(u)=cq$ and $u\notin\eD{c}$.
\end{definition}

With the deletion order in \cref{def:blue}, the vertices that a rail
skips at its own reference offset lie to its right in the grid
drawing; see \cref{fig:rail_separators}(a).  This is why the
definition counts them as right of the rail.  Two consecutive proper
rails delimit a strip, which contains both rails and everything drawn
between them.

\begin{definition}[Closed strips]
\label{def:rail_strips}
For consecutive proper indices $b$ and $b+1$, the \emph{closed
strip between $\eD{b}$ and $\eD{b+1}$} is the subgraph of
$\eAG(P,T)$ with vertex set
\[
    \bigl\{u\in V(\eAG(P,T)):
        bq\le \mathrm{off}(u)<(b+1)q\bigr\}
        \cup V(\eD{b+1}).
\]
It contains every expanded edge between these vertices, except that
an edge whose endpoints both belong to $V(\eD{b+1})$ is
included only if it belongs to $\eD{b+1}$.
\end{definition}

Unfolding the definitions, the vertices of the closed strip between
$\eD{b}$ and $\eD{b+1}$ are exactly those that are neither left of
$\eD{b}$ nor right of $\eD{b+1}$: the lower bound
$bq\le\mathrm{off}(u)$ excludes the vertices left of $\eD{b}$, and the
strict upper bound, together with $V(\eD{b+1})$, excludes exactly the
vertices right of $\eD{b+1}$, including those that $\eD{b+1}$ skips.
The edge condition excludes diagonal edges cutting across the deletion
corners of $\eD{b+1}$.  Hence, the closed strip consists of both rails
and the vertices and edges drawn between them, and every vertex in it
has reference offset in $\fragmentcc{bq}{(b+1)q}$.

A rail is proper if the whole pattern reference fits into the text
reference at shift $bq$.  Proper rails traverse the whole pattern
reference, so they run from the top row to the bottom row.  This makes
them ordered
separators of the expanded graph: a path between two distant rails
must meet all rails in between; see \cref{fig:rail_separators}(b).
Hence, cutting a path at its rail contacts yields pieces that lie
between neighboring rails.

\begin{lemma}[Rails are ordered separators]
\dglabel{lem:canonical_rails}[def:blue]
The proper rails are pairwise disjoint directed paths from the top
row to the bottom row, and their endpoints occur in index order on
both rows.  For $c\in\prD$, every path between a vertex left of
$\eD{c}$ and a vertex right of $\eD{c}$ has a rail contact on
$\eD{c}$.  In particular, a path with no internal rail contact whose
endpoints lie on proper rails $\eD{b}$ and $\eD{b'}$ satisfies
$|b-b'|\le1$.
\end{lemma}
\begin{proof}
For a proper rail, $\hD{b}$ has a vertex at every pattern-reference
coordinate, so $\eD{b}$ traverses all steps of $\A_P$ and is a directed
path from the top row to the bottom row.  These paths are
vertex-disjoint by \cref{obs:diagonals_disjoint}, and every vertex of
$\eD{b}$ has reference offset $bq$, so their order on both rows agrees
with their index order.

Along every edge of $\eAG(P,T)$, each of $\phP$ and $\phT$ increases
by $0$ or $1$, so the reference offset changes by at most $1$.  Let
$c\in\prD$, and suppose that an edge $u\to v$ joins a vertex left of
$\eD{c}$ and a vertex right of $\eD{c}$.  If $u$ is left of $\eD{c}$,
then $\mathrm{off}(u)=cq-1$ and $\mathrm{off}(v)=cq$, so the edge
advances $\phT$ but not $\phP$.  Hence, $v$ lies in the left column of
the rectangle of \cref{def:blue} over $\phPT(v)\in\hD{c}$, and
$\eD{c}$ contains this column.  If $u$ is right of $\eD{c}$, then $\mathrm{off}(u)=cq$ and
$\mathrm{off}(v)=cq-1$, so the edge advances $\phP$ but not $\phT$.
Hence, $u$ lies in the bottom row of the rectangle over
$\phPT(u)\in\hD{c}$, and $\eD{c}$ contains this row.  In both cases,
the endpoint at offset $cq$ lies on $\eD{c}$, so it is not right of
$\eD{c}$, a contradiction.  Thus, no edge joins the two sides, and
every path between them has a vertex on neither side, that is, a rail
contact on $\eD{c}$.  For the final assertion,
suppose that $|b-b'|>1$; by symmetry, $b<b'$.  Since $\prD$ is an integer interval,
$b+1\in\prD$.  The endpoint on $\eD{b}$ lies left of $\eD{b+1}$, and
the endpoint on $\eD{b'}$ lies right of $\eD{b+1}$, so the path has a
rail contact on $\eD{b+1}$.  This contact is internal, because the
rails are disjoint.
\end{proof}

Along a single rail, the suffix distance changes by at most the
canonical cost of the traversed segment, because the reference
projection of the rail costs nothing.  Later sections use this in two
ways: the suffix distances of two vertices on a common rail differ by at
most the number of edits of $\A_P$ and $\A_T$ between them, and, since
$\tdtob$ vanishes on the bottom row, following a proper rail down to the
bottom row bounds $\tdtob(u)$ by the canonical cost of the rest of the
rail.

\begin{lemma}[Distance stability along a rail]\label{lem:rail_stability}
Let $a,a'$ be vertices on the same rail such that $a'$ is reachable from $a$.
Then,
\[
    |\tdtob(a)-\tdtob(a')|\le \cancost(a,a').
\]
\end{lemma}
\begin{proof}
Let $\mathcal R$ be the segment of the rail from $a$ to $a'$.  Its
input projection $\pPT(\mathcal R)$ is a path in $\AG(P,T)$ from
$\pPT(a)$ to $\pPT(a')$, and it has the same cost as $\mathcal R$ by
\cref{obs:expanded_projection}.  Since $\tdtob(a)=\dtob(\pPT(a))$ and
$\tdtob(a')=\dtob(\pPT(a'))$ by definition,
\cref{lem:semiglobal_path_stability}, applied to $\pPT(\mathcal R)$,
gives $|\tdtob(a)-\tdtob(a')|\le\cost(\mathcal R)$.  The reference
projection of $\mathcal R$ has cost zero, so \cref{lem:canonical_cost}
gives $\cost(\mathcal R)\le\cancost(a,a')$.
\end{proof}

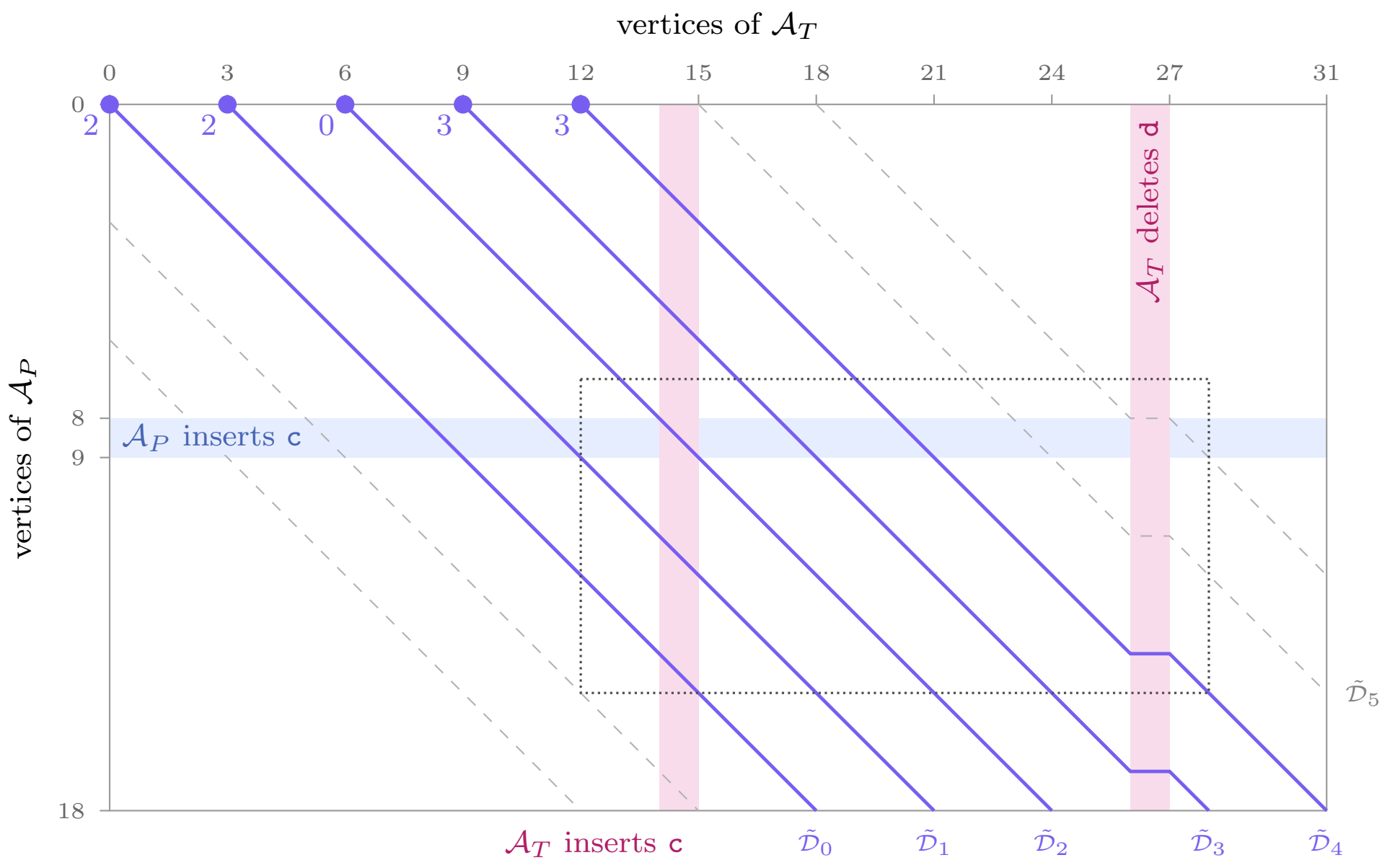
\begin{figure}[t]
\centering
\begin{tikzpicture}[x=0.36cm,y=-0.36cm]
  \fill[color1!16] (0,8) rectangle (31,9);
  \fill[color3!16] (14,0) rectangle (15,18);
  \fill[color3!16] (26,0) rectangle (27,18);
  \draw[black!40] (0,0) rectangle (31,18);
  \foreach \t in {0,3,...,27,31} {
    \draw[black!40] (\t,0) -- (\t,-0.25);
    \node[font=\tiny,anchor=south,black!60] at (\t,-0.25) {$\t$};
  }
  \foreach \p in {0,8,9,18} {
    \draw[black!40] (0,\p) -- (-0.25,\p);
    \node[font=\tiny,anchor=east,black!60] at (-0.25,\p) {$\p$};
  }
  \node[font=\scriptsize,anchor=south] at (15.5,-1.3) {vertices of $\A_T$};
  \node[font=\scriptsize,anchor=south,rotate=90] at (-1.5,9) {vertices of $\A_P$};
  \foreach \b in {-2,-1} {\draw[black!30,dashed] (0,-3*\b) -- (18+3*\b,18);}
  \draw[black!30,dashed] (15,0) -- (26,11) -- (27,11) -- (31,15);
  \draw[black!30,dashed] (18,0) -- (26,8) -- (27,8) -- (31,12);
  \draw[color2,line width=0.9pt] (0,0) -- (18,18);
  \draw[color2,line width=0.9pt] (3,0) -- (21,18);
  \draw[color2,line width=0.9pt] (6,0) -- (24,18);
  \draw[color2,line width=0.9pt] (9,0) -- (26,17) -- (27,17) -- (28,18);
  \draw[color2,line width=0.9pt] (12,0) -- (26,14) -- (27,14) -- (31,18);
  \draw[black!70,densely dotted,line width=0.6pt] (12,7) rectangle (28,15);
  \foreach \b/\v in {0/2,1/2,2/0,3/3,4/3} {
    \fill[color2] (3*\b,0) circle (2.4pt);
    \node[font=\scriptsize,anchor=north east,color2,inner sep=1pt] at (3*\b-0.15,0.15) {$\v$};
  }
  \foreach \b/\x in {0/18,1/21,2/24,3/28,4/31} {\node[font=\tiny,color2,anchor=north] at (\x,18.1) {$\eD{\b}$};}
  \node[font=\tiny,black!50,anchor=west] at (31.1,15) {$\eD{5}$};
  \node[font=\scriptsize,color1!70!black,anchor=west,fill=color1!16,inner sep=1pt] at (0.2,8.5) {$\A_P$ inserts $\texttt{c}$};
  \node[font=\scriptsize,color3!80!black,anchor=north east] at (15,18.15) {$\A_T$ inserts $\texttt{c}$};
  \node[font=\scriptsize,color3!80!black,rotate=90,anchor=east,inner sep=1pt] at (26.5,0.3) {$\A_T$ deletes $\texttt{d}$};
\end{tikzpicture}
\caption{The running example of \cref{ex:running_rails} in expanded
coordinates: rows are the vertices of $\A_P$ and columns those of
$\A_T$, with edits shaded blue and magenta.  The proper rails
$\eD{0},\ldots,\eD{4}$ (purple) are labelled with $\tdtob$ at their
earliest vertices; $\eD{2}$ crosses both insertions at the same point,
where they cancel.  Dashed lines are rails that are not proper, and the
dotted rectangle is the window of \cref{fig:expanded_graph}.}
\label{fig:running_reference}
\end{figure}

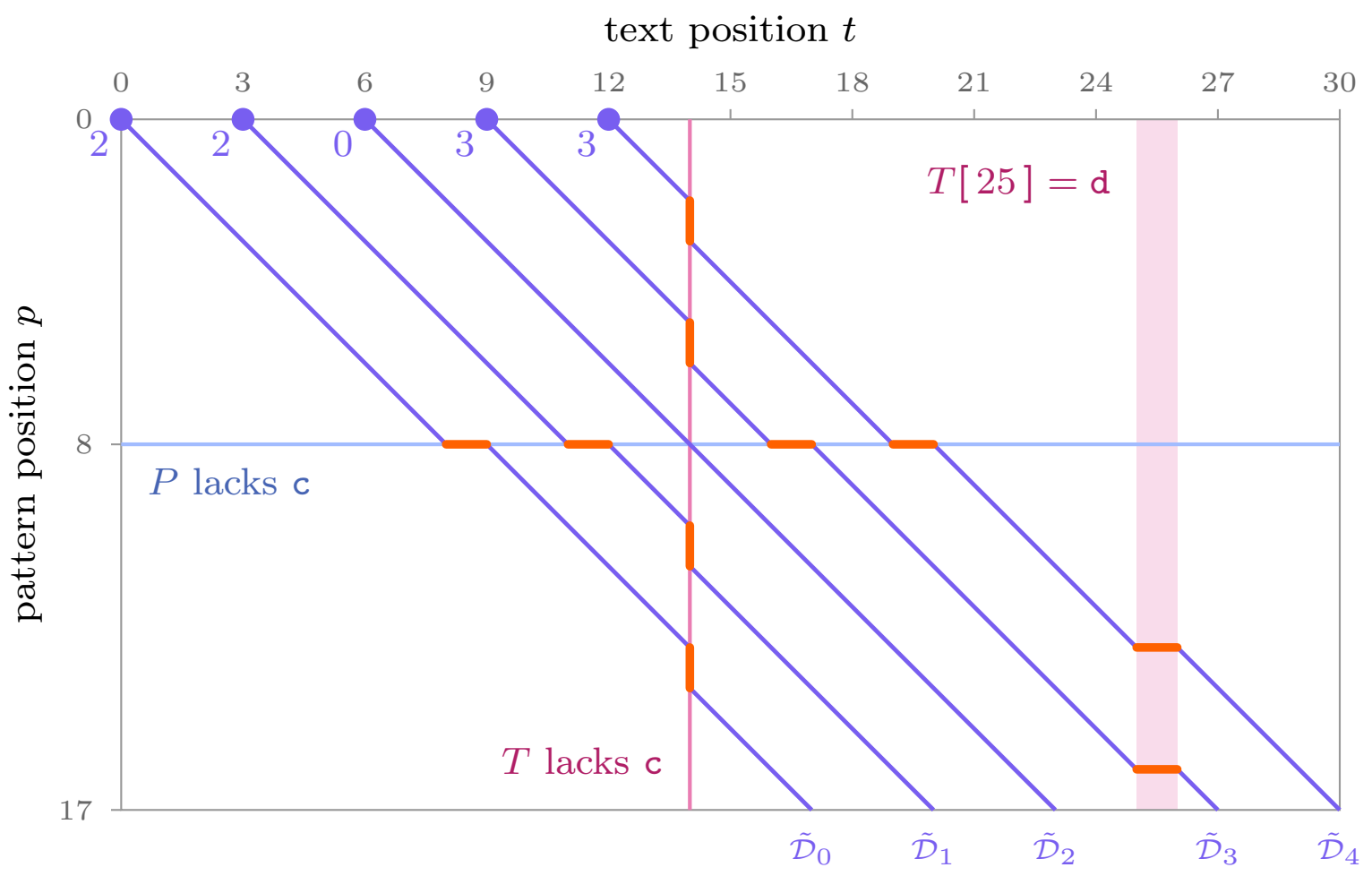
\begin{figure}[t]
\centering
\begin{tikzpicture}[x=0.3cm,y=-0.3cm]
  \fill[color3!16] (25,0) rectangle (26,17);
  \draw[black!40] (0,0) rectangle (30,17);
  \foreach \t in {0,3,...,30} {
    \draw[black!40] (\t,0) -- (\t,-0.25);
    \node[font=\tiny,anchor=south,black!60] at (\t,-0.25) {$\t$};
  }
  \foreach \p in {0,8,17} {
    \draw[black!40] (0,\p) -- (-0.25,\p);
    \node[font=\tiny,anchor=east,black!60] at (-0.25,\p) {$\p$};
  }
  \node[font=\scriptsize,anchor=south] at (15,-1.3) {text position $t$};
  \node[font=\scriptsize,anchor=south,rotate=90] at (-1.5,8.5) {pattern position $p$};
  \draw[color1!60,line width=0.8pt] (0,8) -- (30,8);
  \draw[color3!60,line width=0.8pt] (14,0) -- (14,17);
  \node[font=\scriptsize,color1!70!black,anchor=north west] at (0.2,8.15) {$P$ lacks $\texttt{c}$};
  \node[font=\scriptsize,color3!80!black,anchor=east] at (13.8,15.8) {$T$ lacks $\texttt{c}$};
  \node[font=\scriptsize,color3!80!black,anchor=east] at (24.8,1.6) {$T\position{25}=\texttt{d}$};
  \draw[color2,line width=0.9pt] (0,0) -- (8,8) -- (9,8) -- (14,13) -- (14,14) -- (17,17);
  \draw[color2,line width=0.9pt] (3,0) -- (11,8) -- (12,8) -- (14,10) -- (14,11) -- (20,17);
  \draw[color2,line width=0.9pt] (6,0) -- (23,17);
  \draw[color2,line width=0.9pt] (9,0) -- (14,5) -- (14,6) -- (16,8) -- (17,8) -- (25,16) -- (26,16) -- (27,17);
  \draw[color2,line width=0.9pt] (12,0) -- (14,2) -- (14,3) -- (19,8) -- (20,8) -- (25,13) -- (26,13) -- (30,17);
  \foreach \a/\b/\c/\d in {8/8/9/8,14/13/14/14,11/8/12/8,14/10/14/11,16/8/17/8,14/5/14/6,19/8/20/8,14/2/14/3,25/16/26/16,25/13/26/13} {
    \draw[color4,line width=1.8pt,line cap=round] (\a,\b) -- (\c,\d);
  }
  \foreach \b/\v in {0/2,1/2,2/0,3/3,4/3} {
    \fill[color2] (3*\b,0) circle (2.4pt);
    \node[font=\scriptsize,anchor=north east,color2,inner sep=1pt] at (3*\b-0.15,0.15) {$\v$};
  }
  \foreach \b/\x in {0/17,1/20,2/23,3/27,4/30} {\node[font=\tiny,color2,anchor=north] at (\x,17.1) {$\eD{\b}$};}
\end{tikzpicture}
\caption{The proper rails of \cref{ex:running_rails} in input
coordinates, labelled with their costs.  The blue and magenta lines mark
the missing $\texttt{c}$ of $P$ and of $T$, and the magenta column is the
extra $\texttt{d}$.  The rail $\eD{2}$ meets both lines at one point,
where the edits cancel, and projects to the exact occurrence
$T\fragmentco{6}{23}$.  The other rails pay $1$ at each line, and $\eD{3},\eD{4}$
also $1$ at the $\texttt{d}$ (orange), so each costs its canonical cost.}
\label{fig:running_input}
\end{figure}

\begin{example}[Running example, continued]\label{ex:running_rails}
In \cref{ex:overview_periodic}, the alignment $\A_P$ only inserts
characters, so its vertices are indexed by pattern-reference
coordinates: $\ell_P=18$ and $\phP(i)=i$ for all $i\in\fragment{0}{18}$.  As in
\cref{ex:breakpoint_representation}, $\pP(i)=i$ for $i\le8$ and
$\pP(i)=i-1$ for $i\ge9$.  The alignment $\A_T$ also deletes the
$\texttt{d}$, so it has $\ell_T=31$ steps, one more than the length of
$\hT$: we have $\phT(i)=i$ for $i\le26$ and $\phT(i)=i-1$ for $i\ge27$,
whereas $\pT(i)=i$ for $i\le14$ and $\pT(i)=i-1$ for $i\ge15$.  The
edits of the supplied alignments are the step $8\to9$ of $\A_P$, which
inserts the missing $\texttt{c}$ of $P$, and the steps $14\to15$ and
$26\to27$ of $\A_T$, which insert the missing $\texttt{c}$ of $T$ and
delete $T\position{25}=\texttt{d}$.  We have $\canD=\fragment{-6}{10}$
and $\prD=\fragment{0}{4}$.  For $b\in\fragment{0}{2}$, the proper rail
$\eD{b}$ consists of the vertices $(\hat p,\hat p+3b)$ with
$\hat p\in\fragment{0}{18}$.  For $b\in\{3,4\}$, the rail $\eD{b}$
takes a horizontal edge where $\A_T$ deletes the $\texttt{d}$, so it
consists of the vertices $(\hat p,\hat p+3b)$ with $\hat p\le26-3b$ and
the vertices $(\hat p,\hat p+3b+1)$ with $\hat p\ge26-3b$.

On the rail $\eD{2}$, both missing characters fall on the same edge:
the edge $(8,14)\to(9,15)$ covers empty fragments of $P$ and $T$, so it
costs $0$ and its input projection is a single vertex.  All other edges
of $\eD{2}$ are matches, so $\eD{2}$ projects to the diagonal of
$\AG(P,T)$ from $(0,6)$ to $(17,23)$, that is, to the exact occurrence
of $P$ at position $6$.  On the rail $\eD{0}$, the two missing
characters fall on different edges: the edge $(8,8)\to(9,9)$ covers
the character $T\position{8}$ but no character of $P$, the edge
$(14,14)\to(15,15)$ covers $P\position{13}$ but no character of~$T$,
and each of them costs $1$.  The rails $\eD{3}$ and $\eD{4}$ also pass
the deletion: their horizontal edges $(17,26)\to(17,27)$ and
$(14,26)\to(14,27)$ cover $T\position{25}=\texttt{d}$ and cost $1$.
Thus, the canonical costs $\cancost(u_b,z_b)$ from the earliest vertex
$u_b=(0,3b)$ of $\eD{b}$, on the top row, to its latest vertex $z_b$,
on the bottom row,
are $2,2,2,3,3$ for $b=0,\ldots,4$, whereas
the suffix distances $\tdtob(u_b)$ are $2,2,0,3,3$.  This agrees with
\cref{lem:rail_stability}, because $\tdtob$ vanishes on the bottom row.
Only on $\eD{2}$ is the cost of the rail smaller than its canonical
cost.
\Cref{fig:running_reference,fig:running_input} show these rails in
expanded and in input coordinates.
\end{example}

The reference offsets of consecutive rails differ by $q$, so, by
\cref{lem:offset_change}, a cheap path cannot meet two distant rails.

\begin{lemma}[Cheap paths cannot meet distant rails]
    \dglabel{lem:blue_jump_limit}[def:blue]
    Consider a path $\eA$ in $\eAG(P,T)$.
    If $\eA$ shares vertices with rails $\eD{b}$ and
    $\eD{b'}$, then
    \[
        \cost(\eA)\ge |b-b'|q-(d_P+d_T)\ge |b-b'|q-4d.
    \]
\end{lemma}
\begin{proof}
Choose vertices $u\in\eA\cap\eD{b}$ and $u'\in\eA\cap\eD{b'}$, ordered
as on $\eA$.  Their reference offsets are $bq$ and $b'q$.  Hence,
\cref{lem:offset_change}, applied to the subpath of $\eA$ from $u$ to
$u'$, and the standing bounds $d_P\le d$ and $d_T\le3d$ give
\[
    \cost(\eA)
      \ge |b-b'|q-\cancost(u,u')
      \ge |b-b'|q-(d_P+d_T)
      \ge |b-b'|q-4d.\qedhere
\]
\end{proof}

\begin{figure}[t]
\centering
\begin{tikzpicture}[x=0.6cm,y=-0.6cm]
  \foreach \i in {0,...,8} {\foreach \j in {0,...,11} {\draw[black!11] (\j,\i) -- (\j+1,\i+1);}}
  \foreach \i in {0,...,9} {\draw[black!18] (0,\i) -- (12,\i);}
  \foreach \j in {0,...,12} {\draw[black!18] (\j,0) -- (\j,9);}
  \foreach \c [count=\i from 0] in {a,a,b,a,a,b,a,a,b} {\node[font=\ttfamily\scriptsize,anchor=mid] at (-0.35,\i+0.5) {\c};}
  \foreach \c [count=\j from 0] in {a,a,b,a,a,b,a,a,b,a,a,b} {\node[font=\ttfamily\scriptsize,anchor=mid] at (\j+0.5,-0.35) {\c};}
  \foreach \o in {0,3,6,9} {\draw[color2!70,dashed,line width=0.8pt] (\o,0) -- ({min(12,\o+9)},{min(9,12-\o)});}
  \foreach \o in {3,6} {\draw[color2!70,dashed,line width=0.8pt] (0,\o) -- ({9-\o},9);}
  \foreach \a/\b/\c/\d/\col in {
     0/1/1/2/black, 1/2/2/3/color4, 2/3/3/4/color4,
     3/4/4/5/black, 4/5/4/6/color4, 4/6/5/7/black, 5/7/6/8/color4,
     6/8/7/9/color4, 7/9/8/10/black, 8/10/9/11/color4} {
    \draw[line width=1.7pt,\col,line cap=round] (\b,\a) -- (\d,\c);
  }
  \foreach \a/\b in {0/1,1/2,2/3,3/4,4/5,4/6,5/7,6/8,7/9,8/10,9/11} {\fill (\b,\a) circle (1.6pt);}
  \node[font=\scriptsize,color2!80!black,anchor=north] at (9,9.05) {$\hat t-\hat p\in q\mathbb Z$};
  \node[font=\small,anchor=east] at (-0.8,4.5) {$\hF$};
  \node[font=\small,anchor=south] at (6,-0.9) {$\hG$};
\end{tikzpicture}
\caption{The idea behind \cref{lem:noblue}, in the reference alignment
graph for $Q=\texttt{aab}$ and $q=3$.  The path (thick) never uses an
edge of an in-phase diagonal (dashed).  Some of its diagonal edges are
free (black), as $\texttt{a}=\texttt{a}$ also occurs out of phase, and
the others are edits (orange).  No $q$ consecutive edges are free: they
would match two out-of-phase length-$q$ fragments of $Q^\infty$, which
the synchronization property forbids.}
\label{fig:noblue}
\end{figure}
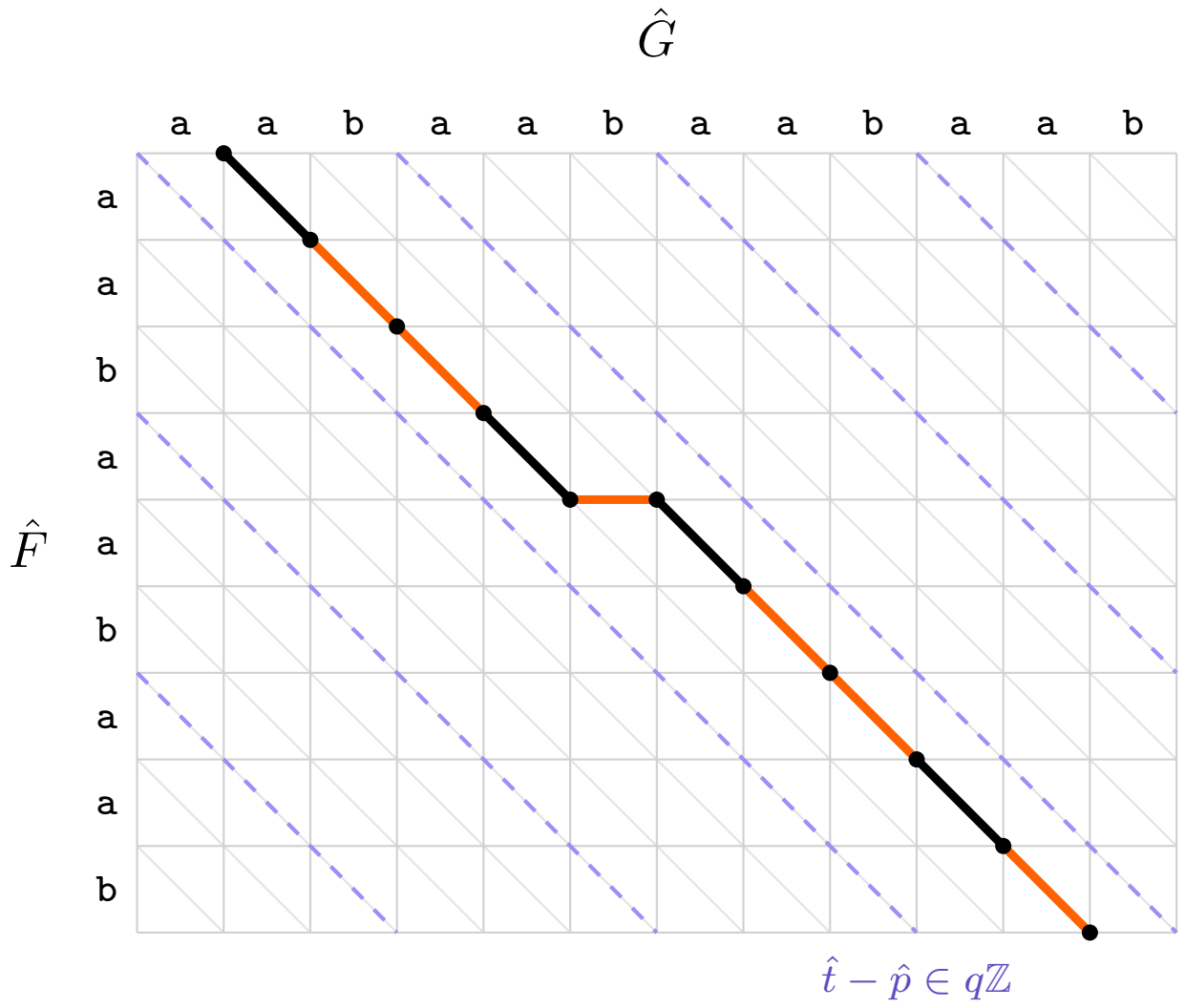

The next lemma is the key consequence of periodicity.  Consider a
path that never uses a reference-diagonal edge of a rail.  Its reference
projection then never matches $q$ consecutive characters: such a run
would be in phase (\cref{fact:matched_period}), so its edges would lie
on a reference rail and lift to edges of a rail
(\cref{obs:expansion_lifts}).  Hence,
every $q$ consecutive edges of the reference projection include an
edit, and up to the canonical cost, these edits carry over to the path
itself; see \cref{fig:noblue}.

\begin{lemma}[Paths avoiding the diagonal edges of rails are expensive]\dglabel{lem:noblue}(Minimum cost of paths avoiding reference-diagonal edges of rails)
    Let $\eA$ be a path from $u$ to $u'$ in $\eAG(P,T)$.
    Set $\hF=\phP(u,u')$ and $\hG=\phT(u,u')$.
    If $\eA$ shares no reference-diagonal edge with any rail, then
    \[\cost(\eA)+\cancost(u,u')\ge \max\left(\floor{\tfrac1q|\hF|},\floor{\tfrac1q|\hG|}\right).\]
\end{lemma}
\begin{proof}
    Let $\hA:\hF\onto\hG$ be the projection of $\eA$ under $\phPT$, and
    let $E\coloneqq\ed^{\hA}(\hF,\hG)$ be its cost.
    By \cref{lem:canonical_cost}, it suffices to prove that
    $E\ge\floor{|\hF|/q}$ and $E\ge\floor{|\hG|/q}$.

    Suppose, to the contrary, that $\max(|\hF|,|\hG|)\ge(E+1)q$.  By
    \cref{fact:matched_period}, $\hA$ contains $q$ consecutive edges of
    cost $0$, starting at a vertex $(\qp,\qt)$ with $\qt-\qp=bq$ for an
    integer $b$; moreover, $b\in\canD$ because $\qp\in\fragment{x_P}{y_P}$
    and $\qt\in\fragment{x_T}{y_T}$.  Hence, this portion of $\hA$ follows the reference
    diagonal $\hD{b}$.  Every diagonal edge of $\hA$ is the projection of
    a reference-diagonal edge of $\eA$ (otherwise, the projected path
    would contain a horizontal or vertical step instead of that diagonal
    edge).  By \cref{obs:expansion_lifts}, such an edge is the unique
    lift of the corresponding edge of $\hD{b}$, so it belongs to
    $\eD{b}$.  Thus, $\eA$ shares a reference-diagonal edge with the
    rail $\eD{b}$, a contradiction.
\end{proof}

\subsection{Localization of Tight Paths}
\label{sec:tight_path_localization}

Finally, we show that tight paths starting at the occurrence starts
assigned to a block index $j$ stay close to the rail of index $b_j$.
We combine three quantities into a potential: the remaining distance
$\tdtob$, the deviation of the reference offset from that of the target
rail, and the canonical cost $\cancost(u)$ of the rest of the supplied
alignments.  Along a tight path, $\tdtob$ decreases by the cost of the
path.  By \cref{lem:offset_change}, the deviation increases by at most
this cost plus the canonical cost of the path, and the latter is exactly
the decrease of $\cancost(u)$ (\cref{obs:cancost_factorizes}).  Hence,
the potential does not increase along tight paths.

\begin{definition}[Relevant vertices]
\label{def:relevant_vertices}
\let\qed\relax %
For $b\in\canD$ and a vertex $u$ of $\eAG(P,T)$, define the
\emph{penalty} of $u$ with respect to $b$ by
\[
    \tpena_b(u) \coloneqq |\mathrm{off}(u)-bq| + \cancost(u).
\]
We call $\tdtob(u)+\tpena_b(u)$ the \emph{potential} of $u$ with respect
to $b$.  For every threshold $W>0$, define the set of \emph{relevant
vertices}
\[
    \tClose_{W,b}
    \coloneqq
    \{u\in \eAG(P,T) : \tdtob(u)+\tpena_b(u)\le W\}.
    \tag*{\lipicsEnd}
\]
\end{definition}

In particular, every relevant vertex $u$ satisfies
$|\mathrm{off}(u)-bq|\le W$: its reference offset differs from $bq$, the
reference offset of the rail $\eD{b}$, by at most $W$.

\begin{lemma}[Relevance is preserved along tight paths]
\label{lem:relevant_path}
Let $\eA$ be a tight path from $u$ to $v$ in $\eAG(P,T)$.
For every $b\in \canD$ and $W>0$, if $u\in\tClose_{W,b}$, then every
vertex of $\eA$ belongs to $\tClose_{W,b}$.
\end{lemma}
\begin{proof}
Fix a vertex $x$ of $\eA$ and let $\mathcal B$ be the prefix of $\eA$
from $u$ to $x$; it is tight because $\eA$ is.  By
\cref{lem:offset_change}, the triangle inequality, and the additivity of
canonical costs (\cref{obs:cancost_factorizes}),
\begin{align*}
    \cost(\mathcal B)
      &\ge |\mathrm{off}(x)-\mathrm{off}(u)|-\cancost(u,x)\\
      &\ge |\mathrm{off}(x)-bq|-|\mathrm{off}(u)-bq|-\cancost(u)+\cancost(x)\\
      &=\tpena_b(x)-\tpena_b(u).
\end{align*}
Tightness of $\mathcal B$ means $\cost(\mathcal B)=\tdtob(u)-\tdtob(x)$,
so
\[
    \tdtob(x)+\tpena_b(x)
       \le\tdtob(u)+\tpena_b(u)\le W.
\]
Since $x$ was arbitrary, the whole path is relevant.
\end{proof}

The occurrence starts assigned to a block index $j$ are relevant with
respect to the rail of index $b_j=j\tau/q$: their potential is at most
$3\kappa$, which is the threshold $W$ used throughout the algorithm
(\cref{def:common_factorization_parameters}).

\begin{lemma}[Relevance of assigned occurrence starts]
\label{lem:ckw_start_relevance}
Let $j\in J$ (\cref{def:jdef}), let
$v\in\OccE_d(P,T)\cap\fragmentco{r_j}{r_{j+1}}$, and let $u$ be a
vertex of the top row of $\eAG(P,T)$ with $\pT(u)=v$.  Then,
$b_j\in\canD$ and
\[
    \tdtob(u)+\tpena_{b_j}(u)\le3\kappa.
\]
In particular, $u\in\tClose_{3\kappa,b_j}$.
\end{lemma}
\begin{proof}
The promises of \SM give $m\ge8dq\ge8d$ and $d_P\le d$, so
$y_P\ge m-d_P\ge7d>\kappa$, where the last inequality uses $d\ge1$ and
$\kappa\le5d$.  Hence, the
definition of $J$ gives $x_T-y_P<j\tau<y_T$, and $b_jq=j\tau$ shows
$b_j\in\canD$.

By \cref{lem:semiglobal_expanded_transfer}, $\tdtob(u)\le d$.  Let
$\hat v$ be maximal such that $(v,\hat v)\in\A_T$; by
\cref{lem:working_budget_ckw_slicing},
$j\tau-\kappa\le\hat v\le j\tau+2\kappa$.  The reference coordinates
that $\A_T$ pairs with $v$ form an interval
$\fragment{\hat v-\ell}{\hat v}$ with $\ell\le d_T\le\kappa$, because
every step of $\A_T$ inside it is an edit.  Since $\phT(u)$ is one of them,
$|\phT(u)-j\tau|\le2\kappa$.  Since $u$ lies in the top row,
$\phP(u)=0$, so $\mathrm{off}(u)=\phT(u)$, and
$\cancost(u)\le d_P+d_T$.  Therefore,
\[
    \tdtob(u)+\tpena_{b_j}(u)
    \le d+|\phT(u)-j\tau|+\cancost(u)
    \le d+2\kappa+d_P+d_T
    =3\kappa.\qedhere
\]
\end{proof}

\begin{figure}[t]
\centering
\begin{tikzpicture}[x=0.3cm,y=-0.3cm]
  \begin{scope}
    \clip (0,0) rectangle (24,18);
    \foreach \p/\a/\b in {0/0/0,1/1/1,2/2/2,3/3/3,4/4/4,5/5/5,6/6/6,7/7/7,8/8/9,9/8/10,10/9/11,11/10/13,12/11/15,13/12/16,14/13/17,15/13/18,16/14/19,17/15/20,18/15/22} {
      \fill[black!14] (\a-0.5,\p-0.5) rectangle (\b+0.5,\p+0.5);
    }
  \end{scope}
  \fill[color1!30,fill opacity=0.5] (0,8) rectangle (24,9);
  \fill[color3!30,fill opacity=0.5] (14,0) rectangle (15,18);
  \draw[black!40] (0,0) rectangle (24,18);
  \foreach \t in {0,3,...,24} {
    \draw[black!40] (\t,0) -- (\t,-0.25);
    \node[font=\tiny,anchor=south,black!60] at (\t,-0.25) {$\t$};
  }
  \foreach \p in {0,8,9,18} {
    \draw[black!40] (0,\p) -- (-0.25,\p);
    \node[font=\tiny,anchor=east,black!60] at (-0.25,\p) {$\p$};
  }
  \node[font=\scriptsize,anchor=south] at (12,-1.3) {vertices of $\A_T$};
  \node[font=\scriptsize,anchor=south,rotate=90] at (-1.6,9) {vertices of $\A_P$};
  \draw[black!40,dashed] (0,3) -- (15,18);
  \draw[color2,line width=0.9pt] (0,0) -- (18,18);
  \draw[color2,line width=0.9pt] (3,0) -- (21,18);
  \draw[color2,line width=0.9pt] (6,0) -- (24,18);
  \node[font=\tiny,black!50,anchor=north] at (15,18.1) {$\eD{-1}$};
  \foreach \b/\x in {0/18,1/21,2/24} {\node[font=\tiny,color2,anchor=north] at (\x,18.1) {$\eD{\b}$};}
  \draw[color4,line width=1.8pt,line cap=round,line join=round]
     (0,0) -- (8,8) -- (8,9) -- (9,9) -- (14,14) -- (15,14) -- (15,15) -- (18,18);
  \fill[color4] (0,0) circle (2.6pt);
  \foreach \x/\y in {8/8,8/9,9/9,14/14,15/14,15/15,18/18} {\fill[color4!70!black] (\x,\y) circle (1.4pt);}
  \foreach \x/\y/\v in {4.3/3.7/5, 11.8/11.2/3, 16.8/16.2/1} {
    \node[font=\tiny,color4!80!black,anchor=south west,inner sep=1.5pt] at (\x,\y) {$\v$};
  }
\end{tikzpicture}
\caption{The relevant vertices $\tClose_{5,0}$ (shaded) of the running
example in expanded coordinates, for the least threshold $W=5$ at
which the start $0$ is relevant.  The proper rails are purple, and
$\eD{-1}$ is dashed.  The relevant vertices form a corridor around
$\eD{0}$ that widens downward.  The tight path of
\cref{ex:running_localization} (orange) stays inside it, and the potential
$\tdtob+\tpena_0$ (numbers) never increases along its diagonal stretches.}
\label{fig:running_relevance}
\end{figure}
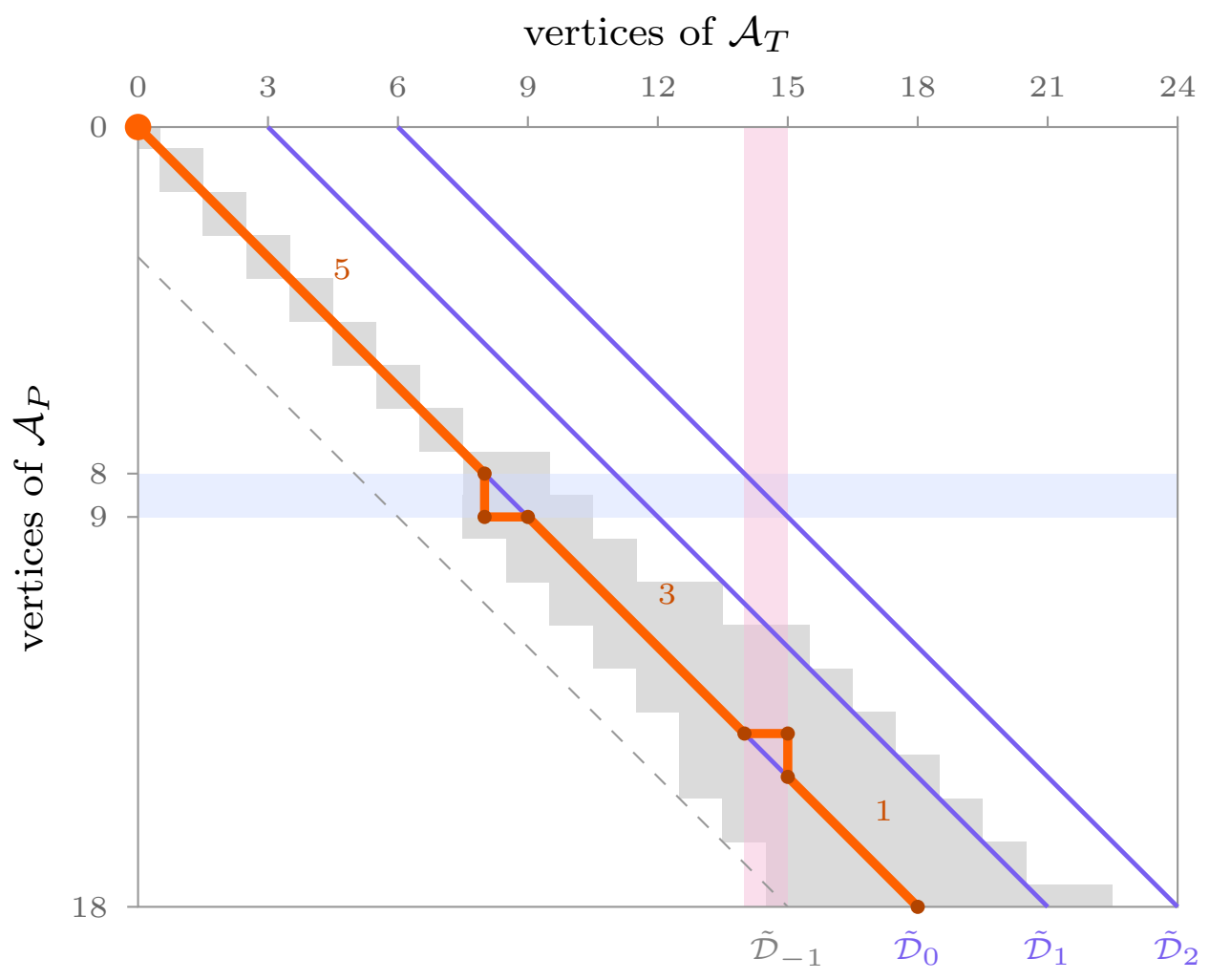

\begin{example}[Running example, continued]\label{ex:running_localization}
In \cref{ex:running_blocks}, the start $0$ has cost $2=d$ and is
assigned to $j=0$, whose rail has index $b_0=0$.  The vertex of
$\eAG(P,T)$ with input projection $(0,0)$ is $u=(0,0)$.  An optimal alignment of $P$ with $T\fragmentco{0}{17}$
inserts $T\position{8}$, deletes $P\position{13}$, and matches all
other characters.  The rail $\eD{0}$ is a tight path that realizes this
alignment, but so is the path that takes each of the two missing
characters with two edges.  The table lists the vertices of the latter
path at which its direction changes, together with the three summands
of the potential $\tdtob+\tpena_0$, where
$\tpena_0(z)=|\mathrm{off}(z)|+\cancost(z)$.
\begin{center}
\begin{tabular}{@{}lcccccccc@{}}
\toprule
vertex $z$ & $(0,0)$ & $(8,8)$ & $(9,8)$ & $(9,9)$ & $(14,14)$ & $(14,15)$ & $(15,15)$ & $(18,18)$\\
$\pPT(z)$ & $(0,0)$ & $(8,8)$ & $(8,8)$ & $(8,9)$ & $(13,14)$ & $(13,14)$ & $(14,14)$ & $(17,17)$\\
\midrule
$\tdtob(z)$ & $2$ & $2$ & $2$ & $1$ & $1$ & $1$ & $0$ & $0$\\
$|\mathrm{off}(z)|$ & $0$ & $0$ & $1$ & $0$ & $0$ & $1$ & $0$ & $0$\\
$\cancost(z)$ & $3$ & $3$ & $2$ & $2$ & $2$ & $1$ & $1$ & $1$\\
\midrule
$\tdtob(z)+\tpena_0(z)$ & $5$ & $5$ & $5$ & $3$ & $3$ & $3$ & $1$ & $1$\\
\bottomrule
\end{tabular}
\end{center}
The edge $(8,8)\to(9,8)$ is the step at which $\A_P$ inserts the
missing $\texttt{c}$.  It covers no character of $P$ or $T$, so it
costs $0$, but it moves the path to the reference offset $-1$, one
position away from the rail $\eD{0}$.  The canonical cost drops by one
and pays for this move, so the potential stays $5$.  The next edge
inserts $T\position{8}$ at cost $1$ and returns to the rail, so both
$\tdtob$ and the deviation of the offset decrease.  The same happens on
the other side of the rail at the missing character of $T$: the edge
$(14,14)\to(14,15)$, at which $\A_T$ inserts the missing $\texttt{c}$,
costs $0$ and moves the path to the offset $1$, and the next edge
deletes $P\position{13}$ at cost $1$.  On the bottom row, the canonical
cost $1$ still counts the deletion of $\A_T$, which lies to the right
of the path.  The potential never increases, as
\cref{lem:relevant_path} guarantees, so the path stays among the
relevant vertices shown in \cref{fig:running_relevance}.  Its initial
value $5\le3\kappa=15$ illustrates the bound of
\cref{lem:ckw_start_relevance}, although this small instance violates
the promise $q\le m/(8d)$ of \SM.  The bound also covers starts whose
alignments follow another rail: the start $8$ is assigned to $j=2$, but
its optimal alignment deletes $P\fragmentco{0}{2}$ and then follows
$\eD{2}$, so the potential of $(0,8)$ with respect to $b_2=4$ is
$2+4+3=9\le3\kappa$.
\end{example}

\section{Dense Covers}\label{sec:dense_covers_marks}

By \cref{lem:noblue}, a path that avoids the reference-diagonal edges
of rails pays roughly one edit per period that it traverses, minus the edits of the
supplied alignments in the traversed range.  Hence, such a path can be
as cheap as its canonical cost only where $\A_P$ or $\A_T$ has many
edits in a short range.  This section makes this precise, and
\cref{sec:mark_savings} turns it into a number attached to each strip
between two consecutive reference rails.

We first define, for a single alignment, the \emph{dense cover}: a
family of fragments, no two of which overlap or even touch, such that
every fragment whose length is small compared with the number of edits
inside it is contained in one of them (\cref{sec:dense_cover_def}).  It
consists of few fragments of small total length
(\cref{sec:dense_cover_size}), and it can be computed in time quadratic
in the number of edits (\cref{sec:dense_cover_algorithm}).  Applying
this construction to $\A_P$ and $\A_T$ yields the reference covers.  If
a path between two rail vertices uses no reference-diagonal edge of
any rail (such as the detours of \cref{def:detours}) and costs at most
its canonical cost, then one of the two reference fragments spanned by
its endpoints lies in a fragment of the corresponding cover
(\cref{sec:canonical_covers}).  The informal counterpart of this
section is the first part of \cref{sec:overview_marks}, and
\cref{fig:overview_reference} shows the two covers in reference
coordinates.

\subsection{The Dense Cover of a String}\label{sec:dense_cover_def}

Throughout
\cref{sec:dense_cover_def,sec:dense_cover_size,sec:dense_cover_algorithm},
$X=S\fragmentco{s_X}{t_X}$ and $Y=S'\fragmentco{s_Y}{t_Y}$ are
fragments of strings or bi-infinite strings $S$ and $S'$, and
$\A:X\onto Y$ is an alignment.  Moreover, $\rho_*$ and $\rho_+$ always
denote integers with $\rho_*\ge2$ and $\rho_+\ge0$.  The vertices of
$\A$ are positions in $S$ and $S'$ (\cref{def:alignment_graph}), and
the fragments of $X$ are the fragments $S\fragmentco{\ell}{r}$ with
$s_X\le\ell\le r\le t_X$ (\cref{def:fragments}).  In our applications,
$S=Q^\infty$, $X$ is $\hP$ or $\hT$, $\rho_*=5q$, and
$\rho_+\in\{0,2d\}$.

An alignment may insert characters between two consecutive
positions of $X$.  Such an insertion is located at a single boundary
position $x$ of $X$, which we represent by the empty fragment
$S\fragmentco{x}{x}$.  To handle such empty fragments consistently, we
distinguish fragments that share a position from fragments that merely
touch.

\begin{definition}[Fragment overlap conventions]
\label{def:fragment_overlap}
Two fragments $S\fragmentco{\ell}{r}$ and
$S\fragmentco{\ell'}{r'}$ of a string or bi-infinite string $S$ are
\emph{disjoint} if
$r\le\ell'$ or $r'\le\ell$; otherwise, they \emph{overlap}.  They are
\emph{strictly disjoint} if $r<\ell'$ or $r'<\ell$; otherwise, they
\emph{weakly overlap}.  Finally,
$S\fragmentco{\ell}{r}$ is \emph{contained} in
$S\fragmentco{\ell'}{r'}$, written
$S\fragmentco{\ell}{r}\subseteq S\fragmentco{\ell'}{r'}$, if
$\ell'\le\ell$ and $r\le r'$.
\end{definition}

In other words, two fragments weakly overlap if the closed intervals
$\fragment{\ell}{r}$ and $\fragment{\ell'}{r'}$ intersect, and a
fragment is contained in another if it is one of its subfragments
(\cref{def:fragments}).  For
example, $S\fragmentco{2}{5}$ and $S\fragmentco{5}{7}$ are disjoint but
weakly overlap.  The empty fragment $S\fragmentco{x}{x}$ overlaps
$S\fragmentco{\ell'}{r'}$ if $\ell'<x<r'$.  If $x=\ell'$ or $x=r'$,
the two fragments are disjoint but still weakly overlap.
Containment is not an inclusion of position sets: the empty fragment
$S\fragmentco{x}{x}$ is contained in $S\fragmentco{\ell'}{r'}$ only if
$\ell'\le x\le r'$.  In particular, $S\fragmentco{x}{x}$ is contained
in both $S\fragmentco{\ell}{x}$ and $S\fragmentco{x}{r}$.  Hence, an
insertion located at $x$ is counted in both of these disjoint
fragments (\cref{def:one_argument_alignment_cost}), which is why later
statements about edits require strictly disjoint fragments.

Next, we assign each edit of an alignment a fragment of $X$, its
\emph{support}, and count the edits of an alignment inside a given
fragment.

\begin{definition}[Supports and fragment alignment costs]
\dglabel{def:one_argument_alignment_cost}[def:alignment_restriction_cost]
The \emph{support} of an edge $e=(x,y)\to(x',y')$ of $\A$ is the
fragment $X_e\coloneqq S\fragmentco{x}{x'}$ of $X$.  For a fragment $F$
of $X$, the \emph{fragment alignment cost} $\ed^\A(F)$ is the number of
edits $e$ of $\A$ with $X_e\subseteq F$.
\end{definition}

The support of a substitution or a deletion is the single character
of $X$ that it modifies, whereas the support of an insertion is the
empty fragment at its location.  Hence, an empty fragment
$S\fragmentco{x}{x}$ is the support of the insertions located at $x$.
Insertions at either
endpoint of $F$ contribute to $\ed^\A(F)$.  The next observation
shows that $\ed^\A(F)$ is the largest cost of a restriction of $\A$
that aligns $F$ with a fragment of $Y$.

\begin{observation}[Fragment cost and alignment restrictions]
\label{obs:fragment_cost_restriction}
For every fragment $F$ of $X$,
\[
    \ed^\A(F)
    =
    \max\{\ed^\A(F,G):G\text{ is a fragment of }Y
                         \text{ and }F\agn{\A}G\}.
\]
\end{observation}
\begin{proof}
Write $F=S\fragmentco{\ell}{r}$.  Consider the subpath of $\A$
from its earliest vertex with $X$-coordinate $\ell$ to its latest vertex
with $X$-coordinate $r$.  Its edits are exactly the edits whose
supports lie in $F$, including insertions at either endpoint.
Every other restriction of $\A$ whose $X$-projection is $F$ is a
subpath of this one.  Since edge costs are nonnegative, the maximum
equals the number of these edits.
\end{proof}

The quantity $\ed^\A(F)$ is monotone with respect to containment
of fragments.  It is not additive, because an insertion between two
adjacent fragments is counted by both, but it is superadditive on
strictly disjoint fragments.

\begin{observation}[Superadditivity of fragment costs]\label{obs:ed_superadditive}
    Let $F$ be a fragment of $X$, and let $F_1,\ldots,F_s$ be pairwise
    strictly disjoint fragments of $X$ contained in $F$.
    Then,
    \[
        \ed^\A(F)\ge \sum_{i=1}^s \ed^\A(F_i).
    \]
\end{observation}
\begin{proof}
    Every edit counted by $\ed^\A(F_i)$ is also counted by
    $\ed^\A(F)$ because $F_i\subseteq F$.  No support, including an
    empty one, is contained in two strictly disjoint fragments.  Thus,
    the sets of edits counted by $\ed^\A(F_1),\ldots,\ed^\A(F_s)$ are
    disjoint subsets of the set counted by $\ed^\A(F)$.
\end{proof}

We are interested in fragments that are short compared with the
number of edits inside them, that is, fragments with more than one edit
per $\rho_*$ positions on average.  Prefixes and suffixes also
receive an additive slack $\rho_+$.

\begin{definition}[Dense fragments]
    \dglabel{def:dense}
\let\qed\relax %
    For the \emph{multiplicative slack} $\rho_*$ and the
    \emph{additive slack} $\rho_+$, a fragment $X'$ of $X$ is
    \emph{\((\rho_*,\rho_+)\)-dense} with respect to \(\A\) if
    \[
        |X'| < \rho_* \cdot (\ed^\A(X') + \sigma),
        \qquad\text{where }
        \sigma\coloneqq
        \begin{cases}
            \rho_+ & \text{if $X'$ is a prefix or a suffix of $X$},\\
            0 & \text{otherwise}.
        \end{cases}
        \tag*{\lipicsEnd}
    \]
\end{definition}

A maximal $(\rho_*,0)$-dense fragment $F$ that is a proper fragment of
$X$ satisfies $|F|=\rho_*\cdot \ed^\A(F)-1$.  Indeed, extending $F$ by one
character on an available side yields a fragment $F^+$ that, by
maximality, is not $(\rho_*,0)$-dense, so monotonicity gives
$|F|+1=|F^+|\ge\rho_*\cdot \ed^\A(F^+)\ge\rho_*\cdot \ed^\A(F)$.  On the other
hand, the density of $F$ gives $|F|<\rho_*\cdot\ed^\A(F)$.  The
\emph{locked fragments} of \cite{CKW20}, which \cite{CKW22} uses to
localize the edits of an alignment relative to a periodic reference,
play a similar role; dense covers replace them in this paper.

Dense fragments may overlap in complicated ways.  We merge them into a
family of strictly disjoint fragments, so that every dense fragment is
contained in a single member of the family.

\begin{definition}[Dense cover]
\dglabel{def:dense_cover}[def:dense]
Join two $(\rho_*,\rho_+)$-dense
fragments of $X$ when they weakly overlap.  For each connected
component, take its \emph{hull}, namely
the smallest fragment of $X$ containing all its members.  The resulting
set of fragments is the
\emph{$(\rho_*,\rho_+)$-dense cover of $X$ with respect to $\A$},
denoted by $\Es_{\A}(\rho_*,\rho_+)$.
\end{definition}

If no fragment of $X$ is $(\rho_*,\rho_+)$-dense, then
$\Es_{\A}(\rho_*,\rho_+)=\emptyset$.  If $\rho_+>0$, then the empty
prefix and the empty suffix of $X$ are dense, so the dense cover
contains a prefix and a suffix of $X$.

\begin{observation}[Structure of the dense cover]
\dglabel{obs:dense_cover_structure}[def:dense_cover]
For every component of $(\rho_*,\rho_+)$-dense fragments, the closed
interval of its hull is the union of the closed intervals of its
members.  The fragments of $\Es_{\A}(\rho_*,\rho_+)$ are pairwise
strictly disjoint, and every $(\rho_*,\rho_+)$-dense fragment is
contained in one of them.
\end{observation}
\begin{proof}
The members of a component are linked by weak overlaps, that is, by
intersections of their closed intervals.  Hence, the union of these
closed intervals is an interval from the smallest left endpoint to the
largest right endpoint, which is the closed interval of the hull.  If
the hulls of two components weakly overlapped, then so would two of
their members, and the components would coincide.  Finally, every
dense fragment belongs to a component and is contained in its hull.
\end{proof}

\begin{figure}[t]
\centering
\begin{tikzpicture}[x=0.47cm,y=-0.47cm]
  \foreach \a/\b/\r in {4/6/0,9/14/0,18/20/0,5/7/1,10/15/1,11/16/2} {
    \draw[line width=2.4pt,color1] (\a+0.08,\r*0.55) -- (\b-0.08,\r*0.55);
  }
  \foreach \a/\b/\r in {0/2/0,19/24/1} {
    \draw[line width=2.4pt,color1!45] (\a+0.08,\r*0.55) -- (\b-0.08,\r*0.55);
  }
  \node[font=\scriptsize,anchor=east] at (-0.3,0.55) {dense};
  \foreach \i in {0,...,23} {\draw[black!45] (\i,1.9) rectangle (\i+1,2.6);}
  \foreach \i in {5,11,13} {\fill[color4] (\i,1.9) rectangle (\i+1,2.6);}
  \draw[line width=2pt,color4] (20,1.75) -- (20,2.75);
  \node[font=\scriptsize,anchor=east] at (-0.3,2.25) {$X$};
  \foreach \i in {0,4,...,24} {\node[font=\tiny,anchor=base] at (\i,3.2) {$\i$};}
  \foreach \a/\b/\c in {0/2/0,4/7/1,9/16/2,18/24/1} {
    \fill[black!14] (\a,3.45) rectangle (\b,4.05);
    \node[font=\scriptsize,anchor=mid] at ({(\a+\b)/2},3.75) {$\c$};
  }
  \node[font=\scriptsize,anchor=mid east] at (-0.3,3.75) {cover};
\end{tikzpicture}
\caption{The dense cover of \cref{ex:dense_cover} with
$\rho_*=3$ and $\rho_+=1$.  The alignment substitutes the characters
at positions $5$, $11$, and $13$ (orange cells) and inserts one at
$20$ (orange bar).  The top rows show the maximal $(3,1)$-dense
fragments; the light ones, a prefix and a suffix, are dense only
because of the additive slack.  The bottom row shows the cover fragments
with their fragment alignment costs.}
\label{fig:covers_example}
\end{figure}
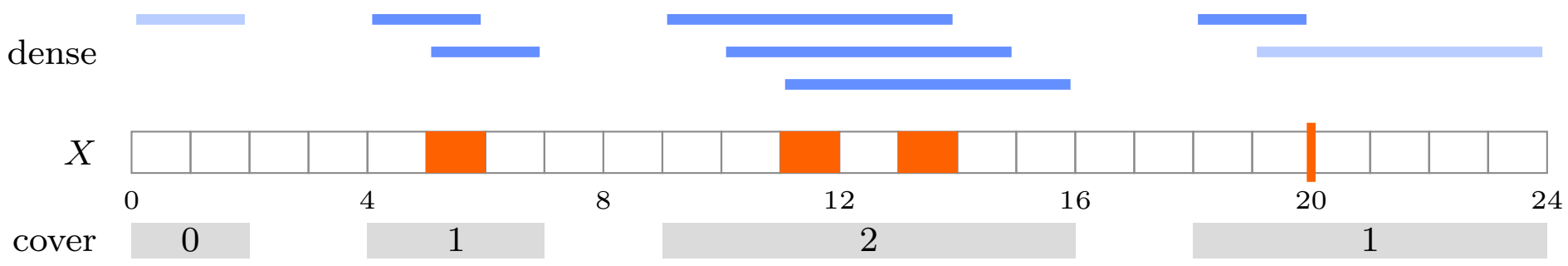

\begin{example}[A dense cover]\label{ex:dense_cover}
Let $X$ be a string of length $24$, so that $S=X$, let $\rho_*=3$
and $\rho_+=1$, and suppose that $\A$ has
four edits: substitutions with supports $X\fragmentco{5}{6}$,
$X\fragmentco{11}{12}$, and $X\fragmentco{13}{14}$, and an insertion
with support $X\fragmentco{20}{20}$; see \cref{fig:covers_example}.  A
fragment containing exactly one edit is $(3,0)$-dense if and only if
its length is at most $2$, and a fragment containing two edits is
$(3,0)$-dense if and only if its length is at most $5$.  Thus, for
example, $X\fragmentco{5}{7}$, $X\fragmentco{9}{14}$, and
$X\fragmentco{18}{20}$ are $(3,1)$-dense, whereas $X\fragmentco{8}{14}$ is
not.  The prefix $X\fragmentco{0}{2}$ contains no edit but is
$(3,1)$-dense, and so is the suffix $X\fragmentco{19}{24}$, which
contains the insertion.  The $(3,1)$-dense fragments form four components,
and the dense cover is
\[
    \Es_{\A}(3,1)=\bigl\{X\fragmentco{0}{2},\ X\fragmentco{4}{7},\
       X\fragmentco{9}{16},\ X\fragmentco{18}{24}\bigr\}.
\]
Note that $X\fragmentco{18}{20}$ and $X\fragmentco{19}{24}$ belong to
the same component, and that the cover fragment
$X\fragmentco{0}{2}$ contains no edits.
\end{example}

The cover lets us assign each edit of $\A$ to a unique cover fragment.
This holds even for an insertion, whose support is empty.

\begin{observation}[Edited edges are covered by the dense cover]
    \dglabel{fct:edits_only_in_cover}[def:dense_cover]
    Put $\Es\coloneqq\Es_{\A}(\rho_*,\rho_+)$.
    For every edit $e$ of $\A$, its support is contained in a
    unique fragment $\E\in\Es$, and $\ed^\A(\E)>0$.
    Consequently, every fragment $F$ of $X$ with $\ed^\A(F)>0$
    weakly overlaps some $\E\in\Es$ with $\ed^\A(\E)>0$.
\end{observation}
\begin{proof}
    The support $X_e$ counts the edit $e$, so
    $|X_e|\le1<\rho_*\le\rho_*\cdot \ed^\A(X_e)$.  Hence, $X_e$ is
    $(\rho_*,0)$-dense and, by \cref{obs:dense_cover_structure}, is
    contained in a cover fragment $\E$.  This fragment also counts $e$,
    so $\ed^\A(\E)>0$.  Uniqueness follows because the cover fragments
    are strictly disjoint.  Finally, if $\ed^\A(F)>0$, choose an edit counted
    by $F$.  Its support is contained in both $F$ and its cover
    fragment, which therefore weakly overlap.
\end{proof}

\subsection{Size of the Dense Cover}\label{sec:dense_cover_size}

Although there can be many dense fragments, the dense cover has few
fragments, and their total length is proportional to the number of
edits.  The cardinality bound controls the size of the representation
of the cover; the length bound will control the number of shifts at
which a pattern cover fragment and a text cover fragment can meet.

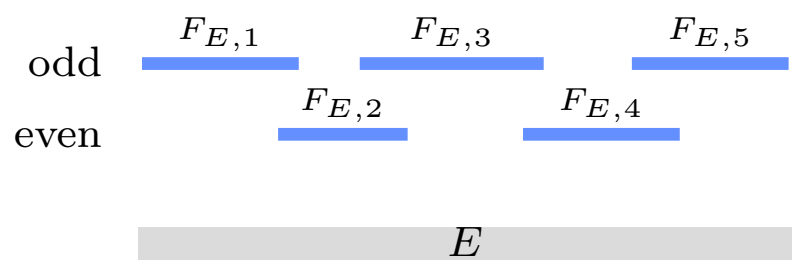
\begin{figure}[t]
\centering
\begin{tikzpicture}[x=0.36cm,y=-0.36cm]
  \foreach \a/\b/\i/\r in {0/3/1/0,4/7.5/3/0,9/12/5/0,2.5/5/2/1,7/10/4/1} {
    \draw[line width=2.4pt,color1] ({\a+0.06},{\r*1.3}) -- ({\b-0.06},{\r*1.3});
    \node[font=\tiny,anchor=base] at ({(\a+\b)/2},{\r*1.3-0.45}) {$F_{E,\i}$};
  }
  \fill[black!14] (0,3.0) rectangle (12,3.6);
  \node[font=\scriptsize,anchor=mid] at (6,3.3) {$E$};
  \node[font=\scriptsize,anchor=mid east] at (-0.3,0) {odd};
  \node[font=\scriptsize,anchor=mid east] at (-0.3,1.3) {even};
\end{tikzpicture}
\caption{Step~1 of the proof of \cref{lem:dense_decomposition}: a
minimal chain $F_{E,1},\ldots,F_{E,5}$ with hull $E$.  Consecutive
members weakly overlap, and members two apart are strictly disjoint,
so the even-indexed members are pairwise strictly disjoint, and so are
the odd-indexed ones.}
\label{fig:dense_cover_proofs}
\end{figure}

\begin{lemma}[Size of the dense cover]
\dglabel{lem:dense_decomposition}[def:dense]
The dense cover $\Es=\Es_{\A}(\rho_*,\rho_+)$ satisfies
\[
    |\Es|\le \ed^{\A}(X,Y)+2
    \qquad\text{and}\qquad
    \sum_{E\in\Es}|E|
       \le2\rho_*\cdot\bigl(\ed^{\A}(X,Y)+\rho_+\bigr).
\]
\end{lemma}

\begin{proof}
The idea is to cover each fragment of the cover by a chain of dense fragments in
which only consecutive fragments weakly overlap.  The odd-indexed and
the even-indexed fragments of all chains then form two families of
strictly disjoint fragments, and \cref{obs:ed_superadditive} bounds the
number of edits in each family.

\emph{Step 1: chains.}  For every $E\in\Es$, choose an
inclusion-minimal subfamily of its component whose hull is $E$ and
whose weak-overlap graph is connected.  List its members as
$F_{E,1},\ldots,F_{E,k_E}$ by increasing left endpoint; see
\cref{fig:dense_cover_proofs}.  By minimality, no member is contained
in another, so the right endpoints increase as well.  Consecutive
members weakly overlap: if $F_{E,i}$ and $F_{E,i+1}$ did not, then, as
both endpoints increase along the list, no member up to $F_{E,i}$
would weakly overlap a later member, and the weak-overlap graph would be
disconnected.  Minimality
also implies that $F_{E,i}$ and $F_{E,i+2}$ are strictly disjoint:
otherwise, they would weakly overlap, and removing $F_{E,i+1}$ would
preserve both the hull and connectivity, contradicting minimality.
Since the fragments of the cover are strictly disjoint
(\cref{obs:dense_cover_structure}), the
odd-indexed members of all chains are pairwise strictly disjoint, and
so are the even-indexed members.

\emph{Step 2: total length.}  In this step, $F$ ranges over the
members of all chains.  Let $\sigma_F$ be the value of $\sigma$ for $F$
in \cref{def:dense}, so that
\[
    |F|<\rho_*\cdot \bigl(\ed^{\A}(F)+\sigma_F\bigr)
\]
and $\sigma_F$ can be nonzero only if $F$ is a prefix or a suffix of $X$.  The
closed intervals of all prefixes of $X$ contain~$s_X$, so the dense
prefixes weakly overlap and lie in a single component, and they are
nested, so a minimal chain contains at most one of them.  Hence, the chains contain at most one prefix in total,
and similarly at most one suffix.  Thus, $\sum_F\sigma_F\le2\rho_+$.  Moreover,
\cref{obs:ed_superadditive}, applied separately to the odd-indexed and
the even-indexed fragments, gives $\sum_F\ed^{\A}(F)\le2\cdot\ed^{\A}(X,Y)$.
Since the chain of each $E\in\Es$ covers $E$,
\[
    \sum_{E\in\Es}|E|
       \le\sum_F|F|
       \le\rho_*\sum_F\bigl(\ed^{\A}(F)+\sigma_F\bigr)
       \le2\rho_*\cdot\bigl(\ed^{\A}(X,Y)+\rho_+\bigr).
\]

\emph{Step 3: number of fragments.}  At most two components contain
a prefix or a suffix of $X$.  Every other component contains a
$(\rho_*,0)$-dense fragment $F$, which satisfies $\ed^{\A}(F)>0$.
Choosing one such fragment per component yields strictly disjoint
fragments, so \cref{obs:ed_superadditive} bounds their number by
$\ed^{\A}(X,Y)$.  This proves the cardinality bound.
\end{proof}

\subsection{Computing the Dense Cover}\label{sec:dense_cover_algorithm}

Next, we compute dense covers from breakpoint representations.  Our
algorithm takes time quadratic in the number of edits, which suffices
for our applications.

\begin{lemma}[{\tt DenseCover($X$, $Y$, $\brp{\A}$, $\rho_*$, $\rho_+$)}]
\dglabel{lem:dense}[lem:dense_decomposition,def:breakpoint_representation]
Given $X$, $Y$, the breakpoint representation $\brp{\A}$, $\rho_*$,
and $\rho_+$, one can compute the dense cover $\Es_{\A}(\rho_*,\rho_+)$
in $\Oh((c+1)^2)$ time, where $c\coloneqq\ed^{\A}(X,Y)$.
\end{lemma}
\begin{proof}
Let $e_1,\ldots,e_c$ be the edits of $\A$ in path order, and let
$S\fragmentco{s_i}{t_i}$ be the support of $e_i$.  Thus,
$t_i\in\{s_i,s_i+1\}$, and the sequences $(s_i)$ and $(t_i)$ are
nondecreasing.  A fragment $F=S\fragmentco{\ell}{r}$ \emph{counts} $e_i$
if it contains the support of $e_i$, that is, if $\ell\le s_i$ and
$t_i\le r$; by \cref{def:one_argument_alignment_cost}, $\ed^{\A}(F)$
is the number of edits counted by $F$.  Since both sequences are
nondecreasing, the edits counted by a fragment form a possibly empty
range $e_a,\ldots,e_b$.

\emph{Overview.}  We compute a family $\mathcal U$ of fragments of
$X$, called \emph{candidates}: the windows defined below, and the
longest dense prefix and suffix of $X$ if they exist.  The candidates
satisfy two conditions:
\begin{enumerate}[label=(\roman*),ref=(\roman*)]
    \item\label{it:dense_cover_candidates_inside} every candidate is
    contained in a fragment of the cover, and
    \item\label{it:dense_cover_candidates_cover} every dense fragment is
    contained in a candidate.
\end{enumerate}
Then, we return the hulls of the components of the weak-overlap graph
on $\mathcal U$.  The following claim shows that these hulls form the
cover.

\begin{claim}[Merging candidates]\label{clm:dense_cover_merging}
Let $\mathcal U$ be a finite family of fragments of $X$ that satisfies
\ref{it:dense_cover_candidates_inside} and
\ref{it:dense_cover_candidates_cover}.  Then, the hulls of the components of the weak-overlap
graph on $\mathcal U$ are exactly the fragments of
$\Es_{\A}(\rho_*,\rho_+)$.
\end{claim}
\begin{claimproof}
For each fragment $E$ of the cover, let $\mathcal U_E$ consist of the
members of $\mathcal U$ contained in $E$.  By \ref{it:dense_cover_candidates_inside}, these families
partition $\mathcal U$, and members of distinct families are strictly
disjoint, because the fragments of the cover are
(\cref{obs:dense_cover_structure}).  Hence, every
component of $\mathcal U$ lies within a single family $\mathcal U_E$.

Fix a fragment $E$ of the cover, and let $K$ be the component of dense
fragments whose hull is $E$.  We use two facts.
\begin{enumerate}[label=(\alph*),ref=(\alph*)]
    \item\label{it:dense_cover_merging_contained} Every $F\in K$ is
    contained in a member $U_F$ of $\mathcal U_E$.  Indeed,
    by~\ref{it:dense_cover_candidates_cover}, $F$ is contained in a
    member $U_F$ of $\mathcal U$, which weakly overlaps $E\supseteq F$;
    by~\ref{it:dense_cover_candidates_inside} and the strict
    disjointness of the fragments of the cover, $U_F\subseteq E$.
    \item\label{it:dense_cover_merging_overlap} Every
    $U\in\mathcal U_E$ weakly overlaps a member of $K$.  Indeed, the
    closed interval of $E$ is the union of the closed intervals of the
    members of $K$ (\cref{obs:dense_cover_structure}), and it contains
    the left endpoint of $U$.
\end{enumerate}
Weak overlap is preserved when either fragment is enlarged.  Hence,
by~\ref{it:dense_cover_merging_contained}, weakly overlapping members
$F,F'$ of $K$ yield weakly overlapping members $U_F,U_{F'}$, so the
members $U_F$ with $F\in K$ lie in one component, because $K$ is
connected.  By~\ref{it:dense_cover_merging_overlap}
and~\ref{it:dense_cover_merging_contained}, every $U\in\mathcal U_E$
weakly overlaps some $F\in K$ and hence $U_F$, so $\mathcal U_E$ forms
a single component.  By~\ref{it:dense_cover_merging_contained}, its
hull contains every member of $K$, and thus~$E$; it is also contained
in $E$, so it is~$E$.
\end{claimproof}

\emph{Windows.}  For $1\le a\le c$, let $B_a$ be the set of indices
$b\in\fragment{a}{c}$ with $t_b-s_a<\rho_*\cdot(b-a+1)$; it contains $a$
because $t_a-s_a\le1<\rho_*$.  For $b\in B_a$,
put $h\coloneqq b-a+1$ and define the \emph{window}
$W_{ab}\coloneqq S\fragmentco{\ell_{ab}}{r_{ab}}$, where
\[
    \ell_{ab}\coloneqq\max(s_X,\ t_b-\rho_*h+1),
    \qquad
    r_{ab}\coloneqq\min(t_X,\ s_a+\rho_*h-1).
\]
Since $t_b-s_a\le\rho_*h-1$, we have
$\ell_{ab}\le s_a\le t_b\le r_{ab}$.  Windows have two properties.

First, every $(\rho_*,0)$-dense fragment $F$ that counts exactly
$e_a,\ldots,e_b$ satisfies $b\in B_a$ and $F\subseteq W_{ab}$.
Indeed, write $F=S\fragmentco{\ell}{r}$ and $h\coloneqq b-a+1$.  Since
$F$ counts $e_a$ and $e_b$, we have $\ell\le s_a$ and $t_b\le r$, and
density gives $r-\ell<\rho_*h$.  Hence, $t_b-s_a\le r-\ell<\rho_*h$,
that is, $b\in B_a$.  Moreover, $\ell>r-\rho_*h\ge t_b-\rho_*h$ and
$r<\ell+\rho_*h\le s_a+\rho_*h$.  Together with $s_X\le\ell$ and
$r\le t_X$, this gives $\ell_{ab}\le\ell$ and $r\le r_{ab}$.

Second, every window $W_{ab}$ is contained in a fragment of the
cover.  Indeed, the fragments
$L_{ab}\coloneqq S\fragmentco{\ell_{ab}}{t_b}$ and
$R_{ab}\coloneqq S\fragmentco{s_a}{r_{ab}}$ count $e_a,\ldots,e_b$, so
each of them counts at least $h$ edits, and each has length at most
$\rho_*h-1$ by the definitions of $\ell_{ab}$ and $r_{ab}$.  Hence,
$L_{ab}$ and $R_{ab}$ are $(\rho_*,0)$-dense and thus dense.  Both
contain $S\fragmentco{s_a}{t_b}$, so they weakly overlap and belong to
the same component of dense fragments.  The hull of this component is
a fragment of the cover; it contains $L_{ab}$ and $R_{ab}$ and hence
also their hull, which is $W_{ab}$.

\emph{Candidates.}  The candidates are the windows $W_{ab}$ with
$a\in\fragment{1}{c}$ and $b\in B_a$, together with the longest dense
prefix and the longest dense suffix of $X$, whenever they exist.  The
windows satisfy~\ref{it:dense_cover_candidates_inside} by their second
property, and the longest dense prefix and suffix
satisfy~\ref{it:dense_cover_candidates_inside} because they are dense.

For~\ref{it:dense_cover_candidates_cover}, let $F$ be a dense fragment.  If $F$ is a prefix of $X$,
then it is contained in the longest dense prefix, because the prefixes
of $X$ are nested; suffixes are handled symmetrically.  Otherwise,
$F$ is $(\rho_*,0)$-dense by \cref{def:dense}.  Then,
$\ed^{\A}(F)>|F|/\rho_*\ge0$, so $F$ counts a nonempty range
$e_a,\ldots,e_b$, and the first property of windows gives $b\in B_a$
and $F\subseteq W_{ab}$.

\emph{Prefixes and suffixes.}  Put $t_0\coloneqq s_X$ and, for
$b\in\fragment{0}{c}$,
\[
    \bar r_b\coloneqq\min(t_X,\ s_X+\rho_*\cdot(b+\rho_+)-1).
\]
If $t_b\le \bar r_b$, then the prefix $S\fragmentco{s_X}{\bar r_b}$ counts
$e_1,\ldots,e_b$, so it counts at least $b$ edits, and its length is
less than $\rho_*\cdot(b+\rho_+)$; hence, it is dense.  Conversely, a dense
prefix $S\fragmentco{s_X}{r}$ that counts exactly $b$ edits satisfies
$t_b\le r$ and $r-s_X<\rho_*\cdot(b+\rho_+)$, so $r\le \bar r_b$.  Therefore,
the longest dense prefix ends at the largest $\bar r_b$ with $t_b\le \bar r_b$,
and it does not exist if there is no such $b$.  Symmetrically, put
$s_{c+1}\coloneqq t_X$ and, for $a\in\fragment{1}{c+1}$,
\[
    \bar\ell_a\coloneqq\max(s_X,\ t_X-\rho_*\cdot(c-a+1+\rho_+)+1).
\]
The longest dense suffix starts at the smallest $\bar\ell_a$ with
$\bar\ell_a\le s_a$, and it does not exist if there is no such $a$.

\emph{Merging.}  There may be $\Theta(c^2)$ windows, so we do not
sort them directly.  For a fixed $a$, the closed interval of every
window $W_{ab}$ contains $s_a$, so these windows pairwise weakly
overlap, and the closed interval of their hull
\[
    H_a\coloneqq
    S\fragmentco{\min_{b\in B_a}\ell_{ab}}{\max_{b\in B_a}r_{ab}}
\]
is the union of their closed intervals.  Hence, a fragment weakly
overlaps $H_a$ if and only if it weakly overlaps one of these windows,
so replacing them by $H_a$ changes neither the components of the
weak-overlap graph nor their hulls.  We sort the at most $c+2$
fragments $H_1,\ldots,H_c$ and the longest dense prefix and suffix by
left endpoint and merge them in one scan: a fragment starts a new
component if its left endpoint exceeds the right endpoints of all
earlier fragments, and it joins the current component otherwise.

\emph{Running time.}  We read the supports from $\brp{\A}$ in
$\Oh(c+1)$ time.  For each $a$, one scan over $b\in\fragment{a}{c}$
determines $B_a$, the windows $W_{ab}$, and $H_a$ in $\Oh(c+1)$ time,
so all of them take $\Oh(({c+1})^2)$ time.  The longest dense prefix and
suffix take $\Oh(c+1)$ time, and sorting and merging take
$\Oh(({c+1})\lg c)$ time.
\end{proof}

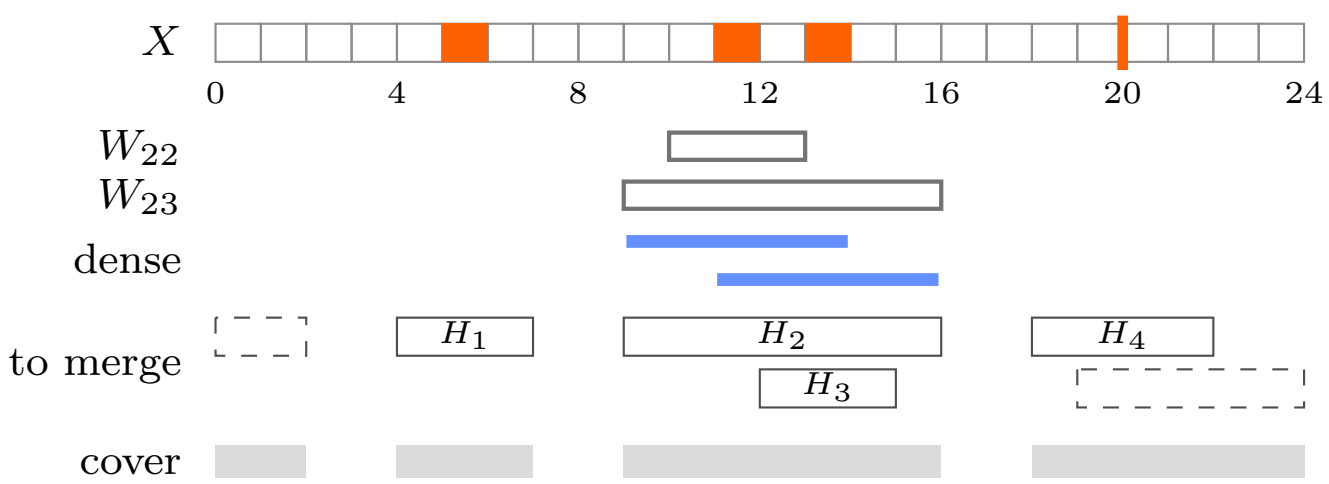
\begin{figure}[t]
\centering
\begin{tikzpicture}[x=0.3cm,y=-0.36cm]
  \foreach \i in {0,...,23} {\draw[black!45] (\i,0) rectangle (\i+1,0.7);}
  \foreach \i in {5,11,13} {\fill[color4] (\i,0) rectangle (\i+1,0.7);}
  \draw[line width=2pt,color4] (20,-0.15) -- (20,0.85);
  \node[font=\scriptsize,anchor=mid east] at (-0.3,0.35) {$X$};
  \foreach \i in {0,4,...,24} {\node[font=\tiny,anchor=base] at (\i,1.45) {$\i$};}
  \draw[black!55,thick] (10,2.0) rectangle (13,2.5);
  \draw[black!55,thick] (9,2.9) rectangle (16,3.4);
  \node[font=\scriptsize,anchor=mid east] at (-0.3,2.25) {$W_{22}$};
  \node[font=\scriptsize,anchor=mid east] at (-0.3,3.15) {$W_{23}$};
  \draw[line width=2.4pt,color1] (9.06,4.0) -- (13.94,4.0);
  \draw[line width=2.4pt,color1] (11.06,4.7) -- (15.94,4.7);
  \node[font=\scriptsize,anchor=mid east] at (-0.3,4.35) {dense};
  \foreach \a/\b/\r/\l in {4/7/0/1,9/16/0/2,12/15/1/3,18/22/0/4} {
    \draw[black!70] (\a,{5.4+0.95*\r}) rectangle (\b,{6.1+0.95*\r});
    \node[font=\tiny,inner sep=0pt,anchor=mid] at ({(\a+\b)/2},{5.63+0.95*\r}) {$H_{\l}$};
  }
  \foreach \a/\b/\r in {0/2/0,19/24/1} {
    \draw[black!70,dashed] (\a,{5.4+0.95*\r}) rectangle (\b,{6.1+0.95*\r});
  }
  \node[font=\scriptsize,anchor=mid east] at (-0.3,6.225) {to merge};
  \foreach \a/\b in {0/2,4/7,9/16,18/24} {\fill[black!14] (\a,7.75) rectangle (\b,8.35);}
  \node[font=\scriptsize,anchor=mid east] at (-0.3,8.05) {cover};
\end{tikzpicture}
\caption{The run of \cref{ex:dense_cover_algorithm}, from top to
bottom: the windows $W_{22}=X\fragmentco{10}{13}$ and
$W_{23}=X\fragmentco{9}{16}$ for $a=2$; the dense fragments
$X\fragmentco{9}{14}$ and $X\fragmentco{11}{16}$, which weakly overlap
with hull $W_{23}$; the six sorted and merged fragments, namely the hulls
$H_1,\ldots,H_4$ of the windows and the longest dense prefix and suffix
(dashed); and the hulls of their weak-overlap components, forming the
dense cover.}
\label{fig:dense_cover_windows}
\end{figure}

\begin{example}[Computing the dense cover of \cref{ex:dense_cover}]
\label{ex:dense_cover_algorithm}
We run the proof of \cref{lem:dense} on \cref{ex:dense_cover}, where
$s_X=0$, $t_X=24$, $\rho_*=3$, $\rho_+=1$, and $c=4$.  The supports
are given by $(s_1,t_1)=(5,6)$, $(s_2,t_2)=(11,12)$,
$(s_3,t_3)=(13,14)$, and $(s_4,t_4)=(20,20)$.  For $a=2$, the index
$b=3$ belongs to $B_2$ because $t_3-s_2=3<3\cdot2$, and its window
is $W_{23}=X\fragmentco{9}{16}$; the index $b=4$ does not, because
$t_4-s_2=9\ge3\cdot3$.  The windows and their hulls $H_a$ are as
follows.
\begin{center}
\begin{tabular}{@{}cccc@{}}
\toprule
$a$ & $B_a$ & windows $W_{ab}$, $b\in B_a$ & $H_a$\\
\midrule
$1$ & $\{1\}$ & $X\fragmentco{4}{7}$ & $X\fragmentco{4}{7}$\\
$2$ & $\{2,3\}$ & $X\fragmentco{10}{13}$, $X\fragmentco{9}{16}$ & $X\fragmentco{9}{16}$\\
$3$ & $\{3\}$ & $X\fragmentco{12}{15}$ & $X\fragmentco{12}{15}$\\
$4$ & $\{4\}$ & $X\fragmentco{18}{22}$ & $X\fragmentco{18}{22}$\\
\bottomrule
\end{tabular}
\end{center}
For the prefixes, $\bar r_0=2\ge t_0=0$, whereas $\bar r_b<t_b$ for all $b\ge1$
(for example, $\bar r_1=5<t_1=6$), so the longest dense prefix is
$X\fragmentco{0}{2}$, which counts no edit.

The suffixes are more interesting.  Here, $s_5=t_X=24$ and
$\bar\ell_a=24-3\cdot(6-a)+1=3a+7$ for $a\in\fragment{1}{5}$.  For $a=5$, the
suffix $X\fragmentco{\bar\ell_5}{24}=X\fragmentco{22}{24}$ counts no edit
and is dense.  For $a=4$, we have $\bar\ell_4=19\le s_4=20$: the suffix
$X\fragmentco{19}{24}$ counts the insertion $e_4$, whose support
$X\fragmentco{20}{20}$ is empty, and it is dense because
$5<3\cdot(1+1)$.  For $a\le3$, we have $\bar\ell_a>s_a$.  For example, a
suffix that counts exactly $e_3$ and $e_4$ starts at or before
$s_3=13$, so its length is at least $11$, whereas density requires
length less than $3\cdot(2+1)=9$, that is, a start at or after
$\bar\ell_3=16$.  Hence, the longest dense suffix is
$X\fragmentco{19}{24}$; without the insertion, it would be
$X\fragmentco{22}{24}$.

The longest dense prefix and suffix and the five windows are the
candidates.  After the windows of each $a$ are replaced by
$H_a$, the remaining $c+2=6$ fragments form four weak-overlap
components with hulls $X\fragmentco{0}{2}$,
$X\fragmentco{4}{7}$, $X\fragmentco{9}{16}$, and $X\fragmentco{18}{24}$,
which are exactly the fragments of the dense cover in
\cref{ex:dense_cover}; see also \cref{fig:dense_cover_windows}.  In
the last component, $H_4=X\fragmentco{18}{22}$ supplies the left
endpoint and the longest dense suffix supplies the right endpoint.
No window reaches beyond position $22$, so without the suffix
candidate, the algorithm would miss the dense fragment
$X\fragmentco{19}{24}$.
\end{example}

\subsection{Reference Covers and Detours}\label{sec:canonical_covers}

We apply the construction to the supplied alignments, viewed from
the reference side.

\begin{definition}[Dense covers of the references]
\dglabel{def:canonical_dense_covers}[def:dense_cover]
The \emph{reference covers} $\Es_P$ and $\Es_T$ are the
$(5q,2d)$- and $(5q,0)$-dense covers of $\hP$ and $\hT$ with
respect to $\A_P^{-1}$ and $\A_T^{-1}$, respectively.
\end{definition}

By \cref{lem:dense_decomposition} and $d_P\le d$, $d_T\le 3d$, both
covers consist of $\Oh(d)$ fragments of total length $\Oh(qd)$.  The
additive slack $2d$ of $\Es_P$ captures every short prefix and suffix
of $\hP$.

\begin{observation}[Short prefixes and suffixes of the pattern reference]
\dglabel{obs:short_boundary_fragments}[def:canonical_dense_covers]
Every prefix of $\hP$ of length less than $10dq$ is contained in a
fragment of $\Es_P$ that is a prefix of $\hP$.  Symmetrically, every
suffix of $\hP$ of length less than $10dq$ is contained in a fragment
of $\Es_P$ that is a suffix of $\hP$.
\end{observation}
\begin{proof}
Such a fragment $\hF$ satisfies $|\hF|<5q\cdot2d$, so it is
$(5q,2d)$-dense by \cref{def:dense} and hence contained in a fragment
of $\Es_P$ (\cref{obs:dense_cover_structure}).  A fragment of $\hP$
that contains a prefix of $\hP$ is itself a prefix of $\hP$, and
symmetrically for suffixes.
\end{proof}

The covers are defined through the inverse alignments, whose supports
are fragments of the references.  Viewed from $\A_P$ and $\A_T$, we
call these supports reference supports.

\begin{definition}[Reference supports]
\dglabel{def:reference_supports}[def:one_argument_alignment_cost]
Let $\A$ be an alignment of a fragment of a string onto a fragment of
$Q^\infty$, for example, $\A_P$ or $\A_T$.  The \emph{reference support}
of an edit $(p,\hat p)\to(p',\hat p')$ of $\A$ is the fragment
$\Q\fragmentco{\hat p}{\hat p'}$, that is, the support of the
corresponding edit $(\hat p,p)\to(\hat p',p')$ of $\A^{-1}$.
\end{definition}

Thus, for a fragment $F$ of $\hP$, the cost $\ed^{\A_P^{-1}}(F)$ counts
the edits of $\A_P$ whose reference supports lie in $F$.  By
\cref{fct:edits_only_in_cover}, the reference support of every edit of
$\A_P$ lies in a unique fragment of $\Es_P$, and the reference support
of every edit of $\A_T$ lies in a unique fragment of $\Es_T$.

A path in the expanded alignment graph $\eAG(P,T)$ may follow a rail
for a while, leave it, and later return to the same or another rail.
We call the parts of the path between its reference-diagonal rail
edges its detours.

\begin{definition}[Detours]
\dglabel{def:detours}[def:blue,def:reference_diagonal_edge]
Let $\eA$ be a path in $\eAG(P,T)$.  Remove from $\eA$ every
reference-diagonal edge that belongs to a rail.  The remaining
components that contain at least one edge are the \emph{detours} of
$\eA$.
\end{definition}

A detour may also stay on a rail.  Where $\A_P$ or $\A_T$ deletes a
character, the rail takes a vertical or a horizontal edge.  Such an
edge is not reference-diagonal, so it is not removed.

The next corollary is the reason for considering dense covers.  It
applies to every path between two rail vertices that uses no
reference-diagonal edge of any rail, in particular, to every detour
whose endpoints lie on rails.  The proof compares two estimates.  If
neither of the two reference fragments spanned by the endpoints lies in
a fragment of the corresponding cover, then both contain at most one
edit per $5q$ positions, whereas, by \cref{lem:noblue}, the path pays
roughly one edit per $q$ positions.

\begin{corollary}[Cover containment below the canonical cost]
    \dglabel{cor:nobluesame}[lem:noblue,lem:canonical_cost,def:canonical_dense_covers]
    Let $u\ne v$ be vertices on rails, and let $\eA$ be a path
    from $u$ to $v$ in $\eAG(P,T)$ with
    $\cost(\eA)\le\cancost(u,v)$.
    If $\eA$ shares no reference-diagonal edge with any rail,
    then $\phP(u,v)$ is contained in a fragment of $\Es_P$ or
    $\phT(u,v)$ is contained in a fragment of $\Es_T$.
\end{corollary}
\begin{proof}
    Set $F=\pP(u,v)$, $\hF=\phP(u,v)$, $G=\pT(u,v)$, and $\hG=\phT(u,v)$.
    Let $f\coloneqq \ed^{\A_P}(F,\hF)$ and $g\coloneqq \ed^{\A_T}(G,\hG)$,
    so that $\cancost(u,v)=f+g$.
    Suppose, for a contradiction, that neither reference fragment is
    contained in a fragment of the corresponding cover.  Then, neither
    $\hF$ nor $\hG$ is $(5q,0)$-dense, because every $(5q,0)$-dense
    fragment is also $(5q,2d)$-dense and hence contained in a fragment
    of the corresponding cover.  By \cref{obs:fragment_cost_restriction}, we
    have $\ed^{\A_P^{-1}}(\hF)\ge f$ and $\ed^{\A_T^{-1}}(\hG)\ge g$,
    and therefore $|\hF|\ge 5qf$ and $|\hG|\ge 5qg$.
    By \cref{lem:noblue},
    \[
        \cost(\eA)
        \ge \max\left(\floor{\tfrac{|\hF|}{q}},\floor{\tfrac{|\hG|}{q}}\right) - f - g
        \ge \tfrac12\left(\floor{\tfrac{|\hF|}{q}}+\floor{\tfrac{|\hG|}{q}}\right) - f - g
        \ge \tfrac52(f+g)-f-g
        = \tfrac32(f+g).
    \]
    Together with $\cost(\eA)\le f+g$, this implies $f=g=0$.

    It remains to rule out $f=g=0$.  In this case, the steps of $\A_P$
    between the rows of $u$ and $v$ are matches, and so are the steps
    of $\A_T$ between their columns, so each of these steps advances
    both its input coordinate and its reference coordinate.  Hence, a
    vertical or horizontal edge of $\eA$ advances exactly one of $\pP$
    and $\pT$, so it aligns a character with the empty string and costs
    $1$.  Since $\cost(\eA)\le f+g=0$, all edges of $\eA$ are diagonal.
    As $u\ne v$, the path $\eA$ has an earliest edge.  This edge
    advances both $\phP$ and $\phT$, so its reference projection is a
    diagonal edge of the reference rail through $\phPT(u)$, which is
    the reference rail of the rail through $u$.  That rail contains
    a lift of every diagonal edge of its reference rail, and, by
    \cref{obs:expansion_lifts}, this lift is unique.  Hence, the
    earliest edge of $\eA$ is a reference-diagonal edge of the rail
    through $u$, contradicting the assumption that $\eA$ shares no
    reference-diagonal edge with any rail.
\end{proof}

The constants in \cref{def:canonical_dense_covers} are chosen with a
margin.  The multiplicative slack $5q$ corresponds to one edit per five
periods.  A path between rail vertices that avoids the
reference-diagonal edges of rails and whose reference projections are
not contained in fragments of the corresponding covers costs at least
$3/2$ times its canonical cost (proof of \cref{cor:nobluesame}), and,
if its endpoints lie on distinct rails, at least $q/5$ more than its
canonical cost (see \cref{eq:few_mark_hop_lower} in the proof of
\cref{lem:blocks_with_few_marks}).  The additive slack $2d$ yields
\cref{obs:short_boundary_fragments}, which covers the paths that start
or end at the boundary of the pattern without using a
reference-diagonal edge of a rail
(\cref{lem:terminal_defect,lem:blocks_with_few_marks}).

\begin{example}[Running example, continued]\label{ex:running_covers}
In \cref{ex:overview_periodic}, take $d=2$ as in
\cref{ex:running_blocks}.  Then, $\Es_P$ and $\Es_T$ are the $(15,4)$-
and $(15,0)$-dense covers.  The inverse alignment $\A_P^{-1}$ deletes
the reference character at coordinate $8$, with support
$\Q\fragmentco{8}{9}$.  The inverse alignment $\A_T^{-1}$ deletes the
reference character at coordinate $14$ and inserts the $\texttt{d}$ at
coordinate $26$, with supports $\Q\fragmentco{14}{15}$ and
$\Q\fragmentco{26}{26}$.  These are the three edits shaded in
\cref{fig:running_reference}.
Since $|\hP|=18<10dq=60$, \cref{obs:short_boundary_fragments} shows
that $\hP$ itself lies in a fragment of $\Es_P$, so $\Es_P=\{\hP\}$.
The text cover has no additive slack: a fragment
of $\hT$ that contains exactly one of the two supports is
$(15,0)$-dense if and only if its length is at most $14$, and a
fragment that contains both supports is $(15,0)$-dense if and only if
its length is at most $29$.  Hence, $\Q\fragmentco{0}{29}$ and
$\Q\fragmentco{1}{30}$ are $(15,0)$-dense, and their hull yields
$\Es_T=\{\hT\}$.  The fragment $\hT$ itself
is not $(15,0)$-dense, because $|\hT|=30=15\cdot2$; it belongs to the
cover only as the hull of these two dense fragments.
\end{example}

\section{Mark Loads and Path Savings}\label{sec:mark_savings}

This section attaches a mark load to every strip between two
consecutive reference rails (\cref{sec:marks}) and uses these
loads to bound how much a path can save relative to the canonical
costs.  The saving of a path inside a strip is charged to the edits
selected for this strip, whose number is at most the mark load of the
strip (\cref{sec:marking_charging}).  Along an optimal path, these
charges add up without counting an edit twice.  Consequently, if the
load is small compared with the period, optimal paths leave the rails
only inside the dense covers (\cref{sec:detour_containment}).

The algorithm for eligible block indices (\cref{sec:powered_ckw_sweep})
uses this section, and the load partition of
\cref{sec:ckw_load_partition} uses the mark loads.  The planar
variation theorem of \cref{sec:mark_distance_variation}, which only the
algorithm for high-load block indices uses, builds on the charging
lemmas of \cref{sec:marking_charging}.  The informal counterpart of this
section is \cref{sec:overview_marks}.

\subsection{Marks and Mark Loads}\label{sec:marks}

By \cref{cor:nobluesame}, if a detour between rail vertices costs at
most its canonical cost, then one of the two reference fragments
spanned by its endpoints lies in a fragment of the corresponding cover.
We attach loads to the strips between consecutive reference rails
rather than to single diagonals, so that one load accounts for detours
whose endpoints lie on either of the two bounding rails.

\begin{definition}[Reference strips and weak meeting]
\dglabel{def:reference_strips}[def:canonical_dense_covers]
For $b\in\canD$, the \emph{reference strip} of index $b$, or simply
\emph{strip~$b$}, consists of the vertices $(\hat p,\hat t)$ of
$\AG(\hP,\hT)$ with $bq\le\hat t-\hat p\le(b+1)q$.
Fragments $\E_P=\Q\fragmentco{\hat x}{\hat y}$ of $\Es_P$ and
$\E_T=\Q\fragmentco{\hat x'}{\hat y'}$ of $\Es_T$ \emph{weakly meet strip~$b$}
if strip~$b$ contains a vertex $(\hat p,\hat t)$ with
$\hat p\in\fragmentcc{\hat x}{\hat y}$ and $\hat t\in\fragmentcc{\hat x'}{\hat y'}$.
\end{definition}

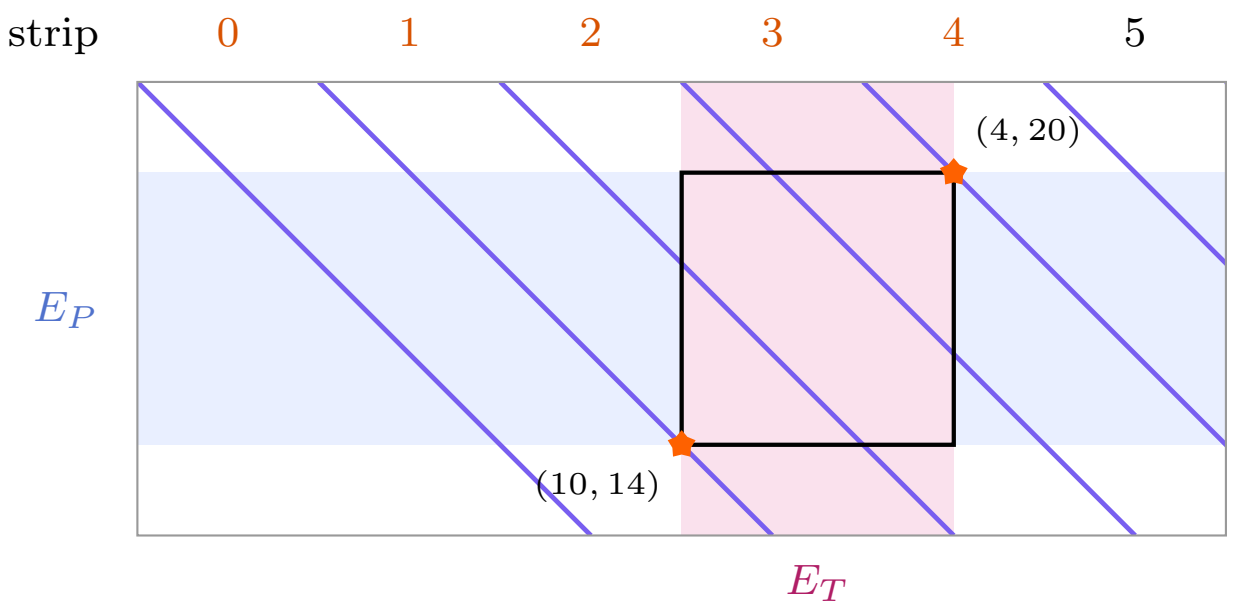
\begin{figure}[t]
\centering
\begin{tikzpicture}[x=0.3cm,y=-0.3cm]
  \fill[color1!14] (2,4) rectangle (26,10);
  \fill[color3!14] (14,2) rectangle (20,12);
  \begin{scope}
    \clip (2,2) rectangle (26,12);
    \foreach \b in {0,...,6} {
      \draw[color2,line width=0.9pt] ({2+4*\b},2) -- ({12+4*\b},12);
    }
  \end{scope}
  \draw[black!40] (2,2) rectangle (26,12);
  \draw[thick] (14,4) rectangle (20,10);
  \foreach \b in {0,...,4} {\node[font=\scriptsize,color4!85!black,anchor=base] at ({4+4*\b},1.2) {$\b$};}
  \node[font=\scriptsize,anchor=base] at (24,1.2) {$5$};
  \node[font=\scriptsize,anchor=base east] at (1.6,1.2) {strip};
  \node[star,star points=5,fill=color4,inner sep=1.3pt] at (14,10) {};
  \node[star,star points=5,fill=color4,inner sep=1.3pt] at (20,4) {};
  \node[font=\scriptsize,color1!80!black,anchor=mid east] at (1.6,7) {$\E_P$};
  \node[font=\scriptsize,color3!80!black,anchor=north] at (17,12.2) {$\E_T$};
  \node[font=\tiny,anchor=north east] at (14,10.1) {$(10,14)$};
  \node[font=\tiny,anchor=south west] at (20,3.9) {$(4,20)$};
\end{tikzpicture}
\caption{Weak meeting in reference coordinates for $q=4$,
$\E_P=\Q\fragmentco{4}{10}$ (blue rows), and $\E_T=\Q\fragmentco{14}{20}$
(magenta columns).  The purple diagonals $\hat t-\hat p=bq$ bound the
strips.  The closed rectangle $\fragment{4}{10}\times\fragment{14}{20}$
spans the offsets $4,\ldots,16$, so it weakly meets the strips
$b=0,\ldots,4$ (orange), attaining the bound $2+(|\E_P|+|\E_T|)/q=5$;
the strips $0$ and $4$ meet it only at its corners (stars).}
\label{fig:reference_strips}
\end{figure}

If $b+1\in\canD$, then strip~$b$ consists of the vertices between the
reference rails $\hD{b}$ and $\hD{b+1}$.  The differences
$\hat t-\hat p$ with $\hat p\in\fragmentcc{\hat x}{\hat y}$ and
$\hat t\in\fragmentcc{\hat x'}{\hat y'}$ form the interval
$\fragmentcc{\hat x'-\hat y}{\hat y'-\hat x}$.  Hence, $\E_P$ and $\E_T$ weakly meet strip
$b$ if and only if this interval intersects $\fragmentcc{bq}{(b+1)q}$,
that is, if and only if
\begin{equation}\label{eq:covers_weak_intersection}
    \hat x'-\hat y-q\le bq\le\hat y'-\hat x.
\end{equation}
Thus, each pair of cover fragments weakly meets a contiguous range of
at most $2+(|\E_P|+|\E_T|)/q$ strips; see \cref{fig:reference_strips}.
In \cref{fig:overview_reference}, these are the strips that meet the
intersection of the shaded rows of $\E_P$ and the shaded columns of
$\E_T$.

The marks measure how much a detour can save relative to its
canonical cost.  Such a saving requires edits of $\A_P$ and $\A_T$
that cancel each other, as the two missing characters do on the rail
$\eD{2}$ of the running example (\cref{ex:running_rails}).  Each
cancellation uses an edit of each alignment, so the side with fewer
edits limits the saving.  Accordingly, a pair of cover fragments
contributes to each strip that it weakly meets the number of edits in
the pattern fragment or in the text fragment, whichever is smaller.  At
the ends of the pattern, a path can also save without cancellations.
For example, compare a path from a rail vertex to the bottom row with
the canonical route that follows the rail to its endpoint on the
bottom row.  If the path reaches the bottom row before this endpoint,
it does not pay for the edits of $\A_T$ that the canonical route
passes between the two endpoints.  There, only the text edits
limit the saving, so the boundary fragments of the pattern cover
receive infinite weight, and their pair marks equal the text weights.

\begin{definition}[Marking weights and mark loads]
\dglabel{def:marking}[def:canonical_dense_covers,def:reference_strips,def:one_argument_alignment_cost]
For $\E_P\in\Es_P$ and $\E_T\in\Es_T$, define their
\emph{marking weights} by
\begin{align*}
    w_P(\E_P)&\coloneqq
    \begin{cases}
        \infty & \text{if $\E_P$ is a prefix or suffix of $\hP$},\\
        \ed^{\A_P^{-1}}(\E_P) & \text{otherwise};
    \end{cases}\\
    w_T(\E_T)&\coloneqq\ed^{\A_T^{-1}}(\E_T),
\end{align*}
and their \emph{pair mark} by $\min(w_P(\E_P),w_T(\E_T))$.  For
$b\in\canD$, the \emph{mark load} $\mu(b)$ of strip $b$ is the sum of
the pair marks of all pairs $(\E_P,\E_T)\in\Es_P\times\Es_T$ that
weakly meet strip~$b$.
\end{definition}

Every text weight is finite, so all pair marks and mark loads are
nonnegative integers.  In fact, all marking weights are positive:
every fragment of $\Es_T$ and every interior fragment of $\Es_P$
contains a $(5q,0)$-dense fragment and hence an edit.  Thus,
$\mu(b)>0$ holds if and only if some fragment of $\Es_P$ weakly meets
strip $b$ with some fragment of $\Es_T$.

\begin{example}[Running example, continued]\label{ex:running_marks}
In \cref{ex:running_covers}, the pattern cover $\Es_P=\{\hP\}$ consists
of a prefix of $\hP$, so $w_P(\hP)=\infty$, and $w_T(\hT)=2$ because
$\hT$ contains the reference supports of both edits of $\A_T$.  By
\cref{eq:covers_weak_intersection}, the cover fragments $\hP$ and
$\hT$ weakly meet strip $b$ if and only if $0-18-3\le3b\le30-0$, which
holds for every $b\in\canD=\fragment{-6}{10}$.  Thus,
$\mu(b)=\min(\infty,2)=2$ for all $b\in\canD$.  In this small
instance, the cover fragments are much longer than the distance $q$
between consecutive rails, so the loads do not single out the rail
$\eD{2}$ on which the two missing characters cancel; they only show
that every proper rail may be affected.
\end{example}

Although the number of strips can be very large, the total load is
small.  In particular, only $\Oh(d^2)$ strips have positive load.

\begin{lemma}[Total mark load]\label{lem:few_marks}
The total mark load satisfies
\[
    \sum_{b\in\canD}\mu(b)\le318d^2.
\]
\end{lemma}
\begin{proof}
We sum each pair's contribution over all strips, then use the bounds
on cover lengths and weights.
Put $L_P\coloneqq\sum_{\E_P\in\Es_P}|\E_P|$ and
$L_T\coloneqq\sum_{\E_T\in\Es_T}|\E_T|$.  By
\cref{lem:dense_decomposition}, we have
$L_P\le10q(d_P+2d)\le30qd$ and $L_T\le10qd_T\le30qd$.
By \cref{eq:covers_weak_intersection}, each pair
$(\E_P,\E_T)$ contributes $\min(w_P(\E_P),w_T(\E_T))$ to at most
$2+(|\E_P|+|\E_T|)/q$ mark loads.  Hence,
\[
    \sum_{b\in\canD}\mu(b)
    \le\sum_{\E_P\in\Es_P}\sum_{\E_T\in\Es_T}
        \min(w_P(\E_P),w_T(\E_T))
        \Bigl(2+\tfrac{|\E_P|}{q}+\tfrac{|\E_T|}{q}\Bigr).
\]
The cover fragments are strictly disjoint
(\cref{obs:dense_cover_structure}), so \cref{obs:ed_superadditive},
applied to $\hP$ and $\hT$, shows that
the interior pattern weights sum to at most $d_P$ and the text weights
sum to at most $d_T$.  We bound the three resulting sums separately.

\emph{The constant term.}  Using $\min(w_P,w_T)\le w_T$ and
$|\Es_P|\le d_P+2$ (\cref{lem:dense_decomposition}), it is at most
$2(d_P+2)d_T$.

\emph{The term with $|\E_P|$.}  Using $\min(w_P,w_T)\le w_T$ again,
it is at most $\sum_{\E_P\in\Es_P}|\E_P|\cdot d_T/q=d_TL_P/q$.

\emph{The term with $|\E_T|$.}  For a fixed $\E_T\in\Es_T$, the
interior pattern fragments contribute at most $d_P$, and the at most
two boundary pattern fragments contribute $w_T(\E_T)\le d_T$ each.
Hence,
\[
    \sum_{\E_P\in\Es_P}\min(w_P(\E_P),w_T(\E_T))\le d_P+2d_T,
\]
and the term is at most $(d_P+2d_T)L_T/q$.

Summing up, and using $d_P\le d$, $d_T\le3d$, $L_P,L_T\le30qd$, and
$d_P+2\le3d$ (because $d\ge1$), we obtain
\[
    \sum_{b\in\canD}\mu(b)
       \le2(d_P+2)d_T+
          \frac{d_TL_P+(d_P+2d_T)L_T}{q}
       \le18d^2+90d^2+210d^2=318d^2.\qedhere
\]
\end{proof}

\subsection{Charging Savings to Marks}\label{sec:marking_charging}

The \emph{saving} of a path $\mathcal A$ from $u$ to $v$ is
$\cancost(u,v)-\cost(\mathcal A)$, the number of edits of $\A_P$ and
$\A_T$ that the path passes minus its cost.  If $u$ and $v$ lie on one
rail, this is how much cheaper the path is than following the supplied
alignments along the rail.  Endpoints on different rails admit no such
comparison, because a route that follows both alignments keeps the
reference offset fixed.  Nevertheless, we charge the savings of paths
between neighboring rails, and of paths to the bottom row, to the marks
of the strips that these paths traverse.

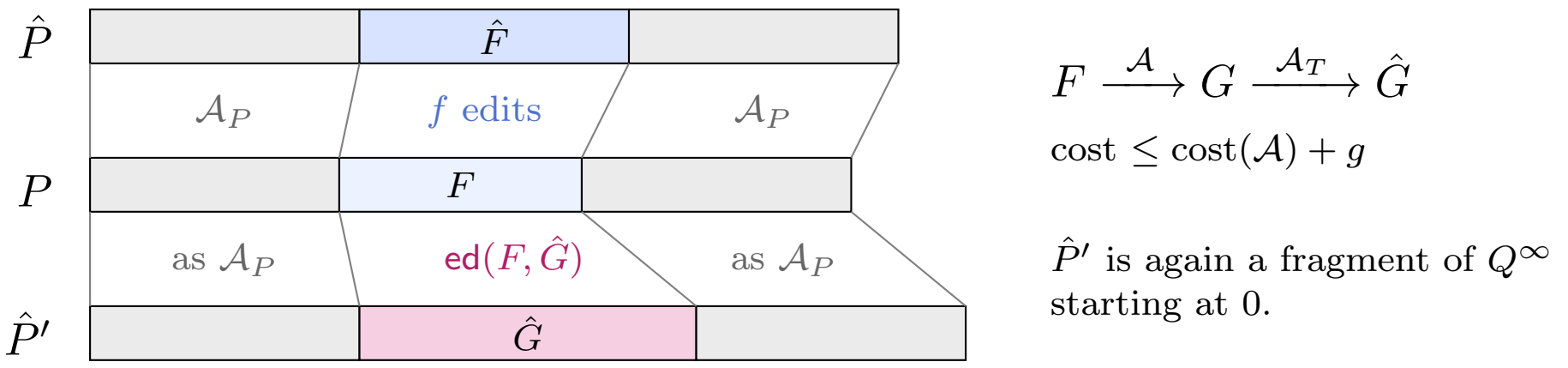
\begin{figure}[t]
\centering
\begin{tikzpicture}[x=0.52cm,y=0.52cm]
  \node[font=\small,anchor=east] at (-0.3,0) {$\hP\vphantom{\hP'}$};
  \draw[fill=black!8] (0,-0.4) rectangle (4,0.4);
  \draw[fill=color1!25] (4,-0.4) rectangle (8,0.4); \node[font=\scriptsize] at (6,0) {$\hF$};
  \draw[fill=black!8] (8,-0.4) rectangle (12,0.4);
  \node[font=\small,anchor=east] at (-0.3,-2.2) {$P\vphantom{\hP'}$};
  \draw[fill=black!8] (0,-2.6) rectangle (3.7,-1.8);
  \draw[fill=color1!12] (3.7,-2.6) rectangle (7.3,-1.8); \node[font=\scriptsize] at (5.5,-2.2) {$F$};
  \draw[fill=black!8] (7.3,-2.6) rectangle (11.3,-1.8);
  \node[font=\small,anchor=east] at (-0.3,-4.4) {$\hP'$};
  \draw[fill=black!8] (0,-4.8) rectangle (4,-4.0);
  \draw[fill=color3!22] (4,-4.8) rectangle (9,-4.0); \node[font=\scriptsize] at (6.5,-4.4) {$\hG$};
  \draw[fill=black!8] (9,-4.8) rectangle (13,-4.0);
  \foreach \a/\b in {0/0,4/3.7,8/7.3,12/11.3} {\draw[black!50] (\a,-0.4) -- (\b,-1.8);}
  \foreach \a/\b in {0/0,3.7/4,7.3/9,11.3/13} {\draw[black!50] (\a,-2.6) -- (\b,-4.0);}
  \node[font=\scriptsize,color1!80!black,anchor=mid] at (5.85,-1.1) {$f$ edits};
  \node[font=\scriptsize,black!60,anchor=mid] at (2,-1.1) {$\A_P$};
  \node[font=\scriptsize,black!60,anchor=mid] at (10,-1.1) {$\A_P$};
  \node[font=\scriptsize,black!60,anchor=mid] at (2,-3.3) {as $\A_P$};
  \node[font=\scriptsize,black!60,anchor=mid] at (10.3,-3.3) {as $\A_P$};
  \node[font=\scriptsize,color3!80!black,anchor=mid] at (6.3,-3.3) {$\ed(F,\hG)$};
  \node[font=\small,anchor=west] at (14,-0.5) {$F\xrightarrow{\ \mathcal A\ }G\xrightarrow{\ \A_T\ }\hG$};
  \node[font=\scriptsize,anchor=west] at (14,-1.7) {cost $\le\cost(\mathcal A)+g$};
  \node[font=\scriptsize,anchor=west,text width=4.4cm,align=flush left] at (14,-3.6) {$\hP'$ is again a fragment of $Q^\infty$ starting at~$0$.};
\end{tikzpicture}
\caption{The exchange argument of \cref{lem:exchange}.  A path
$\mathcal A$ from $u$ to $v$ aligns $F=\pP(u,v)$ with $G=\pT(u,v)$, which
the supplied alignments map onto $\hF$ with $f$ edits and onto $\hG$ with
$g$ edits.  As $u$ and $v$ lie on rails (or $v$ on the bottom row),
replacing $\hF$ by $\hG$ in $\hP$ yields another fragment of $Q^\infty$.
Aligning $F$ with $\hG$ through $G$ costs at most $\cost(\mathcal A)+g$,
so the optimality of $\A_P$ gives $f\le\cost(\mathcal A)+g$.}
\label{fig:exchange}
\end{figure}

Both kinds of paths are compared with the supplied alignments by the
same exchange argument, which uses the optimality of $\A_P$ and $\A_T$;
see \cref{fig:exchange}.

\begin{lemma}[Exchange bound]\label{lem:exchange}
Let $u,v$ be vertices of $\eAG(P,T)$ such that $v$ is reachable from
$u$, the vertex $u$ lies on a rail, and $v$ lies on a rail or on the
bottom row.  Put
\[
    f\coloneqq\ed^{\A_P}(\pP(u,v),\phP(u,v))
    \qquad\text{and}\qquad
    g\coloneqq\ed^{\A_T}(\pT(u,v),\phT(u,v)),
\]
so that $\cancost(u,v)=f+g$.  Then, every path $\mathcal A$ from $u$ to
$v$ satisfies $f\le\cost(\mathcal A)+g$, that is,
$\cancost(u,v)-\cost(\mathcal A)\le2g$.  If $v$ lies on a rail, then
also $g\le\cost(\mathcal A)+f$, and hence
$\cancost(u,v)-\cost(\mathcal A)\le2\min(f,g)$.
\end{lemma}
\begin{proof}
Put $F\coloneqq\pP(u,v)$ and $G\coloneqq\pT(u,v)$, and write
$\phP(u,v)=\Q\fragmentco{\hat p}{\hat p'}$ and $\phT(u,v)=\Q\fragmentco{\hat t}{\hat t'}$.
Since $u$ lies on a rail, $\mathrm{off}(u)=\hat t-\hat p$ is a multiple of $q$.
If $v$ lies on a rail, then $\hat t'-\hat p'$ is a multiple of $q$ as well; if
$v$ lies on the bottom row, then $\hat p'=y_P$.  In both cases, since $\Q$
has period $q$, replacing $\Q\fragmentco{\hat p}{\hat p'}$ by $\Q\fragmentco{\hat t}{\hat t'}$
in $\hP=\Q\fragmentco{0}{y_P}$ yields the fragment
$\Q\fragmentco{0}{y_P+(\hat t'-\hat t)-(\hat p'-\hat p)}$ of~$Q^\infty$.  Keeping $\A_P$
outside $F$, which costs $d_P-f$, and aligning $F$ with
$\Q\fragmentco{\hat t}{\hat t'}$ optimally gives an alignment of $P$ with this
fragment of cost $d_P-f+\ed(F,\Q\fragmentco{\hat t}{\hat t'})$.  Since
$d_P=\edl{P}{Q}$ is the minimum cost of aligning $P$ with a fragment of
$Q^\infty$, this cost is at least $d_P$, that is,
$f\le\ed(F,\Q\fragmentco{\hat t}{\hat t'})$.  Moreover,
$\ed(F,G)\le\cost(\mathcal A)$ by \cref{obs:expanded_projection}, and
the restriction of $\A_T$ aligns $G$ with $\Q\fragmentco{\hat t}{\hat t'}$ at
cost $g$, so the triangle inequality gives $f\le\cost(\mathcal A)+g$.
If $v$ lies on a rail, then replacing $\Q\fragmentco{\hat t}{\hat t'}$ by
$\Q\fragmentco{\hat p}{\hat p'}$ in $\hT$ and using $d_T=\edl{T}{Q}$ gives
$g\le\cost(\mathcal A)+f$ in the same way.  These two bounds are equivalent to
$\cancost(u,v)-\cost(\mathcal A)\le2g$ and
$\cancost(u,v)-\cost(\mathcal A)\le2f$, respectively.
\end{proof}

\paragraph*{Selected edits.}  By \cref{lem:exchange}, a path between
two rails saves, relative to its canonical cost, at most twice the
number of edits that it passes on the side with fewer edits.  For
every strip, we therefore fix in advance,
for every pair of cover fragments that weakly meets the strip, the side
with the smaller marking weight, and we charge savings only to edits on
the fixed sides.

\begin{definition}[Selected edits]
\label{def:selected_edits}
Let $b\in\canD$.  For every pair $(\E_P,\E_T)\in\Es_P\times\Es_T$ that
weakly meets strip~$b$, the pair \emph{selects} all edits of $\A_P$ with
reference support in $\E_P$ if $w_P(\E_P)<w_T(\E_T)$, and all edits of
$\A_T$ with reference support in $\E_T$ otherwise.  The set $S_b$ of
edits \emph{selected for strip~$b$} consists of the edits selected by
at least one such pair.
\end{definition}

\begin{observation}[Number of selected edits]\label{obs:selected_count}
For every $b\in\canD$, we have $|S_b|\le\mu(b)$.
\end{observation}
\begin{proof}
A pair $(\E_P,\E_T)$ that weakly meets strip~$b$ selects
$\ed^{\A_P^{-1}}(\E_P)=w_P(\E_P)$ edits if $w_P(\E_P)<w_T(\E_T)$, and
$\ed^{\A_T^{-1}}(\E_T)=w_T(\E_T)$ edits otherwise; in both cases, it
selects as many edits as its pair mark $\min(w_P(\E_P),w_T(\E_T))$.
Summing over these pairs counts every edit of $S_b$ at least once,
which gives $|S_b|\le\mu(b)$.  The inequality can be strict, because an
edit may be selected by many pairs.
\end{proof}

\paragraph*{Paths between neighboring rails.}
\Cref{lem:rail_path_savings} below charges the saving of a path between
neighboring rails to the selected edits that it passes.

\begin{lemma}[Charging a rail-to-rail path]
\label{lem:rail_path_savings}
Let $b,b+1\in\prD$, and let $\mathcal A$ be a path in $\eAG(P,T)$
from $u$ to $v$ without internal rail contacts such that each of $u$
and $v$ lies on $\eD{b}$ or $\eD{b+1}$.  Then,
$\cancost(u,v)-\cost(\mathcal A)$ is at most twice the number of edits
of $S_b$ that $\mathcal A$ passes.
\end{lemma}
\begin{proof}
Put $\hF\coloneqq\phP(u,v)$ and $\hG\coloneqq\phT(u,v)$, and let
$f\coloneqq\ed^{\A_P}(\pP(u,v),\hF)$ and
$g\coloneqq\ed^{\A_T}(\pT(u,v),\hG)$ be the numbers of edits of
$\A_P$ and of $\A_T$ that $\mathcal A$ passes, so that
$\cancost(u,v)=f+g$.  Let
$\delta\coloneqq f+g-\cost(\mathcal A)$ denote the saving of
$\mathcal A$.  If $\delta\le0$, there is nothing to prove, so we assume
that $\delta>0$.  Both $u$ and $v$ lie on rails, so
$\delta\le2\min(f,g)$ by \cref{lem:exchange}; in particular, $f,g>0$.
We show that all edits of $\A_P$ that $\mathcal A$ passes or all edits
of $\A_T$ that $\mathcal A$ passes belong to $S_b$.
Then, $\mathcal A$ passes at least $\min(f,g)$ edits of $S_b$, and
the bound $\delta\le2\min(f,g)$ yields the lemma.

\begin{claim}[Cover containment]\label{clm:ordered_cover}
The fragment $\hF$ is contained in a fragment of $\Es_P$, or $\hG$ is
contained in a fragment of $\Es_T$.
\end{claim}
\begin{claimproof}
Since $f+g>0$, we have $u\ne v$.  Suppose first that $\mathcal A$
contains a reference-diagonal edge of a rail.  Both endpoints of this
edge lie on the rail, and $\mathcal A$ has no internal rail contact,
so this edge is all of $\mathcal A$.  The step of $\A_P$ on this edge
advances the pattern-reference coordinate by one, so its reference
support is $\hF$, and it is an edit because $f\ge1$.  Hence, $\hF$ is contained in a
fragment of $\Es_P$ by \cref{fct:edits_only_in_cover}.
Otherwise, $\mathcal A$ shares no reference-diagonal edge with a
rail, and $\delta>0$ means $\cost(\mathcal A)<\cancost(u,v)$, so
\cref{cor:nobluesame} gives the containment.
\end{claimproof}

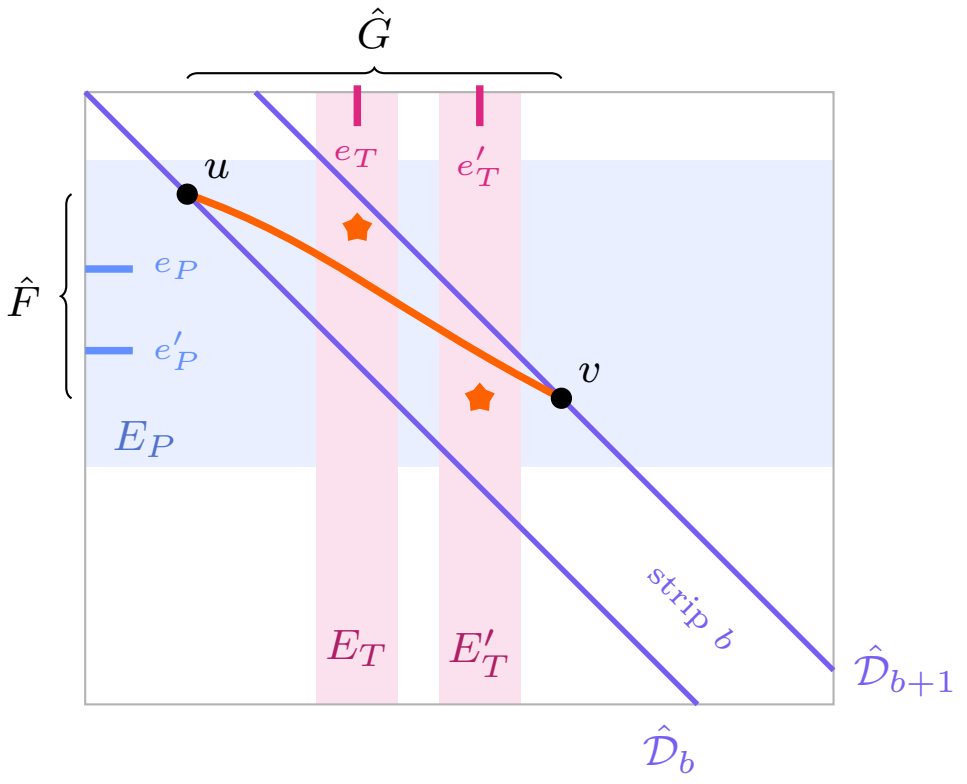
\begin{figure}[t]
\centering
\begin{tikzpicture}[x=0.45cm,y=-0.45cm]
  \fill[color1!14] (0,1) rectangle (11,5.5);
  \fill[color3!14] (3.4,0) rectangle (4.6,9);
  \fill[color3!14] (5.2,0) rectangle (6.4,9);
  \draw[black!30] (0,0) rectangle (11,9);
  \draw[color2,line width=1pt] (0,0) -- (9,9);
  \draw[color2,line width=1pt] (2.5,0) -- (11,8.5);
  \node[font=\scriptsize,color2,anchor=north] at (8.6,9) {$\hD{b}$};
  \node[font=\scriptsize,color2,anchor=west] at (11,8.5) {$\hD{b+1}$};
  \node[font=\tiny,color2,rotate=-45] at (8.9,7.6) {strip $b$};
  \draw[color4,line width=1.4pt] (1.5,1.5) .. controls (3.5,2.2) and (4.5,3.2) .. (7,4.5);
  \fill (1.5,1.5) circle (2pt); \fill (7,4.5) circle (2pt);
  \node[font=\scriptsize,anchor=south west,inner sep=1pt] at (1.65,1.35) {$u$};
  \node[font=\scriptsize,anchor=south west,inner sep=1pt] at (7.15,4.35) {$v$};
  \draw[decorate,decoration={brace,amplitude=2pt,mirror}] (-0.2,1.5) -- (-0.2,4.5) node[midway,anchor=east,font=\scriptsize,xshift=-2pt] {$\hF$};
  \draw[decorate,decoration={brace,amplitude=2pt}] (1.5,-0.2) -- (7,-0.2) node[midway,anchor=south,font=\scriptsize,yshift=2pt] {$\hG$};
  \node[font=\scriptsize,color1!80!black,anchor=west] at (0.1,5.1) {$\E_P$};
  \node[font=\scriptsize,color3!80!black,anchor=north] at (4,7.6) {$\E_T$};
  \node[font=\scriptsize,color3!80!black,anchor=north] at (5.8,7.6) {$\E_T'$};
  \draw[color3,line width=1.4pt] (4,-0.1) -- (4,0.5);
  \node[font=\tiny,color3,anchor=north] at (4,0.5) {$e_T$};
  \draw[color3,line width=1.4pt] (5.8,-0.1) -- (5.8,0.5);
  \node[font=\tiny,color3,anchor=north] at (5.8,0.5) {$e_T'$};
  \draw[color1,line width=1.4pt] (0,2.6) -- (0.7,2.6);
  \node[font=\tiny,color1,anchor=west] at (0.7,2.6) {$e_P$};
  \draw[color1,line width=1.4pt] (0,3.8) -- (0.7,3.8);
  \node[font=\tiny,color1,anchor=west] at (0.7,3.8) {$e_P'$};
  \node[star,star points=5,fill=color4,inner sep=1.4pt] at (4,2) {};
  \node[star,star points=5,fill=color4,inner sep=1.4pt] at (5.8,4.5) {};
\end{tikzpicture}
\caption{Selected edits in the proof of \cref{lem:rail_path_savings}
for $\hF\subseteq\E_P$ (blue), in reference coordinates.  The orange
curve, projecting a path from $u\in\eD{b}$ to $v\in\eD{b+1}$, passes
edits $e_P,e_P'$ of $\A_P$ (left) and $e_T,e_T'$ of $\A_T$ (top),
supported in $\E_T$ and $\E_T'$ (magenta).  So $(\E_P,\E_T)$ and
$(\E_P,\E_T')$ weakly meet strip $b$ (stars); if one of them selects the
pattern side, $e_P,e_P'$ are selected, and otherwise $e_T,e_T'$ are.}
\label{fig:charging_cases}
\end{figure}

Suppose, for a contradiction, that $\mathcal A$ passes an edit $e_P$
of $\A_P$ and an edit $e_T$ of $\A_T$, neither of which is selected
for strip~$b$.  Let $\E_P=\Q\fragmentco{\hat x}{\hat y}$ and
$\E_T=\Q\fragmentco{\hat x'}{\hat y'}$ be the fragments of $\Es_P$ and $\Es_T$
that contain their reference supports (\cref{fct:edits_only_in_cover}),
and write $\hF=\Q\fragmentco{\hat p}{\hat p'}$ and $\hG=\Q\fragmentco{\hat t}{\hat t'}$, so
that $\mathrm{off}(u)=\hat t-\hat p$ and $\mathrm{off}(v)=\hat t'-\hat p'$ lie in
$\fragmentcc{bq}{(b+1)q}$.  Both $\E_P$ and $\hF$ contain the reference
support of $e_P$, so they weakly overlap, that is, $\hat x\le\hat p'$ and
$\hat p\le\hat y$.  Symmetrically, $\hat x'\le\hat t'$ and $\hat t\le\hat y'$.  By
\cref{clm:ordered_cover} and the uniqueness in
\cref{fct:edits_only_in_cover}, $\hF\subseteq\E_P$ or
$\hG\subseteq\E_T$.  In the first case, $\hat x\le\hat p$ and $\hat p'\le\hat y$, so
\[
    \hat x'-\hat y\le\hat t'-\hat p'\le(b+1)q
    \qquad\text{and}\qquad
    \hat y'-\hat x\ge\hat t-\hat p\ge bq.
\]
In the second case, $\hat x'\le\hat t$ and $\hat t'\le\hat y'$, so
\[
    \hat x'-\hat y\le\hat t-\hat p\le(b+1)q
    \qquad\text{and}\qquad
    \hat y'-\hat x\ge\hat t'-\hat p'\ge bq.
\]
In both cases, $\E_P$ and $\E_T$ weakly meet strip~$b$ by
\cref{eq:covers_weak_intersection}; see \cref{fig:charging_cases}.
Hence, the pair $(\E_P,\E_T)$ selects all edits of $\A_P$ with
reference support in $\E_P$, including $e_P$, or all edits of $\A_T$
with reference support in $\E_T$, including $e_T$
(\cref{def:selected_edits}), a contradiction.
\end{proof}

\begin{example}[Running example, continued]\label{ex:running_charging}
We trace the proof of \cref{lem:rail_path_savings} in the running
example.  Take $b=2$, and let $\mathcal A$ be the single edge from
$u=(8,14)$ to $v=(9,15)$ of the rail $\eD{2}$ from
\cref{ex:running_rails}.  It covers no character of $P$ or $T$, so it
costs $0$, whereas $f=g=1$, because it passes an edit of each
supplied alignment.  Its saving is therefore $2=2\min(f,g)$, so the
bound of \cref{lem:exchange} is attained.  As in the proof of
\cref{clm:ordered_cover}, $\hF=\Q\fragmentco{8}{9}$ is the reference
support of the edit of $\A_P$ on this edge, and it lies in the only
pattern cover fragment $\E_P=\hP$ of \cref{ex:running_covers}.  The
only pair of cover fragments, $(\hP,\hT)$, weakly meets every strip
(\cref{ex:running_marks}), and $w_P(\hP)=\infty>2=w_T(\hT)$, so both
edits of $\A_T$ are selected for every strip, and no edit of $\A_P$
is.  The edge passes one of them, the insertion of $\A_T$ at reference
coordinate~$14$, so it passes one edit of $S_2$, and the bound of
\cref{lem:rail_path_savings} is attained as well.
\end{example}

\paragraph*{Paths to the bottom row.}  Optimal paths end on the
bottom row, generally not at the end of a rail.  We call the ends of
the rails their bottom endpoints, and we let moves along the bottom
row be free, because an alignment of the whole pattern may end
anywhere in the text.

\begin{definition}[Bottom endpoints and free bottom moves]\label{def:rail_bottom_endpoints}
For $b\in\prD$, let $z_b$ denote the latest vertex of the rail $\eD{b}$;
we call it the \emph{bottom endpoint} of $\eD{b}$.
For a path $\mathcal A$ in $\eAG(P,T)$, let $\cost^\circ(\mathcal A)$
denote its cost when every edge inside the bottom row has cost $0$.
\end{definition}

By \cref{lem:canonical_rails}, the vertex $z_b$ lies on the bottom row,
and the vertices $z_b$ occur on the bottom row in index order.  The
reference projection of $z_b$ is the latest vertex $(y_P,y_P+bq)$ of the
reference rail~$\hD{b}$, so $\mathrm{off}(z_b)=bq$.

Along the bottom row, only edits of $\A_T$ are passed.  They are
selected because their cover fragments form pairs with the suffix
fragment of $\Es_P$, whose marking weight is infinite.  More generally,
this holds for the edits of $\A_T$ next to every short suffix of $\hP$.

\begin{observation}[Text edits next to a short suffix of the pattern]
\label{obs:suffix_pays}
Let $b\in\canD$, and let $u,v$ be vertices of $\eAG(P,T)$ such that $v$
is reachable from $u$,
$\mathrm{off}(u),\mathrm{off}(v)\in\fragmentcc{bq}{(b+1)q}$, and
$\phP(u,v)$ is a suffix of $\hP$ of
length less than $10dq$.  Then, every edit of $\A_T$ with reference
support in $\phT(u,v)$ is selected for strip~$b$.
\end{observation}
\begin{proof}
By \cref{obs:short_boundary_fragments}, $\phP(u,v)$ lies in a fragment
$\E_P=\Q\fragmentco{\hat x}{\hat y}$ of $\Es_P$ that is a suffix of $\hP$, so
$w_P(\E_P)=\infty$ and $\hat x\le\phP(u)\le\phP(v)\le\hat y$.  Consider an edit
of $\A_T$ with reference support in $\phT(u,v)$.  By
\cref{fct:edits_only_in_cover}, its reference support lies in a
fragment $\E_T=\Q\fragmentco{\hat x'}{\hat y'}$ of $\Es_T$.  Both $\E_T$ and
$\phT(u,v)$ contain this support, so they weakly overlap, that is,
$\hat x'\le\phT(v)$ and $\phT(u)\le\hat y'$.  Hence,
\[
    \hat x'-\hat y\le\phT(v)-\phP(v)=\mathrm{off}(v)\le(b+1)q
    \qquad\text{and}\qquad
    \hat y'-\hat x\ge\phT(u)-\phP(u)=\mathrm{off}(u)\ge bq.
\]
By \cref{eq:covers_weak_intersection}, $\E_P$ and $\E_T$ weakly meet
strip $b$.  Since $w_P(\E_P)=\infty$, this pair selects all edits of
$\A_T$ with reference support in $\E_T$ (\cref{def:selected_edits}),
including the considered edit.
\end{proof}

A path $\eA$ from a rail vertex $x$ to a vertex $y$ of the bottom row
generally ends off the rails, so \cref{lem:exchange} bounds its saving
$\cancost(x,y)-\cost(\eA)$ only by twice the number of edits of $\A_T$
that $\eA$ passes.  A
positive saving forces the remaining pattern reference to be a short
suffix of $\hP$, and then \cref{obs:suffix_pays} shows that these text
edits are selected.

\begin{figure}[t]
\centering
\begin{tikzpicture}[x=0.5cm,y=-0.5cm]
  \fill[color1!14] (-1,3) rectangle (17,8);
  \fill[color3!16] (9.2,0) rectangle (11.4,8);
  \fill[color3!16] (11.8,0) rectangle (14.6,8);
  \draw[thick] (-1,8) -- (17,8);
  \node[font=\scriptsize,anchor=east] at (-1,8) {bottom row};
  \node[font=\scriptsize,anchor=east] at (-1,3) {row of $x$};
  \draw[color2,line width=1.1pt] (4,0) -- (12,8);
  \draw[color2,line width=1.1pt] (7,0) -- (15,8);
  \node[font=\scriptsize,color2,anchor=north] at (12,8.05) {$z_b\vphantom{z_{b+1}y}$};
  \node[font=\scriptsize,color2,anchor=north] at (15,8.05) {$z_{b+1}\vphantom{z_{b+1}y}$};
  \node[font=\scriptsize,color2,anchor=south] at (4,0) {$\hD{b}$};
  \node[font=\scriptsize,color2,anchor=south] at (7,0) {$\hD{b+1}$};
  \fill (7,3) circle (2.2pt); \node[font=\scriptsize,anchor=east] at (6.9,3) {$x$};
  \draw[color4,line width=1.6pt,rounded corners=3pt] (7,3) -- (8.6,3.6) -- (10.6,5.2) -- (12.4,6.6) -- (13.6,8);
  \fill[color4] (13.6,8) circle (2.2pt); \node[font=\scriptsize,anchor=north,color4] at (13.6,8.05) {$y\vphantom{z_{b+1}y}$};
  \foreach \t in {10.2,10.8,12.8} {\draw[color3,line width=1.4pt] (\t,-0.2) -- (\t,0.4);}
  \node[font=\tiny,color3,anchor=south] at (12,-0.3) {edits of $\A_T$ counted by $g$};
  \node[star,star points=5,fill=color4,inner sep=1.2pt] at (10.4,5.75) {};
  \node[star,star points=5,fill=color4,inner sep=1.2pt] at (12.6,7.4) {};
  \node[font=\scriptsize,color1!80!black,anchor=west] at (15.2,4.2) {suffix of $\hP$ in $\E_P$, $w_P=\infty$};
  \node[font=\scriptsize,color3!80!black,anchor=north] at (10.3,9.1) {$\E_T$};
  \node[font=\scriptsize,color3!80!black,anchor=north] at (13.2,9.1) {$\E_T$};
\end{tikzpicture}
\caption{The last-piece comparison in \cref{lem:terminal_defect}, in
reference coordinates.  A path from $x\in\eD{b}$ to $y$ on the bottom row,
with $bq\le\mathrm{off}(y)\le(b+1)q$, avoids reference-diagonal edges of
rails.  A positive saving over
$\cancost(x,y)$ makes the blue part of $\hP$ a short suffix inside a
cover fragment $\E_P$ of infinite weight; its pairs with the text cover
fragments (magenta) meet strip $b$ (stars) and select the text
edits.}
\label{fig:terminal_defect}
\end{figure}
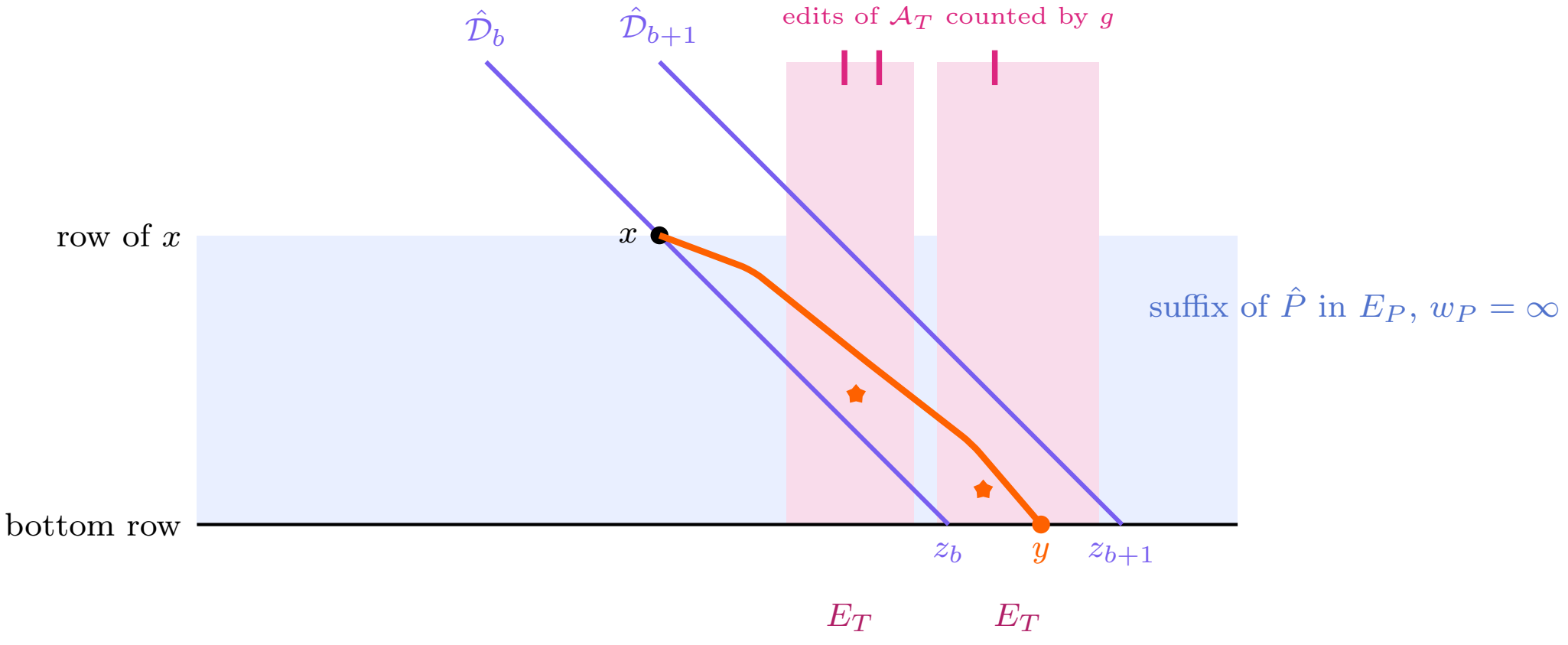

\begin{lemma}[Savings of a last piece]\label{lem:terminal_defect}
Let $b,b+1\in\prD$, and let $\eA$ be a path from a vertex $x$ of
$\eD{b}$ or $\eD{b+1}$ to a vertex $y$ of the bottom row
with $bq\le\mathrm{off}(y)\le(b+1)q$.  Suppose that $\eA$ shares no
reference-diagonal edge with a rail.  Then, $\cancost(x,y)-\cost(\eA)$
is at most twice the number of edits of $S_b$ that $\eA$ passes.
\end{lemma}
\begin{proof}
We bound the saving by the text edits passed and then show that these
edits are selected; see \cref{fig:terminal_defect}.  Put
\[
    w\coloneqq\cost(\eA),\quad
    f\coloneqq\ed^{\A_P}(\pP(x,y),\phP(x,y)),\quad
    g\coloneqq\ed^{\A_T}(\pT(x,y),\phT(x,y)).
\]
By \cref{lem:exchange}, the saving $f+g-w$ is at most $2g$.  Suppose that this saving is positive;
otherwise, there is nothing to prove.  By \cref{lem:noblue},
\[
    \floor{\frac{|\phP(x,y)|}{q}}
    \le w+f+g
    <2(f+g)
    \le2(d_P+d_T)\le8d.
\]
Since $y$ lies on the bottom row, $\phP(x,y)$ is a suffix of $\hP$
of length less than $8dq$.  The offsets of $x$ and $y$ belong to
$\fragmentcc{bq}{(b+1)q}$, so \cref{obs:suffix_pays} shows that all
$g$ edits of $\A_T$ that $\eA$ passes belong to $S_b$.  The saving is
at most $2g$, as required.
\end{proof}

\paragraph*{Paths inside a strip.}  The two lemmas combine into one
bound for a path inside a closed strip (\cref{def:rail_strips}) whose
endpoints lie on the two rails or the bottom row.  Moves inside that row
are free, as in the $\cost^\circ$ of
\cref{def:rail_bottom_endpoints}.

\begin{lemma}[Charging a path inside a strip]\label{lem:strip_path_savings}
Let $b,b+1\in\prD$, and let $\mathcal A$ be a path from $u$ to $v$
inside the closed strip between $\eD{b}$ and $\eD{b+1}$.  Suppose that
each of $u$ and $v$ lies on one of these rails or on the bottom row.  Then,
$\cancost(u,v)-\cost^\circ(\mathcal A)$ is at most twice the number of
edits of $S_b$ that $\mathcal A$ passes.
\end{lemma}
\begin{proof}
Split $\mathcal A$ into pieces at its rail contacts and at its
vertices on the bottom row.  Every edit counted by $\cancost(u,v)$ is
passed by exactly one edge of $\mathcal A$, so, by additivity
(\cref{obs:cancost_factorizes}), it suffices to prove the bound for
every piece.

A piece outside the bottom row starts on $\eD{b}$ or $\eD{b+1}$, has no
internal rail contact, and costs the same in $\cost^\circ$ as
originally.  If it ends on a rail, then \cref{lem:rail_path_savings}
applies.  Otherwise, it ends on the bottom row and contains no rail
edge, because its start is its only rail contact; hence, it shares no
reference-diagonal edge with a rail, and \cref{lem:terminal_defect}
applies.

Every other piece is a single edge inside the bottom row; if $u$ lies
on the bottom row, then all pieces are of this kind.  Such an edge
costs $0$ in $\cost^\circ$ and passes at most one edit, of $\A_T$.  Its
endpoints have reference offsets in $\fragmentcc{bq}{(b+1)q}$, and
$\phP$ spans the empty suffix of $\hP$ between them, so
\cref{obs:suffix_pays} shows that this edit belongs to $S_b$.
\end{proof}

\subsection{Detours of Optimal Paths}\label{sec:detour_containment}

We now derive the consequence of the strip charging lemma that the
algorithm for eligible block indices (\cref{sec:powered_ckw_sweep})
uses.  Consider an optimal alignment of the pattern, viewed as a path
in $\eAG(P,T)$, and its detours (\cref{def:detours}).  If the period is
large or the load around the path is small, then one reference
projection of each detour lies in a fragment of its cover; the side
may differ between detours.

\begin{figure}[t]
\centering
\begin{tikzpicture}[x=0.5cm,y=-0.5cm]
  \draw[thick] (5.6,9) -- (14.2,9);
  \node[font=\scriptsize,anchor=east] at (5.6,9) {bottom row};
  \foreach \x/\l in {0/{b-1},2/{b},4/{b+1}} {
    \draw[color2,line width=1.1pt] (\x,0) -- (\x+9,9);
    \node[font=\scriptsize,color2,anchor=south] at (\x,0) {$\eD{\l}$};
    \fill[color2] (\x+9,9) circle (2.2pt);
    \node[font=\scriptsize,color2,anchor=north] at (\x+9,9.15) {$z_{\l}\vphantom{\bar y}$};
  }
  \draw[color4,line width=1.8pt,line cap=round] (2.3,0.3) -- (2.7,0.7);
  \draw[color4,line width=1.8pt] (2.7,0.7) .. controls (3.6,0.75) and (4,1.3) .. (4,2);
  \draw[color4,line width=1.8pt,line cap=round] (4,2) -- (4.4,2.4);
  \draw[color4,line width=1.8pt] (4.4,2.4) .. controls (5.4,2.5) and (6.2,2.7) .. (6.9,2.9);
  \draw[color4,line width=1.8pt] (6.9,2.9) .. controls (7.1,3.6) and (7.2,4.5) .. (7.3,5.3);
  \draw[color4!55!black,line width=1.8pt] (7.3,5.3) .. controls (7.35,6) and (7.45,6.8) .. (7.5,7.5);
  \draw[color4!55!black,line width=1.8pt,line cap=round] (7.5,7.5) -- (7.9,7.9);
  \draw[color4!55!black,line width=1.8pt,dash pattern=on 2pt off 1pt] (7.9,7.9) .. controls (8.8,8) and (9.6,8.4) .. (10,9);
  \draw[->,color4,line width=1.1pt] (10,8.7) -- (14.1,8.7);
  \foreach \p in {(2.3,0.3),(2.7,0.7),(4,2),(4.4,2.4),(6.9,2.9),(7.3,5.3),(7.5,7.5),(7.9,7.9)} {\fill \p circle (1.7pt);}
  \node[font=\scriptsize,anchor=east] at (2.2,0.3) {$y$};
  \fill (10,9) circle (1.7pt);
  \node[font=\scriptsize,anchor=north] at (10,9.15) {$\bar y\vphantom{z_{b+1}}$};
  \draw[color4,line width=1.8pt] (9.4,0.6) -- (10.4,0.6);
  \node[font=\scriptsize,anchor=west] at (10.5,0.6) {pieces between $\eD{b}$ and $\eD{b+1}$};
  \draw[color4!55!black,line width=1.8pt] (9.4,1.7) -- (10.4,1.7);
  \node[font=\scriptsize,anchor=west] at (10.5,1.7) {pieces between $\eD{b-1}$ and $\eD{b}$};
  \draw[color4!55!black,line width=1.8pt,dash pattern=on 2pt off 1pt] (9.4,2.8) -- (10.4,2.8);
  \node[font=\scriptsize,anchor=west] at (10.5,2.8) {last piece};
  \draw[->,color4,line width=1.1pt] (9.4,3.9) -- (10.4,3.9);
  \node[font=\scriptsize,anchor=west] at (10.5,3.9) {free moves toward $z$};
\end{tikzpicture}
\caption{Charging the rest of an optimal path after a detour
(\cref{lem:blocks_with_few_marks}).  The rails met by the path lie
strictly inside a proper rail interval $I$, and $M=\sum_{c\in I}\mu(c)$.
From the latest vertex $y$ of the detour, we follow the path to the
bottom row and then move freely to $z=z_{\max I}$.  Cutting at rail
contacts (dots) gives pieces with disjoint edit sets, so by
\cref{lem:strip_path_savings}, their total saving is at most
$2\sum_{c\in I}|S_c|\le2M$.}
\label{fig:single_path_savings}
\end{figure}
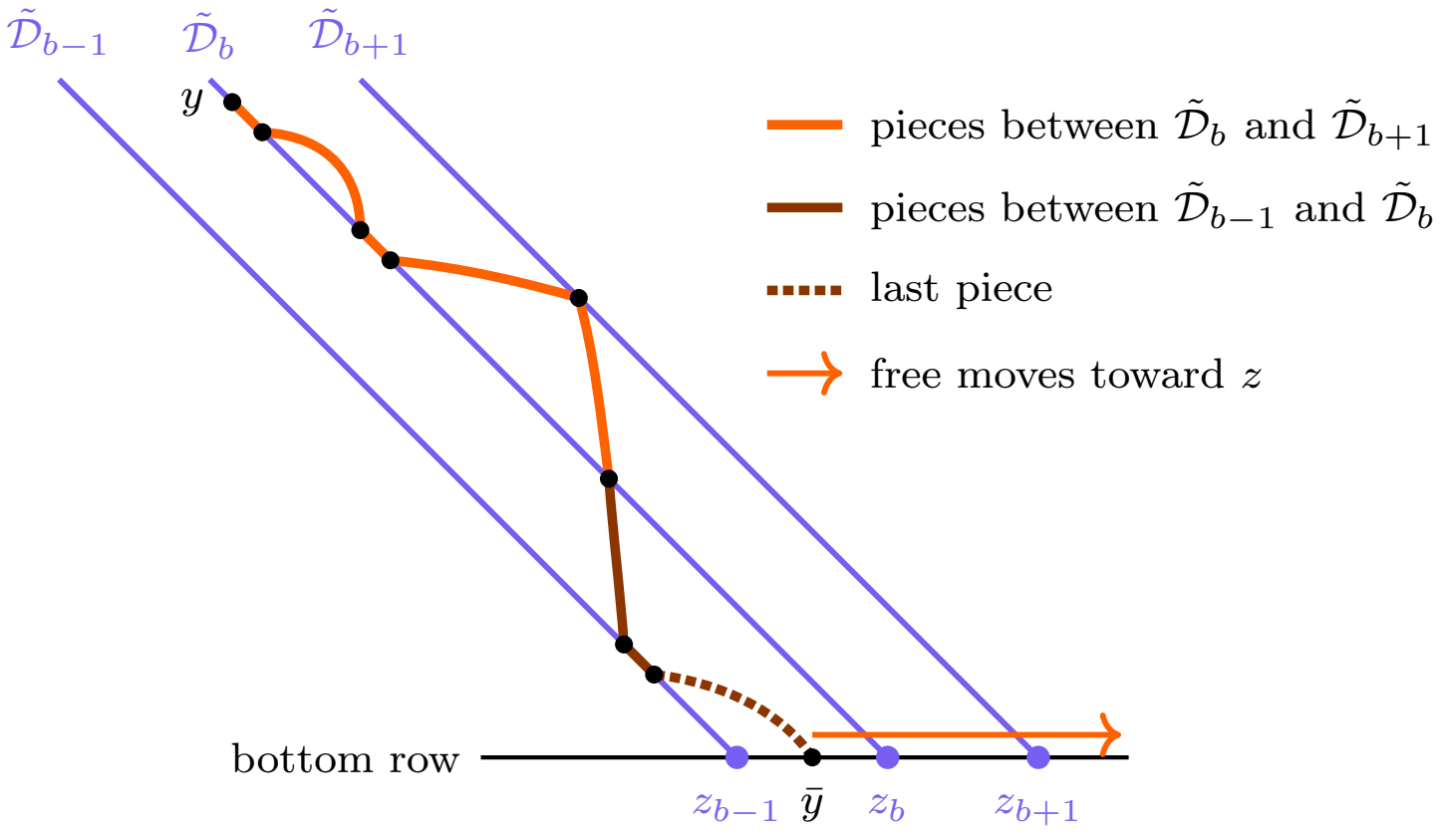

The difficult detours are those between distinct rails.  If neither
reference projection of such a detour lies in a cover fragment, the
detour costs at least $q/5$ more than its canonical cost.  We show that
the rest of the path cannot make up for this excess.  To compare the
rest of the path with the route that follows the rail from the start
of the detour, we extend both by free moves along the bottom row to
the same vertex, the bottom endpoint of a rail to the right of the
path.  \Cref{lem:strip_path_savings} charges the saving of the extended
rest of the path, relative to its canonical cost, to the selected edits
that it passes, so this saving is at most twice the load of the strips
around the path; see \cref{fig:single_path_savings}.  If this load is
less than $q/10$, the saving is less than $q/5$, so the route along the
rail is cheaper than the path from the start of the detour,
contradicting optimality.  The informal counterpart is the third case
in \cref{sec:overview_marks}.

\begin{lemma}[Dense-cover containment from a large period or small load]
\label{lem:blocks_with_few_marks}
Let $\eA$ be a shortest path from a vertex of the top row of
$\eAG(P,T)$ to the bottom row, with $\cost(\eA)\le d$.
Suppose that $q>\kappa$ or that there is an integer interval
$I\subseteq\prD$ such that every rail met by $\eA$ has index strictly
between $\min I$ and $\max I$, and $\sum_{c\in I}\mu(c)<q/10$.
Then, for every detour of $\eA$, with earliest vertex $x$ and latest
vertex $y$, $\phP(x,y)$ is contained in a fragment of $\Es_P$ or
$\phT(x,y)$ is contained in a fragment of $\Es_T$.
\end{lemma}
\begin{proof}
We handle detours at the ends of $\eA$ by their length and the other
detours by their rail contacts.  Fix a detour $\mathcal X$ of $\eA$
with earliest vertex $x$ and latest vertex $y$, and put
$\hF\coloneqq\phP(x,y)$ and $\hG\coloneqq\phT(x,y)$.  Let $f$ and $g$
be the numbers of edits of $\A_P$ and of $\A_T$ that $\mathcal X$
passes, so that $\cancost(x,y)=f+g$.  The detour is tight and shares no
reference-diagonal edge with a rail.

\emph{Case 1: $\mathcal X$ contains the earliest or the latest edge of
$\eA$.}  Then, $\hF$ is a prefix or a suffix of $\hP$.
By \cref{lem:noblue},
$\floor{|\hF|/q}\le\cost(\mathcal X)+f+g\le\kappa$, so
\[
    |\hF|<(\kappa+1)q\le(5d+1)q\le10dq,
\]
using $d\ge1$.  Hence, \cref{obs:short_boundary_fragments} puts $\hF$
in a fragment of $\Es_P$.

Otherwise, $\mathcal X$ is preceded and followed by removed rail
edges (\cref{def:detours}), so $x\in\eD{b}$ and $y\in\eD{b'}$ for some
$b,b'\in\canD$.

\emph{Case 2: $b=b'$.}  Tightness and \cref{lem:rail_stability} give
\[
    \cost(\mathcal X)
    =\tdtob(x)-\tdtob(y)
    \le\cancost(x,y).
\]
Since $\mathcal X$ contains an edge, $x\ne y$, so
\cref{cor:nobluesame} puts one of the two reference projections in a
fragment of its cover.

\emph{Case 3: $b\ne b'$.}  By \cref{lem:blue_jump_limit},
\[
    d\ge\cost(\eA)\ge q-(d_P+d_T),
\]
so $q\le\kappa$.  Hence, the second alternative of the assumption
holds.  Put $M\coloneqq\sum_{c\in I}\mu(c)$, and suppose, for a
contradiction, that neither $\hF$ nor $\hG$ is contained in a fragment
of its cover.  Then, neither fragment is $(5q,0)$-dense
(\cref{obs:dense_cover_structure}), so
\cref{obs:fragment_cost_restriction} gives
$\floor{|\hF|/q}\ge5f$ and $\floor{|\hG|/q}\ge5g$.  Since the
endpoints lie on distinct rails, \cref{lem:offset_change} gives
$q\le|b-b'|q\le\cost(\mathcal X)+f+g$.
Together with \cref{lem:noblue}, this yields
\[
    \cost(\mathcal X)+f+g
      \ge\max(q,5f,5g)
      \ge\tfrac15\cdot q+\tfrac25\cdot5f+\tfrac25\cdot5g
      =q/5+2f+2g,
\]
where the second inequality bounds the maximum from below by a
weighted average.  Therefore,
\begin{equation}\label{eq:few_mark_hop_lower}
    \cost(\mathcal X)\ge\cancost(x,y)+q/5.
\end{equation}

It remains to show that the rest of $\eA$ cannot save enough to pay
for this excess.  Let $\bar y$ be the last vertex of $\eA$, which lies
on the bottom row, and put $z\coloneqq z_{\max I}$.  The vertex $y$
lies on $\eD{b'}$, strictly between the extreme rails of $I$, and $\eA$
meets neither extreme rail, so, by \cref{lem:canonical_rails}, the
suffix of $\eA$ from $y$ stays strictly between them.  In particular,
$\bar y$ precedes $z$ on the bottom row.  Let $\mathcal B$ be this
suffix followed by the path along the bottom row from $\bar y$ to $z$.
The suffix is a shortest path from $y$ to the bottom row, and its
edges inside the bottom row cost $0$, because otherwise, stopping
earlier would be cheaper.  Hence, $\cost^\circ(\mathcal B)=\tdtob(y)$.

\begin{claim}[Savings of the remaining suffix]
\label{clm:remaining_suffix_savings}
The path $\mathcal B$ satisfies $\cancost(y,z)-\cost^\circ(\mathcal B)\le2M$.
\end{claim}
\begin{claimproof}
Cut $\mathcal B$ at its rail contacts into pieces.  A piece that
starts at a rail contact of $\eA$, on a rail $\eD{c}$, lies in the
closed strip between $\eD{c}$ and one of its neighbors
(\cref{lem:canonical_rails}), and
$c-1,c,c+1\in I$.  Every other piece lies inside the bottom row,
between two neighboring bottom endpoints $z_c$ and $z_{c+1}$ with
$c,c+1\in I$.  Hence, \cref{lem:strip_path_savings} applies to every
piece for some strip $c$ with $c,c+1\in I$, and the piece saves at
most twice the number of edits of $S_c$ that it passes.  Both
coordinates are nondecreasing along $\mathcal B$, so every edit is
passed by at most one piece.  Additivity
(\cref{obs:cancost_factorizes}) and \cref{obs:selected_count} now give
\[
    \cancost(y,z)-\cost^\circ(\mathcal B)
       \le2\sum_{c\in I}|S_c|
       \le2M.\claimqedhere
\]
\end{claimproof}

Tightness, \cref{eq:few_mark_hop_lower}, the claim, and additivity
(\cref{obs:cancost_factorizes}) yield
\[
    \tdtob(x)
      =\cost(\mathcal X)+\tdtob(y)
      \ge\cancost(x,z)+q/5-2M
      >\cancost(x,z),
\]
where the last step uses $M<q/10$.  On the other hand,
\cref{lem:rail_stability}, applied to $x$ and the bottom endpoint $z_b$
of its rail, gives $\tdtob(x)\le\cancost(x,z_b)$, and additivity gives
$\cancost(x,z_b)\le\cancost(x,z)$, a contradiction.
\end{proof}

\begin{example}[Running example, continued]\label{ex:running_single_path}
In the setting of \cref{ex:running_rails,ex:running_charging}, consider
the top vertex $u_2=(0,6)$ of the rail $\eD{2}$, whose bottom endpoint
is $z_2=(18,24)$, and let $z=z_4$ be the bottom endpoint of the last of
the five proper rails.  Following $\eD{2}$ passes the two missing
characters but not the deletion of $\A_T$, so $\cancost(u_2,z_2)=2$,
and the bottom row from $z_2$ to $z_4$ passes this deletion, so
$\cancost(u_2,z)=3$.  In contrast, $\tdtob(u_2)=0$ because $P$ occurs
exactly at position $6$.  To illustrate the suffix charging in the proof of
\cref{lem:blocks_with_few_marks}, follow a shortest path from $u_2$ along
$\eD{2}$ to $z_2$ and then move freely along the bottom row to $z_4$.
Cut at its rail contacts, this path consists of the single edges of
$\eD{2}$ and of the bottom-row pieces from $z_2$ to $z_3$ and from $z_3$
to $z_4$.  Only two pieces have a positive saving: the edge
$(8,14)\to(9,15)$, with saving $2$, and the piece from $z_2$ to $z_3$,
with saving $1$, because it passes the deletion of $\A_T$ at no cost.
Each of them passes one edit of $S_2$ (\cref{ex:running_charging}), so
the total saving $\cancost(u_2,z_4)-\tdtob(u_2)=3$ is at most
$2\cdot2=4$.  For $I=\fragmentcc{0}{4}$, the load is $10\ge q/10$, so this
instance illustrates only the charging argument, not the small-load
hypothesis of \cref{lem:blocks_with_few_marks}.
\end{example}

\section{The Planar Variation Theorem}\label{sec:mark_distance_variation}

This section proves the main structural result of the paper.  Fix a row of
the expanded alignment graph and pick one vertex $u_b$ on each rail
$\eD{b}$ of an interval of rails.  The canonical costs $\cancost(u_b)$,
which count the edits of the two supplied alignments that remain after
$u_b$, form a simple baseline: their \emph{total variation}, the sum of
absolute differences between consecutive values, is the canonical cost
between the two extreme vertices.  The suffix distances $\tdtob(u_b)$
deviate from this baseline where edits of $\A_P$ and $\A_T$ cancel on a
rail and where shortest paths leave their rails.  The theorem bounds
their total variation by that of the baseline plus a charge to the mark
loads near the rails.  Bounding every suffix distance separately with
the charging lemmas of \cref{sec:mark_savings} would charge these loads
once for every rail; the theorem charges them once for the entire
row.

Shortest paths from a rail can visit nearby rails, so the charge
involves the loads in the following enlargement of the interval, whose
extreme rails are too far away for these paths to reach
(\cref{lem:comparison_graph_faithfulness}).

\begin{definition}[Enlargement of a rail interval]
\label{def:rail_interval_enlargement}
\let\qed\relax %
Set $\eta^\star\coloneqq\floor{8d/q}+1$.
For a nonempty integer interval
$C=\fragmentcc{c_1}{c_2}\subseteq\canD$, its \emph{enlargement} is
\[
    C^+\coloneqq
        \fragmentcc{c_1-\eta^\star}{c_2+\eta^\star}.
        \tag*{\lipicsEnd}
\]
\end{definition}

\begin{restatable}[Planar variation theorem]{theoremq}{planarvariation}\label{thm:combi}
Let $C=\fragmentcc{c_1}{c_2}\subseteq\canD$ be a nonempty integer
interval with $C^+\subseteq\prD$, and let $\ep\in\fragment{0}{\ell_P}$.
For every $b\in C$, pick a vertex $u_b$ of $\eD{b}$ in row $\ep$ of
$\eAG(P,T)$.
Then,
\[
    \sum_{b=c_1+1}^{c_2}|\tdtob(u_b)-\tdtob(u_{b-1})|
    \le
    \cancost(u_{c_1},u_{c_2})
        +4\sum_{b\in C^+}\mu(b).\qedhere
\]
\end{restatable}

\Cref{thm:overview_variation} in the technical overview is an informal
version of this theorem; \cref{sec:overview_variation} sketches its
proof.  Only the
high-load algorithm uses the theorem: in the exit row of a slab, every
pair of neighboring rails whose suffix distances differ by more than
$q/2$ contributes more than $q/2$ to the total variation, so there are
few such pairs (\cref{sec:block_geometry}).  For eligible block
indices (\cref{sec:powered_ckw_sweep}), the bounds of \cref{sec:mark_savings} suffice.

\begin{example}[Running example, continued]\label{ex:running_variation}
In the setting of \cref{ex:running_rails}, fix a row $\ep$, let
$C=\fragment{0}{4}$, and let $u_b$ be the earliest vertex of $\eD{b}$
in row $\ep$; this is $(\ep,\ep+3b)$ unless $b\ge3$ and $\ep>26-3b$,
in which case it is $(\ep,\ep+3b+1)$.  The table lists the suffix
distances $\tdtob(u_b)$ for selected rows, together with their
variation $V\coloneqq\sum_{b=1}^{4}|\tdtob(u_b)-\tdtob(u_{b-1})|$ and
the canonical cost $\cancost(u_0,u_4)$.  The latter counts the insertion of
$\A_T$ at the step $14\to15$ for $3\le\ep\le14$ and the deletion at the
step $26\to27$ for $\ep\ge15$.
\begin{center}
\begin{tabular}{@{}cccc@{}}
\toprule
row $\ep$ & $\tdtob(u_0),\ldots,\tdtob(u_4)$ & $V$ & $\cancost(u_0,u_4)$\\
\midrule
$0$ & $2,2,0,3,3$ & $5$ & $0$\\
$3$ & $2,2,0,3,2$ & $6$ & $1$\\
$6$ & $2,2,0,2,2$ & $4$ & $1$\\
$9$ & $1,1,0,1,1$ & $2$ & $1$\\
$12$ & $1,0,0,1,1$ & $2$ & $1$\\
$15$ & $0,0,0,1,0$ & $2$ & $1$\\
$18$ & $0,0,0,0,0$ & $0$ & $1$\\
\bottomrule
\end{tabular}
\end{center}
In every row, following $\eD{b}$ yields a shortest path from $u_b$ to the
bottom row.  To explain the excess $V-\cancost(u_0,u_4)$, compare
$\tdtob(u_b)$ with the canonical cost $\cancost(u_b,z)$ to the common
target $z\coloneqq z_4$.  By additivity, these canonical costs do not
increase with $b$, and their variation is exactly $\cancost(u_0,u_4)$.
The suffix distances fall short of them for two reasons: the rails
$\eD{0}$, $\eD{1}$, and $\eD{2}$ reach the bottom row before the
deletion of $\A_T$, which saves $1$, and, above row $9$, the two missing
characters cancel on $\eD{2}$, which saves another $2$.  In row $0$,
for example, the canonical costs are $3,3,3,3,3$.  The first shortfall,
on the rails $0,1,2$, adds $1$ to the variation, and the second, on
rail $2$ alone, adds $4$, which accounts for $V=5$.  The proof of the
theorem pays for such shortfalls with the marks of
\cref{ex:running_marks}.  The instance is too small for the hypothesis
$C^+\subseteq\prD$, since $\eta^\star=\floor{16/3}+1=6$ while only five
rails are proper; it only illustrates the quantities in the theorem.
\end{example}

\begin{figure}[t]
\centering
\begin{tikzpicture}[x=0.42cm,y=-0.42cm,>=stealth]
  \fill[black!6] (0,0) -- (5,0) -- (11,6) -- (6,6) -- cycle;
  \draw[black!30] (-0.6,0) -- (12,0);
  \draw[black!30] (-0.6,6) -- (12,6);
  \node[font=\scriptsize,anchor=east] at (-0.6,0) {row $\ep$};
  \node[font=\scriptsize,anchor=east] at (-0.6,6) {bottom row};
  \draw[line width=2pt,black!55] (6,6) -- (11,6);
  \foreach \s/\l in {0/{b-1},2.5/{b},5/{b+1}} {
    \draw[line width=1.3pt,color2] (\s,0) -- (\s+6,6);
    \fill (\s,0) circle (2pt);
    \fill (\s+6,6) circle (2pt);
    \node[font=\scriptsize,anchor=south] at (\s,0.05) {$u_{\l}$};
    \node[font=\scriptsize,anchor=north] at (\s+6,6.05) {$z_{\l}$};
  }
  \draw[->,thick,color4] (3.3,0.8) .. controls (5.5,1.6) and (6.8,2.6) .. (7.2,3.0)
     .. controls (7.8,3.6) and (8.6,4.3) .. (9.5,4.55);
  \draw[color3,line width=1.4pt] (5.5,3) -- (8,3);
  \fill[color3] (5.5,3) circle (1.8pt);
  \fill[color3] (7.2,3) circle (1.8pt);
  \fill[color3] (8,3) circle (1.8pt);
  \node[font=\scriptsize,color3,anchor=west] at (8.2,3) {cut};
  \node[font=\small] at (6,8.0) {(a)};
\end{tikzpicture}
\hspace{0.8cm}
\begin{tikzpicture}[x=0.62cm,y=-0.62cm,>=stealth]
  \draw[thick] (-0.4,0) -- (7.4,0);
  \draw[thick,color1] (0,0) .. controls (0.2,2.6) and (1.8,3.9) .. (3.5,4.4);
  \draw[thick,color1] (1,0) .. controls (1.0,1.5) and (0.95,2.4) .. (1.188,2.988);
  \draw[thick,color1] (7,0) .. controls (6.8,2.6) and (5.2,3.9) .. (3.5,4.4);
  \draw[thick,color1] (6,0) .. controls (6.0,1.5) and (6.05,2.4) .. (5.812,2.988);
  \foreach \x in {2.5,3.5,4.5} {\draw[black!40] (\x,2.45) -- (3.5,4.4);}
  \draw[line width=1.6pt,color3] (2.5,1.9) -- (4.5,1.9);
  \draw[thick,color4] (2,0) .. controls (2,0.8) and (2.5,0.9) .. (2.5,1.4);
  \draw[thick,color4] (3,0) .. controls (3,0.8) and (3.5,0.9) .. (3.5,1.4);
  \draw[thick,color4] (4,0) .. controls (4,0.8) and (3.5,0.9) .. (3.5,1.4);
  \draw[thick,color4] (5,0) .. controls (5,0.8) and (4.5,0.9) .. (4.5,1.4);
  \foreach \x in {2.5,3.5,4.5} {
    \fill (\x,1.4) circle (1.5pt);
    \draw[->,line width=1.6pt,color4] (\x,1.4) -- (\x,2.42);
    \fill[color3] (\x,1.9) circle (1.7pt);
  }
  \node[font=\scriptsize,anchor=east] at (2.35,2.2) {$S$};
  \node[font=\scriptsize,color3,anchor=west] at (4.6,1.95) {$\Pi$};
  \foreach \i in {0,...,7} {\fill (\i,0) circle (2pt); \node[font=\scriptsize,anchor=south] at (\i,0.05) {$s_{\i}$};}
  \fill (3.5,4.4) circle (2pt); \node[font=\scriptsize,anchor=north] at (3.5,4.45) {$z$};
  \node[font=\small] at (3.5,5.6) {(b)};
\end{tikzpicture}
\caption{(a)~A band graph $G_{I,\ep}$ for $I=\fragment{b-1}{b+1}$
between $R_{b-1}$ and $R_{b+1}$ (purple).  The cut of an edit
selected for strip $b$ (magenta) crosses the edges of strip $b$ that pass
the edit.  (b)~Raising the weights of $S$ (thick)
crossed by $\Pi$ (magenta) affects $s_2,\ldots,s_5$; their path prefixes
(orange) and $\Pi$ form a connected $X$ disjoint from the paths of the
other sources (blue), so the affected sources are consecutive.}
\label{fig:marking_comparison}
\end{figure}
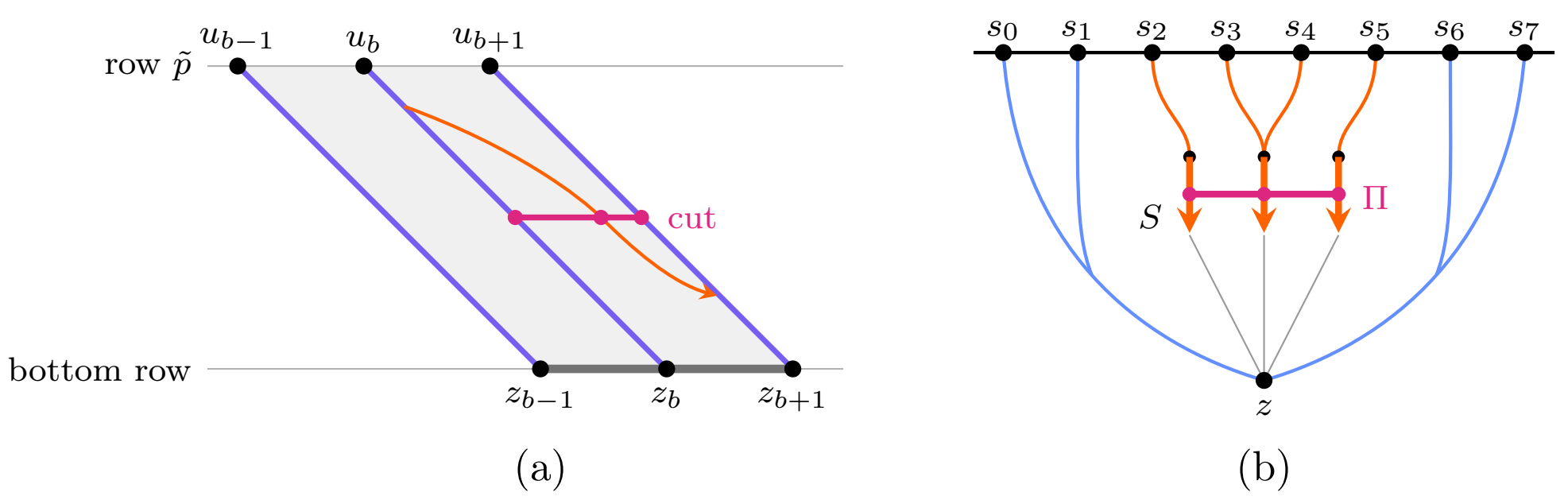

\paragraph*{Outline of the proof.}  The proof has three ingredients.
\begin{enumerate}
    \item The charging lemma of \cref{sec:marking_charging}: a path
    inside the closed strip between $\eD{b}$ and $\eD{b+1}$, with
    endpoints on these rails or the bottom row and free bottom moves,
    saves at most twice the number of edits of $S_b$ that it passes
    (\cref{lem:strip_path_savings}), and $|S_b|\le\mu(b)$
    (\cref{obs:selected_count}).
    \item A \emph{band graph} (\cref{sec:comparison_graph}): the part of
    $\eAG(P,T)$ between the extreme rails of $C^+$, in which moving
    along the bottom row is free; see
    \cref{fig:marking_comparison}(a).  It is a plane DAG that preserves
    the distances $\tdtob(u_b)$ to a common target on the bottom row,
    and the vertices $u_b$ lie in order on its outer face.
    \item A planar lemma (\cref{sec:planar_elimination}): let the
    sources and the target lie on the outer face of a plane DAG, and
    raise by one the weights of a set of edges such that every path
    uses at most one of them and, if there are at least two of them,
    they form a path in the dual graph through inner faces.  Then,
    the distances to the target grow by one for the sources of a block
    that is consecutive along the outer face and stay the same for all
    other sources; see \cref{fig:marking_comparison}(b).
\end{enumerate}
These ingredients combine as follows.  For every edit of $S_b$, the
edges of the closed strip between $\eD{b}$ and $\eD{b+1}$ that pass
this edit, including those of both rails, cross a cut through the strip
(\cref{sec:eliminate_hops}).  A unit increase raises their weights by
one, except that the weight of an edge of a rail or of the bottom row
never exceeds the canonical cost of the edge.  Such edges cross the cut
only at its ends, so the edges whose weights are raised, if there are
at least two of them, still form a path in the dual graph, and the
planar lemma applies.  After two unit
increases for every selected edit, at most $2\sum_{b\in C^+}\mu(b)$ in
total, the charging lemma shows that no path is cheaper than its
canonical cost, and following a rail and then the bottom row attains
this cost.  Hence, the final distances are the canonical costs to the
target, whose total variation telescopes to $\cancost(u_{c_1},u_{c_2})$,
and they exceed the initial distances by one indicator of a block of
sources per unit increase.  Each indicator has total variation at most
$2$, which yields the factor $4$.  \Cref{ex:running_proof} runs the
whole argument on the running example.

\subsection{The Band Graph}\label{sec:comparison_graph}

We now build the plane graph in which we compare distances.  Fix an
integer interval $I\subseteq\prD$ with at least two indices and a row
$\ep\in\fragmentco{0}{\ell_P}$; we exclude the bottom row
$\ep=\ell_P$, where \cref{thm:combi} is trivial because all suffix
distances vanish.  Put
\[
    \mathsf H_{\ep}
      \coloneqq
      \fragment{\ep}{\ell_P}\times\fragment{0}{\ell_T}
    \qquad\text{and}\qquad
    R_b\coloneqq\eD{b}\cap\mathsf H_{\ep}
      \quad\text{for}\quad b\in I.
\]
Call $\mathsf H_{\ep}$ the \emph{suffix rectangle}; it consists of the
rows $\ep,\ldots,\ell_P$ and contains every vertex reachable from
row $\ep$.  By \cref{lem:canonical_rails}, every $R_b$ is a directed
path from the top row of $\mathsf H_{\ep}$, which is row $\ep$, to the
bottom row, and it ends at the vertex $z_b$ of
\cref{def:rail_bottom_endpoints}.

\begin{definition}[Band graph]\label{def:comparison_graph}
The \emph{band graph} $G_{I,\ep}$ has target $z\coloneqq z_{\max I}$
and is the union of the closed strips between consecutive rails of
$I$, restricted to $\mathsf H_{\ep}$.
Every edge inside the bottom row costs $0$; every other edge keeps its
cost in $\eAG(P,T)$.
\end{definition}

Adjacent closed strips (\cref{def:rail_strips}) share exactly their
common rail, with its edges.  Free moves along the bottom row reflect that an alignment of
the whole pattern may end anywhere in the text.  From every vertex of
$R_b$, following $R_b$ and then the bottom row leads to $z$.  We will raise edge weights until
these routes cost exactly their canonical costs and no other route is
cheaper.

\begin{lemma}[Plane band graph]
\label{lem:comparison_graph_planarity}
The graph $G_{I,\ep}$ is a plane DAG.  The target lies on its outer
face, and the vertices of the rails in row $\ep$ occur in rail-index
order on an arc of the outer-face boundary that avoids the target.
\end{lemma}
\begin{proof}
Draw $\eAG(P,T)$ as a rectangular grid, with pattern coordinates
increasing downwards and text coordinates increasing to the right.
Every edge is a grid edge or the diagonal from the top-left to the
bottom-right corner of a grid square.  Since each square contains only
this one diagonal, the drawing is planar, and so is its subgraph
$G_{I,\ep}$.  Every edge increases at least one coordinate and decreases
neither, so the graph is acyclic.  Finally, the
vertices of the rails in row $\ep$ lie in rail-index order along the
top row of $\mathsf H_{\ep}$, whereas the target $z$ lies on the bottom
row strictly below it, on the complementary arc of the outer boundary.
\end{proof}

Restricting the graph to a finite interval of rails could cut off parts
of shortest paths.  The following lemma shows that this does not
happen for shortest paths from rails of $C$: they are too cheap to
reach the extreme rails of $C^+$.

\begin{lemma}[Distance preservation in an enlarged rail interval]
\label{lem:comparison_graph_faithfulness}
Let $C\subseteq\canD$ be a nonempty integer interval with
$C^+\subseteq\prD$, and fix
$\ep\in\fragmentco{0}{\ell_P}$.
For every $b\in C$ and every
$u\in\eD{b}\cap\mathsf H_{\ep}$,
\[
    \dist_{G_{C^+,\ep}}(u,z_{\max C^+})=\tdtob(u).
\]
\end{lemma}
\begin{proof}
Set $I=C^+$ and $z=z_{\max I}$; since $\eta^\star\ge1$, the interval
$I$ has at least three indices, so $G_{I,\ep}$ is defined.  Fix $b\in C$
and $u\in R_b$.

\emph{Lower bound.}  Every edge $x\to y$ costs at least
$\tdtob(x)-\tdtob(y)$: outside the bottom row, this is the triangle
inequality, and inside it, both suffix distances vanish.  Telescoping
along a path to $z$, where $\tdtob(z)=0$, gives the lower bound.

\emph{Upper bound.}  Let $\eA$ be a shortest path in $\eAG(P,T)$ from
$u$ to the bottom row that ends at its earliest vertex $\bar x$ on the
bottom row.  Its cost is $\tdtob(u)\le\cancost(u,z_b)\le d_P+d_T\le4d$
by \cref{lem:rail_stability}, because $\tdtob(z_b)=0$.  If $\eA$ met
an extreme rail $R_c$ of $I$, then $|c-b|\ge\eta^\star$ because
$b\in C$, so \cref{lem:blue_jump_limit} would give
\[
    \cost(\eA)\ge\eta^\star q-(d_P+d_T)
       \ge\eta^\star q-4d>4d,
\]
a contradiction.  Hence, by \cref{lem:canonical_rails}, all
its vertices lie, like $u$, right of $\eD{\min I}$ and left of
$\eD{\max I}$.  Thus, $\eA$ consists of edges of $G_{I,\ep}$ outside the
bottom row, and the bottom row leads from $\bar x$ to $z$ along
edges of cost $0$.  Together, these paths cost $\tdtob(u)$.
\end{proof}

\subsection{Planar Unit Increases}\label{sec:planar_elimination}

Increasing edge weights may increase the distances from many sources.
In a plane graph, however, if every path uses at most one of the
increased edges and a curve through the inner faces crosses these edges
one after another, then the sources whose distances increase are
consecutive along the outer face.  The proof adapts the classical
argument showing that, for roots on the outer face of a plane graph,
the roots whose shortest-path trees contain a fixed edge form an
interval of the outer face~\cite{Klein05}.

\begin{lemma}[Planar unit increase]\label{lem:planar_unit_increase}
Let $G$ be a finite plane DAG with nonnegative integer edge weights and
a target $z$ on its outer face, and let $s_0,\ldots,s_\sigma$ be
vertices with finite distances to $z$ that occur in this order on an
arc of the outer face avoiding $z$.  Let $S$ be a set of edges such
that every path uses at most one of them, and, if $|S|\ge2$, let $\Pi$
be a simple curve that meets $G$ only in one interior point of each
edge of $S$ and otherwise runs inside the inner faces of $G$.  Then, increasing the weights of the edges of $S$ by
one increases the distances to $z$ by one for a possibly empty block
$s_i,\ldots,s_j$ of consecutive sources and leaves the other distances
unchanged.
\end{lemma}
\begin{proof}
Every path uses at most one edge of $S$, so its cost increases by zero
or one, and so does the distance from every source.  We call a source
\emph{affected} if its distance increases, that is, if every shortest
path from it uses an edge of $S$.  It suffices to show that no
unaffected source lies between two affected ones; see
\cref{fig:marking_comparison}(b).

If $S$ is empty, no source is affected, so assume otherwise.
Subdivide every edge of $S$, at its point on $\Pi$ if $|S|\ge2$.  If
$|S|\ge2$, the parts of $\Pi$ between consecutive new vertices form a
path of new edges, which we also call $\Pi$; if $|S|=1$, let $\Pi$
consist of the single new vertex.  The resulting graph $G^\circ$ is
plane, and since all new edges run inside inner faces, the sources and
$z$ lie on its outer face in the same order as in $G$.  For every affected source, choose
a shortest path in $G$; it uses an edge of $S$, and we keep its prefix
in $G^\circ$ that ends at the new vertex on this edge.  Let $X$ be the
union of these prefixes and $\Pi$.  For every unaffected source, choose
a shortest path in $G$ that uses no edge of $S$, and let $Y$ be the
union of these paths.  Then, $X$ is connected because all prefixes end
on $\Pi$, and $Y$ is connected because all its paths end at $z$.

We claim that $X$ and $Y$ are vertex-disjoint.  The new vertices do not
lie on $Y$, because $Y$ uses no edge of $S$.  Suppose that the path in
$Y$ from an unaffected source $s$ meets the prefix in $X$ from an
affected source $s'$ at a vertex $v$ of $G$.  The part of the latter
from $s'$ to $v$ uses no edge of $S$, and neither does the part of the
former from $v$ to $z$.  Both parts are parts of shortest paths through
$v$, so their concatenation is a shortest path from $s'$ that uses no
edge of $S$, contradicting the choice of $s'$.

Finally, if an unaffected source $s_h$ lay between affected sources
$s_i$ and $s_j$, then $s_i$, $s_h$, $s_j$, and $z$ would occur in this
cyclic order around the outer face of $G^\circ$, alternating between
$X$ and $Y$.  This is impossible, because two vertex-disjoint connected
subgraphs of a plane graph cannot contain alternating vertices of one
face.
\end{proof}

\subsection{Cuts through the Strips}\label{sec:eliminate_hops}

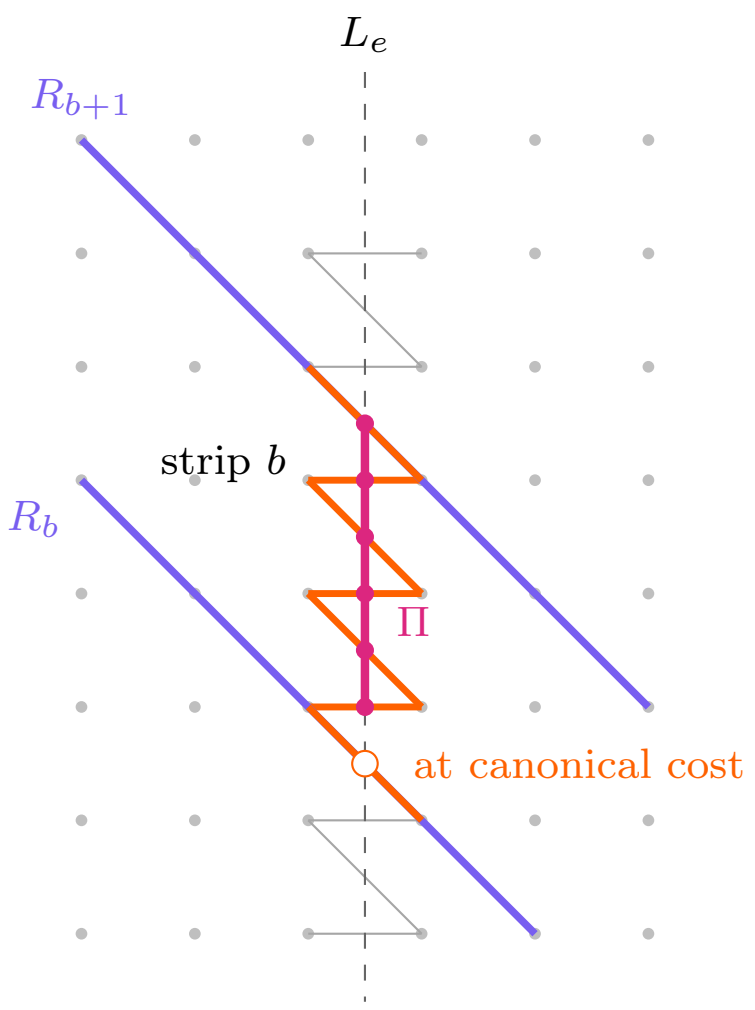
\begin{figure}[t]
\centering
\begin{tikzpicture}[x=0.75cm,y=-0.75cm,>=stealth]
  \foreach \x in {2,...,7} {\foreach \y in {1,...,8} {\fill[black!25] (\x,\y) circle (1.1pt);}}
  \draw[black!35] (4,2) -- (5,2);
  \draw[black!35] (4,2) -- (5,3);
  \draw[black!35] (4,3) -- (5,3);
  \draw[black!35] (4,7) -- (5,7);
  \draw[black!35] (4,7) -- (5,8);
  \draw[black!35] (4,8) -- (5,8);
  \draw[black!60,dashed] (4.5,0.4) -- (4.5,8.6);
  \node[font=\scriptsize,anchor=south] at (4.5,0.4) {$L_e$};
  \foreach \a/\b/\c/\d in {4/4/5/4, 4/4/5/5, 4/5/5/5, 4/5/5/6, 4/6/5/6} {
    \draw[color4,line width=1.3pt] (\a,\b) -- (\c,\d);
  }
  \draw[color2,line width=1.4pt] (2,4) -- (6,8);
  \draw[color2,line width=1.4pt] (2,1) -- (7,6);
  \node[font=\scriptsize,color2,anchor=north east] at (2,4) {$R_b$};
  \node[font=\scriptsize,color2,anchor=south] at (2,1) {$R_{b+1}$};
  \draw[color4,line width=1.3pt] (4,3) -- (5,4);
  \draw[color4,line width=1.3pt] (4,6) -- (5,7);
  \draw[color3,line width=1.6pt] (4.5,3.5) -- (4.5,6);
  \foreach \y in {3.5,4,4.5,5,5.5,6} {\fill[color3] (4.5,\y) circle (1.7pt);}
  \node[font=\scriptsize,color3,anchor=west] at (4.6,5.25) {$\Pi$};
  \draw[color4,fill=white] (4.5,6.5) circle (2.4pt);
  \node[font=\scriptsize,color4,anchor=west] at (4.75,6.5) {at canonical cost};
  \node[font=\scriptsize] at (3.25,3.85) {strip $b$};
\end{tikzpicture}
\caption{The cut set $K_e$ of \cref{lem:cut_sets} for an edit $e\in S_b$
of $\A_T$, in expanded coordinates.  The orange edges of the closed
strip, including those of both rails, cross the dashed line $L_e$
between the columns of $e$ and form $K_e$.  The edge of $R_b$ already has
its canonical cost and is left unchanged (hollow); the other crossings stay
consecutive along $L_e$, and the part $\Pi$ of $L_e$ between them
(magenta) meets no other edge.}
\label{fig:cut_sets}
\end{figure}

Let $b,b+1\in I$, and let $e\in S_b$ be an edit selected for strip $b$
(\cref{def:selected_edits}).  If $e$ is an edit of $\A_P$, between its
vertices $i$ and $i+1$, let $L_e$ be the horizontal line between the
rows $i$ and $i+1$ of the drawing of $G_{I,\ep}$; if $e$ is an edit of
$\A_T$, between its vertices $i$ and $i+1$, let $L_e$ be the vertical
line between the columns $i$ and $i+1$.  An edge $x\to y$
crosses $L_e$ if and only if it passes $e$, that is, if
$\cancost(x,y)$ counts $e$.  Let $K_e$ be the set of edges in the
closed strip between $R_b$ and $R_{b+1}$ that pass $e$, including
edges of both rails; see
\cref{fig:marking_comparison}(a) and \cref{fig:cut_sets}.

For this pair $b,e$, a \emph{unit increase} raises the weight of every
edge of $K_e$ by one, except that an edge of a rail or inside the
bottom row keeps its weight once this weight equals its canonical
cost.  Initially, such edges cost at most their canonical costs: by
\cref{lem:canonical_cost}, a rail edge does, because its reference
projection costs zero, and an edge inside the bottom row costs $0$ by \cref{def:comparison_graph}.
Hence, their weights never exceed their canonical costs.  An edge of
a rail $R_c$ with $\min I<c<\max I$ lies in two closed strips, so unit
increases for both strips may raise its weight, but never above its
canonical cost.

\begin{lemma}[Cut sets]\label{lem:cut_sets}
Consider a unit increase, applied after any number of other unit
increases.  The set of edges whose weights it raises satisfies the
hypotheses on $S$ in \cref{lem:planar_unit_increase} for $G_{I,\ep}$.
\end{lemma}
\begin{proof}
Consider the unit increase for $b$ and $e$.  Both coordinates are
nondecreasing along every edge, so every path crosses $L_e$ at most
once and hence uses at most one edge of $K_e$.

It remains to find the curve $\Pi$.  The edges of $K_e$ cross $L_e$
where it runs through the part of the band bounded by $R_b$,
$R_{b+1}$, row $\ep$, and the bottom row.  Both rails are monotone, so
this part of $L_e$ is a segment.  Every edge that crosses this segment lies in the
closed strip and passes $e$, so it lies in $K_e$.  An edge of a rail
or inside the bottom row can cross $L_e$ only at an end of the
segment, so the edges of $K_e$ whose weights stay unchanged do not
separate the crossings of the edges whose weights are raised.  If at
least two weights are raised, let $\Pi$ be the part of $L_e$ from the
first to the last of these crossings.  It passes through no vertex, it
meets $G_{I,\ep}$ only in one interior point of each edge whose weight
is raised, and otherwise it runs inside the band, hence inside inner
faces.
\end{proof}

Two unit increases per selected edit pay for all savings.  Since the
weights of edges of rails and of the bottom row are never raised above
the canonical costs of these edges, the route along a rail and then the bottom row does not become
more expensive than its canonical cost.

\begin{lemma}[Final distances are canonical]\label{lem:canonical_lower_bound}
Let $G'$ be obtained from $G_{I,\ep}$ by applying two unit increases
for every $b$ with $b,b+1\in I$ and every $e\in S_b$.  Then, for every
vertex $v$ of a rail $R_c$, $c\in I$, the distance from $v$ to $z$ in $G'$ equals
$\cancost(v,z)$.
\end{lemma}
\begin{proof}
We first show that every edge of a rail or inside the bottom row ends
with its canonical cost, and then we cut an arbitrary path into
pieces.

\emph{Edges of rails and of the bottom row.}  Let $x\to y$ be such an
edge, and let the closed strip between $R_b$ and $R_{b+1}$, with
$b,b+1\in I$, contain it.  By \cref{lem:strip_path_savings}, applied to
the edge alone, $\cancost(x,y)$ exceeds the initial weight of the edge
by at most twice the number of edits of $S_b$ that the edge passes.
For each such edit $e$, the edge lies in $K_e$, so each of the two
unit increases for $b$ and $e$ raises its weight by one until this
weight reaches $\cancost(x,y)$.  Hence, the final weight is
$\cancost(x,y)$; increases for the other closed strip containing the
edge, if any, do not change this.

\emph{Arbitrary paths.}  Consider a path from $v$ to $z$.  Split it at
its rail contacts and at its earliest vertex on the bottom row, and
split the part after that vertex into single edges.  A piece that is a
single edge of a rail or inside the bottom row costs its canonical
cost.  Every other piece $\mathcal A$ starts on a rail, ends on a rail
or the bottom row, and has no internal rail contact, so, by
\cref{lem:canonical_rails}, it lies in the closed strip between $R_b$
and $R_{b+1}$ for some $b$ with $b,b+1\in I$, and none of its edges
lies on a rail or inside the bottom row.  Every edit
$e\in S_b$ that $\mathcal A$ passes is passed by exactly one edge of
$\mathcal A$; this edge lies in $K_e$ and receives both unit
increases.  Hence, the weight of $\mathcal A$ in $G'$ exceeds its
initial cost by twice the number of edits of $S_b$ that it passes, and
\cref{lem:strip_path_savings} shows that this weight is at least the
canonical cost between the endpoints of $\mathcal A$.  By additivity
(\cref{obs:cancost_factorizes}), the path costs at least
$\cancost(v,z)$ in~$G'$.  Following $R_c$ and then the bottom
row attains this bound, because every edge on this route has its
canonical cost.
\end{proof}

\subsection{Proof of the Planar Variation Theorem}
\label{sec:mark_distance_bounds_proof}

\planarvariation*
\begin{proof}
If $\ep=\ell_P$, then all suffix distances $\tdtob(u_b)$ vanish, and
there is nothing to prove.  Otherwise, construct $G_{C^+,\ep}$, and
put $M^+\coloneqq\sum_{b\in C^+}\mu(b)$.
By \cref{lem:comparison_graph_faithfulness}, the distances from the
vertices $u_b$ to the target $z$ in this graph are $\tdtob(u_b)$, and
\cref{lem:comparison_graph_planarity} places these vertices and $z$ on
the outer face as required by \cref{lem:planar_unit_increase}.  For
every $b$ with $b,b+1\in C^+$ and every $e\in S_b$, apply two unit
increases, and let $r$ be their total number.  By
\cref{obs:selected_count}, we have $|S_b|\le\mu(b)$ for every such $b$,
so
\[
    r=2\sum_{b=\min C^+}^{\max C^+-1}|S_b|\le2\sum_{b\in C^+}\mu(b)=2M^+.
\]
By \cref{lem:cut_sets,lem:planar_unit_increase}, the $i$-th unit
increase increases the distances from the vertices $u_b$ with $b$ in some
integer interval $J_i$ by one and leaves the others unchanged.
The final distances are $\cancost(u_b,z)$ by
\cref{lem:canonical_lower_bound}.  Write $[b\in J_i]$ for $1$ if
$b\in J_i$ and $0$ otherwise.  Then,
\[
    \cancost(u_b,z)=\tdtob(u_b)+\sum_{i=1}^{r}[b\in J_i]
    \qquad\text{for every }b\in C.
\]
For every $b\in\fragmentcc{c_1+1}{c_2}$, the vertex $u_b$ lies to the
right of $u_{b-1}$ in row $\ep$, so additivity
(\cref{obs:cancost_factorizes}) gives
\[
    \cancost(u_{b-1},z)-\cancost(u_b,z)
       =\cancost(u_{b-1},u_b)\ge0.
\]
The variation of the final distances therefore telescopes to
$\cancost(u_{c_1},u_{c_2})$.  The indicator of an interval has total
variation at most $2$, so the variation of the initial distances is
at most this canonical cost plus $2r\le4M^+$.
\end{proof}

\begin{example}[Running example: the proof at work]\label{ex:running_proof}
We run the proof on the instance of \cref{ex:running_rails}, with
$I=\fragment{0}{4}$, the top row $\ep=0$, and $u_b=(0,3b)$.  (The
instance is too small for $C^+\subseteq\prD$, but the steps below do
not need this assumption.)  Following its rail is a shortest path from
every $u_b$ (\cref{ex:running_variation}), so
the distances to $z=z_4$ in $G_{I,0}$ are still
$\tdtob(u_b)=2,2,0,3,3$, with total variation $5$.  By
\cref{ex:running_charging}, both edits of $\A_T$, the insertion at the
step $14\to15$ and the deletion at the step $26\to27$, are selected for
every strip, and no edit of $\A_P$ is.  Hence, for each
$b\in\fragment{0}{3}$, the closed strip between $R_b$ and $R_{b+1}$ has
two cut sets: its edges that cross the vertical line between the
columns $14$ and $15$, and those that cross the line between the
columns $26$ and $27$.  In \cref{fig:running_reference},
these lines run through the middle of the magenta shading.

The rail edge $(8,14)\to(9,15)$ of $R_2$, on which the two missing
characters cancel, lies in the closed strips on both sides of $R_2$
and crosses the first line; in \cref{fig:running_reference}, it is
the diagonal of the cell where the blue shading crosses the left
magenta shading.  Its weight
rises from $0$ to its canonical cost $2$ and stays there, although the
insertion is selected for both strips.  Every detour around it,
through strip $1$ or strip $2$, crosses the same line with an edge off
the rails, so it costs at least $2$ as well.  The free bottom edge
$(18,26)\to(18,27)$ between $z_2$ and $z_3$ passes the deletion and
lies only in the closed strip between $R_2$ and $R_3$.  Its weight
rises from $0$ to its canonical cost $1$ in the first unit increase for
the deletion and this strip, and it stays there in the second.  The final distances are the
canonical costs $3,3,3,3,3$, and their variation is
$0=\cancost(u_0,u_4)$.

To see the blocks of affected sources, apply the unit increases strip
by strip in increasing order and, within each strip, those for the
insertion before those for the deletion.  Then, only three unit
increases change a distance from a source: each of the two increases
for the insertion in strip $2$ raises only the distance from $u_2$, and
the first increase for the deletion in strip $2$ raises the distances
from $u_0$, $u_1$, and~$u_2$.  The increases for strip $1$ already
bring the rail edge $(8,14)\to(9,15)$ to its canonical cost, but they
do not change the distance from $u_2$, because the detour
$(8,14)\to(8,15)\to(9,15)$ through strip $2$ still costs $0$.  On $\fragment{0}{4}$, the indicator of
$\{2\}$ has total variation $2$, and that of $\fragment{0}{2}$ has
total variation $1$.  These three indicators account exactly for the
drop in variation: $2+2+1=5$.
\end{example}

\section{Monge Matrices, Distance Matrices, and Ferns}\label{sec:fern_toolbox}

The algorithms in the remaining sections compute and multiply
distance matrices of pairs of fragments of $P$ and $T$.  This section
collects the necessary tools, most of which are specializations of
results of \cite{GK25,gk24}.
\Cref{sec:monge_toolbox} recalls the compact representation of
bounded-difference Monge matrices, which allows multiplying
$(\Delta+1)\times(\Delta+1)$ distance matrices in
$\tilde\Oh(\Delta)$ time.  \Cref{sec:ferns} adds backward edges to
alignment graphs, so that all distances between their vertices are
finite, and introduces \emph{ferns}, which are distance matrices that
are correct only up to a threshold.  \Cref{sec:factor_products} shows
that the distance matrix of a pair of strings agrees up to the
threshold with the product of the matrices of its factors, provided
that the cuts between the factors are chosen with enough slack.
Finally,
\cref{sec:fern_tables} constructs ferns for all pairs of strings from
two families in which consecutive strings differ by few edits; this is
how the fern \emph{catalogue} of \cref{sec:ckw_common} is built.  The
informal counterpart of this section is the discussion of distance
matrices in \cref{sec:overview_blocks}.

\subsection{Core-Based Monge Matrices}\label{sec:monge_toolbox}

We compose distance matrices with the min-plus product, which is associative.

\begin{definition}[Min-plus product]
\label{def:min_plus_product}
The \emph{min-plus product} of a $p\times s$ matrix $A$ and an
$s\times r$ matrix $B$ with entries in
$\mathbb R\cup\{+\infty\}$ is the $p\times r$ matrix $A\oplus B$ with
entries
\[
    (A\oplus B)[i,j]
      \coloneqq
      \min_{h\in\fragmentco{0}{s}}
          \bigl(A[i,h]+B[h,j]\bigr),
\]
where rows and columns are indexed from zero.

For a square matrix $A$
and an integer $u\ge1$, $A^{\oplus u}$ denotes the product of $u$
copies of $A$.
\end{definition}

\begin{definition}[Monge matrices]
\label{def:monge}
A real-valued matrix $A$ is \emph{Monge} if
\[
    A[i,j]+A[i',j']\le A[i,j']+A[i',j]
    \qquad\text{for } i<i' \text{ and } j<j'.
\]
For a constant $\beta\ge0$, it is \emph{$\beta$-bounded-difference}
if horizontally or vertically adjacent entries differ in absolute
value by at most $\beta$.
\end{definition}

A Monge matrix is determined by its topmost row, its leftmost column, and
the second-order differences of its entries, which are nonnegative.
For integer-valued bounded-difference Monge matrices, few of these differences are
nonzero, which yields a space-efficient representation.

\begin{definition}[Condensed representation]
\label{def:condensed_representation}
For a $p\times s$ matrix $A$, its \emph{density matrix} has entries
\[
    \dens{A}[i,j]
      =A[i,j+1]+A[i+1,j]-A[i,j]-A[i+1,j+1]
      \qquad\text{for }i\in \fragmentco{0}{p-1}\text{ and }j\in \fragmentco{0}{s-1}.
\]
The \emph{core} of $A$ lists the nonzero entries of $\dens{A}$ together
with their positions.  The core, the topmost row, and the leftmost
column form the \emph{condensed representation} of $A$.
\end{definition}

Algorithms access Monge matrices through a data structure built from
this representation.

\begin{fact}[{Core-based matrix oracle; \cite[Lemma~4.4]{GK25} and \cite[Definition~5.3]{gk24}}]
\label{fact:cmo}
Every integer-valued $p\times s$ Monge matrix $A$ admits a data
structure, called a \emph{core-based matrix oracle} (CMO) of $A$,
that supports entry queries in $\Oh(\lg \max(p,s))$ time
and lists the core of $A$ in time linear in the size of the core plus one.
\end{fact}

For integer-valued bounded-difference Monge matrices, the core is
small, and CMOs support the following operations.

\begin{lemma}[CMO interface for $2$-bounded-difference Monge matrices; \cite{GK25,gk24}]
\label{lem:bd_monge_cmo_operations}
Nonnegative integer $2$-bounded-difference Monge matrices are closed
under compatible min-plus products and contiguous submatrix extraction.
A $p\times s$ matrix in this class has core size
$\Oh(\min(p,s))$ and admits a CMO of size $\Oh(p+s)$.
For matrices in this class with positive dimensions at most $n$,
the following operations are supported.
\begin{description}[leftmargin=!,labelwidth=\widthof{\bfseries Multiplication.}]
    \item[Construction.]
        Build a CMO from a condensed representation in
        $\Oh(n\lg n)$ time, or from an explicit $p\times s$
        matrix in $\Oh(ps+n\lg n)$ time.
    \item[Multiplication.]
        Given CMOs of compatible matrices $A$ and $B$, construct
        a CMO of $A\oplus B$ in $\Oh(n\lg n)$ time.
    \item[Cropping.]
        Given a CMO of $A$ and contiguous row and column intervals,
        construct a CMO of the submatrix of $A$ that they select in
        $\Oh(n\lg n)$ time.
    \item[Vector multiplication.]
        Given a CMO of $A$, nonempty contiguous row and column intervals
        $R,C$, and a finite vector $x$ indexed by $C$, compute
        $A[R,C]\oplus x$ in
        $\Oh((|R|+|C|)\lg n)$ time and $\Oh(|R|+|C|)$
        working space, without constructing the submatrix $A[R,C]$.
\end{description}
\end{lemma}
\begin{proof}
The core bound and the closure properties are
\cite[Lemmas~4.9 and~4.10]{GK25}, and the construction,
multiplication, and cropping algorithms are
\cite[Lemmas~4.4, 4.6, and~4.7]{GK25}; see also~\cite{Tiskin15} and~\cite{GGK25}
for the multiplication of unit-Monge and core-sparse Monge matrices,
respectively.  For an explicit input matrix,
we first scan it to obtain its condensed representation, as in
\cite[Lemma~5.4]{gk24}.  For vector multiplication, consider the
implicit matrix with entries $A[r,c]+x[c]$ for $r\in R$ and $c\in C$.
It is Monge, because restriction preserves the Monge inequalities and
the column offsets $x[c]$ cancel in them.  Each of its entries can be
retrieved in $\Oh(\lg n)$ time.  The SMAWK
algorithm~\cite{AggarwalKMSW87} computes all
its row minima using $\Oh(|R|+|C|)$ entry queries and working space
\cite[Theorem~3.12]{gk24}, which gives the claimed bounds; see also
\cite[Lemma~5.35]{gk24}.
\end{proof}

\subsection{Restricted Distance Matrices and Ferns}\label{sec:ferns}

We use the restricted distance matrices and ferns of
\cite[Definitions~5.16 and~5.18]{gk24}, specialized to unit edit costs
and equal input and output widths.  A restricted distance matrix
records the distances from the leftmost $\Delta+1$ vertices of the top row
of the alignment graph to the rightmost $\Delta+1$ vertices of its bottom
row; see \cref{fig:fern_example}.

In an alignment graph, a vertex can reach only the vertices to its
bottom-right.  To make all entries of our distance matrices finite,
we add backward edges.  Their cost is high enough that they do not
change the distance from a vertex to any vertex that it can already
reach.

\begin{definition}[Augmented alignment graph; unit-weight specialization
of {\cite[Definition 3.14]{gk24}}]
\label{def:augmented_ag}
For fragments $F$ and $G$, the \emph{augmented alignment graph}
$\oAG(F,G)$ is obtained from $\AG(F,G)$ by adding, for every edge
$u\to v$ of $\AG(F,G)$, a reverse edge $v\to u$ of cost $2$.
\end{definition}

In $\oAG(F,G)$, every vertex can reach every other vertex, so all
distances are finite.

\begin{lemmaq}[Distance preservation for reachable vertices;
unit-weight specialization of {\cite[Lemma 3.16]{gk24}}]
\label{lem:augmented_distance_preservation}
\let\qed\relax %
For fragments $F$ and $G$, if a vertex $v$ is reachable from a vertex
$u$ in $\AG(F,G)$, then
\[
    \dist_{\oAG(F,G)}(u,v)=\dist_{\AG(F,G)}(u,v).
    \tag*{\lipicsEnd}
\]
\end{lemmaq}

The \emph{offset} of a vertex $(x,y)$ of $\oAG(F,G)$ is $y-x$.

\begin{observation}[Offset changes]
\label{obs:augmented_offset_change}
Along every walk in $\oAG(F,G)$, the offset changes by at most the
cost of the walk.
\end{observation}
\begin{proof}
Diagonal edges keep the offset, and the other edges of $\AG(F,G)$
change it by one and cost one.  Reverse edges change it by at most one
and cost two.
\end{proof}

\begin{definition}[Restricted distance matrix]\label{def:restricted_distance_matrix}
Let fragments $F,G$ start at $x,y$ and end at $x',y'$, respectively.
For an integer $\Delta\in\fragmentcc{0}{|G|}$, the
\emph{$\Delta$-restricted distance matrix} is
the $(\Delta+1)\times(\Delta+1)$ matrix defined by
\[
    \udmat{F}{G}{\Delta}[i,j]
      \coloneqq\dist_{\oAG(F,G)}
        ((x,y+i),(x',y'-\Delta+j))
      \qquad\text{for }i,j\in \fragmentcc{0}{\Delta}.
\]
Its \emph{input states} are the vertices $(x,y+i)$, and its
\emph{output states} are the vertices $(x',y'-\Delta+j)$.
\end{definition}

The matrix $\udmat{F}{G}{\Delta}$ depends only on the strings that $F$
and $G$ spell, not on the locations of these fragments: translating
every vertex by $(-x,-y)$ maps $\oAG(F,G)$ onto the augmented alignment
graph of these strings and preserves edge costs.  If $F$ and
$G$ start at position~$0$, the input and output states are
$(0,i)$ and $(|F|,|G|-\Delta+j)$.

\begin{fact}[{Properties of restricted distance matrices; \cite[Section~5.1]{GK25} and \cite[Fact~5.15]{gk24}}]
\label{fact:restricted_matrix_properties}
Every restricted distance matrix is nonnegative, integer-valued,
Monge, and $2$-bounded-difference.
\end{fact}

By \cref{lem:bd_monge_cmo_operations}, the core and the CMO of a
$\Delta$-restricted distance matrix therefore have size
$\Oh(\Delta+1)$.

We only need distances up to a threshold; larger distances may be
replaced by any values above it.  The following relation identifies
values that agree up to a threshold $k$.

\begin{definition}[$k$-equivalence; {\cite[Definition~4.11]{GK25}}]
\label{def:threshold_equivalence}
For $k\ge0$ and $a,b\in\mathbb R_{\ge0}\cup\{+\infty\}$,
write $a\meq{k}b$ if
$a=b$ or $\min(a,b)>k$.
We extend this relation entrywise to matrices and vectors.
\end{definition}

Thus, $k$-equivalent values agree if one of them is at most $k$.
For example, $1\meq{1}1$ and $2\meq{1}5$, but $1\nmeq{1}2$.

\begin{observation}[Min-plus congruence]
\label{obs:threshold_congruence}
With the usual conventions for $+\infty$, $k$-equivalence is a
congruence for minimum and addition of nonnegative values, and hence
for compatible min-plus products, including matrix-vector products.
In particular, whenever $A\meq{k}B$, we have
$A\oplus x\meq{k}B\oplus x$ for every compatible nonnegative vector
$x$, and $A^{\oplus u}\meq{k}B^{\oplus u}$ for square matrices $A,B$ and
integers $u\ge1$.
\end{observation}
\begin{proof}
This extends \cite[Observation~4.12]{GK25} and \cite[Remark~5.8]{gk24}
to nonnegative extended reals.  Let $a\meq{k}a'$ and $b\meq{k}b'$.  If
$a+b\le k$, then $a,b\le k$, so $a'=a$ and $b'=b$; symmetrically, if
$a'+b'\le k$, then $a=a'$ and $b=b'$.  Hence, $a+b\meq{k}a'+b'$, and
minima are handled similarly.
\end{proof}

\begin{definition}[Fern]\label{def:fern}
For fragments $F,G$, an integer $\Delta\in\fragmentcc{0}{|G|}$, and $k\ge0$,
a \emph{$(\Delta,k)$-fern} of $(F,G)$ is a nonnegative real matrix
$\Phi$ satisfying $\Phi\meq{k}\udmat{F}{G}{\Delta}$.
\end{definition}

\begin{figure}[t]
\centering
\begin{tikzpicture}[x=0.75cm,y=-0.75cm]
  \foreach \i in {0,1} {\foreach \j in {0,...,3} {\draw[black!15] (\j,\i) -- (\j+1,\i+1);}}
  \foreach \i in {0,1,2} {\draw[black!25] (0,\i) -- (4,\i);}
  \foreach \j in {0,...,4} {\draw[black!25] (\j,0) -- (\j,2);}
  \foreach \j/\c in {0/a,1/b,2/a,3/b} {\node[font=\small,anchor=mid] at (\j+0.5,-0.85) {\texttt{\c}};}
  \foreach \i/\c in {0/a,1/b} {\node[font=\small,anchor=mid] at (-0.35,\i+0.5) {\texttt{\c}};}
  \foreach \j in {0,1,2} {\fill[color1] (\j,0) circle (2.6pt);}
  \foreach \j in {2,3,4} {\fill[color3] (\j,2) circle (2.6pt);}
  \foreach \j in {0,1,2} {\node[font=\scriptsize,anchor=south,color1] at (\j,0.05) {$\j$};}
  \foreach \j in {0,1,2} {\node[font=\scriptsize,anchor=north,color3] at (\j+2,2.05) {$\j$};}
  \draw[line width=1.5pt,color2] (0,0) -- (2,2);
  \draw[line width=1.5pt,color2] (2,0) -- (4,2);
  \node[anchor=west,font=\small] at (5,1) {$\udmat{\texttt{ab}}{\texttt{abab}}{2}=
    \begin{pmatrix}0&1&2\\1&2&1\\2&1&0\end{pmatrix}$};
\end{tikzpicture}
\caption{The restricted distance matrix of \cref{ex:fern}.  Rows
of the matrix correspond to the input states on the top row (blue),
and columns correspond to the output states on the bottom row
(magenta).  The two purple paths realize the zero entries.}
\label{fig:fern_example}
\end{figure}

\begin{example}\label{ex:fern}
Let $F=\texttt{ab}$, $G=\texttt{abab}$, and $\Delta=2$.  The
input states are the vertices $(0,i)$ for $i\in\fragment{0}{2}$, and
the output states are the vertices $(2,2+j)$ for $j\in\fragment{0}{2}$.
Since $i\le2\le2+j$, the output state $(2,2+j)$ is reachable from the
input state $(0,i)$ in $\AG(F,G)$, so, by
\cref{lem:augmented_distance_preservation}, the entry
$\udmat{F}{G}{2}[i,j]$ is the edit distance between $F$ and
$G\fragmentco{i}{2+j}$, which yields
the matrix in \cref{fig:fern_example}.  For example, the entry in row $1$
and column $1$ is $\ed(\texttt{ab},\texttt{ba})=2$.  The density
matrix of $\udmat{F}{G}{2}$ has two nonzero entries, both equal to $2$, at positions
$(0,1)$ and $(1,0)$.  Replacing all entries equal to $2$ by $5$ yields
a $(2,1)$-fern of $(F,G)$, but not a $(2,2)$-fern.
\end{example}

Restricting a fern to its top rows and right columns gives a fern of
smaller width, because the leftmost $\Delta'+1$ input states and the rightmost
$\Delta'+1$ output states are among the original ones.

\begin{lemmaq}[Trimming; {\cite[Lemma~5.20]{gk24}}]
\label{lem:trim}
If $\Phi$ is a $(\Delta,k)$-fern of $(F,G)$, then, for an integer
$\Delta'\in\fragmentcc{0}{\Delta}$ and a threshold $0\le k'\le k$,
\[
    \Phi'\coloneqq
      \Phi[\fragmentcc{0}{\Delta'},\fragmentcc{\Delta-\Delta'}{\Delta}]
\]
is a $(\Delta',k')$-fern of $(F,G)$.
If $\Phi$ is integer-valued, $2$-bounded-difference, and Monge, so is $\Phi'$.
\end{lemmaq}

\subsection{Factor Products}\label{sec:factor_products}

\begin{figure}[t]
\centering
\begin{tikzpicture}[x=0.55cm,y=-0.55cm]
  \fill[color1!12] (0,0) rectangle (5.5,3);
  \fill[color3!12] (1.5,3) rectangle (9.5,7);
  \fill[color2!12] (5.5,7) rectangle (11,10);
  \draw[color1!70] (0,0) rectangle (5.5,3);
  \draw[color3!70] (1.5,3) rectangle (9.5,7);
  \draw[color2!70] (5.5,7) rectangle (11,10);
  \draw[black!40] (0,0) rectangle (11,10);
  \draw[line width=3pt,color4!50] (1.5,3) -- (5.5,3);
  \draw[line width=3pt,color4!50] (5.5,7) -- (9.5,7);
  \node[font=\tiny,color4!80!black,anchor=south east] at (1.5,2.9) {$y_1$};
  \node[font=\tiny,color4!80!black,anchor=south west] at (5.5,2.9) {$y_1+\Delta$};
  \node[font=\tiny,color4!80!black,anchor=north east] at (5.5,7.1) {$y_2$};
  \node[font=\tiny,color4!80!black,anchor=north west] at (9.5,7.1) {$y_2+\Delta$};
  \draw[densely dotted,black!50] (2,0) -- (2,3) (5,0) -- (5,3);
  \draw[densely dotted,black!50] (6,10) -- (6,7) (9,10) -- (9,7);
  \draw[|-|,thick,black!70] (2,3) -- (5,3);
  \draw[|-|,thick,black!70] (6,7) -- (9,7);
  \node[font=\tiny,anchor=south] at (2,-0.1) {$x_1-k$};
  \node[font=\tiny,anchor=south] at (5,-0.1) {$x_1+|Y|-|X|+k$};
  \node[font=\tiny,anchor=north] at (6,10.1) {$x_2-k$};
  \node[font=\tiny,anchor=north] at (9,10.1) {$x_2+|Y|-|X|+k$};
  \draw[color4,line width=1.5pt,line cap=round] (0.5,0) -- (2.5,2) -- (3,2) -- (6,5) -- (6,5.5) -- (10.5,10);
  \fill[color4] (4,3) circle (2.2pt); \fill[color4] (7.5,7) circle (2.2pt);
  \foreach \j in {0,0.5,...,4} {\fill[color1] (\j,0) circle (1.3pt);}
  \foreach \j in {7,7.5,...,11} {\fill[color2] (\j,10) circle (1.3pt);}
  \node[font=\scriptsize,anchor=east] at (-0.1,3) {$x_1$};
  \node[font=\scriptsize,anchor=east] at (-0.1,7) {$x_2$};
  \node[font=\scriptsize,anchor=east] at (-0.1,0) {$0$};
  \node[font=\scriptsize,anchor=east] at (-0.1,10) {$|X|$};
  \node[font=\scriptsize,color1!80!black] at (3.6,0.8) {$(X_0,Y_0)$};
  \node[font=\scriptsize,color3!80!black] at (7.9,4.3) {$(X_1,Y_1)$};
  \node[font=\scriptsize,color2!80!black] at (7.5,9.1) {$(X_2,Y_2)$};
\end{tikzpicture}
\caption{Factor products through covered seams
(\cref{thm:natural_factor_fern_product}) with $z=3$ factors, whose
alignment graphs are shaded.  The orange bars are the shared $\Delta+1$
states, and the dots on the top and bottom are the input and output
states.  For paths of cost at most $k$, the crossing columns on row $x_i$
lie in the black interval $\fragmentcc{x_i-k}{{x_i+|Y|-|X|+k}}$; the
highlighted path crosses the seams within the shared states.}
\label{fig:covered_seam}
\end{figure}
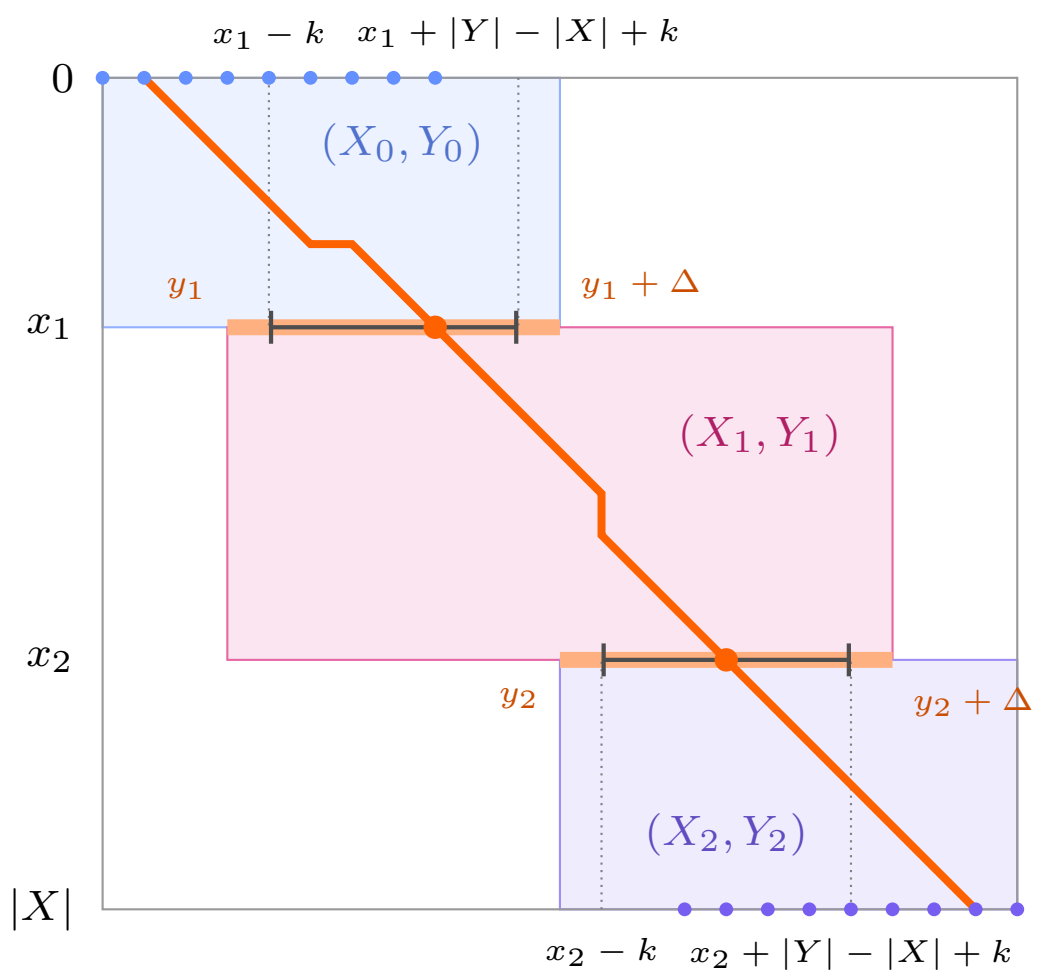

We now compose ferns.  We cut strings $X$ and $Y$ into factors $X_i$
and $Y_i$, respectively, so that consecutive factors $Y_i$ overlap in
$\Delta$ characters, as in \cref{fig:overview_factorization}.  Since alignment
graphs of fragments keep the positions of the underlying strings, each
$\AG(X_i,Y_i)$ is a rectangle in $\AG(X,Y)$, and consecutive rectangles
share $\Delta+1$ states on the row of their common cut; see
\cref{fig:covered_seam}.  Hence, paths in consecutive graphs
$\AG(X_i,Y_i)$ that meet at shared states form a path in $\AG(X,Y)$.
Conversely, a path in $\AG(X,Y)$ splits in this way if it crosses each
cut row within the shared states.  The following condition on the cuts
guarantees this for all paths of cost at most $k$.

\begin{definition}[Cuts and covered seams]\label{def:covered_seam}
\let\qed\relax %
For strings $X,Y$ and integers $\Delta\in\fragmentcc{1}{|Y|}$ and $z\ge1$, take
integer cuts $x_i$ and $y_i$ for $i\in\fragmentcc{0}{z}$, nondecreasing
in $i$, with $x_0=y_0=0$, $x_z=|X|$, and $y_z=|Y|-\Delta$.
They induce the factors
\[
    X_i\coloneqq X\fragmentco{x_i}{x_{i+1}}
    \qquad\text{and}\qquad
    Y_i\coloneqq Y\fragmentco{y_i}{y_{i+1}+\Delta}
    \qquad\text{for }i\in\fragmentco{0}{z}.
\]
For a nonnegative integer $k$, an internal seam $(x_i,y_i)$, where
$i\in \fragmentoo{0}{z}$, is \emph{covered at threshold $k$} if
\[
    \fragmentcc{0}{|Y|}
       \cap\fragmentcc{x_i-k}{x_i+|Y|-|X|+k}
       \subseteq\fragmentcc{y_i}{y_i+\Delta}.
       \tag*{\lipicsEnd}
\]
\end{definition}

A path of cost at most $k$ from an input state to an output state can
cross row $x_i$ only in
$\fragmentcc{x_i-k}{\allowbreak {x_i+|Y|-|X|+k}}$, because its offset
changes by at most $k$ (\cref{obs:augmented_offset_change}).  Coverage places all valid columns from this
interval among the columns $\fragmentcc{y_i}{y_i+\Delta}$ of the shared
states.

\begin{theorem}[Factor products through covered seams]
\label{thm:natural_factor_fern_product}
Let $(X_i,Y_i)_{i=0}^{z-1}$ be induced by the cuts of
\cref{def:covered_seam}, and let $k\in\fragmentcc{0}{|X|}$ be an integer.
If every internal seam is covered at threshold $k$, then
\[
    \udmat{X}{Y}{\Delta}
    \meq{k}
    \bigoplus_{i=0}^{z-1}\udmat{X_i}{Y_i}{\Delta}.
\]
Consequently, the product of any $(\Delta,k)$-ferns of these factors
is a $(\Delta,k)$-fern of $(X,Y)$.
\end{theorem}
\begin{proof}
We follow the path-splitting argument of \cite[Theorem~5.25]{gk24}; see
\cref{fig:covered_seam}.  Write $D\coloneqq\udmat{X}{Y}{\Delta}$ and
$E\coloneqq\bigoplus_{i=0}^{z-1}\udmat{X_i}{Y_i}{\Delta}$.

\emph{Step 1: $D\le E$.}  The augmented alignment graph
$\oAG(X_i,Y_i)$ is the subgraph of $\oAG(X,Y)$ induced by the rectangle
$\fragmentcc{x_i}{x_{i+1}}\times\fragmentcc{y_i}{y_{i+1}+\Delta}$.
Its input state $g_i$ is the vertex $(x_i,y_i+g_i)$, and its output
state $g_{i+1}$ is the vertex $(x_{i+1},y_{i+1}+g_{i+1})$, which is the
input state $g_{i+1}$ of the next factor.  Hence, local paths concatenate into paths in $\oAG(X,Y)$, and
$D\le E$ holds entrywise.

\emph{Step 2: $E\le D$ at all entries where $D\le k$.}  Fix
$g_0,g_z\in\fragmentcc{0}{\Delta}$ with $c\coloneqq D[g_0,g_z]\le k$.
The input state $g_0$ of $D$ is $u_0\coloneqq(0,g_0)$, with offset
$g_0$, and the output state $g_z$ is $u_z\coloneqq(|X|,|Y|-\Delta+g_z)$,
with offset $|Y|-|X|-\Delta+g_z$.  Applying \cref{obs:augmented_offset_change} to a
cheapest walk from $u_0$ to $u_z$ in $\oAG(X,Y)$ shows that
\[
    |Y|-|X|-\Delta+g_z\ge g_0-c.
\]
As $c\le k\le|X|$, the column of $u_z$ is
$|Y|-\Delta+g_z\ge g_0+|X|-c\ge g_0$,
so $u_z$ is reachable from $u_0$ in $\AG(X,Y)$.  By
\cref{lem:augmented_distance_preservation}, some path $\mathcal A$ from
$u_0$ to $u_z$ in $\AG(X,Y)$ has cost $c$.

We now split $\mathcal A$ at the cut rows.  For $i\in\fragmentoo{0}{z}$, let
$u_i=(x_i,v_i)$ be the earliest vertex of $\mathcal A$ in row $x_i$.
The prefix of $\mathcal A$
ending at $u_i$ and the suffix starting at $u_i$ cost at most $c$ each,
so \cref{obs:augmented_offset_change} gives
\[
    v_i-x_i\ge g_0-c\ge-k
    \qquad\text{and}\qquad
    v_i-x_i\le|Y|-|X|-\Delta+g_z+c\le|Y|-|X|+k.
\]
Since also $v_i\in\fragmentcc{0}{|Y|}$, coverage of the seam $(x_i,y_i)$ yields
$v_i\in\fragmentcc{y_i}{y_i+\Delta}$.  Hence, $u_i$ is the output state
$g_i\coloneqq v_i-y_i$ of $\udmat{X_{i-1}}{Y_{i-1}}{\Delta}$ and the
input state $g_i$ of $\udmat{X_i}{Y_i}{\Delta}$.  Similarly, because
$x_0=y_0=0$, $x_z=|X|$, and $y_z=|Y|-\Delta$, the vertex
$u_0$ is the input state $g_0$ of $\udmat{X_0}{Y_0}{\Delta}$, and
$u_z$ is the output state $g_z$ of $\udmat{X_{z-1}}{Y_{z-1}}{\Delta}$.

Since the cut rows are nondecreasing, the vertices $u_0,\ldots,u_z$
appear on $\mathcal A$ in this order.  For $i\in\fragmentco{0}{z}$, let
$\mathcal A_i$ be the part of $\mathcal A$ from $u_i$ to $u_{i+1}$,
which is empty if $u_i=u_{i+1}$.  Since the edges of
$\AG(X,Y)$ do not decrease rows or columns, the vertices of
$\mathcal A_i$ lie in rows $\fragmentcc{x_i}{x_{i+1}}$ and columns
$\fragmentcc{v_i}{v_{i+1}}\subseteq\fragmentcc{y_i}{y_{i+1}+\Delta}$.
Thus, $\mathcal A_i$ is a path in $\AG(X_i,Y_i)$ from the input state
$g_i$ to the output state $g_{i+1}$, and
\[
    E[g_0,g_z]
    \le\sum_{i=0}^{z-1}\udmat{X_i}{Y_i}{\Delta}[g_i,g_{i+1}]
    \le\sum_{i=0}^{z-1}\cost(\mathcal A_i)
    =c=D[g_0,g_z].
\]

\emph{Conclusion.}  By Steps~1 and~2, each entry of $E$ equals the
corresponding entry of $D$ if the latter is at most $k$, and both
exceed $k$ otherwise; thus, $D\meq{k}E$.  Finally, let $\Phi_i$ be a $(\Delta,k)$-fern of $(X_i,Y_i)$
for $i\in\fragmentco{0}{z}$.  By \cref{obs:threshold_congruence},
$\bigoplus_{i=0}^{z-1}\Phi_i\meq{k}E\meq{k}D$, so the product is a
$(\Delta,k)$-fern of $(X,Y)$.
\end{proof}

\subsection{Constructing Tables of Ferns}\label{sec:fern_tables}

In \cref{sec:ckw_common}, we need ferns for all pairs of a pattern
factor and a text factor.  There are $\Oh(d)$ distinct factors on
each side, but they can be arranged in \emph{string sequences}, in
which consecutive factors differ by few edits in total.  Our
construction adapts the common factorization and the all-pairs assembly of
\cite[Lemma~9.10]{CKW22} and \cite[Theorem~9.3]{gk24}, and
\cref{thm:sequence_fern_construction} generalizes the unit-weight case
of the latter in two ways.  First, \cite{gk24} considers pairs from a
single family; like \cite{CKW22}, which constructs seaweed matrices
instead of ferns, we allow two independent families.  Second, both
works measure similarity by the median edit distance, the minimum total
edit distance from the family to a single string.  Listing a family of
median edit distance $\delta$ yields consecutive differences of at most
$2\delta$ edits in total, whereas $\texttt{a},\texttt{a}^2,\ldots,\texttt{a}^s$
differ by $s-1$ edits in total but have median edit distance
$\Theta(s^2)$.  After recalling the primitives and formalizing
string sequences, we prove \cref{thm:sequence_fern_construction}.

\subsubsection{Construction Primitives}

The first primitive maintains two strings under character edits and
reports ferns of the current pair.  As in \cite[Lemma~5.36]{gk24}, it
extracts ferns from the dynamic data structure of
\cite[Theorem~5.10]{GK25}; we use the update and retrieval times that
follow directly from the latter.

\begin{lemma}[Dynamic ferns; based on {\cite[Theorem~5.10]{GK25}}]
    \dglabel{lem:etdyn}
There is a data structure supporting the following operations.
\begin{description}[leftmargin=!,labelwidth=\widthof{\bfseries Character edits.}]
    \item[Initialization.]
        Given nonempty strings $X,Y$ of total length $n$,
        build the data structure in $\Oh(n^2\lg n)$ time.
    \item[Fern retrieval.]
        Given integers $\Delta \in \fragmentcc{1}{|Y|}$ and $k\ge0$, return an
        integer-valued $2$-bounded-difference Monge $(\Delta,k)$-fern
        of the maintained pair $(X,Y)$,
        represented by a CMO of size $\Oh(\Delta)$, in
        $\Oh(n\lg n)$ time.
    \item[Character edits.]
        Apply one character edit to either maintained string, provided
        both strings remain nonempty, in
        $\Oh(n\lg^2 n)$ time.
\end{description}
\end{lemma}
\begin{proof}
We maintain the data structure of \cite[Theorem~5.10]{GK25}, with unit
weights, for the pair $(Y,X)$.  It is built in $\Oh(n^2\lg n)$ time,
returns a CMO of the boundary distance matrix of $(Y,X)$
\cite[Definition~5.6]{GK25} in constant time, and handles a character
edit, a special case of a mega-edit \cite[Definition~5.9]{GK25}, in
$\Oh(n\lg^2 n)$ time.  The entries of this matrix are distances in the
augmented alignment graph of $(Y,X)$ \cite[Definition~5.1]{GK25}, from
its input vertices, which include $(i,0)$ for $i\in\fragmentcc{0}{|Y|}$ in
this order, to its output vertices, which include $(j,|X|)$ for
$j\in\fragmentcc{0}{|Y|}$ in this order.  Swapping the two coordinates
of every vertex maps $\oAG(X,Y)$ onto this graph and preserves edge
costs.  Hence, $\udmat{X}{Y}{\Delta}$ is the contiguous submatrix of
the boundary distance matrix with rows $(i,0)$ for
$i\in\fragmentcc{0}{\Delta}$ and columns $(j,|X|)$ for
$j\in\fragmentcc{|Y|-\Delta}{|Y|}$.  Fern retrieval crops it with
\cref{lem:bd_monge_cmo_operations} in $\Oh(n\lg n)$ time.  The result
is a CMO of $\udmat{X}{Y}{\Delta}$ itself, which is a $(\Delta,k)$-fern
for every $k$, is integer-valued and $2$-bounded-difference Monge by
\cref{fact:restricted_matrix_properties}, and has a CMO of size
$\Oh(\Delta)$ by \cref{lem:bd_monge_cmo_operations}.
\end{proof}

For long strings that are not part of a sequence, we use a static
construction.

\begin{lemma}[Static ferns; unit-weight specialization of
    {\cite[Lemma~6.14]{gk24}}]
\label{lem:static_fern_construction}
Let $X,Y$ be nonempty strings of maximum length $n$.
For integers $\Delta\in\fragmentcc{1}{|Y|}$ and $k\in\fragmentcc{1}{\Delta}$, an integer-valued
$2$-bounded-difference Monge $(\Delta,k)$-fern of $(X,Y)$, represented
by a CMO, can be constructed in
$\Oh(\Delta^2\lg^2 n)$ time in the \modelname model.
\end{lemma}
\begin{proof}
Apply the unit-weight case of \cite[Lemma~6.14]{gk24} with threshold
$\Delta$.  Since $\Delta\le |Y|\le n$, it constructs an integer-valued
$2$-bounded-difference Monge $(\Delta,\Delta)$-fern in
$\Oh(\Delta^2\lg^2(n\Delta))=\Oh(\Delta^2\lg^2 n)$ time.
This is also a $(\Delta,k)$-fern because $k\le\Delta$.  By
\cref{lem:bd_monge_cmo_operations}, its CMO can be constructed from its
$\Oh(\Delta^2)$ entries within the same bound.
\end{proof}

If two strings differ greatly in length, all entries of their
restricted distance matrix are large, and a constant matrix serves as
a fern.

\begin{observation}[Trivial ferns outside the length band]
\label{obs:trivial_length_gap_fern}
Let $X,Y$ be strings, let $\Delta\in\fragmentcc{0}{|Y|}$, and let $k\ge0$ be an
integer.  If every entry of $\udmat{X}{Y}{\Delta}$ exceeds $k$, then
the constant $(\Delta+1)\times(\Delta+1)$ matrix with value $k+1$ is a
$2$-bounded-difference Monge $(\Delta,k)$-fern of $(X,Y)$.  In
particular, this holds whenever
\[
   |Y|-|X|<-k
   \qquad\text{or}\qquad
   |Y|-|X|>k+2\Delta.
\]
A CMO of this matrix can be constructed in $\Oh((\Delta+1)\lg\Delta)$ time.
\end{observation}
\begin{proof}
The constant matrix is Monge, is $0$-bounded-difference, and has an
empty core, so \cref{lem:bd_monge_cmo_operations} constructs its CMO
from its condensed representation in $\Oh((\Delta+1)\lg\Delta)$
time.  It is $k$-equivalent to any matrix whose entries all exceed
$k$, which proves the first assertion.

For $i,j\in\fragmentcc{0}{\Delta}$, every walk defining
$\udmat{X}{Y}{\Delta}[i,j]$ goes from offset $i$ to offset
$|Y|-|X|-\Delta+j$, so, by \cref{obs:augmented_offset_change}, it costs
at least
\[
   \bigl||Y|-|X|-\Delta+j-i\bigr|.
\]
Since $j-i\in\fragment{-\Delta}{\Delta}$, this lower bound exceeds $k$
for all $i,j$ under either of the two length inequalities.
\end{proof}

The construction for long strings cuts every text into blocks and
extends each block by the leftmost characters of the next block, as in
\cref{def:covered_seam}.  A character edit then changes at most two of
the resulting factors.

\begin{observation}[Edits of overlapping factors]
\label{obs:overlapping_factor_edits}
Let $\Gamma$ be a positive integer, and let $S=B_0B_1\cdots B_{z-1}$ be
a string partitioned into blocks of length more than $\Gamma$.
Consider the factors $F_h\coloneqq B_h\cdot B_{h+1}\fragmentco{0}{\Gamma}$
for $h\in\fragmentco{0}{z-1}$ and $F_{z-1}\coloneqq B_{z-1}$.  Suppose
that one character edit is applied to a block $B_h$ at position~$p$,
that is, it substitutes or deletes $B_h[p]$ or inserts a character
that becomes the new $B_h[p]$.  Then, at most three character edits
transform the factors of $S$ into those of the edited string.  They
modify only $F_h$ and, if $h>0$ and $p<\Gamma$, also $F_{h-1}$.
\end{observation}
\begin{proof}
One character edit transforms $F_h$, which contains $B_h$, and the
factors other than $F_{h-1}$ and $F_h$ do not change.  If
$h>0$, then $F_{h-1}$ ends with $B_h\fragmentco{0}{\Gamma}$, which
changes only if $p<\Gamma$.  In this case, a substitution substitutes
one character of $F_{h-1}$.  An insertion inserts a character into
$F_{h-1}$ and pushes its rightmost character out, which we delete.  A
deletion deletes a character of $F_{h-1}$ and brings $B_h[\Gamma]$ into
its suffix, which we append; this character exists because
$|B_h|>\Gamma$.  In each case, at most two character edits transform
$F_{h-1}$.
\end{proof}

\subsubsection{Ferns for String Sequences}

We arrange similar strings in sequences of small total edit distance.

\begin{definition}[String sequence]
\label{def:string_sequence}
A \emph{string sequence} is a sequence
$\boldsymbol S=(S_0,\ldots,S_s)$ of nonempty strings in which
consecutive strings are distinct.  The \emph{drift} of $\boldsymbol S$ is
$\sum_{j=1}^{s}\ed(S_{j-1},S_j)$.
\end{definition}

We now construct the table.  For short strings, we walk through all
pairs of the two sequences, maintaining the current pair in the dynamic
data structure of \cref{lem:etdyn}.  For long strings, we cut all
strings of both sequences at common positions, chosen so that every
seam is covered; the cuts split the strings into parts, and every
string has one factor in each part.  Most parts are far from all edits
between consecutive strings, so the factors of such a part are the
same within each sequence, and we construct only one fern for the
resulting pair.  In the remaining $\Oh(\Lambda)$ parts, the factors
are short, and we maintain the current pair of factors of each of them
in its own instance of \cref{lem:etdyn}.
Walking through all pairs of the two sequences, we replay the edits between consecutive strings on these factors
and maintain the product of the factor
ferns in a balanced tree.

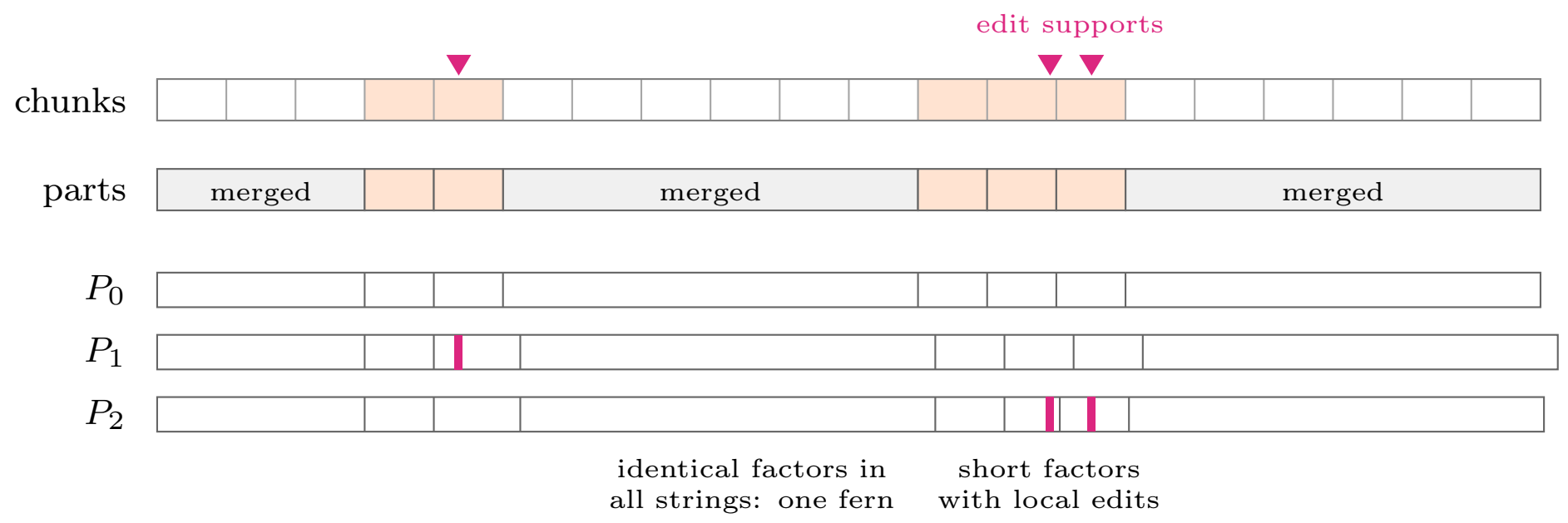
\begin{figure}[t]
\centering
\begin{tikzpicture}[x=0.55cm,y=-0.55cm]
  \foreach \c in {3,4,11,12,13} {\fill[color4!18] (\c,0) rectangle (\c+1,0.6);}
  \foreach \c in {0,...,20} {\draw[black!35] (\c,0) -- (\c,0.6);}
  \draw[black!50] (0,0) rectangle (20,0.6);
  \node[font=\scriptsize,anchor=east] at (-0.2,0.3) {chunks};
  \foreach \e in {4.36,12.91,13.51} {\fill[color3] (\e,-0.05) -- ++(-0.18,-0.3) -- ++(0.36,0) -- cycle;}
  \node[font=\tiny,color3,anchor=south] at (13.2,-0.35) {edit supports};
  \foreach \a/\b in {0/3,5/11,14/20} {\draw[black!60,fill=black!6] (\a,1.3) rectangle (\b,1.9);}
  \foreach \a in {3,4,11,12,13} {\draw[black!60,fill=color4!18] (\a,1.3) rectangle (\a+1,1.9);}
  \node[font=\scriptsize,anchor=mid east] at (-0.2,1.6) {parts};
  \node[font=\tiny,anchor=mid] at (1.5,1.6) {merged};
  \node[font=\tiny,anchor=mid] at (8,1.6) {merged};
  \node[font=\tiny,anchor=mid] at (17,1.6) {merged};
  \foreach \y/\l/\sh/\st in {2.8/P_0/0/0, 3.7/P_1/0.25/0, 4.6/P_2/0.25/-0.2} {
    \draw[black!60] (0,\y) rectangle (20+\sh+\st,\y+0.5);
    \node[font=\scriptsize,anchor=east] at (-0.2,\y+0.25) {$\l$};
    \foreach \c in {3,4,5,11,12,13,14} {
      \pgfmathsetmacro{\cc}{\c+(\c>4.3?\sh:0)+(\c>12.6?\st:0)}
      \draw[black!60] (\cc,\y) -- (\cc,\y+0.5);
    }
  }
  \fill[color3] (4.3,3.7) rectangle (4.42,4.2);
  \fill[color3] (12.85,4.6) rectangle (12.97,5.1);
  \fill[color3] (13.45,4.6) rectangle (13.57,5.1);
  \node[font=\tiny,anchor=north,text width=2.6cm,align=center] at (8.6,5.25) {identical factors in all strings: one fern};
  \node[font=\tiny,anchor=north,text width=2.3cm,align=center] at (12.9,5.25) {short factors with local edits};
\end{tikzpicture}
\caption{The common factorization in Steps~2 and~3 of the proof of
\cref{thm:sequence_fern_construction}.  The root string's coordinates are cut into chunks of length $4\Gamma$.  Chunks whose block, or
the next chunk's block, contains the support of an edit between
consecutive strings (magenta triangles) are marked and kept separate;
maximal runs of the others are merged.  Text cuts lie
$\alpha=k+\Lambda$ positions left of the pattern cuts, a shift not drawn.}
\label{fig:sequence_factorization}
\end{figure}

\begin{theorem}[Ferns from two string sequences]
\label{thm:sequence_fern_construction}
Let $\boldsymbol P=(P_0,\ldots,P_{s_P})$ and
$\boldsymbol T=(T_0,\ldots,T_{s_T})$ be string sequences of maximum
string length $n$ and total drift at most a nonnegative integer $\Lambda$.
For a positive integer $k$ and an integer $\Delta\in\fragmentcc{1}{\min_j|T_j|}$, put
$L\coloneqq \Lambda+k+\Delta+1$.
In $\Oh(L^3\lg^2(nL))$ time in the \modelname model, one can construct
a table $\mathsf H$ whose $(i,j)$-entry is a CMO of an integer-valued
$2$-bounded-difference Monge $(\Delta,k)$-fern of $(P_i,T_j)$.
\end{theorem}
\begin{proof}
Since consecutive strings are distinct, every edit distance between
consecutive strings is at least one, so $s_P+s_T\le\Lambda$ and
$\mathsf H$ has $\Oh((\Lambda+1)^2)$ entries.  We compute optimal
alignments between consecutive strings in both sequences: for two
strings at edit distance $\delta\ge1$, \cite[Lemma~3.20]{gk24} with
thresholds $1,2,4,\ldots$ yields an optimal alignment using
$\Oh(\delta^2)$ \modelname operations.  These distances sum to at
most $\Lambda$, so we use $\Oh((\Lambda+1)^2)$ operations in total.
For $i\in\fragmentoc{0}{s_P}$, let $\mathcal B^P_i$ be the computed alignment from
$P_{i-1}$ to $P_i$, and, for $j\in\fragmentoc{0}{s_T}$, let $\mathcal B^T_j$ be the
computed alignment from $T_{j-1}$ to $T_j$.  The drifts of
$\boldsymbol P$ and $\boldsymbol T$ are the total costs of these
alignments, so their sum is at most $\Lambda$.

Both constructions below \emph{walk} through all pairs $(P_i,T_j)$ in
the following order: for $i=0,\ldots,s_P$, they visit
$T_0,\ldots,T_{s_T}$ if $i$ is even and $T_{s_T},\ldots,T_0$ if $i$ is
odd.  Between consecutive visited pairs, a \emph{move} replaces one
string by its neighbor in its sequence, following the computed
alignment between them (or its reverse).  The walk passes once through
the pattern sequence and $s_P+1\le\Lambda+1$ times through the text
sequence, so it follows every computed alignment at most $\Lambda+1$
times, and its moves follow alignments of total cost
$\Oh((\Lambda+1)^2)$.

\emph{Step 1: large length differences and short strings.}
Put $d_{i,j}\coloneqq|T_j|-|P_i|$ and $d_0\coloneqq d_{0,0}$.  Each edit
changes the length of a string by at most one, so
$\bigl||P_i|-|P_0|\bigr|+\bigl||T_j|-|T_0|\bigr|\le\Lambda$ and, in
particular, $|d_{i,j}-d_0|\le\Lambda$.  If $d_0<-k-\Lambda$, then
$d_{i,j}<-k$ for all $i,j$; if $d_0>k+2\Delta+\Lambda$, then
$d_{i,j}>k+2\Delta$ for all $i,j$.  In both cases, we fill $\mathsf H$
with CMOs of the constant fern of \cref{obs:trivial_length_gap_fern}
in $\Oh((\Lambda+1)^2(\Delta+1)\lg\Delta)$ time and return.  Otherwise, put $\Gamma\coloneqq4L$; then
\begin{equation}\label{eq:sequence_pair_length_bound}
    -k-2\Lambda\le d_{i,j}\le k+2\Delta+2\Lambda<\Gamma.
\end{equation}

If $n\le8\Gamma$, we initialize \cref{lem:etdyn} with $(P_0,T_0)$
and walk through all pairs.  At each move, we apply the edits of the
corresponding alignment, insertions before deletions, so both strings
stay nonempty and have length at most $2n$.  At each visited pair
$(P_i,T_j)$, we retrieve a $(\Delta,k)$-fern, which is possible because
$\Delta\le|T_j|$, and store its CMO in $\mathsf H[i,j]$.  Since the maintained strings
have total length $\Oh(n)$ and $n\le8\Gamma=\Oh(L)$, the
initialization, the $\Oh((\Lambda+1)^2)$ edits,
and the $\Oh((\Lambda+1)^2)$ retrievals take
$\Oh(n^2\lg n+(\Lambda+1)^2n\lg^2 n)=\Oh(L^3\lg^2 L)$ time.

Henceforth, we assume $n>8\Gamma$.  Then, every string in both
sequences has length greater than $n-\Gamma>7\Gamma$.  Indeed, let $S$
be a string of length $n$ in either sequence.  The strings in the same
sequence as $S$ differ in length from $S$ by at most $\Lambda<\Gamma$,
and, by \cref{eq:sequence_pair_length_bound}, the strings in the other
sequence differ in length from $S$ by less than $\Gamma$.

\emph{Step 2: a common factorization; see \cref{fig:sequence_factorization}.}
For an alignment $\mathcal B$ from a string $X$ to a string $Y$ and a
coordinate $a\in\fragmentcc{0}{|X|}$, the \emph{transfer} of $a$ across
$\mathcal B$ is the second coordinate of the earliest vertex of
$\mathcal B$ with first coordinate $a$.  By
\cref{obs:augmented_offset_change}, applied to the prefix of
$\mathcal B$ ending at this vertex, the transfer differs from $a$ by at
most $\cost(\mathcal B)$.  Let $\mathsf{tr}^P_i$ map each coordinate of
$P_0$ to its transfer across $\mathcal B^P_1,\ldots,\mathcal B^P_i$ in
turn, and define $\mathsf{tr}^T_j$ analogously.  Thus, $\mathsf{tr}^P_i$
and $\mathsf{tr}^T_j$ displace coordinates by at most the drifts of
$\boldsymbol P$ and $\boldsymbol T$, respectively; in particular, the
two displacements sum to at most $\Lambda$.

Put $\alpha\coloneqq k+\Lambda$ and $M\coloneqq|P_0|>7\Gamma$.
Partition $\fragmentco{0}{M}$ into chunks of length $4\Gamma$, merging
a shorter remainder at the right end into its left neighbor, so that
every chunk has length in $\fragmentco{4\Gamma}{8\Gamma}$.  For every
chunk boundary $c\in\fragmentoo{0}{M}$, put
\[
    x_i(c)\coloneqq\mathsf{tr}^P_i(c),
    \qquad
    y_j(c)\coloneqq\mathsf{tr}^T_j(c-\alpha),
\]
and put $x_i(0)=y_j(0)=0$, $x_i(M)=|P_i|$, and $y_j(M)=|T_j|$.  The
\emph{block} of a chunk $\fragmentco{c}{c'}$ in $P_i$ is
$P_i\fragmentco{x_i(c)}{x_i(c')}$, and its block in $T_j$ is
$T_j\fragmentco{y_j(c)}{y_j(c')}$.  For each edit of a computed
alignment from a string $S$ to its successor, we mark every chunk
whose block in $S$ contains the support of the edit
(\cref{def:one_argument_alignment_cost}), and the left neighbor of each
such chunk.  Only the empty support of an insertion at a boundary
between two blocks lies in two blocks, so each edit marks at most three
chunks.  The computed alignments have at most $\Lambda$ edits, so at
most $3\Lambda$ chunks are marked.  Keep each marked chunk as a separate
part, and merge each maximal run of unmarked chunks into a single part.
This partitions $\fragmentco{0}{M}$ into $z\le6\Lambda+1$ parts; let
$0=u_0<u_1<\cdots<u_z=M$ be their boundaries, so that part $h$ is
$\fragmentco{u_h}{u_{h+1}}$ for $h\in\fragmentco{0}{z}$.  A part is
\emph{marked} if it is a marked chunk.

For $h\in\fragmentcc{0}{z}$, put $x_{i,h}\coloneqq x_i(u_h)$,
$y_{j,h}\coloneqq y_j(u_h)$ for $h\in\fragmentco{0}{z}$ and
$y_{j,z}\coloneqq|T_j|-\Gamma$.  As in \cref{def:covered_seam}, with
$\Gamma$ in place of $\Delta$, these cuts induce the factors
$P_i\fragmentco{x_{i,h}}{x_{i,h+1}}$ and
$T_j\fragmentco{y_{j,h}}{\allowbreak {y_{j,h+1}+\Gamma}}$ of part $h$, for
$h\in\fragmentco{0}{z}$.

\begin{claim}[Factor lengths]\label{clm:sequence_factor_lengths}
For every pair $(P_i,T_j)$ and every part $h\in\fragmentco{0}{z}$,
the lengths of both factors of part $h$, namely
$P_i\fragmentco{x_{i,h}}{x_{i,h+1}}$ and
$T_j\fragmentco{y_{j,h}}{y_{j,h+1}+\Gamma}$, lie in
$\fragmentoo{\ell_h-\Gamma}{\ell_h+2\Gamma}$, where
$\ell_h\coloneqq u_{h+1}-u_h\ge4\Gamma$.  Consequently, consecutive cuts in $P_i$ and in $T_j$ differ by more
than $2\Gamma$, the cuts satisfy the requirements of
\cref{def:covered_seam} for $X=P_i$ and $Y=T_j$, with $\Gamma$ in
place of $\Delta$, and every factor of a marked part has length
less than $10\Gamma$.
\end{claim}
\begin{claimproof}
The internal cuts of $P_i$ lie within $\Lambda$ of $u_h$, and those of
$T_j$ lie within $\Lambda$ of $u_h-\alpha$.  The outer cuts are
$x_{i,0}=y_{j,0}=0$ and the string ends, which satisfy
$|P_i|\in\fragmentcc{M-\Lambda}{M+\Lambda}$ and
$|T_j|\in\fragmentco{M-\alpha-\Lambda}{M-\alpha+\Gamma+\Lambda}$,
because $|T_0|=M+d_0$ and $0\le d_0+\alpha<\Gamma$ by
\cref{eq:sequence_pair_length_bound}.  The pattern factor of part $h$
spans from the cut at $u_h$ to the cut at $u_{h+1}$, and the text
factor extends $\Gamma$ characters further, except in part $z-1$, where
it ends at $|T_j|$.  Since $\alpha+2\Lambda<\Gamma$, the
length bounds follow.

Consecutive pattern cuts differ by the length of a pattern factor,
which exceeds $\ell_h-\Gamma\ge3\Gamma$.  Consecutive text cuts,
including $y_{j,z}=|T_j|-\Gamma$, differ by the length of a text factor
minus $\Gamma$, which exceeds $\ell_h-2\Gamma\ge2\Gamma$.  Hence, the
cuts increase, and $\Gamma\le|T_j|$ holds because $|T_j|>7\Gamma$.  For
a marked part, $\ell_h<8\Gamma$, so every factor has length less
than $10\Gamma$.
\end{claimproof}

\begin{claim}[Covered seams]\label{clm:sequence_coverage}
For every pair $(P_i,T_j)$, every internal seam is covered at
threshold $k$.
\end{claim}
\begin{claimproof}
Fix an internal boundary $u_h$, where $h\in\fragmentoo{0}{z}$.  The
distances of $x_{i,h}$ from $u_h$ and of $y_{j,h}$ from $u_h-\alpha$
sum to at most $\Lambda$, so
\[
    k=\alpha-\Lambda\le x_{i,h}-y_{j,h}\le\alpha+\Lambda=k+2\Lambda.
\]
Together with \cref{eq:sequence_pair_length_bound}, this yields
\[
    y_{j,h}\le x_{i,h}-k,
    \qquad
    x_{i,h}+d_{i,j}+k-y_{j,h}\le3k+4\Lambda+2\Delta<4L = \Gamma.
\]
Thus,
$\fragmentcc{x_{i,h}-k}{x_{i,h}+d_{i,j}+k}\subseteq\fragmentcc{y_{j,h}}{y_{j,h}+\Gamma}$,
so the seam $(x_{i,h},y_{j,h})$ is covered.
\end{claimproof}

\emph{Step 3: local edits.}
The block of part $h$ in a string is the union of the blocks of its
chunks.  For a pattern string, the blocks of the parts are its
factors.  For a text string, every factor except the rightmost one
consists of the block of its part followed by the $\Gamma$ leftmost
characters of the next block, as in \cref{obs:overlapping_factor_edits}.
By \cref{clm:sequence_factor_lengths}, all blocks of parts have length
more than $2\Gamma$.

Consider consecutive strings $S$ and $S'$ in one of the sequences and
the computed alignment $\mathcal B$ from $S$ to $S'$.  We \emph{replay}
the edits of $\mathcal B$ in their order along $\mathcal B$, so that
each intermediate string consists of a prefix of $S'$ and a suffix of
$S$.  Every internal cut of $S$ moves with the replayed edits until the
replay reaches the earliest vertex of $\mathcal B$ whose first
coordinate is the cut; afterwards, the cut stays fixed, and an
insertion at the cut is applied to the block on its right.  The outer
cuts follow the ends of the string.  Since the internal cuts of $S'$
are the transfers of those of $S$, the replay ends with the blocks of~$S'$,
and each edit is applied to a block that contains its support
in~$S$.  During the replay, every block of a part changes its length by
at most $\cost(\mathcal B)\le\Lambda$, so it remains longer than
$\Gamma$.

\begin{claim}[Local edits]\label{clm:sequence_local_edits}
Replaying the edits of $\mathcal B$ transforms the factors of $S$ into
those of $S'$ by at most $3\cost(\mathcal B)$ character edits, all
applied to factors of marked parts.  Consequently, all strings of
each sequence have the same factor of every unmarked part.
\end{claim}
\begin{claimproof}
An edit is applied to a block that contains its support, so this block
belongs to a marked chunk, which forms a
part $h$ on its own; if $h>0$, the left neighbor of this chunk is also
marked and forms part $h-1$.  By \cref{obs:overlapping_factor_edits},
the edit translates into at most three character edits of the factors
of parts $h$ and $h-1$; for pattern strings, whose factors do not
overlap, it translates into one character edit of the factor of
part $h$.
\end{claimproof}

We implement Steps~2 and~3 without listing the chunks: to find the
chunk of an edit, we binary-search over the chunk boundaries, evaluating $x_i(c)$ or
$y_j(c)$ through the breakpoint representations.  These searches, the
transfers of the $\Oh(\Lambda+1)$ part boundaries, and the translation
of the edits into factor edits take $\Oh((\Lambda+1)^2\lg^2(nL))$
\modelname operations in total.  Every character that a factor edit
inserts or appends is read from $S$ or from the breakpoint
representation.

\emph{Step 4: assembling the table.}
For every unmarked part, we construct a $(\Gamma,k)$-fern of the
common pair of factors (\cref{clm:sequence_local_edits}) using
\cref{lem:static_fern_construction}, which applies because every text
factor has length greater than $\Gamma\ge k$
(\cref{clm:sequence_factor_lengths}).  For every marked part, we
initialize an instance of \cref{lem:etdyn} with the factors of $P_0$
and $T_0$ of this part, both of length less than $10\Gamma$, and
retrieve a $(\Gamma,k)$-fern.  We store these factor ferns at the
leaves of a balanced binary tree, ordered by part, and every
internal node stores the product of its children.

We walk through all pairs.  At each move along a computed alignment, we
apply the factor edits of \cref{clm:sequence_local_edits} to the
instances of the affected marked parts; at a move along the reverse
of a computed alignment, we undo these factor edits in reverse order.
For every part whose factors changed, we then \emph{refresh} its
leaf: we retrieve a new $(\Gamma,k)$-fern, replace the leaf, and
recompute the $\Oh(\lg\Lambda)$ products on its path to the root.

When the walk visits $(P_i,T_j)$, every leaf is a $(\Gamma,k)$-fern of
the current factors of its part: in unmarked parts, these
factors never change, and the other leaves are refreshed after every
change.  By \cref{clm:sequence_factor_lengths,clm:sequence_coverage}
and since $k<|P_i|$, \cref{thm:natural_factor_fern_product} shows that
the root is a $(\Gamma,k)$-fern of $(P_i,T_j)$.  We crop it to rows
$\fragment{0}{\Delta}$ and columns $\fragment{\Gamma-\Delta}{\Gamma}$
using \cref{lem:bd_monge_cmo_operations} and store the resulting CMO
in $\mathsf H[i,j]$; by \cref{lem:trim}, it is the required
$(\Delta,k)$-fern.  Each crop is a separate CMO, so it remains
available while the tree is updated.

Since $z=\Oh(\Lambda+1)$ and $\Lambda+1\le\Gamma$, the initialization
takes $\Oh((\Lambda+1)\Gamma^2\lg^2 n)=\Oh(L^3\lg^2 n)$ time.  Since
the walk follows every computed alignment at most $\Lambda+1$ times,
\cref{clm:sequence_local_edits} bounds the number of factor edits by
$\Oh((\Lambda+1)^2)$; there are at most as many refreshes, and there
are $\Oh((\Lambda+1)^2)$ crops.  The factors of every instance have
length $\Oh(\Gamma)$ throughout, so, by \cref{lem:etdyn,lem:bd_monge_cmo_operations},
a factor edit takes $\Oh(\Gamma\lg^2\Gamma)$ time, a crop takes
$\Oh(\Gamma\lg\Gamma)$ time, and a refresh takes
$\Oh(\Gamma\lg\Gamma\lg\Lambda)$ time.  Altogether,
the walk takes $\Oh((\Lambda+1)^2\Gamma\lg^2(nL))=\Oh(L^3\lg^2(nL))$
time, which proves the theorem.
\end{proof}

\section{Common Factorization and Preprocessing}
\label{sec:ckw_common}

This section builds a common factorization and preprocessing for
\cref{sec:powered_ckw_sweep,sec:block_geometry}.
Its informal counterpart is \cref{sec:overview_blocks}, and
\cref{fig:overview_factorization} illustrates the factorization of a
single block index.  We proceed in four steps.
\begin{enumerate}
    \item \Cref{sec:ckw_exact_factorization} cuts $P$ into consecutive
    pattern factors and $T$ into overlapping text factors at the
    positions that the supplied alignments $\A_P$ and $\A_T$ pair with the multiples
    of~$\tau$.  For each \emph{padded} block index $j$, that is, for all
    but $\Oh(1)$ block indices, it pairs the $i$th pattern factor with
    the $(j+i)$th text factor and adds one pair for the suffix of $P$
    after the last cut.  These \emph{factor pairs of $j$} factorize $P$
    and a fragment of $T$, and each of them receives its restricted
    distance matrix (\cref{sec:ferns}).
    \item \Cref{sec:ckw_exceptional_lists} lists as \emph{exceptional}
    the $\Oh(d)$ factor indices whose reference intervals are near a
    fragment of a dense cover (\cref{sec:dense_covers_marks}), and shows that every other factor is
    \emph{plain}, that is, equal to a fixed periodic string.
    \item \Cref{sec:ckw_catalogues} builds, in $\Oh(d^3\lg^2m)$ time, a
    \emph{catalogue} that returns in $\Oh(\lg m)$ time a fern
    (\cref{sec:fern_toolbox}) of the matrix of any pair.  Most pairs consist
    of plain factors, so $\Oh(d^2)$ precomputed ferns serve all pairs.
    \item \Cref{sec:ckw_load_partition} splits the padded block indices,
    by the mark load (\cref{sec:mark_savings}) near them, into \emph{eligible} ones, handled in
    \cref{sec:powered_ckw_sweep}, and \emph{high-load} ones, handled in
    \cref{sec:block_geometry}.  It also collects the start intervals,
    the exceptional lists, and the high-load indices in a
    \emph{preprocessing record}.
\end{enumerate}

\subsection{Factorization at Padded Block Indices}\label{sec:ckw_exact_factorization}

Recall that the supplied alignments map $P$ and $T$ to the
periodic reference intervals $\Q\fragmentco{0}{y_P}$ and
$\Q\fragmentco{x_T}{y_T}$, respectively.  We place cuts at the
multiples of $\tau$ in the reference and use the alignments to locate
the corresponding cuts in $P$ and $T$.  Since $\tau$ is a multiple of
$q=|Q|$, all reference cuts have the same periodic phase.  We first fix
the parameters; $\kappa$ and $\tau$ are those of
\cref{def:structural_ckw_parameters}.

\begin{definition}[Common factorization and fern parameters]
\label{def:common_factorization_parameters}
\let\qed\relax %
Set
\[
    f\coloneqq21d,
    \qquad
    \Delta\coloneqq2f=42d,
    \qquad
    W\coloneqq3\kappa,
    \qquad
    w\coloneqq\floor{W/q}+1,
    \qquad
    R_*\coloneqq w+\floor{f/q}+1,
\]
and call $f$ the \emph{text margin}.  Let
$t\coloneqq\floor{(y_P-\kappa)/\tau}$.  Using the origins $b_j$ of
\cref{def:jdef}, define the set of \emph{padded} block indices by
\[
    J_{\mathrm{pad}}
      \coloneqq
      \left\{j\in J:
        \fragmentcc{b_j-R_*}{b_j+R_*}\subseteq\prD
      \right\}.
      \tag*{\lipicsEnd}
\]
\end{definition}

\begin{remark}[Roles of the parameters]
\label{rem:margins_and_padding}
An alignment of $P$ with $T$ need
not cross a text cut where it crosses the corresponding pattern cut, so
each text factor extends by the \emph{text margin} $f$ beyond its cuts,
and consecutive text factors overlap in $\Delta=2f$ positions.  The
factor pairs get $\Delta$-restricted distance matrices
(\cref{def:restricted_distance_matrix}), and $f=21d$ is large enough
that all paths considered later cross every seam within an overlap.  The
later algorithms compare distances up to $W=3\kappa$, which bounds the
potential of the assigned occurrence starts and of the tight paths from
them (\cref{lem:ckw_start_relevance,lem:relevant_path}), so the
catalogue stores $(\Delta,W)$-ferns (\cref{def:fern}).  Since $wq>W$,
the vertices relevant to $\eD{b_j}$ at threshold $W$ lie strictly
between the rails $\eD{b_j-w}$ and $\eD{b_j+w}$
(\cref{def:relevant_vertices}), and the exceptional lists widen the
windows of the text factors by $wq$ accordingly.  The algorithms for
$j$ use the rails within distance $R_*$ of $\eD{b_j}$, with
$\floor{f/q}+1$ extra rails for the text margin, and the mark loads
(\cref{def:marking}) of the strips between them; $j$ is padded if all
these rails are proper (\cref{def:proper_rails}).  Finally, the $t$
multiples of $\tau$ in $\fragmentcc{\tau}{y_P-\kappa}$ will be the
internal cuts.
\end{remark}

Several proofs below use the following bounds.

\begin{observation}[Parameter bounds]
\label{obs:common_parameter_bounds}
We have $2\kappa<f<R_*q<38\tau$ and $\tau\le wq$.
\end{observation}
\begin{proof}
Since $\kappa\le5d$, we have $2\kappa<21d=f$.  Moreover,
$R_*q\ge(\floor{f/q}+1)q>f$, and, since $wq\le W+q=3\kappa+q$,
\[
    R_*q\le wq+f+q\le3\kappa+21d+2q\le36d+2q<38\tau,
\]
because $d\le\kappa<\tau$ and $q\le\tau$.  Finally, $wq$ and $\tau$
are multiples of $q$, and $wq>W>\kappa$, whereas $\tau$ is the least
multiple of $q$ exceeding $\kappa$; hence, $\tau\le wq$.
\end{proof}

Padding removes only a constant number of block indices at the two
ends of $J$, and a padded block index exists only if $P$ is long enough
for two internal cuts.

\begin{lemma}[Padded block indices]
\label{lem:padded_ckw_shift_interval}
A block index $j\in J$ is padded if and only if
\[
    x_T+R_*q\le j\tau\le y_T-y_P-R_*q.
\]
Hence, $J_{\mathrm{pad}}$ is an integer interval.  Moreover,
$|J\setminus J_{\mathrm{pad}}|\le78$, and if $J_{\mathrm{pad}}\ne\emptyset$,
then $t\ge2$.
\end{lemma}
\begin{proof}
Since $x_P=0$, we have
$\prD=\{b\in\Z:x_T\le bq\le y_T-y_P\}$ (\cref{def:proper_rails}),
and $b_j=j\tau/q$ (\cref{def:jdef}).  Hence,
$\fragmentcc{b_j-R_*}{b_j+R_*}\subseteq\prD$ holds if and only if
$x_T\le j\tau-R_*q$ and $j\tau+R_*q\le y_T-y_P$, which is the displayed
condition.  The definition of $J$ gives
$x_T-\kappa\le j\tau\le y_T-y_P+\kappa$.  Hence, every excluded index
satisfies $j\tau\in\fragmentco{x_T-\kappa}{x_T+R_*q}$ or
$j\tau\in\fragmentoc{y_T-y_P-R_*q}{y_T-y_P+\kappa}$.  Both intervals
have length $\kappa+R_*q<39\tau$ (\cref{obs:common_parameter_bounds}),
so each of them contains at most $39$ multiples of $\tau$, and
$|J\setminus J_{\mathrm{pad}}|\le78$.

Now, suppose that some $j\in J_{\mathrm{pad}}$ exists.  The displayed
condition gives $y_T-x_T-y_P\ge2R_*q$.  The following steps use
$y_T-x_T\le n+d_T$ and $m\le y_P+d_P$, which follow from the costs of
the supplied alignments, and $n<3m/2+k$, which is part of the input
specification of \cref{prob:sm}:
\[
    y_T-x_T-y_P
       \le n+d_T-y_P
        <\tfrac32m+k+d_T-y_P
       \le\tfrac32(y_P+d_P)+k+d_T-y_P
        =\tfrac{y_P}2+k+\tfrac{3d_P}2+d_T
        \le\tfrac{y_P}2+\kappa.
\]
The last step uses $k\le d/2$ and $d_P\le d$, which give
$k+3d_P/2+d_T\le d+d_P+d_T=\kappa$.  Hence, $y_P>4R_*q-2\kappa$.  Since $R_*q\ge wq\ge\tau$
and $R_*q>f>2\kappa$ (\cref{obs:common_parameter_bounds}), the bound
$t=\floor{(y_P-\kappa)/\tau}\ge2$ follows from
\[
    y_P-\kappa>4R_*q-3\kappa=2R_*q+(2R_*q-3\kappa)>2\tau+(4\kappa-3\kappa)>2\tau.\qedhere
\]
\end{proof}

In the rest of this section and in
\cref{sec:powered_ckw_sweep,sec:block_geometry}, we assume
$J_{\mathrm{pad}}\ne\emptyset$, so that $t\ge2$ by
\cref{lem:padded_ckw_shift_interval}; otherwise, $J$ has at most $78$
indices, and \cref{fct:verifyRj} with $J_{\mathrm{sel}}=J$ solves the
instance directly in $\Oh(d^2)$ time.

The cuts in $P$ and $T$ give one family of pattern factors and one
family of overlapping text factors.

\begin{definition}[Cuts and internal factors]
\label{def:internal_cuts}
The \emph{internal cuts} are the reference coordinates $i\tau$ with
$i\in\fragmentcc{1}{t}$.  The \emph{pattern cuts} $a_i$ and the
\emph{text cuts} $c_u$ are
\begin{align*}
    a_i&\coloneqq\min\{v:(v,i\tau)\in\A_P\}
       &&\text{for }i\in\fragmentcc{0}{t},\\
    c_u&\coloneqq\min\{v:(v,u\tau)\in\A_T\}
       &&\text{for }u\in\Z\text{ with }u\tau\in\fragmentcc{x_T}{y_T}.
\end{align*}
The \emph{factor-index domains} are
\[
    \mathcal I_P\coloneqq\fragmentco{0}{t}
    \qquad\text{and}\qquad
    \mathcal I_T\coloneqq
       \fragmentco{\min J_{\mathrm{pad}}}{\max J_{\mathrm{pad}}+t}.
\]
The \emph{internal pattern factors} are
$P_i\coloneqq P\fragmentco{a_i}{a_{i+1}}$ for $i\in\mathcal I_P$, and
the \emph{internal text factors} are
$T_u\coloneqq T\fragmentco{c_u-f}{c_{u+1}+f}$ for $u\in\mathcal I_T$;
these fragments are well-defined by \cref{lem:centered_ckw_factor_containment}.
\end{definition}

Each cut is the leftmost position of $P$ or $T$ paired with its
reference coordinate; in particular, $a_0=0$.  The
internal pattern factors are consecutive, whereas consecutive internal
text factors overlap in exactly $2f=\Delta$ positions.  At block index
$j$, the $i$th pattern factor is paired with the $(j+i)$th text factor,
so $\mathcal I_T$ consists of the indices $j+i$ with
$j\in J_{\mathrm{pad}}$ and $i\in\mathcal I_P$.

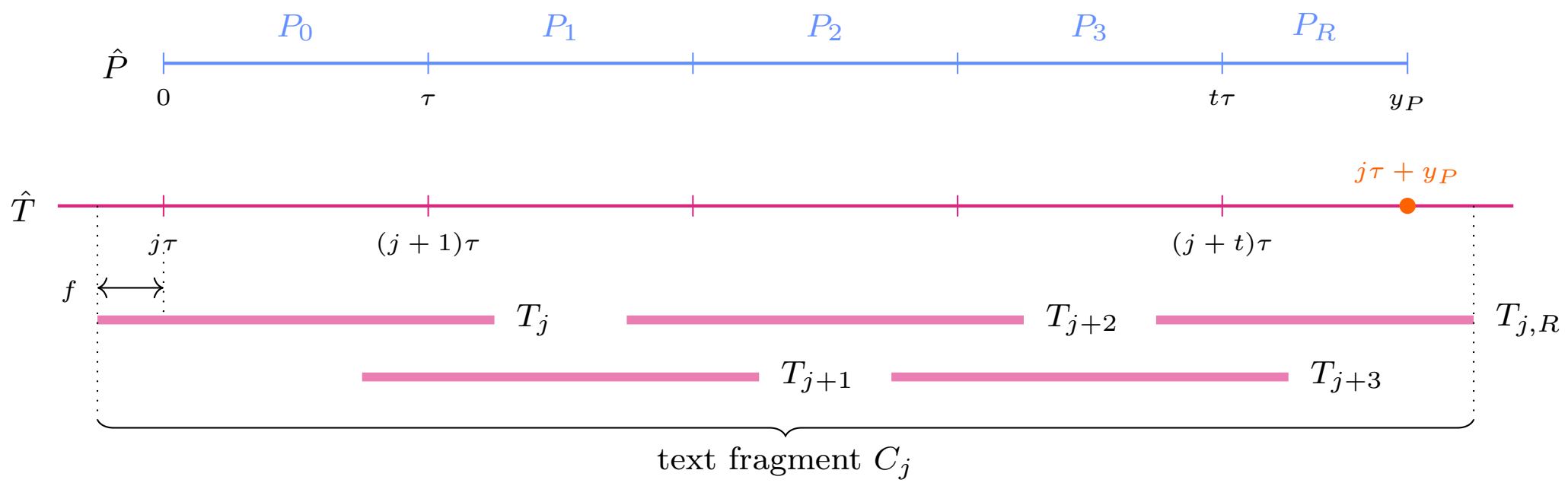
\begin{figure}[t]
\centering
\begin{tikzpicture}[x=1.15cm,y=-0.62cm]
  \draw[thick,color1] (0,0) -- (9.4,0);
  \foreach \x in {0,2,4,6,8} {\draw[color1] (\x,-0.15) -- (\x,0.15);}
  \draw[color1] (9.4,-0.15) -- (9.4,0.15);
  \foreach \x/\l in {1/P_0,3/P_1,5/P_2,7/P_3,8.7/P_R} {\node[font=\scriptsize,anchor=south,color1] at (\x,-0.1) {$\l$};}
  \node[font=\scriptsize,anchor=east] at (-0.15,0) {$\hP$};
  \node[font=\tiny,anchor=north] at (0,0.15) {$0\vphantom{t0}$};
  \node[font=\tiny,anchor=north] at (2,0.15) {$\tau\vphantom{t0}$};
  \node[font=\tiny,anchor=north] at (8,0.15) {$t\tau\vphantom{t0}$};
  \node[font=\tiny,anchor=north] at (9.4,0.15) {$y_P\vphantom{t0}$};
  \draw[thick,color3] (-0.8,2) -- (10.2,2);
  \foreach \x in {0,2,4,6,8} {\draw[color3] (\x,1.85) -- (\x,2.15);}
  \node[font=\scriptsize,anchor=east] at (-0.85,2) {$\hT$};
  \node[font=\tiny,anchor=north] at (0,2.15) {$j\tau\vphantom{(}$};
  \node[font=\tiny,anchor=north] at (2,2.15) {$(j+1)\tau$};
  \node[font=\tiny,anchor=north] at (8,2.15) {$(j+t)\tau$};
  \fill[color4] (9.4,2) circle (2pt);
  \node[font=\tiny,anchor=south,color4] at (9.4,1.9) {$j\tau+y_P$};
  \foreach \a/\b/\r/\l in {-0.5/2.5/3.6/T_j,1.5/4.5/4.4/T_{j+1},3.5/6.5/3.6/T_{j+2},5.5/8.5/4.4/T_{j+3},7.5/9.9/3.6/T_{j,R}} {
    \draw[line width=2.2pt,color3!60] (\a,\r) -- (\b,\r);
    \node[font=\scriptsize,anchor=west] at (\b+0.05,\r) {$\l$};
  }
  \draw[decorate,decoration={brace,amplitude=4pt,mirror}] (-0.5,5.0) -- (9.9,5.0) node[midway,below=3pt,font=\scriptsize] {text fragment $C_j$};
  \draw[dotted] (-0.5,2) -- (-0.5,5.0); \draw[dotted] (9.9,2) -- (9.9,5.0);
  \draw[dotted] (0,2.65) -- (0,3.6);
  \draw[<->] (-0.5,3.15) -- (0,3.15); \node[font=\tiny,anchor=mid east] at (-0.53,3.15) {$f$};
\end{tikzpicture}
\caption{The common factorization for a padded block index $j$
(schematic, reference coordinates).  The pattern reference is cut at
multiples of $\tau$, the text reference at the same points shifted by
$j\tau$, and each text factor extends by $f$ beyond its cuts, so
consecutive ones overlap in $\Delta=2f$ positions; $T_{j,R}$ ends $f$
positions after $j\tau+y_P$ (orange).  The factors are the fragments of
$P$ and $T$ aligned with these intervals.}
\label{fig:ckw_carrier}
\end{figure}

One endpoint pair completes the
factorization: the pattern factor $P_R$ covers the suffix of $P$ after
the rightmost cut; see \cref{fig:ckw_carrier}.
The text partner of $P_R$ depends on $j$.  We map the end
$j\tau+y_P$ of the shifted pattern reference
$\fragmentco{j\tau}{j\tau+y_P}$ to a position $c_j^+$ of $T$ with
$\A_T$.  Extending $T\fragmentco{c_j}{c_j^+}$ by the text margin $f$ on
both sides gives the text fragment $C_j$ of $j$
(\cref{fig:ckw_carrier}), and the factor pairs of $j$ will describe
the alignments of $P$ with $C_j$.

\begin{definition}[Text fragments and factor pairs]
\label{def:centered_carrier}
For $j\in J_{\mathrm{pad}}$, define
\(
    c_j^+\coloneqq
       \min\{v:(v,j\tau+y_P)\in\A_T\}
\)
and the \emph{text fragment of $j$}
\[
    C_j\coloneqq T\fragmentco{c_j-f}{c_j^+ +f}.
\]
The \emph{endpoint factors} are
\(
    P_R\coloneqq P\fragmentco{a_t}{|P|}
\)
and
\(
    T_{j,R}\coloneqq T\fragmentco{c_{j+t}-f}{c_j^+ +f}
\).
The $t+1$ \emph{factor pairs of $j$} are $(P_i,T_{j+i})$ with index
$i$ for $i\in\fragmentco{0}{t}$, and $(P_R,T_{j,R})$ with index $t$.
\end{definition}

We next check that all factor pairs of $j$ exist.  The lemma also
relates the start interval of $j$ to $C_j$: its first position $r_j$
lies in $C_j$, at most $f$ positions after the start of $C_j$.

\begin{lemma}[Factors inside the text fragments]
\label{lem:centered_ckw_factor_containment}
Every internal text factor $T_u$ with $u\in\mathcal I_T$ is
well-defined.  For every $j\in J_{\mathrm{pad}}$, the text fragment
$C_j$ and all factor pairs of $j$ are well-defined, their text
factors have length at least $\Delta$, the first position
$r_j$ of the start interval of $j$ (\cref{def:jdef}) is the leftmost
text position paired with ${j\tau-\kappa}$, and
\[
    c_j-f\le r_j\le c_j\le c_{j+1}\le\cdots\le c_{j+t}\le c_j^+.
\]
\end{lemma}
\begin{proof}
We use two properties of the supplied alignments.  The leftmost
positions paired with increasing reference coordinates are
nondecreasing, and between any two vertices of $\A_T$, which runs from
$(0,x_T)$ to $(n,y_T)$ at cost $d_T$, the text and reference advances
differ by at most $d_T$.  Since $t\ge2$, its definition gives
$0<t\tau\le y_P-\kappa$, so the pattern cuts $a_0\le\cdots\le a_t$ are
defined.  By the definition of $R_*$, we have $R_*q>wq+f>f+d_T$,
because $wq>W = 3\kappa \ge d_T$.  Fix $j\in J_{\mathrm{pad}}$.

\emph{The text cuts.}  By the padding inequalities
(\cref{lem:padded_ckw_shift_interval}), $x_T+R_*q\le j\tau$ and
$j\tau+y_P+R_*q\le y_T$.  Since $t\tau\le y_P-\kappa$, the reference
coordinates $j\tau<(j+1)\tau<\cdots<(j+t)\tau<j\tau+y_P$ lie in
$\fragmentcc{x_T}{y_T}$, so $c_j\le c_{j+1}\le\cdots\le c_{j+t}\le
c_j^+$ are defined.  Moreover, the reference advances by at least
$R_*q$ from $(0,x_T)$ to $(c_j,j\tau)$ and from $(c_j^+,j\tau+y_P)$ to
$(n,y_T)$, so $c_j\ge R_*q-d_T>f$ and $n-c_j^+\ge R_*q-d_T>f$.

\emph{The factors.}  By these bounds, $C_j$ is well-defined, and the
text factors of the factor pairs of $j$ are well-defined and lie
in~$C_j$; moreover, each of them has length at least $2f=\Delta$.  Since
$J_{\mathrm{pad}}$ is an integer interval
(\cref{lem:padded_ckw_shift_interval}), every $u\in\mathcal I_T$ is of
the form $j+i$ with $j\in J_{\mathrm{pad}}$ and $i\in\mathcal I_P$, so
$T_u$ is the text factor of a factor pair of $j$.

\emph{The position $r_j$.}  We have $j\tau-\kappa\ge x_T+R_*q-\kappa>x_T$.
The reference coordinate grows by at most one per step of $\A_T$, so
the leftmost vertex $(v,\hat v)$ of $\A_T$ with $\hat v\ge j\tau-\kappa$
satisfies $\hat v=j\tau-\kappa$.  Hence, $r_j$ is the leftmost text
position paired with $j\tau-\kappa$ (\cref{def:jdef}).  From
$(r_j,j\tau-\kappa)$ to $(c_j,j\tau)$, the reference advances by
$\kappa$, so $0\le c_j-r_j\le\kappa+d_T\le2\kappa<f$
(\cref{obs:common_parameter_bounds}).
\end{proof}

The factor pairs of $j$ factorize the pair $(P,C_j)$ as in
\cref{fig:ckw_carrier}.

\begin{observation}[Factorization of the text fragments]
\label{lem:centered_ckw_exact_factorization}
For every $j\in J_{\mathrm{pad}}$, the pattern factors of the factor
pairs of $j$, taken in the order of their indices, concatenate to $P$,
while their text factors cover $C_j$ with consecutive overlaps of
length $\Delta$.
\end{observation}
\begin{proof}
The pattern factors $P_0,\ldots,P_{t-1},P_R$ are
$P\fragmentco{0}{a_1},\ldots,P\fragmentco{a_{t-1}}{a_t},P\fragmentco{a_t}{|P|}$,
so they concatenate to $P$.  The text factors
$T_j,\ldots,T_{j+t-1},T_{j,R}$ start at
$c_j-f,\ldots,c_{j+t}-f$ and end at
$c_{j+1}+f,\ldots,c_{j+t}+f,c_j^+ +f$, respectively.  These positions
are nondecreasing by \cref{lem:centered_ckw_factor_containment}, so the
text factors cover $C_j=T\fragmentco{c_j-f}{c_j^+ +f}$, and
consecutive ones overlap in $T\fragmentco{c_{j+i}-f}{c_{j+i}+f}$, of
length $2f=\Delta$.
\end{proof}

\begin{example}[Running example, continued]\label{ex:running_factors}
In \cref{ex:overview_periodic} with $d=2$, we have $\kappa=5$,
$\tau=6$, and $t=\floor{(18-5)/6}=2$, so the internal cuts are the
reference coordinates $6$ and $12$.  Since $\A_P$ pairs the reference
coordinate $12$ with the pattern position $11$, the pattern cuts are
$a_1=6$ and $a_2=11$, and
\[
    P_0=\texttt{abcabc},
    \qquad
    P_1=P\fragmentco{6}{11}=\texttt{ababc},
    \qquad
    P_R=\texttt{abcabc}.
\]
Of the internal pattern factors, only $P_1$ differs from
$\Q\fragmentco{0}{\tau}=\texttt{abcabc}$, because it contains the
position where $\A_P$ inserts $\texttt{c}$.  The text
cuts are $(c_0,\ldots,c_5)=(0,6,12,17,23,30)$.  This small instance
violates the standing assumption $J_{\mathrm{pad}}\ne\emptyset$, and the margin $f=21d$
exceeds the length of its text, so no text factor is defined.  If we
ignore the margins, the fragments $T\fragmentco{c_u}{c_{u+1}}$ are
equal to $\texttt{abcabc}$ for $u\in\{0,1,3\}$, whereas
$T\fragmentco{12}{17}=\texttt{ababc}$ and
$T\fragmentco{23}{30}=\texttt{abdcabc}$.  At block index $j$, the
pattern factor $P_1$ is paired with the fragment of index $j+1$.
For $j=1$, that is, on the rail $\eD{2}$, it is paired with
$T\fragmentco{12}{17}$, and the two are equal; this is the
factor-level view of the cancellation in \cref{ex:running_rails}.  The
fragment $T\fragmentco{23}{30}$, which contains the extra $\texttt{d}$,
has index $4>j+1$ for every $j\in J$, so it is never paired with~$P_1$.
In a longer instance with the same three edits, all internal
pattern factors far from these edits would equal $\texttt{abcabc}$,
and all internal text factors far from them would equal
$\Q\fragmentco{-f}{\tau+f}$, of length $\tau+2f=90$; see
\cref{lem:plain_outside_exceptional}.
\end{example}

Each factor pair receives its $\Delta$-restricted distance matrix
(\cref{def:restricted_distance_matrix}).  For
$g,g'\in\fragmentcc{0}{\Delta}$, row $g$ corresponds to entering the
text factor at its position $g$, and column $g'$ to leaving it
$\Delta-g'$ positions before its end.  Since consecutive text factors
overlap in $\Delta$ positions
(\cref{lem:centered_ckw_exact_factorization}), the output states of one factor are the
input states of the next, and the min-plus product of their matrices
minimizes over the state at their common seam.

\begin{definition}[Factor matrices and interval products]
\label{def:natural_ckw_factor_matrices}
For $j\in J_{\mathrm{pad}}$, define the exact factor matrices of the
factor pairs of $j$ by
\[
    I_{j,i}\coloneqq\udmat{P_i}{T_{j+i}}{\Delta}
       \quad\text{for }i\in\fragmentco{0}{t}
    \qquad\text{and}\qquad
    I_{j,t}\coloneqq\udmat{P_R}{T_{j,R}}{\Delta}.
\]
For integer cuts $\ell,e\in\fragmentcc{0}{t+1}$ with $\ell\le e$, define
the interval product over the indices in $\fragmentco{\ell}{e}$ by
\[
    I_{j,\fragmentco{\ell}{e}}\coloneqq\bigoplus_{i=\ell}^{e-1}I_{j,i},
\]
where an empty product is the min-plus identity matrix, with zeros on
the main diagonal and $+\infty$ elsewhere.  Finally, $I_j$ denotes the
product of all factor matrices of $j$, that is,
\(I_j\coloneqq I_{j,\fragmentcc{0}{t}}=I_{j,0}\oplus I_{j,1}\oplus\cdots\oplus I_{j,t}\).
\end{definition}

These matrices are well-defined by
\cref{lem:centered_ckw_factor_containment}.

\subsection{Sparse Factor Families and Exceptional Lists}\label{sec:ckw_exceptional_lists}

If $P$ and $T$ were exactly periodic, all internal pattern factors
would be equal to one string, and so would all internal text factors
and all endpoint text factors.  These strings are the following prototypes.

\begin{definition}[Prototypes and plain factors]
\label{def:prototypes}
Define the periodic prototypes by
\[
    Q_P^\circ\coloneqq\Q\fragmentco{0}{\tau},
    \qquad
    Q_T^\circ\coloneqq\Q\fragmentco{-f}{\tau+f},
    \qquad
    Q_R^\circ\coloneqq\Q\fragmentco{t\tau-f}{y_P+f}.
\]
Call $P_i$ \emph{plain} when $P_i=Q_P^\circ$, and call
$T_u$ \emph{plain} when $T_u=Q_T^\circ$.
\end{definition}

Few factors differ from their prototypes.  The factor $P_i$ is the
part of $P$ that $\A_P$ aligns with $\Q\fragmentco{i\tau}{(i+1)\tau}$,
and $T_u$ is the part of $T$ that $\A_T$ aligns with
$\Q\fragmentco{u\tau}{(u+1)\tau}$, extended by $f$ positions on both
sides.  These intervals are translated by $\tau$ from one index to the
next and $f=\Oh(\tau)$, so an edit of $\A_P$ or $\A_T$ affects only
$\Oh(1)$ factors.  The next lemma makes this precise for a generic
family that also covers the endpoint text factor.  It translates a
reference window $\fragmentco{\alpha}{\beta}$ by the multiples of
$\tau$, maps each translate back to the source string $X$ with $\A$,
and extends the result by $h$ positions on both sides; we use $h=0$
for $P$ and $h=f$ for $T$.  The lemma bounds the total distance of the
resulting factors $F_v$ from their common prototype $F^\circ$, and it
shows that $F_v\ne F^\circ$ only if the reference support of an edit
of $\A$ weakly overlaps $E_v$, the translated window widened by $h$.

\begin{lemma}[Charging prototype deviations to alignment edits]
\label{lem:periodic_factor_charging}
Let $\A:X\onto\Q\fragmentco{x}{y}$ be an alignment.  Fix integers $\alpha$,
$\beta\in\fragmentcc{\alpha}{\alpha+2\tau}$, and $h\in\fragmentcc{0}{f}$.
For every integer $v$ with
$x\le v\tau+\alpha\le v\tau+\beta\le y$, let $\ell_v,r_v$
be the least source coordinates paired by $\A$ with
$v\tau+\alpha,v\tau+\beta$, respectively.
Let $\mathcal I$ be any set of these indices $v$ satisfying
$0\le\ell_v-h\le r_v+h\le|X|$.
For $v\in\mathcal I$, define
\[
    F_v\coloneqq X\fragmentco{\ell_v-h}{r_v+h},
    \qquad
    F^\circ\coloneqq\Q\fragmentco{\alpha-h}{\beta+h}.
\]
Then,
\[
    \sum_{v\in\mathcal I}\ed(F_v,F^\circ)=\Oh(\cost(\A)).
\]
Moreover, $F_v=F^\circ$ whenever no reference support of an edit of $\A$
weakly overlaps the interval
\[
    E_v\coloneqq
       \fragmentco{v\tau+\alpha-h}{v\tau+\beta+h}.
\]
\end{lemma}
\begin{proof}
Fix $v\in\mathcal I$, and let $c_v$ be the number of edits of $\A$
whose reference supports weakly overlap~$E_v$.  We first show that
\begin{equation}\label{eq:periodic_factor_local_charge}
    \ed(F_v,F^\circ)\le2c_v.
\end{equation}
If $E_v$ sticks out of $\fragmentco{x}{y}$, we extend $X$ and $\A$ on
that side by matched characters of $\Q$.  This changes neither $F_v$
nor the edits of $\A$, so we may assume $E_v\subseteq\fragmentco{x}{y}$.
Let $(s_L,v\tau+\alpha-h)$ and $(s_R,v\tau+\beta+h)$ be the earliest
vertices of $\A$ with these reference coordinates, and let $\A_v$ be
the subpath between them.  It passes through the earliest vertices
$(\ell_v,v\tau+\alpha)$ and $(r_v,v\tau+\beta)$ with their reference
coordinates, in this order, and they split it into three parts.  We
denote the costs of these parts by $c_1$, $c_2$, and $c_3$.  The reference
supports of all edits of $\A_v$ are contained in $E_v$, so $c_1+c_2+c_3\le c_v$.  Since $q$
divides~$\tau$, the path $\A_v$ aligns $X\fragmentco{s_L}{s_R}$ with
$\Q\fragmentco{v\tau+\alpha-h}{v\tau+\beta+h}=F^\circ$ at cost
$c_1+c_2+c_3$.  The first and the third part each advance the
reference by $h$, and along an alignment the source and the reference
advance differ by at most the cost, so $|s_L-(\ell_v-h)|\le c_1$ and
$|s_R-(r_v+h)|\le c_3$.  Since
$\ed(X\fragmentco{a}{b},X\fragmentco{a'}{b'})\le|a-a'|+|b-b'|$ for all
fragments of $X$, we obtain
\[
    \ed(F_v,F^\circ)
       \le\ed\bigl(F_v,X\fragmentco{s_L}{s_R}\bigr)
          +\ed\bigl(X\fragmentco{s_L}{s_R},F^\circ\bigr)
       \le(c_1+c_3)+(c_1+c_2+c_3)
       \le2c_v.
\]

If no reference support of an edit of $\A$ weakly overlaps $E_v$, then
$c_v=0$, and \cref{eq:periodic_factor_local_charge} gives $F_v=F^\circ$.
For the sum, consider an edit with reference support
$\Q\fragmentco{z}{z'}$, where $z'\le z+1$.  It weakly overlaps $E_v$
only if $v\tau\in\fragmentcc{z-\beta-h}{z'-\alpha+h}$.  This interval
has length at most $1+2\tau+2f$, so it contains at most
$(1+2\tau+2f)/\tau+1<46$ multiples of $\tau$, because $2f=42d<42\tau$.
Hence, every edit is counted by at most $45$ of the numbers $c_v$,
and \cref{eq:periodic_factor_local_charge} gives
\[
    \sum_{v\in\mathcal I}\ed(F_v,F^\circ)
       \le2\sum_{v\in\mathcal I}c_v
       \le90\cost(\A).\qedhere
\]
\end{proof}

\begin{figure}[t]
\centering
\begin{tikzpicture}[x=0.55cm,y=0.5cm]
  \node[font=\small,anchor=east] at (-1.2,0) {$\hP$};
  \draw[black!60] (0,0) -- (17,0);
  \foreach \i/\l in {0/0,1/\tau,2/2\tau,3/3\tau,4/4\tau,5/5\tau,6/6\tau,7/7\tau,8/8\tau} {\draw[black!60] (2*\i,-0.12) -- (2*\i,0.12); \node[font=\tiny,anchor=north,black!60] at (2*\i,-0.12) {$\vphantom{0\tau}\l$};}
  \fill[color1!45] (7.2,-0.13) rectangle (8.6,0.13);
  \node[font=\scriptsize,color1!80!black,anchor=north] at (7.9,-0.8) {$\E_P$};
  \foreach \i in {2,...,5} {
    \pgfmathsetmacro{\yy}{0.6+0.5*mod(\i,3)}
    \pgfmathsetmacro{\red}{(\i==3||\i==4)?1:0}
    \ifdim\red pt=1pt \def\cc{color1}\else\def\cc{black!40}\fi
    \draw[\cc,line width=1.6pt] ({2*\i},\yy) -- ({2*\i+2},\yy);
    \node[font=\tiny,\cc,anchor=east] at ({2*\i-0.05},\yy) {$H^P_{\i}$};
  }
  \node[font=\scriptsize,anchor=west] at (17.3,0.9) {exceptional: $3,4\in\mathcal X_P$};
  \begin{scope}[yshift=-2.9cm]
  \node[font=\small,anchor=east] at (-1.2,0) {$\hT$};
  \draw[black!60] (0,0) -- (17,0);
  \foreach \i/\l in {0/0,1/\tau,2/2\tau,3/3\tau,4/4\tau,5/5\tau,6/6\tau,7/7\tau,8/8\tau} {\draw[black!60] (2*\i,-0.12) -- (2*\i,0.12); \node[font=\tiny,anchor=north,black!60] at (2*\i,-0.12) {$\vphantom{0\tau}\l$};}
  \fill[color3!55] (9.3,-0.13) rectangle (9.8,0.13);
  \node[font=\scriptsize,color3!80!black,anchor=north] at (9.55,-0.8) {$\E_T$};
  \foreach \i in {2,...,6} {
    \pgfmathsetmacro{\yy}{0.6+0.6*(\i-2)}
    \pgfmathsetmacro{\red}{(\i>=3&&\i<=6)?1:0}
    \ifdim\red pt=1pt \def\cc{color3}\else\def\cc{black!40}\fi
    \draw[\cc,line width=0.8pt] ({2*\i-2.8},\yy) -- ({2*\i+4.8},\yy);
    \draw[\cc,line width=2.6pt] ({2*\i-0.8},\yy) -- ({2*\i+2.8},\yy);
    \node[font=\tiny,\cc,anchor=east] at ({2*\i-2.85},\yy) {$H^T_{\i}$};
  }
  \node[font=\scriptsize,anchor=west] at (17.3,1.8) {exceptional: $3,4,5,6\in\mathcal X_T$};
  \end{scope}
  \node[font=\small] at (-3.2,1.2) {(a)};
  \begin{scope}[yshift=-5.5cm]
  \node[font=\small] at (-3.2,2.6) {(b)};
  \node[font=\small,anchor=east] at (-1.2,0) {$\hT$};
  \draw[black!60] (-1.5,0) -- (8.6,0);
  \draw[black!60] (11.4,0) -- (21.6,0);
  \node[black!60] at (10,0) {$\cdots$};
  \foreach \x/\l in {0/j\tau,3/(j+1)\tau,6/(j+2)\tau,14/(j+t-1)\tau,17/(j+t)\tau,20/(j+t+1)\tau}
    {\draw[black!60] (\x,-0.12) -- (\x,0.12);
     \node[font=\tiny,anchor=north,black!60] at (\x,-0.12) {$\vphantom{(j+1)\tau}\l$};}
  \draw[color4,line width=1.4pt] (-0.5,0) -- (0,0);
  \draw[color4] (-0.5,-0.12) -- (-0.5,0.12);
  \node[font=\tiny,anchor=south,color4] at (-0.5,0.12) {$j\tau-\kappa$};
  \fill[color4] (19.6,0) circle (2.2pt);
  \node[font=\tiny,anchor=south,color4] at (19.6,0.12) {$j\tau+y_P$};
  \draw[color3,line width=2.6pt] (15.8,1.05) -- (20.8,1.05);
  \node[font=\tiny,color3,anchor=mid east] at (15.7,1.05) {$E_{j,R}$};
  \draw[black!60,line width=0.8pt,{Stealth[length=3pt]}-] (-1.4,1.9) -- (7.2,1.9);
  \draw[black!60] (7.2,1.75) -- (7.2,2.05);
  \node[font=\tiny,black!70,anchor=south] at (2.9,2.0) {$H^T_{j}$};
  \draw[black!60,line width=0.8pt,-{Stealth[length=3pt]}] (12.8,1.9) -- (21.5,1.9);
  \draw[black!60] (12.8,1.75) -- (12.8,2.05);
  \node[font=\tiny,black!70,anchor=south] at (17.15,2.0) {$H^T_{j+t}$};
  \end{scope}
\end{tikzpicture}
\caption{The exceptional lists of \cref{def:exceptional_lists}
(schematic).  (a)~$H_3^P$ and $H_4^P$ weakly overlap $\E_P$, so $3$ and
$4$ are exceptional; the text windows extend the thick ones by $wq$ per
side, so $3,\ldots,6$ are exceptional, though only the thick windows of
$4$ and $5$ weakly overlap $\E_T$.  (b)~The ends of the text fragment of
$j$: $c_j-r_j=\kappa$ if $j$ is not exceptional, and $T_{j,R}=Q_R^\circ$
if $j+t$ is not (\cref{lem:plain_outside_exceptional}).}
\label{fig:exceptional_lists}
\end{figure}
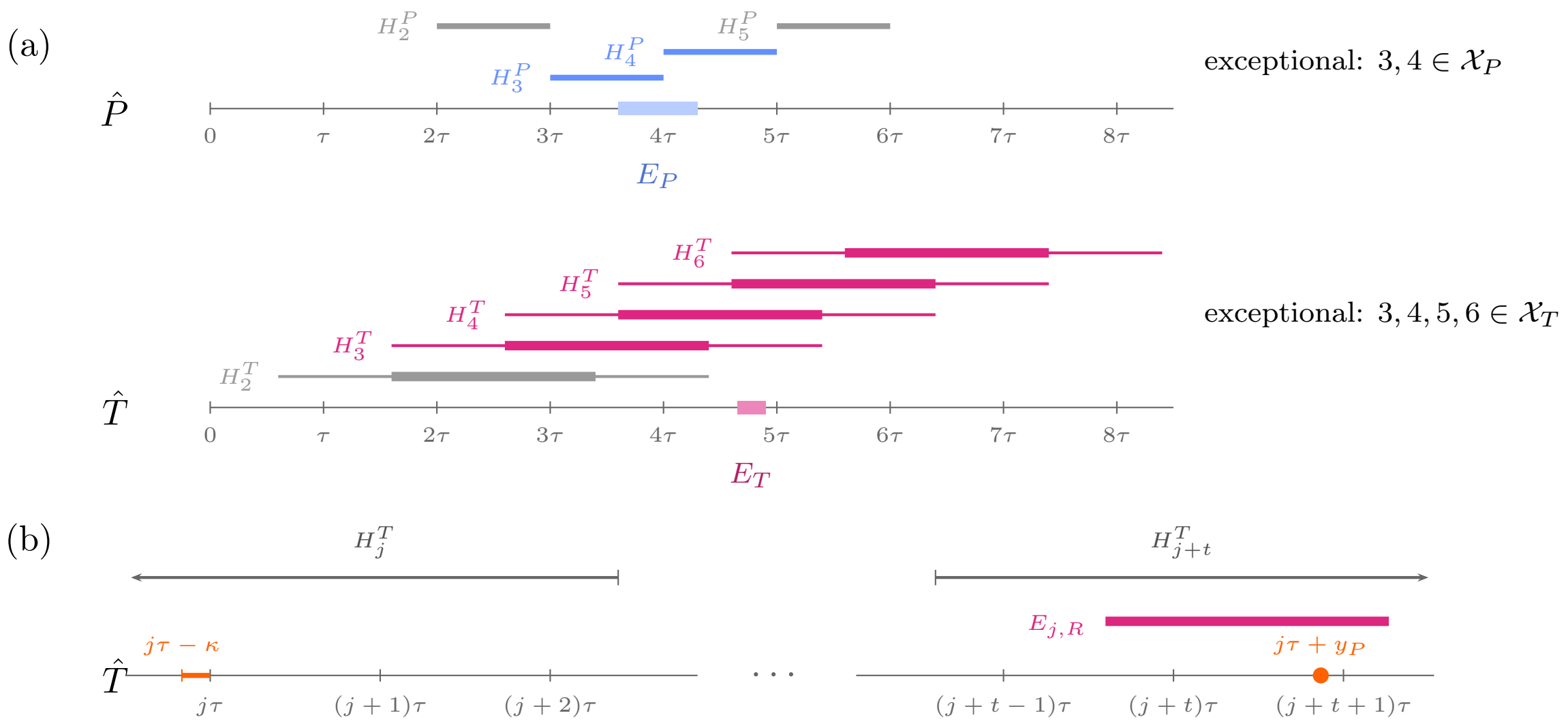

We call a factor index \emph{exceptional} if the reference window of
its factor weakly overlaps a fragment of a reference cover; see
\cref{fig:exceptional_lists}.  The window of $P_i$ is the interval
$E_i=\fragmentco{i\tau}{(i+1)\tau}$ of \cref{lem:periodic_factor_charging}
with $h=0$.  The window of $T_u$ extends the interval
$E_u=\fragmentco{u\tau-f}{(u+1)\tau+f}$ with $h=f$ by $wq>W$ on both
sides, as the vertices relevant to $j$ have offsets within $W$ of $b_jq$
(\cref{rem:margins_and_padding}).  Since all edits lie in cover
fragments (\cref{fct:edits_only_in_cover}), a factor with a
non-exceptional index is plain; the later sections also use that no
cover fragment lies near it.  An exceptional factor may still be plain.

\begin{definition}[Exceptional lists]
\label{def:exceptional_lists}
Let $\Es_P$ and $\Es_T$ be the reference covers
(\cref{def:canonical_dense_covers}).
Define two reference windows for every integer $i$:
\[
    H_i^P\coloneqq\fragmentco{i\tau}{(i+1)\tau},
    \qquad
    H_i^T\coloneqq
       \fragmentco{i\tau-f-wq}{(i+1)\tau+f+wq}.
\]
The two \emph{exceptional lists} are the sorted index sets
\begin{align*}
    \mathcal X_P
       &\coloneqq
       \{i\in\mathcal I_P:
          H_i^P\text{ weakly overlaps a fragment of }\Es_P\},\\
    \mathcal X_T
       &\coloneqq
       \{u\in\fragmentcc{\min J_{\mathrm{pad}}}{\max J_{\mathrm{pad}}+t}:
          H_u^T\text{ weakly overlaps a fragment of }\Es_T\}.
\end{align*}
Call $P_i$ exceptional when $i\in\mathcal X_P$, and call
$T_u$ exceptional when $u\in\mathcal X_T$.
\end{definition}

The range of $\mathcal X_T$ extends $\mathcal I_T$ by the index
$\max J_{\mathrm{pad}}+t$, which has no text factor; for every
$j\in J_{\mathrm{pad}}$, the index $j+t$ classifies the endpoint pair
of $j$ (\cref{lem:plain_outside_exceptional}).

\begin{lemma}[Plain factors outside the exceptional lists]
\label{lem:plain_outside_exceptional}
Every internal factor whose index is outside its exceptional list
equals the corresponding prototype:
\[
    i\notin\mathcal X_P\text{ implies }P_i=Q_P^\circ
       \quad\text{for }i\in\mathcal I_P,
    \qquad
    u\notin\mathcal X_T\text{ implies }T_u=Q_T^\circ
       \quad\text{for }u\in\mathcal I_T.
\]
Moreover, for every $j\in J_{\mathrm{pad}}$,
\[
    j\notin\mathcal X_T\text{ implies }c_j-r_j=\kappa,
    \qquad
    j+t\notin\mathcal X_T\text{ implies }T_{j,R}=Q_R^\circ.
\]
\end{lemma}
\begin{proof}
Apply \cref{lem:periodic_factor_charging} with
$(\alpha,\beta)=(0,\tau)$ to $\A_P$ with $h=0$ and index set
$\mathcal I_P$, and to $\A_T$ with $h=f$ and index set $\mathcal I_T$.
The resulting factors are $P_i$ and $T_u$, with prototypes $Q_P^\circ$
and $Q_T^\circ$, and the corresponding intervals $E_v$ lie inside $H_i^P$ and $H_u^T$,
respectively.  If a factor is not plain, then, by
\cref{lem:periodic_factor_charging}, its interval $E_v$ weakly
overlaps the reference support of an edit.  By
\cref{fct:edits_only_in_cover}, this support lies in a fragment of the
corresponding reference cover.  This fragment therefore weakly overlaps
$H_i^P$ or $H_u^T$, and the factor is exceptional.

For the endpoint factor, fix $j\in J_{\mathrm{pad}}$; note that
$j$ and $j+t$ lie in the range of $\mathcal X_T$.  By \cref{def:common_factorization_parameters},
$t\tau\le y_P-\kappa<(t+1)\tau$, so $y_P-t\tau<\tau+\kappa<2\tau$.  We
apply \cref{lem:periodic_factor_charging} to $\A_T$ with $h=f$, $v=j$,
and $(\alpha,\beta)=(t\tau,y_P)$; by \cref{def:centered_carrier}, the
resulting factor is $T_{j,R}$, with prototype $Q_R^\circ$.  The
corresponding interval is
\[
    E_{j,R}=\fragmentco{(j+t)\tau-f}{j\tau+y_P+f}.
\]
Since $y_P-t\tau<\tau+\kappa$ and $\kappa<W<wq$, the right endpoint
$j\tau+y_P+f$ of $E_{j,R}$ is smaller than the right endpoint
$(j+t+1)\tau+f+wq$ of $H^T_{j+t}$, so $E_{j,R}\subseteq H^T_{j+t}$.
As above, if $T_{j,R}\ne Q_R^\circ$, then $j+t\in\mathcal X_T$.

Finally, $H^T_j$ contains $\fragmentco{j\tau-\kappa}{j\tau}$, because
$f>\kappa$.  If $j\notin\mathcal X_T$, then no fragment of $\Es_T$
weakly overlaps $H^T_j$, so, by \cref{fct:edits_only_in_cover}, no
reference support of an edit of $\A_T$ weakly overlaps this interval.
By \cref{lem:centered_ckw_factor_containment}, $r_j$ and $c_j$ are the
leftmost text positions paired with $j\tau-\kappa$ and $j\tau$, so
$\A_T$ matches $T\fragmentco{r_j}{c_j}$ with
$\Q\fragmentco{j\tau-\kappa}{j\tau}$ without edits.  Hence,
$c_j-r_j=\kappa$.
\end{proof}

The factor-index domains can be very long, but the exceptional lists can be
constructed without scanning them.

\begin{lemma}[Size and construction of the exceptional lists]
\label{lem:exceptional_list_construction}
The exceptional lists (\cref{def:exceptional_lists}) satisfy
$|\mathcal X_P|+|\mathcal X_T|=\Oh(d)$.
Given the endpoints of the fragments of $\Es_P$ and $\Es_T$, both
lists can be constructed in $\Oh(d\lg m)$ time.
\end{lemma}
\begin{proof}
By \cref{lem:dense_decomposition}, each cover consists of $\Oh(d)$
fragments of total length $\Oh(qd)$; recall that $q\le\tau$.

A pattern-cover fragment $E=\fragmentco{a}{b}\in\Es_P$ weakly overlaps
$H_i^P$ if and only if $i\tau\le b$ and $(i+1)\tau\ge a$.  Hence, it
contributes exactly the indices in
$\mathcal I_P\cap\fragmentcc{\ceil{a/\tau}-1}{\floor{b/\tau}}$, and
there are at most $|E|/\tau+2$ of them.  Summing over $\Es_P$ gives
$|\mathcal X_P|=\Oh(d+qd/\tau)=\Oh(d)$.

A text-cover fragment $E=\fragmentco{a}{b}\in\Es_T$ weakly overlaps
$H_u^T$ if and only if $u\tau-f-wq\le b$ and $(u+1)\tau+f+wq\ge a$.
Hence, it contributes exactly the indices in
\[
    \fragmentcc{\min J_{\mathrm{pad}}}{\max J_{\mathrm{pad}}+t}\cap
    \fragmentcc{\ceil{(a-f-wq)/\tau}-1}{\floor{(b+f+wq)/\tau}},
\]
and there are at most $(|E|+2f+2wq)/\tau+2$ of them.  Since
$f+wq=\Oh(d+q)=\Oh(\tau)$, summing over $\Es_T$ gives
$|\mathcal X_T|=\Oh(d+qd/\tau)=\Oh(d)$.

To construct the lists, we note that the endpoints of both index
ranges are explicit (\cref{def:internal_cuts,def:exceptional_lists}); those of
$J_{\mathrm{pad}}$ follow from \cref{lem:padded_ckw_shift_interval}.  For
each cover fragment, we compute the interval of its contributed indices
and enumerate them.  Sorting the $\Oh(d)$ indices of each list and
removing duplicates takes $\Oh(d\lg m)$ time.
\end{proof}

\subsection{Fern Catalogue}\label{sec:ckw_catalogues}

By \cref{lem:plain_outside_exceptional}, most factor pairs consist of
two prototypes, so many factor matrices $I_{j,i}$ coincide and can
share one fern.  A catalogue precomputes $\Oh(d^2)$ ferns, which all
factor pairs reuse, and returns \emph{read-only CMO handles}: pointers to CMOs that remain
unchanged for the lifetime of the catalogue.  Retaining or returning a
handle copies only the pointer, and space bounds count each
constructed CMO once.  The internal factors can form $\Theta(d^2)$
distinct pairs, too many to construct their ferns one by one, so we
construct them in one batch with \cref{thm:sequence_fern_construction}.
The endpoint pairs are fewer: by \cref{lem:plain_outside_exceptional},
only $\Oh(d)$ block indices have an endpoint text factor that may
differ from $Q_R^\circ$, and we construct a fern for each of these
pairs separately.

\begin{lemma}[Fern catalogue]
\label{lem:ckw_fern_catalogue}
Given the breakpoint representations of $\A_P,\A_T$, one can construct
in $\Oh(d^3\lg^2 m)$ time in the \modelname model a \emph{fern
catalogue}: a structure that answers every query
$(j,i)\in J_{\mathrm{pad}}\times\fragmentcc{0}{t}$ in $\Oh(\lg m)$ time
with a read-only CMO handle for an integer-valued, $2$-bounded-difference
Monge matrix that is $W$-equivalent to $I_{j,i}$, that is, for a
$(\Delta,W)$-fern of the factor pair of $j$ with index $i$
(\cref{def:centered_carrier,def:natural_ckw_factor_matrices}).
\end{lemma}
\begin{proof}
\emph{Internal factors.}
Consecutive cuts are $\tau$ apart in the reference, so the costs of
the supplied alignments give $0<\tau-d_P\le|P_i|\le\tau+d_P$ and
$\Delta\le|T_u|\le\tau+\Delta+d_T$ throughout the index domains.  Thus,
all these strings have length $\Oh(\tau+d)$.

We construct the reference covers $\Es_P$ and $\Es_T$
(\cref{def:canonical_dense_covers}) in $\Oh(d^2)$ time by \cref{lem:dense}; the breakpoint representations
of the inverse alignments are obtained by swapping the two components
of every entry.  Then, we construct $\mathcal X_P$ and $\mathcal X_T$
in $\Oh(d\lg m)$ time by \cref{lem:exceptional_list_construction}.

For each of the two families, we list the factors at the exceptional
indices in its domain, followed by a \emph{default} factor at one
non-exceptional index, if there is one; we find this index from the
domain endpoints and the gaps between consecutive exceptional indices.
By \cref{lem:plain_outside_exceptional}, the factors at all
non-exceptional indices, including the default factor, are plain.  Predecessor queries on the
breakpoint representations locate the cuts $a_i$ and $c_u$ of every
listed factor.  We contract consecutive equal strings in each list,
testing equality by length and longest common prefix queries in
$\Oh(1)$ \modelname operations each (\cref{sec:prelims_model}).  Since
consecutive strings are now distinct, this yields two string sequences
(\cref{def:string_sequence}).
\Cref{lem:periodic_factor_charging}, applied with $(\alpha,\beta)=(0,\tau)$
to $\A_P$ with $h=0$ and to $\A_T$ with $h=f$, gives
$\sum_{i\in\mathcal I_P}\ed(P_i,Q_P^\circ)=\Oh(d_P)$ and
$\sum_{u\in\mathcal I_T}\ed(T_u,Q_T^\circ)=\Oh(d_T)$.  Hence, by the
triangle inequality, the drift of a sequence
$(X_0,\ldots,X_s)$ of factors at distinct indices of a family with
prototype $X^\circ$ satisfies
\[
    \sum_{i=1}^s\ed(X_{i-1},X_i)
       \le2\sum_{i=0}^s\ed(X_i,X^\circ)=\Oh(d).
\]
All listed strings are nonempty, have
length $\Oh(\tau+d)$, and the text strings have length at least
$\Delta$ (\cref{lem:centered_ckw_factor_containment}).  We apply \cref{thm:sequence_fern_construction} with a drift
bound $\Lambda=\Oh(d)$, threshold $k=W$, and width $\Delta$; since
$W,\Delta=\Oh(d)$, its parameter $L$ is $\Oh(d)$.  It constructs
$(\Delta,W)$-ferns for all pairs of sequence positions in time
\begin{equation}\label{eq:common_fern_catalogue_time}
    \Oh\left(d^3\lg^2\bigl(d(\tau+d)\bigr)\right)
       =\Oh(d^3\lg^2 m).
\end{equation}
The input promise $q\le m/(8d)$ gives $d,q\le m/8$, while
$\tau\le\kappa+q\le5d+q$, so $d(\tau+d)+2=\Oh(m^2)$.

\emph{Endpoint factors.}
By \cref{lem:plain_outside_exceptional}, $T_{j,R}$ differs from
$Q_R^\circ$ only if $j+t\in\mathcal X_T$.  For every $u\in\mathcal X_T$,
we therefore construct a fern of $(P_R,T_{u-t,R})$ if
$u-t\in J_{\mathrm{pad}}$.  We also construct a \emph{default}
endpoint fern of $(P_R,T_{j,R})$ for one $j\in J_{\mathrm{pad}}$ with
$j+t\notin\mathcal X_T$, if such an index exists; we find it from the
endpoints of $J_{\mathrm{pad}}$ and the sorted list $\mathcal X_T$.
By the same lemma, all $j\in J_{\mathrm{pad}}$ with
$j+t\notin\mathcal X_T$ share the factor $T_{j,R}=Q_R^\circ$.  For each of these
$\Oh(|\mathcal X_T|+1)=\Oh(d)$ pairs, we locate both strings by
predecessor queries on the breakpoint representations and construct a
$(\Delta,W)$-fern by \cref{lem:static_fern_construction}.  This lemma
applies because
$W=3\kappa\le15d<\Delta$, $|T_{j,R}|\ge\Delta$
(\cref{lem:centered_ckw_factor_containment}), and the pattern string
is nonempty:
$|P|-a_t\ge y_P-t\tau-d_P\ge\kappa-d_P>0$.  All strings have length
at most $n=\Oh(m)$, so each construction takes $\Oh(d^2\lg^2m)$ time,
and all of them take $\Oh(d^3\lg^2m)$ time.

\emph{Queries.}  We store the exceptional indices of both families in
sorted arrays with the positions of their strings; a text index $u$
also stores the endpoint fern constructed for $u$.  A query $(j,i)$
with $i\in\fragmentco{0}{t}$ looks up the pattern index $i$ and the
text index $j+i$ by binary search, uses the position of the default
factor for each index that is not stored, and returns the fern of the pair of the two
positions.  The strings at these positions are $P_i$ and $T_{j+i}$, so,
by \cref{def:natural_ckw_factor_matrices}, this is a fern of $I_{j,i}$.
A query $(j,t)$ looks up the text index $j+t$ and returns the endpoint
fern stored with it or, if $j+t\notin\mathcal X_T$, the default
endpoint fern, which exists because $j$ itself qualifies.  In the
latter case, $T_{j,R}$ and the text string of the default pair both
equal $Q_R^\circ$.  Either
way, the returned fern is a fern of $(P_R,T_{j,R})$, that is, of
$I_{j,t}$.  In all cases, a
query performs a constant number of binary searches over $\Oh(d)$
sorted indices, in $\Oh(\lg m)$ time, and returns a handle to a stored
CMO.
\end{proof}

\subsection{Load Partition and Preprocessing Record}\label{sec:ckw_load_partition}

To decide which algorithm handles a padded block index $j$, we sum
the mark loads (\cref{def:marking}) of the strips near the origin
$b_j$.  For periods in an intermediate range, block indices with a
large sum are \emph{high-load} and are handled in
\cref{sec:block_geometry}; all other padded block indices are
\emph{eligible} and are handled in \cref{sec:powered_ckw_sweep}.

\begin{definition}[Block-index load partition]
\label{def:ckw_load_partition}
For each $j\in J_{\mathrm{pad}}$, define its load by
\[
    M_j\coloneqq
      \sum_{b\in \canD\cap\fragmentcc{b_j-R_*}{b_j+R_*}}
          \mu(b).
\]
The corresponding block-index load partition is
\begin{align*}
    J_{\mathrm{high}}
       &\coloneqq
       \left\{j\in J_{\mathrm{pad}}:
          \sqrt d\le q\le\kappa\ \text{and}\ M_j\ge q/10\right\},\\
    J_{\mathrm{elig}}
       &\coloneqq J_{\mathrm{pad}}\setminus J_{\mathrm{high}}.
\end{align*}
Call the indices in $J_{\mathrm{high}}$ \emph{high-load} and those in
$J_{\mathrm{elig}}$ \emph{eligible}.
\end{definition}

The following bounds are used in the running-time analyses: the
total load limits the work charged to marks, and the number of
high-load indices limits the work spent separately on each such index.

\begin{lemma}[Load bounds]
\label{lem:ckw_load_bounds}
The loads satisfy
\[
    \sum_{j\in J_{\mathrm{pad}}}M_j=\Oh(d^2)
    \qquad\text{and}\qquad
    |J_{\mathrm{high}}|=\Oh(d^2/q).
\]
\end{lemma}
\begin{proof}
Each mark load $\mu(b)$ is counted in $M_j$ for all $j$ with
$|b-b_j|\le R_*$.  Consecutive origins differ by $\tau/q$, so there
are at most $1+2R_*q/\tau<77$ such indices $j$
(\cref{obs:common_parameter_bounds}).  Hence,
$\sum_{j\in J_{\mathrm{pad}}}M_j\le77\sum_{b\in\canD}\mu(b)$, and
\cref{lem:few_marks} proves the first bound.  Every
$j\in J_{\mathrm{high}}$ satisfies $M_j\ge q/10$, so
$(q/10)\cdot|J_{\mathrm{high}}|\le\sum_{j\in J_{\mathrm{pad}}}M_j$,
which proves the second bound.
\end{proof}

The common input of the later algorithms consists of the start
intervals, the padded block indices, the exceptional lists, and the
high-load block indices.  We collect these objects in one record.

\begin{definition}[Preprocessing record]
\label{def:preprocessing_record}
For the block-and-start-interval data of \cref{def:jdef},
the exceptional lists of \cref{def:exceptional_lists},
and the high-load indices of \cref{def:ckw_load_partition}, put
\[
    \mathfrak P
       \coloneqq
       \bigl(
          \bigl(J,(r_j)_{j\in J}\bigr),J_{\mathrm{pad}},
          \mathcal X_P,\mathcal X_T,J_{\mathrm{high}}
       \bigr).
\]
Here, the integer intervals $J$ and $J_{\mathrm{pad}}$
(\cref{lem:padded_ckw_shift_interval}) are stored by their endpoints,
$(r_j)_{j\in J}$ as the arithmetic progressions of \cref{fct:rj}, and
$\mathcal X_P$, $\mathcal X_T$, and $J_{\mathrm{high}}$ as sorted lists.
Call $\mathfrak P$ the \emph{preprocessing record}.
\end{definition}

The record does not include the fern catalogue of
\cref{sec:ckw_catalogues}, which is constructed separately.

\begin{lemma}[Constructing the preprocessing record]
\label{lem:preprocessing_record}
Given a valid instance of \SM with $J_{\mathrm{pad}}\ne\emptyset$, the record
$\mathfrak P$ can be constructed in
$\Oh(d^2\lg m)$ time and $\Oh(d^2)$ space.
\end{lemma}
\begin{proof}
\emph{Blocks and exceptional lists.}  We construct the representation of the
start intervals by \cref{fct:rj}.  The padded set is an integer
interval by \cref{lem:padded_ckw_shift_interval}, and we compute its
endpoints by intersecting $J$ with
$\{j:x_T+R_*q\le j\tau\le y_T-y_P-R_*q\}$.  We apply \cref{lem:dense}
to $\brp{\A_P^{-1}}$ with parameters $(5q,2d)$ and to
$\brp{\A_T^{-1}}$ with parameters $(5q,0)$; the breakpoint
representation of an inverse alignment is obtained by swapping the
two components of every entry.  The resulting covers consist of $\Oh(d)$ fragments of
total length $\Oh(qd)$, and we construct the exceptional lists by
\cref{lem:exceptional_list_construction}.

\emph{Marking weights.}  We extract the reference supports of the
edits of $\A_P$ and $\A_T$ from their breakpoint representations.
Both endpoints of the supports are nondecreasing along each alignment,
so two binary searches count the supports contained in a given cover
fragment, including the empty supports at its endpoints.  This
computes all finite marking weights in $\Oh(d\lg m)$ time; boundary
pattern fragments receive weight $\infty$ as in \cref{def:marking}.

\emph{Loads.}  We compute the loads $M_j$ by processing the pairs of
cover fragments.  For a pair $\E_P=\Q\fragmentco{\hat x}{\hat y}$ and
$\E_T=\Q\fragmentco{\hat x'}{\hat y'}$, put
$h\coloneqq\min(w_P(\E_P),w_T(\E_T))$.  By
\cref{eq:covers_weak_intersection}, the strip indices at which the pair places a mark
form the interval
\[
    A\coloneqq\canD\cap
       \fragmentcc{\ceil{(\hat x'-\hat y-q)/q}}{\floor{(\hat y'-\hat x)/q}}.
\]
We skip the pair if $h=0$ or $A=\emptyset$.  Otherwise, we write
$A=\fragmentcc{a}{b}$ and enumerate the block indices
\[
    j\in
    J_{\mathrm{pad}}\cap
       \fragmentcc{\ceil{(a-R_*)q/\tau}}
                  {\floor{(b+R_*)q/\tau}},
\]
which are exactly those with
$\fragmentcc{b_j-R_*}{b_j+R_*}\cap A\ne\emptyset$.  For each of them,
we add the contribution $h\cdot|A\cap\fragmentcc{b_j-R_*}{b_j+R_*}|$ of
the pair to the dictionary entry of $j$, which we keep in a balanced
search tree.  Every enumerated combination of a pair and an index
contributes at least $h\ge1$ to $M_j$, so the
number of combinations is at most $\sum_jM_j=\Oh(d^2)$ by
\cref{lem:ckw_load_bounds}.
There are $\Oh(d^2)$ cover pairs, and each interval computation takes
constant time.  Thus, the sparse dictionary of the positive loads is
constructed in $\Oh(d^2\lg m)$ time and $\Oh(d^2)$ space.

\emph{High-load indices.}  If $q<\sqrt d$ or $q>\kappa$, we set
$J_{\mathrm{high}}=\emptyset$.  Otherwise, we sort the dictionary
entries whose value is at least $q/10$ to obtain $J_{\mathrm{high}}$ in
$\Oh(d^2\lg m)$ time.
\end{proof}

\section{Algorithm for Eligible Block Indices}\label{sec:powered_ckw_sweep}

This section reports, in $\Oh(d^3\lg^2m)$ time, the occurrence starts
assigned to all eligible block indices as $\Oh(d^3)$ arithmetic
progressions of difference $\tau$ (\cref{prop:eligible_output}),
following the strategy of \cref{sec:overview_sweep,sec:overview_marks}.
\begin{enumerate}
    \item The factorization of \cref{sec:ckw_common} gives each padded
    block index $j$ the min-plus product $I_j$ of the matrices of its
    factor pairs (\cref{def:natural_ckw_factor_matrices}).  Most pairs
    have indices outside the exceptional lists
    (\cref{lem:exceptional_list_construction}) and hence the same matrix
    (\cref{lem:plain_outside_exceptional}).  We replace every long run of such pairs
    by one fixed matrix, the \emph{long-run matrix}, and obtain the
    \emph{powered product} $K_j$ (\cref{sec:sweep_powers}).
    \item We show that $K_j$ still determines the occurrence starts
    assigned to $j$ (\cref{sec:sweep_cut_rows,sec:sweep_correctness}).
    A \emph{witness}, an optimal alignment for an occurrence start,
    survives the replacement if it is \emph{in phase} somewhere in every
    long run, that is, if it visits a vertex where the two periodic
    strings are synchronized.  If
    $q<\sqrt d$, staying off phase across a long run costs more than the
    long-run matrix charges, so every witness survives.  Otherwise,
    eligibility means $q>\kappa$ or a low load, and
    \cref{lem:blocks_with_few_marks} shows that the witnesses are
    \emph{clean}: they follow a single rail across every long run.
    \item As $j$ grows, pairs with exceptional text factors shift past
    those with exceptional pattern factors, but the long-run matrix does
    not depend on the length of the run, so $K_j$ changes only
    $\Oh(d^2)$ times.  A sweep maintains $K_j$ and builds a \emph{range
    oracle} (\cref{sec:sweep_oracle}), which \cref{sec:block_geometry}
    also uses.
    \item We split the eligible indices into $\Oh(d^2)$ groups of
    consecutive indices, within which $K_j$ and $r_j-j\tau$ are
    fixed.  For each group, one vector of row minima selects
    the rows that yield occurrence starts, and each such row yields one
    progression of difference $\tau$ across the group
    (\cref{sec:sweep_output}).
\end{enumerate}
\Cref{alg:eligible} summarizes the procedure;
\cref{fig:long_run_matrix,fig:cut_rows,fig:clean_path} illustrate the
long-run matrix, the cut rows, and the clean witnesses.

Earlier algorithms also treat a run of repeated factor pairs as a power
of one matrix \cite[Claim~7.4]{GK25}, \cite[Claim~6.8]{gk24}.  The
sweeps over block indices in \cite[Claim~4.37]{CKW22} and
\cite[Theorem~7.37]{gk24} trim
every run to a common length bound, and \cite[Section~9.3]{gk24}
precomputes the powers for all lengths up to this bound.  Then, a run
changes its matrix whenever its trimmed length changes.  Here, every
long run gets the long-run matrix, which does not depend on the length
of the run, so only runs shorter than a constant cutoff change the
powered product.

Throughout this section, we fix a valid instance of \SM with
$J_{\mathrm{pad}}\ne\emptyset$ and use the parameters of
\cref{def:structural_ckw_parameters,def:common_factorization_parameters}.
We prove correctness up to the threshold $d$, but we compute ferns up to
the larger threshold $W=3\kappa\ge3d$.  All CMOs used or returned in this
section represent nonnegative, integer-valued, $2$-bounded-difference
Monge matrices on $\fragment{0}{\Delta}\times\fragment{0}{\Delta}$.

\subsection{Plain-Run Matrices and Powered Products}\label{sec:sweep_powers}

By \cref{lem:plain_outside_exceptional}, every internal factor pair whose
indices lie outside the exceptional lists consists of the prototypes
$Q_P^\circ$ and $Q_T^\circ$, so a run of such pairs contributes a power
of one matrix.  We keep short runs and replace each long run by a fixed
matrix: the cost of moving to a state in which the two periodic strings
are in phase, matching for free, and moving to the exit state.

\begin{definition}[Plain-run matrices]
\label{def:canonical_run_powers}
\let\qed\relax %
Let $A\coloneqq\udmat{Q_P^\circ}{Q_T^\circ}{\Delta}$ be the matrix of
the prototypes, and fix the cutoff $\chi\coloneqq42$.  The set of
\emph{in-phase states} is
$S\coloneqq\fragment{0}{\Delta}\cap(f+q\mathbb Z)$, and the
\emph{long-run matrix} $\overline A$, with rows and columns indexed by
$\fragment{0}{\Delta}$, is given by
\[
    \overline A[g,h]\coloneqq\min_{c\in S}\bigl(|g-c|+|h-c|\bigr).
    \tag*{\lipicsEnd}
\]
\end{definition}

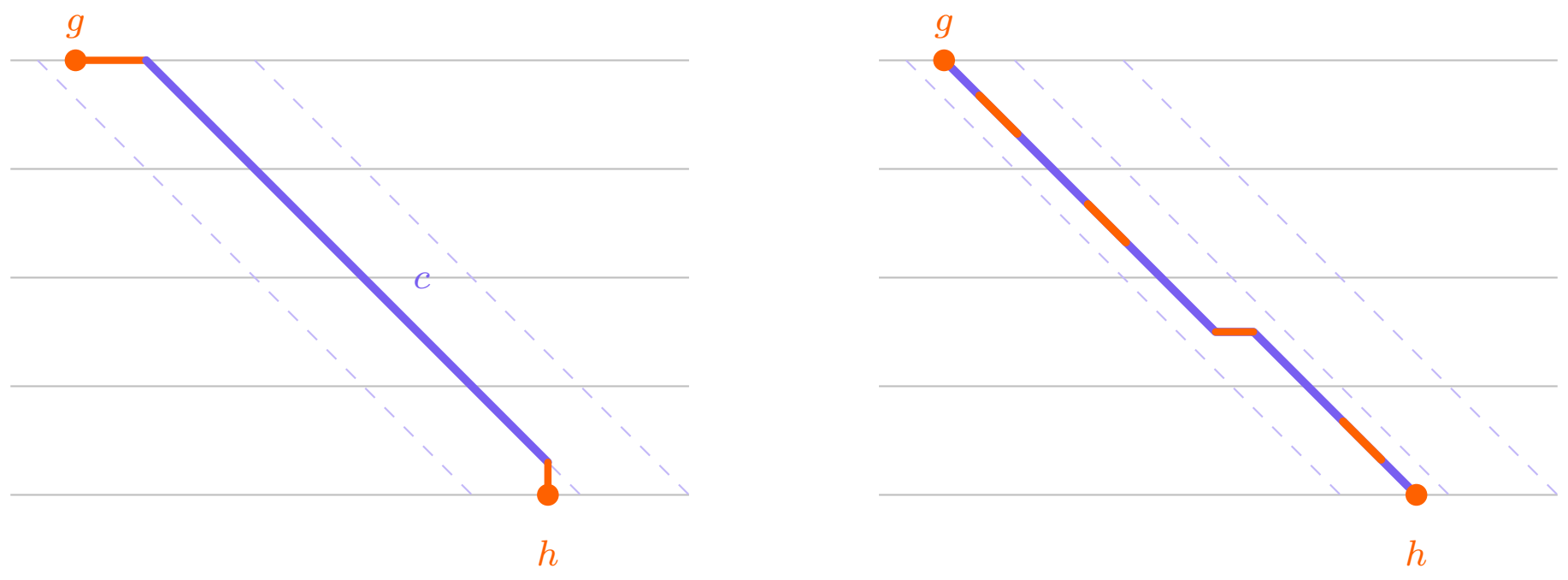
\begin{figure}[t]
\centering
\begin{tikzpicture}[x=0.42cm,y=-0.42cm]
  \foreach \r in {0,2,4,6,8} {\draw[black!25] (0,\r) -- (12.5,\r);}
  \foreach \c in {0.5,2.5,4.5} {\draw[color2!45,dashed] (\c,0) -- ({\c+8},8);}
  \draw[color4,line width=1.6pt,line cap=round] (1.2,0) -- (2.5,0);
  \draw[color2,line width=1.8pt,line cap=round] (2.5,0) -- (9.9,7.4);
  \draw[color4,line width=1.6pt,line cap=round] (9.9,7.4) -- (9.9,8);
  \fill[color4] (1.2,0) circle (2.4pt);
  \fill[color4] (9.9,8) circle (2.4pt);
  \node[font=\scriptsize,color4,anchor=base] at (1.2,-0.55) {$g$};
  \node[font=\scriptsize,color4,anchor=base] at (9.9,9.3) {$h$};
  \node[font=\scriptsize,color2,anchor=base west] at (7.1,4.2) {$c$};
  \node[font=\small,anchor=base] at (6.25,11.2) {(a) in phase};
\end{tikzpicture}
\hspace{1.2cm}
\begin{tikzpicture}[x=0.42cm,y=-0.42cm]
  \foreach \r in {0,2,4,6,8} {\draw[black!25] (0,\r) -- (12.5,\r);}
  \foreach \c in {0.5,2.5,4.5} {\draw[color2!45,dashed] (\c,0) -- ({\c+8},8);}
  \draw[color2,line width=1.8pt,line cap=round,line join=round]
     (1.2,0) -- (6.2,5) -- (6.9,5) -- (9.9,8);
  \draw[color4,line width=1.6pt,line cap=round] (6.2,5) -- (6.9,5);
  \foreach \k in {1,3} {\draw[color4,line width=1.8pt,line cap=round]
     ({0.85+\k},{\k-0.35}) -- ({1.55+\k},{\k+0.35});}
  \draw[color4,line width=1.8pt,line cap=round] (8.55,6.65) -- (9.25,7.35);
  \fill[color4] (1.2,0) circle (2.4pt);
  \fill[color4] (9.9,8) circle (2.4pt);
  \node[font=\scriptsize,color4,anchor=base] at (1.2,-0.55) {$g$};
  \node[font=\scriptsize,color4,anchor=base] at (9.9,9.3) {$h$};
  \node[font=\small,anchor=base] at (6.25,11.2) {(b) off phase};
\end{tikzpicture}
\caption{The long-run matrix in a long plain run (schematic), with
in-phase diagonals dashed.  (a)~A path from $g$ to $h$
meeting the in-phase diagonal of state $c$ costs at least
$|g-c|+|h-c|$, which the orange-purple path attains; the best $c$ gives
$\overline A[g,h]$.  (b)~A path strictly between consecutive in-phase
diagonals has an edit in each of the more than $q\ge\overline A[g,h]$
periods (orange) if $q<\sqrt d$.  Both survive replacing the run by
$\overline A$.}
\label{fig:long_run_matrix}
\end{figure}

In the alignment graph of $(Q_P^\circ,Q_T^\circ)$, the diagonal edge
from a vertex $(x,x+g)$ compares $\Q[x]$ with $\Q[x+g-f]$.  If $g$ is
in phase, then $g-f$ is a multiple of $q$, so these characters are equal
and the edge costs~$0$.  Hence, the path of diagonal edges from the
input state $g$ to the output state $g$ costs~$0$.  The set $S$ contains
$f$, so it is nonempty, and $\overline A[g,h]$ is the cost of moving
from the entry state $g$ to the best $c\in S$, following diagonal edges
at cost $0$, and moving to the exit state $h$; see
\cref{fig:long_run_matrix}.

The next lemma shows that $A^{\oplus s}\le\overline A$ for every
$s\ge\chi$, so replacing a run of at least $\chi$ factor pairs by
$\overline A$ never decreases any entry.  For each $c\in S$, a path
through $s$ copies of the prototype pair moves from state $g$ to $c$,
follows diagonal edges at cost $0$, and moves to $h$, at cost
$|g-c|+|h-c|$.  It deletes at most $\Delta<\chi\tau\le s\tau$ characters
of the pattern, because $\chi=\Delta/d$ and $\tau>d$.

\begin{lemma}[The long-run matrix dominates long powers]
\label{lem:canonical_run_power_monotonicity}
The long-run matrix $\overline A$ is nonnegative, integer-valued,
$1$-bounded-difference, and Monge.  Moreover,
$\overline A\ge A^{\oplus s}$ entrywise for every integer $s\ge\chi$.
\end{lemma}
\begin{proof}
The properties of $\overline A$ are elementary, and domination follows
from an in-phase path.

\emph{Step 1: properties of $\overline A$.}  For fixed $h$, each
function $g\mapsto|g-c|+|h-c|$ changes by at most $1$ between adjacent
integers, and so does their minimum over $c\in S$; by symmetry, the same
holds in~$h$.  For the Monge property, we use that if $a\le a'$,
then $t\mapsto|t-a|-|t-a'|$ is nondecreasing, so
$|x-a|+|x'-a'|\le|x-a'|+|x'-a|$ whenever $x\le x'$.  Fix
$g<g'$ and $h<h'$, and let $c,c'\in S$ satisfy
$\overline A[g,h']=|g-c|+|h'-c|$ and $\overline A[g',h]=|g'-c'|+|h-c'|$.
If $c\le c'$, we bound
$\overline A[g,h]$ using $c$ and $\overline A[g',h']$ using $c'$, and
the inequality for $h<h'$ and $c\le c'$ gives
\[
    \overline A[g,h]+\overline A[g',h']
       \le|g-c|+|h-c|+|g'-c'|+|h'-c'|
       \le\overline A[g,h']+\overline A[g',h].
\]
Otherwise, $c'<c$; we swap the roles of $c$ and $c'$ in these bounds
and use the inequality for $g<g'$.

\emph{Step 2: the in-phase path.}
Let $s\ge\chi$, and consider $\AG(X,Y)$ for the strings
${X\coloneqq\Q\fragmentco{0}{s\tau}}$ and
$Y\coloneqq\Q\fragmentco{-f}{s\tau+f}$.
For $g,h\in\fragment{0}{\Delta}$ and $c\in S$,
we build a path in $\AG(X,Y)$
from $(0,g)$ to $(s\tau,s\tau+h)$; see \cref{fig:long_run_matrix}(a).
It first takes $|g-c|$ deletions if $g>c$ or insertions if $g<c$, then
diagonal edges, and finally $|h-c|$ deletions if $c>h$ or insertions if
$c<h$.  It deletes at most $\Delta$ characters of $X$, and
$|X|=s\tau\ge\chi\tau>\chi d=\Delta$ because $\tau>d$, so the path
exists.  Its diagonal edges compare $X[x]=\Q[x]$ with
$Y[x+c]=\Q[x+c-f]$, which are equal because $q$ divides $c-f$.  Hence,
the path costs $|g-c|+|h-c|$, and its offsets lie between $g$, $c$, and
$h$, hence in $\fragment{0}{\Delta}$.

\emph{Step 3: the path cost bounds the product.}  Since $q$ divides $\tau$,
the $i$th copy of the prototype pair, for $i\in\fragmentco{0}{s}$, is
$\bigl(X\fragmentco{i\tau}{(i+1)\tau},Y\fragmentco{i\tau}{(i+1)\tau+\Delta}\bigr)$.
For $i\in\fragmentoo{0}{s}$, let $g_i$ be the offset of the path's
latest vertex in row $i\tau$, with $g_0\coloneqq g$ and
$g_s\coloneqq h$.  By Step~2, these vertices split the path into one
piece per copy, the $i$th from input state $g_i$ to output state
$g_{i+1}$, which costs at least $A[g_i,g_{i+1}]$.  Hence,
$A^{\oplus s}[g,h]\le|g-c|+|h-c|$, and minimizing over $c\in S$ gives
$A^{\oplus s}\le\overline A$.
\end{proof}

We now apply this replacement to the factor matrices of a block index
$j$ (\cref{def:natural_ckw_factor_matrices}).  Its internal factor pair
with index $i$ is $(P_i,T_{j+i})$, so it consists of the prototypes
unless $i\in\mathcal X_P$ or $j+i\in\mathcal X_T$
(\cref{lem:plain_outside_exceptional}).  We call the maximal intervals
of the remaining indices the \emph{plain runs} of $j$ and replace the factors
of each run with at least $\chi$ elements by $\overline A$.  We do this
for every interval $\fragmentco{\ell}{e}$ of factor indices, not just
for $I_j$, because the range oracle of \cref{sec:sweep_oracle} applies
such partial products to vectors.

\begin{definition}[Plain runs and powered products]
\label{def:powered_ckw_interval_products}
For $j\in J_{\mathrm{pad}}$, a \emph{plain run} of $j$ is a maximal
nonempty integer interval $G\subseteq\fragmentco{0}{t}$ such that
$i\notin\mathcal X_P$ and $j+i\notin\mathcal X_T$ for all $i\in G$; see \cref{def:exceptional_lists}.

For a nonempty integer interval $\fragmentco{\ell}{e}\subseteq \fragmentcc{0}{t}$,
the
\emph{powered product} $K_{j,\fragmentco{\ell}{e}}$ is obtained from
$I_{j,\fragmentco{\ell}{e}}$ by replacing, for each plain run $G$ of $j$
such that $|G\cap\fragmentco{\ell}{e}|\ge \chi$, the
factors $I_{j,i}$ with $i\in G\cap\fragmentco{\ell}{e}$ by the single
matrix $\overline A$.  The powered product of $j$ is
$K_j\coloneqq K_{j,\fragmentcc{0}{t}}$.
\end{definition}

The threshold $\chi$ applies to $G\cap\fragmentco{\ell}{e}$, not to
$G$: if $|G\cap\fragmentco{\ell}{e}|<\chi\le|G|$, the factors $A$ are kept.

\begin{example}[Powered products]\label{ex:powered_word}
Consider illustrative lists $\mathcal X_P=\{41\}$ and
${\mathcal X_T=\{226\}}$, sparser than in any valid instance, with
$t=210$ and $100,101\in J_{\mathrm{pad}}$, and assume that every factor
outside them is plain, as in \cref{lem:plain_outside_exceptional}.  The plain runs of $j=100$ are
$\fragmentco{0}{41}$, $\fragmentco{42}{126}$, and
$\fragmentco{127}{210}$.  For $j=101$, the pair with index $125$ is
$(P_{125},T_{226})$, so the plain runs are $\fragmentco{0}{41}$,
$\fragmentco{42}{125}$, and $\fragmentco{126}{210}$.  In both cases,
only the last two runs have at least $\chi=42$ elements, so
\begin{align*}
    K_{100}&=A^{\oplus 41}\oplus I_{100,41}\oplus\overline A
       \oplus I_{100,126}\oplus\overline A\oplus I_{100,210},\\
    K_{101}&=A^{\oplus 41}\oplus I_{101,41}\oplus\overline A
       \oplus I_{101,125}\oplus\overline A\oplus I_{101,210}.
\end{align*}
The pairs with index $41$ share $P_{41}$, those with indices $126$ and
$125$ share $T_{226}$, and their other factors are plain.  Likewise,
$T_{100,R}=T_{101,R}=Q_R^\circ$, because $310,311\notin\mathcal X_T$.
Hence, $I_{101,41}=I_{100,41}$, $I_{101,125}=I_{100,126}$, and
$I_{101,210}=I_{100,210}$, so $K_{101}=K_{100}$, although the runs
change.
\end{example}

\begin{observation}[Powered products dominate exact products]
\label{obs:powered_product_dominance}
For every $j\in J_{\mathrm{pad}}$ and every nonempty integer interval
$\fragmentco{\ell}{e}\subseteq\fragmentcc{0}{t}$, we have
$K_{j,\fragmentco{\ell}{e}}\ge I_{j,\fragmentco{\ell}{e}}$ entrywise.
\end{observation}
\begin{proof}
For each plain run $G$ of $j$ with $|G'|\ge\chi$, where
$G'\coloneqq G\cap\fragmentco{\ell}{e}$,
the powered product replaces the factors with indices in $G'$ by
$\overline A$, whereas the exact product contains $A^{\oplus|G'|}$ in
their place by \cref{lem:plain_outside_exceptional}; all other factors
coincide.  Hence, \cref{lem:canonical_run_power_monotonicity} and the
monotonicity of min-plus multiplication prove the claim.
\end{proof}

\subsection{Cut Rows and Path Pieces}\label{sec:sweep_cut_rows}

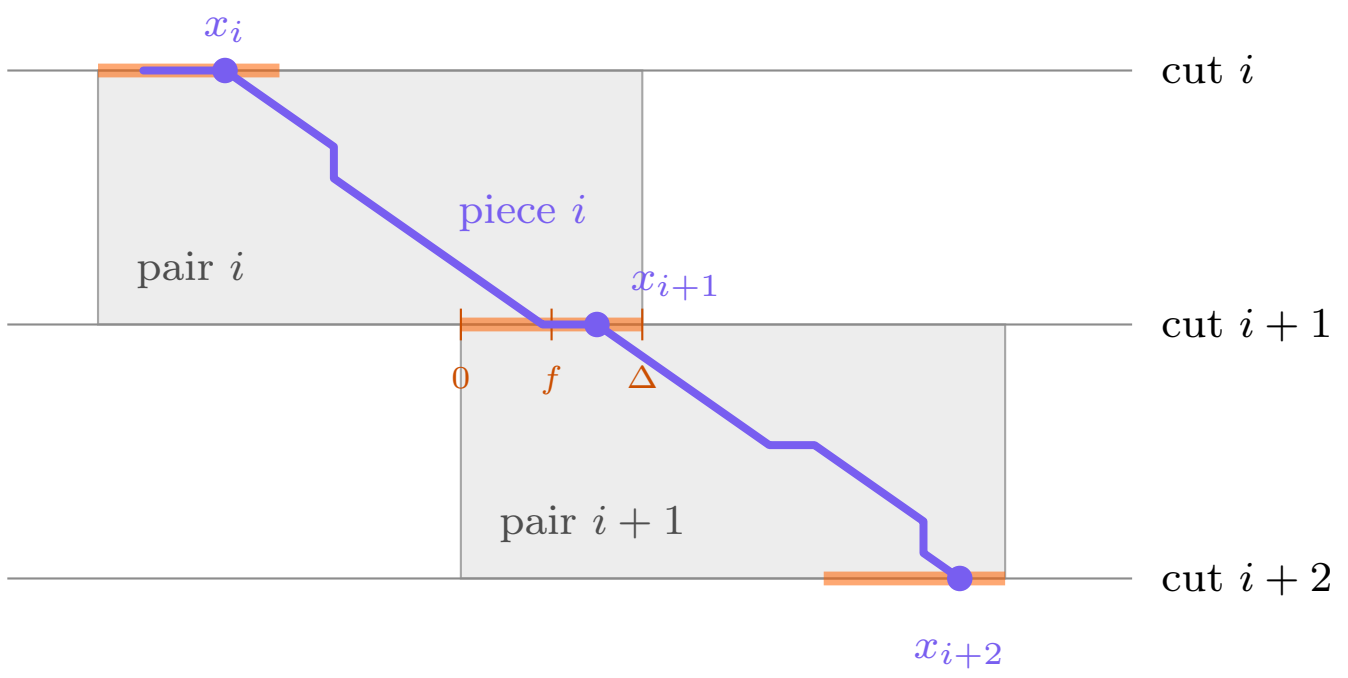
\begin{figure}[t]
\centering
\begin{tikzpicture}[x=0.6cm,y=-0.42cm]
  \fill[black!7] (1,0) rectangle (7,4);
  \fill[black!7] (5,4) rectangle (11,8);
  \draw[black!35] (1,0) rectangle (7,4);
  \draw[black!35] (5,4) rectangle (11,8);
  \node[font=\scriptsize,black!70,anchor=base west] at (1.2,3.3) {pair $i$};
  \node[font=\scriptsize,black!70,anchor=base west] at (5.2,7.3) {pair $i+1$};
  \foreach \y/\l in {0/i,4/i+1,8/i+2} {
    \draw[black!45] (0,\y) -- (12.4,\y);
    \node[font=\scriptsize,anchor=mid west] at (12.5,\y) {cut $\l$};
  }
  \foreach \x/\y in {1/0,5/4,9/8} {
    \draw[color4,line width=2.6pt,opacity=0.55] (\x,\y) -- (\x+2,\y);
  }
  \foreach \x in {5,6,7} {\draw[color4!80!black] (\x,3.75) -- (\x,4.25);}
  \node[font=\tiny,color4!80!black,anchor=base] at (5,5.0) {$0\vphantom{\Delta}$};
  \node[font=\tiny,color4!80!black,anchor=base] at (6,5.0) {$f\vphantom{\Delta}$};
  \node[font=\tiny,color4!80!black,anchor=base] at (7,5.0) {$\Delta$};
  \draw[color2,line width=1.5pt,line cap=round,line join=round]
     (1.5,0) -- (2.4,0) -- (3.6,1.2) -- (3.6,1.7) -- (5.8,3.9) -- (5.9,4)
     -- (6.5,4) -- (8.4,5.9) -- (8.9,5.9) -- (10.1,7.1) -- (10.1,7.6) -- (10.5,8);
  \foreach \x/\y in {2.4/0,6.5/4,10.5/8} {\fill[color2] (\x,\y) circle (2.4pt);}
  \node[font=\scriptsize,color2,anchor=base] at (2.4,-0.55) {$x_i$};
  \node[font=\scriptsize,color2,anchor=base west] at (6.65,3.45) {$x_{i+1}$};
  \node[font=\scriptsize,color2,anchor=base] at (10.5,9.25) {$x_{i+2}$};
  \node[font=\scriptsize,color2,anchor=mid west] at (4.75,2.2) {piece $i$};
\end{tikzpicture}
\caption{Cut rows and pieces (schematic;
\cref{def:cut_rows,def:cut_row_states,def:path_pieces}).  The factor
pairs $i$ and $i+1$ of $j$ (gray) lie between consecutive lifted cut
rows.  Consecutive text factors share the $\Delta+1$ states of their
common cut (orange), from the anchor $z_{j,i+1}$ (state $0$) through the
text cut $c_{j+i+1}$ (state $f$) to state $\Delta$.  The path (purple) is
split at its latest vertices $x_i,x_{i+1},x_{i+2}$ on the cut rows.}
\label{fig:cut_rows}
\end{figure}

To compare path costs with products of factor matrices, we locate the cuts
between the factor pairs in the ordinary and expanded alignment graphs
and split paths at them; see \cref{fig:cut_rows}.

\begin{definition}[Cut rows]
\label{def:cut_rows}
Put $a_{t+1}\coloneqq|P|$, and use the pattern cuts
$0=a_0,a_1,\ldots,a_t$ of \cref{def:internal_cuts}.
Define their \emph{pattern-reference coordinates}, the reference
coordinates that $\A_P$ pairs with them, by
\[
    \widehat a_i\coloneqq i\tau\quad\text{for }i\in\fragmentcc{0}{t}
    \qquad\text{and}\qquad
    \widehat a_{t+1}\coloneqq y_P.
\]
For every $i\in\fragmentcc{0}{t+1}$, call $\{a_i\}\times\fragment{0}{|T|}$
the $i$th \emph{cut row} of $\AG(P,T)$, and let $\widetilde a_i$ be the
unique expansion position with
$\bigl(\pP(\widetilde a_i),\phP(\widetilde a_i)\bigr)=(a_i,\widehat a_i)$.
The row $\widetilde a_i$ of $\eAG(P,T)$ is the $i$th \emph{lifted cut
row}.
\end{definition}

The lifted cut rows satisfy
$0=\widetilde a_0<\widetilde a_1<\cdots<\widetilde a_{t+1}=\ell_P$,
because their defining pairs occur on $\A_P$ in this order.  The cuts
with $i\in\{0,t+1\}$ are the two ends of $P$, and the others are the
seams between consecutive factor pairs.  On each cut row, we measure
text positions from an anchor, so that the vertex $g\in\fragment{0}{\Delta}$
positions to the right of the anchor is both the input state $g$ of
the factor matrix that starts at this cut and the output state $g$ of
the factor matrix that ends there (\cref{def:restricted_distance_matrix}).

\begin{definition}[Cut-row states]
\label{def:cut_row_states}
\let\qed\relax %
For $j\in J_{\mathrm{pad}}$, define the \emph{text anchors}
\[
    z_{j,i}\coloneqq c_{j+i}-f\quad\text{for }i\in\fragmentcc{0}{t}
    \qquad\text{and}\qquad
    z_{j,t+1}\coloneqq c_j^+-f.
\]
For $i\in\fragmentcc{0}{t+1}$, the \emph{state} at cut $i$ of an
expanded vertex $x$ on the $i$th lifted cut row is
\[
    \widetilde\rho_{j,i}(x)\coloneqq\pT(x)-z_{j,i}.
    \tag*{\lipicsEnd}
\]
\end{definition}

An expanded vertex $x$ on the $i$th lifted cut row projects to the
vertex $\pPT(x)=(a_i,z_{j,i}+\widetilde\rho_{j,i}(x))$ of the $i$th cut
row.  At an internal cut $i$, the states $0$, $f$, and $\Delta$
correspond to the text positions $c_{j+i}-f$, $c_{j+i}$, and
$c_{j+i}+f$: the left end of the overlap of the adjacent text
factors, the text cut, and the right end.

To compare a path with a product of factor matrices, we split it at
the lifted cut rows; see \cref{fig:cut_rows}.

\begin{definition}[Pieces of a path]
\label{def:path_pieces}
Fix a nonempty integer interval
$\fragmentco{\ell}{e}\subseteq\fragmentcc{0}{t}$ and a directed path
$\mathcal A$ in $\eAG(P,T)$ from a vertex $x_\ell$ on the $\ell$th
lifted cut row to a vertex $x_e$ on the $e$th lifted cut row.  For
$i\in\fragmentoo{\ell}{e}$, let $x_i$ be the latest vertex of
$\mathcal A$ on the $i$th lifted cut row.  For a nonempty integer
interval $\fragmentco{i_1}{i_2}\subseteq\fragmentco{\ell}{e}$, the
\emph{portion} of $\mathcal A$ between the cuts $i_1$ and $i_2$ is its
subpath from $x_{i_1}$ to $x_{i_2}$.  For $i\in\fragmentco{\ell}{e}$,
the $i$th \emph{piece} of $\mathcal A$ is its portion between the cuts
$i$ and $i+1$.  For
$j\in J_{\mathrm{pad}}$ and $i\in\fragment{\ell}{e}$, the \emph{state}
of $\mathcal A$ at cut $i$ is $\widetilde\rho_{j,i}(x_i)$.
\end{definition}

Since the rows along $\mathcal A$ never decrease,
$x_\ell,\ldots,x_e$ occur on $\mathcal A$ in this order, and
the pieces partition the edges of $\mathcal A$.  The next observation
shows that vertices close to the rail $\eD{b_j}$ have states in
$\fragment{0}{\Delta}$.

\begin{observation}[Cut-row states near the rail]
\label{obs:cut_states_near_rail}
Let $j\in J_{\mathrm{pad}}$ and $i\in\fragmentcc{0}{t+1}$.  Every
expanded vertex $x$ on the $i$th lifted cut row with
$|\mathrm{off}(x)-b_jq|\le W$ satisfies
$\widetilde\rho_{j,i}(x)\in\fragment{0}{\Delta}$.
\end{observation}
\begin{proof}
By \cref{def:internal_cuts,def:centered_carrier}, $\A_T$ pairs the
text position $z_{j,i}+f$ with the reference coordinate
$\widehat a_i+j\tau$.  Since $\phP(x)=\widehat a_i$ and $b_jq=j\tau$, we have
$|\phT(x)-(\widehat a_i+j\tau)|=|\mathrm{off}(x)-b_jq|\le W$.  Along
$\A_T$, the text and reference advances differ by at most $d_T$, so
\[
    |\pT(x)-(z_{j,i}+f)|\le |\phT(x)-(\widehat a_i+j\tau)|+d_T \le 3\kappa+d_T\le15d+3d<21d=f,
\]
where $\kappa\le5d$ and $d_T\le3d$ by \cref{rem:parameter_bounds}.
Hence, $\widetilde\rho_{j,i}(x)=\pT(x)-z_{j,i}\in \fragment{0}{2f}=\fragment{0}{\Delta}$.
\end{proof}

The suffix distance $\dtob$ (\cref{def:semiglobal_suffix_distances})
decides the occurrence starts.  Products of factor matrices bound its
change between cut rows from above, ruling out false positives in
\cref{sec:sweep_correctness,sec:block_geometry}.

\begin{lemma}[Factor products and suffix distances]
\label{lem:factor_product_suffix_bound}
For $j\in J_{\mathrm{pad}}$, a
nonempty integer interval $\fragmentco{\ell}{e}\subseteq\fragmentcc{0}{t}$,
and states $g,g'\in\fragment{0}{\Delta}$,
\[
    \dtob(a_\ell,z_{j,\ell}+g)
      \le I_{j,\fragmentco{\ell}{e}}[g,g']+\dtob(a_e,z_{j,e}+g').
\]
\end{lemma}
\begin{proof}
By \cref{lem:centered_ckw_exact_factorization} and
\cref{def:cut_row_states}, adjacent factors use the same ordinary
vertex for each state of their common cut.  Choose states
$g\eqqcolon g_\ell,\ldots,g_e\coloneqq g'$ attaining the minimum in
$I_{j,\fragmentco{\ell}{e}}[g,g']$ and an optimal path for each
$I_{j,i}[g_i,g_{i+1}]$ with $i\in\fragmentco{\ell}{e}$.  Concatenating these paths yields a walk
$\mathcal C$ in $\oAG(P,T)$ of cost $I_{j,\fragmentco{\ell}{e}}[g,g']$
from $(a_\ell,z_{j,\ell}+g)$ to $(a_e,z_{j,e}+g')$.  The walk may use
reverse edges, because the factor matrices are distances in $\oAG$
(\cref{def:restricted_distance_matrix}).  By
\cref{lem:semiglobal_path_stability}, $\dtob$ changes along every edge
of $\AG(P,T)$ by at most its cost, which is at most $1$, and hence along
every reverse edge by less than its cost $2$.  Summing over the edges
of $\mathcal C$ gives
$\dtob(a_\ell,z_{j,\ell}+g)-\dtob(a_e,z_{j,e}+g')\le\cost(\mathcal C)$.
\end{proof}

\subsection{Surviving Paths and Occurrence Starts}\label{sec:sweep_correctness}

By \cref{obs:powered_product_dominance}, the replacement never
decreases an entry.  We say that a path from the $\ell$th to the $e$th
lifted cut row \emph{survives} the replacement if
$K_{j,\fragmentco{\ell}{e}}[g,g']$ is at most its cost, where $g$ and
$g'$ are its states at the cuts $\ell$ and $e$.  In
\cref{ex:overview_drift}, an optimal alignment stays off phase in a
long run, so its cost grows with the run length, whereas
$\overline A$ does not.  A path whose reference offsets stay within
$W$ of $b_jq$ survives if it is in phase at least once in every long
plain run, that is, if it has a vertex there whose reference
offset is a multiple of $q$.  \emph{Clean} paths, whose detours from the rails
stay near exceptional factors, have this property.  If $q<\sqrt d$,
every such path survives, even when off phase.
\Cref{lem:ckw_clean_powered_exactness} proves both claims, and the
witnesses of eligible block indices are paths of these kinds
(\cref{lem:few_mark_ckw_uniform_instantiation}).

We first consider paths whose edges off the rails lie inside the
reference covers; we call them clean.

\begin{definition}[Clean paths]
\label{def:clean_paths}
A directed path in $\eAG(P,T)$ is \emph{clean} if, for each of its
edges $x\to y$ that does not belong to a rail, $\phP(x,y)$ is
contained in a fragment of $\Es_P$ or $\phT(x,y)$ in a fragment of
$\Es_T$.
\end{definition}

Within plain runs, the alignments $\A_P$ and $\A_T$ have no edits,
and the fragments of the reference covers $\Es_P$ and $\Es_T$ stay
away.  There, the state at each cut equals the reference offset plus
${f-b_jq}$.

\begin{observation}[Offsets in plain runs]
\label{obs:plain_run_offsets}
Let $j\in J_{\mathrm{pad}}$, and let $\fragmentco{\ell}{e}$ be a
nonempty integer interval contained in a plain run of $j$.  Let
$\mathcal A$ be a directed path in $\eAG(P,T)$ from a vertex $x_\ell$
on the $\ell$th lifted cut row to a vertex $x_e$ on the $e$th lifted
cut row such that $|\mathrm{off}(x_i)-b_jq|\le W$ for
$i\in\{\ell,e\}$.  Then,
\[
    \widetilde\rho_{j,i}(x_i)=\mathrm{off}(x_i)-b_jq+f
    \qquad\text{for }i\in\{\ell,e\}.
\]
Moreover, $\cancost(x_\ell,x_e)=0$, $\phP(x_\ell,x_e)$ is strictly
disjoint from all fragments of $\Es_P$, and $\phT(x_\ell,x_e)$ is
strictly disjoint from all fragments of $\Es_T$.
\end{observation}
\begin{proof}
For $i\in\{\ell,e\}$, we have
$\phP(x_i)=i\tau$, so
$|\phT(x_i)-(j+i)\tau|=|\mathrm{off}(x_i)-b_jq|\le W<wq$ because
$b_jq=j\tau$.  Thus, $\phT(x_i)\in\fragmentcc{(j+\ell)\tau-wq}{(j+e)\tau+wq}\eqqcolon R$.  Since $\phP$
and $\phT$ are nondecreasing along $\mathcal A$, we have
$\phP(x_\ell,x_e)=\Q\fragmentco{\ell\tau}{e\tau}$, and
$\phT(x_\ell,x_e)$ is contained in $\Q\fragmentco{\min R}{\max R}$.

The windows $H_i^P$ and $H^T_{j+i}$ with
$i\in\fragmentco{\ell}{e}$ cover $\fragmentco{\ell\tau}{e\tau}$ and
$R$, respectively.  For each such~$i$, we have $i\notin\mathcal X_P$
and $j+i\notin\mathcal X_T$, so, by \cref{def:exceptional_lists}, no
fragment of $\Es_P$ weakly overlaps $H_i^P$, and no fragment of $\Es_T$
weakly overlaps $H^T_{j+i}$.  A fragment that weakly overlaps one of
the two projections also weakly overlaps a window covering it, so
$\phP(x_\ell,x_e)$ and $\phT(x_\ell,x_e)$ are strictly disjoint from all
fragments of $\Es_P$ and $\Es_T$, respectively.  By
\cref{fct:edits_only_in_cover}, $\mathcal A$ passes no edits of $\A_P$
or~$\A_T$, that is, $\cancost(x_\ell,x_e)=0$.

Let $i\in\{\ell,e\}$.  The alignment $\A_T$ pairs the text position
$z_{j,i}+f=c_{j+i}$ with the reference coordinate $(j+i)\tau$.  By
\cref{fct:edits_only_in_cover}, $\A_T$ has no edits in $R$, so
$\pT(x_i)-c_{j+i}=\phT(x_i)-(j+i)\tau$ and
\[
    \widetilde\rho_{j,i}(x_i)
       =\pT(x_i)-c_{j+i}+f
       =\phT(x_i)-\phP(x_i)-j\tau+f
       =\mathrm{off}(x_i)-b_jq+f.\qedhere
\]
\end{proof}

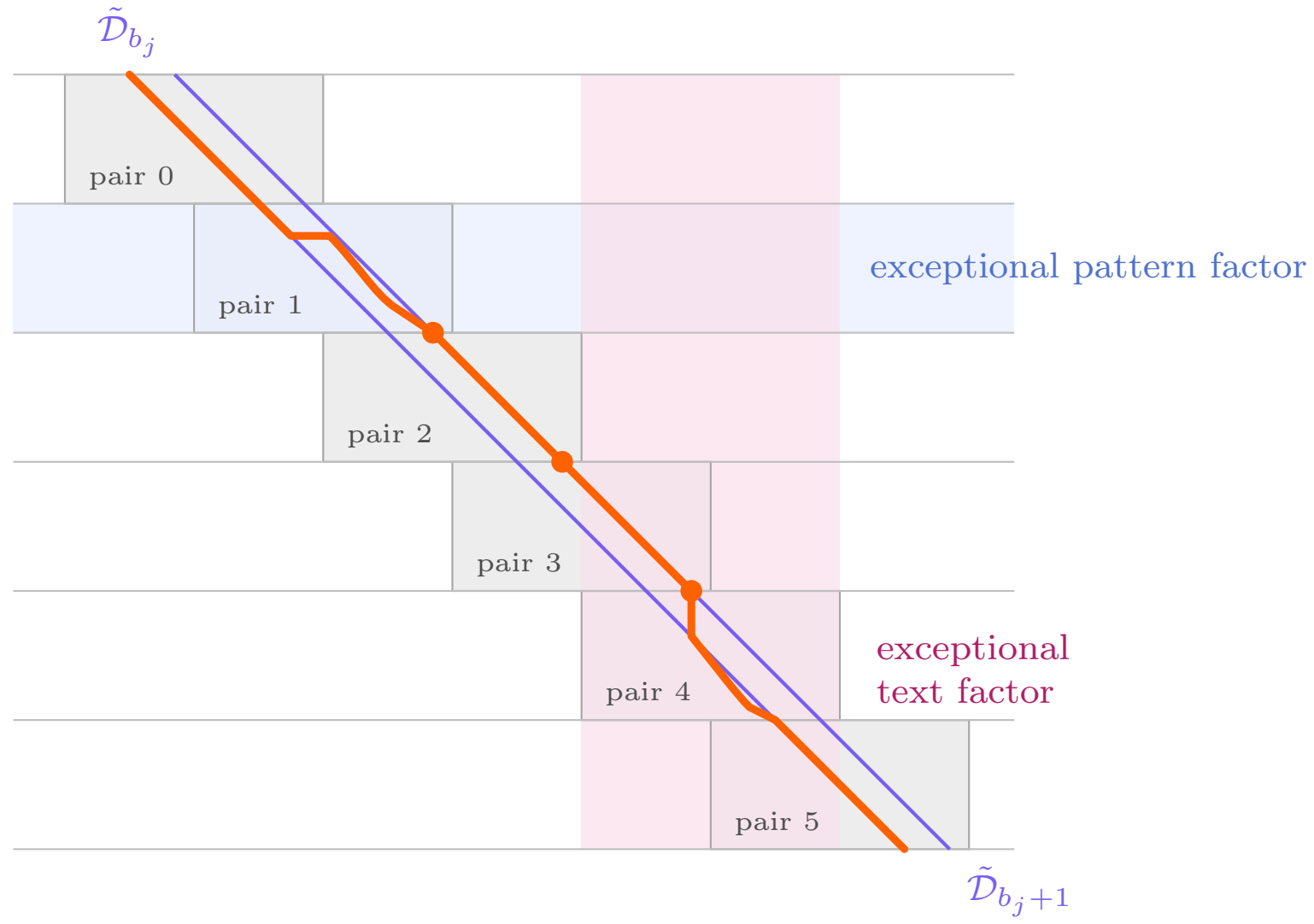
\begin{figure}[t]
\centering
\begin{tikzpicture}[x=0.5cm,y=-0.5cm]
  \foreach \i in {0,...,5} {\fill[black!7] (2*\i-0.2,2*\i) rectangle (2*\i+3.8,2*\i+2);}
  \fill[color1!14,opacity=0.75] (-1,2) rectangle (14.5,4);
  \fill[color3!14,opacity=0.75] (7.8,0) rectangle (11.8,12);
  \foreach \i in {0,...,5} {
    \draw[black!35] (2*\i-0.2,2*\i) rectangle (2*\i+3.8,2*\i+2);
    \node[font=\tiny,black!70,anchor=base west] at (2*\i-0.1,2*\i+1.7) {pair $\i$};
  }
  \foreach \y in {0,2,...,12} {\draw[black!25] (-1,\y) -- (14.5,\y);}
  \foreach \c in {0,0.7} {\draw[color2,line width=0.8pt] (0.8+\c,0) -- (12.8+\c,12);}
  \node[font=\scriptsize,color2,anchor=south] at (0.8,0) {$\eD{b_j}$};
  \node[font=\scriptsize,color2,anchor=north west] at (13.5,12) {$\eD{b_j+1}$};
  \draw[color4,line width=1.7pt,line cap=round,line join=round]
     (0.8,0) -- (2.8,2) -- (3.3,2.5) -- (3.9,2.5) .. controls (4.3,2.9) and (4.6,3.4) .. (4.9,3.6) -- (5.5,4)
     -- (9.5,8) -- (9.5,8.7) .. controls (9.9,9.2) and (10.2,9.6) .. (10.4,9.8) -- (10.8,10) -- (12.8,12);
  \foreach \y in {4,6,8} {\fill[color4] ({\y+1.5},\y) circle (2.4pt);}
  \node[font=\scriptsize,color1!80!black,anchor=west] at (12.0,3) {exceptional pattern factor};
  \node[font=\scriptsize,color3!80!black,anchor=west,text width=1.8cm] at (12.1,9.2) {exceptional text factor};
\end{tikzpicture}
\caption{A clean path (\cref{def:clean_paths}) near the rail
$\eD{b_j}$, in reference coordinates (schematic), with the factor pairs
of $j$ in gray.  Its detours lie inside the pairs with an exceptional
pattern factor (blue rows) or text factor (magenta columns).  Across the
plain run of pairs $2$ and $3$, it follows the rail $\eD{b_j+1}$ and
crosses every cut in the in-phase state $f+q$ (dots), so it survives
replacing the run by $\overline A$.}
\label{fig:clean_path}
\end{figure}

Across a plain run, the covers stay away (\cref{obs:plain_run_offsets}),
so a clean path whose offsets stay within $W$ of $b_jq$ has no edges
off the rails there: it follows a single rail and is in phase; see
\cref{fig:clean_path}.

The next lemma shows that clean paths and, if $q<\sqrt d$, all paths
survive the replacement, provided that their offsets stay within $W$ of
$b_jq$.  In the latter case, a path that stays off phase across a long
run pays for every period (\cref{lem:noblue}), which is more than the
long-run matrix charges.

\begin{lemma}[Paths preserved by powered products]
\label{lem:ckw_clean_powered_exactness}
Fix $j\in J_{\mathrm{pad}}$, a nonempty integer interval
$\fragmentco{\ell}{e}\subseteq\fragmentcc{0}{t}$, and a directed path
$\mathcal A$ in $\eAG(P,T)$ from a vertex $x_\ell$ on the $\ell$th
lifted cut row to a vertex $x_e$ on the $e$th lifted cut row.  Suppose
that $|\mathrm{off}(x')-b_jq|\le W$ for every vertex $x'$ of
$\mathcal A$ and that $q<\sqrt d$ or $\mathcal A$ is clean.  Then, the
endpoint states of $\mathcal A$ lie in $\fragment{0}{\Delta}$ and
satisfy
\[
    K_{j,\fragmentco{\ell}{e}}[\widetilde\rho_{j,\ell}(x_\ell),
                    \widetilde\rho_{j,e}(x_e)]
       \le\cost(\mathcal A).
\]
\end{lemma}
\begin{proof}
We bound every factor of the powered product by the cost of the
corresponding part of $\mathcal A$: a factor $I_{j,i}$ by the $i$th
piece, and a factor $\overline A$ by the portion across the replaced
run.

\emph{Step 1: factors $I_{j,i}$.}  Let $x_\ell,\ldots,x_e$ and
$\mathcal A_\ell,\ldots,\mathcal A_{e-1}$ be the vertices and the
pieces of $\mathcal A$ from \cref{def:path_pieces}, and put $g_i\coloneqq\widetilde\rho_{j,i}(x_i)$
for $i\in\fragmentcc{\ell}{e}$.  By \cref{obs:cut_states_near_rail},
every $g_i$ lies in $\fragment{0}{\Delta}$; this includes the endpoint
states $g_\ell$ and $g_e$.  The projection of
$\mathcal A_i$ under $\pPT$ has the same cost
(\cref{obs:expanded_projection}).  Since $\pP$ and $\pT$ are
nondecreasing along $\mathcal A$, this projection stays between its
endpoints, the vertices of the states $g_i$ and $g_{i+1}$ on the cut
rows $i$ and $i+1$.  Hence, it is a path in the alignment graph of the
factor pair of $j$ with index $i$, and
$I_{j,i}[g_i,g_{i+1}]\le\cost(\mathcal A_i)$.  These bounds cover all
factors of the powered product other than $\overline A$.

\emph{Step 2: replaced runs.}  Consider a plain run of $j$ whose intersection
$G=\fragmentco{i_1}{i_2}$ with $\fragmentco{\ell}{e}$ has
$h\coloneqq|G|\ge\chi$ elements, so that the powered product replaces
its factors by $\overline A$.  Let $\mathcal B$ be the portion of
$\mathcal A$ between the cuts $i_1$ and $i_2$.  We show that
$\overline A[g_{i_1},g_{i_2}]\le\cost(\mathcal B)$.  By
\cref{obs:plain_run_offsets}, applied to $\fragmentco{i_1}{i_2}$,
$\cancost(x_{i_1},x_{i_2})=0$, the reference projections of
$\mathcal B$ are strictly disjoint from all fragments of the
corresponding covers, and $g_i=\mathrm{off}(x_i)-b_jq+f$ for
${i\in\{i_1,i_2\}}$.

\textsc{Case 1:} some vertex $x'$ of $\mathcal B$ has a reference
offset divisible by $q$.  The state $c\coloneqq\mathrm{off}(x')-b_jq+f$ lies
in $f+q\mathbb Z$ and satisfies $|c-f|\le W\le15d<f$, so $c\in S$.
Since $\cancost(x_{i_1},x_{i_2})=0$, \cref{obs:cancost_factorizes}
gives $\cancost(x_{i_1},x')=\cancost(x',x_{i_2})=0$, so applying
\cref{lem:offset_change} to the parts of $\mathcal B$ before and after
$x'$ yields
\[
    \overline A[g_{i_1},g_{i_2}]
       \le|g_{i_1}-c|+|g_{i_2}-c|
       =|\mathrm{off}(x_{i_1})-\mathrm{off}(x')|
          +|\mathrm{off}(x')-\mathrm{off}(x_{i_2})|
       \le\cost(\mathcal B).
\]

\textsc{Case 2:} no vertex of $\mathcal B$ has a reference offset
divisible by $q$.  Then, no edge of $\mathcal B$ belongs to a rail,
because all vertices of a rail have reference offsets divisible by $q$.
If $\mathcal A$ were clean, one reference projection of an edge of
$\mathcal B$ would be contained in a fragment of the corresponding
dense cover, and this fragment would weakly overlap the corresponding
reference projection of $\mathcal B$, a contradiction.  Thus,
$q<\sqrt d$.  By
\cref{lem:offset_change,obs:cancost_factorizes}, each edge of
$\mathcal B$ changes the reference offset by at most one, so $g_{i_1}$
and $g_{i_2}$ lie strictly between two consecutive elements $c<c+q$
of $f+q\mathbb Z$.  These are in-phase states: they lie within
$W+q\le15d+d<f$ of $f$ because $g_{i_1}$ does within $W$ and
$q<\sqrt d$.  As $c<g_{i_1},g_{i_2}<c+q$, the sums
$|g_{i_1}-c|+|g_{i_2}-c|$ and $|g_{i_1}-c-q|+|g_{i_2}-c-q|$ add up to
$2q$, so $\overline A[g_{i_1},g_{i_2}]\le q$.  On the other hand,
$\mathcal B$ shares no edge with a rail, $\cancost(x_{i_1},x_{i_2})=0$,
and $|\phP(x_{i_1},x_{i_2})|=h\tau$, so \cref{lem:noblue} gives
$\cost(\mathcal B)\ge\floor{h\tau/q}>q$,
because $h\ge\chi=42$ and $\tau>d>q^2$, so $h\tau/q>42q$.

\emph{Step 3: combining the bounds.}  In both cases of Step~2,
$\overline A[g_{i_1},g_{i_2}]\le\cost(\mathcal B)
=\sum_{i\in G}\cost(\mathcal A_i)$.  Together with Step~1, choosing the
states $g_i$ at the cuts between consecutive factors of the powered
product shows that
$K_{j,\fragmentco{\ell}{e}}[g_\ell,g_e]\le\sum_{i=\ell}^{e-1}\cost(\mathcal A_i)=\cost(\mathcal A)$.
\end{proof}

We now show that, for an eligible block index $j$, the row minima of
$K_j$ determine the occurrence starts assigned to $j$, that is, those
in its start interval $\fragmentco{r_j}{r_{j+1}}$ (\cref{def:jdef}).
Row $g$ of $K_j$ corresponds to the start vertex $(0,c_j-f+g)$, because
$z_{j,0}=c_j-f$ (\cref{def:cut_row_states}).  For the all-zero vector
$\mathbf0$ on $\fragment{0}{\Delta}$, the entry $(K_j\oplus\mathbf0)[g]$
is the minimum of row $g$, so one matrix--vector product yields all row
minima.  We call optimal paths from the occurrence
starts to the bottom row witnesses.

\begin{definition}[Witnesses]
\label{def:witnesses}
Let $j\in J_{\mathrm{pad}}$, and let $v\in\fragmentco{r_j}{r_{j+1}}$
satisfy $\dtob(0,v)\le d$.  A \emph{witness} for $v$ is a shortest
path in $\eAG(P,T)$ from a vertex $x$ of the top row with
$\pPT(x)=(0,v)$ to the bottom row.
\end{definition}

Every witness for $v$ costs $\dtob(0,v)$
(\cref{lem:semiglobal_expanded_transfer}), the minimum edit distance
between $P$ and a fragment of $T$ starting at $v$.  The next lemma
shows that the row minima of $K_j$ never fall below these costs and
attain them when they are at most $d$.  The lower bound follows from
\cref{lem:factor_product_suffix_bound,obs:powered_product_dominance},
the upper bound from \cref{lem:ckw_clean_powered_exactness} applied to
a witness (clean by \cref{lem:blocks_with_few_marks} if
$q\ge\sqrt d$).

\begin{lemma}[Occurrence starts from row minima]
\label{lem:few_mark_ckw_uniform_instantiation}
For every $j\in J_{\mathrm{elig}}$, a position
$v\in\fragmentco{r_j}{r_{j+1}}$ belongs to $\OccE_k(P,T)$ if and only
if $v=c_j-f+g$ for some $g\in\fragment{0}{\Delta}$ with
${(K_j\oplus\mathbf0)[g]\le k}$.
\end{lemma}
\begin{proof}
By \cref{lem:semiglobal_expanded_transfer}, a position $v$ belongs to
$\OccE_k(P,T)$ if and only if $\dtob(0,v)\le k$.

\emph{No false positives.}  Let $g\in\fragment{0}{\Delta}$, and put
$v\coloneqq c_j-f+g$.  Row $g$ of $K_j$ corresponds to the start vertex
$(0,v)$, and every column $g'$ of $K_j$ to the vertex
$(|P|,c_j^+ -f+g')$ on the bottom row of $\AG(P,T)$, where $\dtob$
vanishes.  By
\cref{lem:factor_product_suffix_bound}, with $\ell=0$ and $e=t+1$, and
by \cref{obs:powered_product_dominance},
\[
    \dtob(0,v)\le I_j[g,g']\le K_j[g,g']
       \qquad\text{for }g'\in\fragment{0}{\Delta}.
\]
Taking the minimum over $g'$ gives $\dtob(0,v)\le(K_j\oplus\mathbf0)[g]$.
Hence, if ${(K_j\oplus\mathbf0)[g]\le k}$, then
$v\in\OccE_k(P,T)$.

\emph{No false negatives.}  Let
$v\in\OccE_k(P,T)\cap\fragmentco{r_j}{r_{j+1}}$.  Then,
$\dtob(0,v)\le k\le d$ because $d\ge2k$, so $v$ has a witness $\eA$, say
from $x$ to $y$.  Since $v\in\OccE_d(P,T)$ and $W=3\kappa$,
\cref{lem:ckw_start_relevance} gives $x\in\tClose_{W,b_j}$, that is,
$x$ is relevant: its potential $\tdtob(x)+\tpena_{b_j}(x)$ is at most
$W$ (\cref{def:relevant_vertices}).  The path $\eA$ is a shortest path
to the bottom row, hence tight, so, by \cref{lem:relevant_path}, all
its vertices belong to $\tClose_{W,b_j}$.  In particular, since
$\tpena_{b_j}(x')\ge|\mathrm{off}(x')-b_jq|$, we have
$|\mathrm{off}(x')-b_jq|\le W$ for every vertex $x'$ of $\eA$.

\emph{Clean witnesses.}  Suppose that $q\ge\sqrt d$.  Since $j$ is
eligible, $q>\kappa$ or $M_j<q/10$ (\cref{def:ckw_load_partition}).  In
the latter case, $I\coloneqq\fragmentcc{b_j-w}{b_j+w}$ satisfies
$I\subseteq\prD$ and $\sum_{c\in I}\mu(c)\le M_j$ because $j$ is
padded and $R_*\ge w$, and every rail $\eD{c}$ met by $\eA$ satisfies
$|c-b_j|q\le W<wq$, so $\min I<c<\max I$.  Thus,
\cref{lem:blocks_with_few_marks} applies to $\eA$: one reference
projection of every detour is contained in a fragment of the
corresponding cover.  Every edge of $\eA$ off the rails lies in a
detour (\cref{def:detours}), so $\eA$ is clean.

\emph{Applying \cref{lem:ckw_clean_powered_exactness}.}  The vertices
$x$ and $y$ lie on the top and bottom rows, the lifted cut rows $0$
and $t+1$.  Since $q<\sqrt d$ or $\eA$ is clean, the lemma applies with
$\ell=0$ and $e=t+1$.  With $g\coloneqq\widetilde\rho_{j,0}(x)=v-(c_j-f)$
and $g'\coloneqq\widetilde\rho_{j,t+1}(y)$, it gives
$g,g'\in\fragment{0}{\Delta}$ and
\[
    (K_j\oplus\mathbf0)[g]\le K_j[g,g']
       \le\cost(\eA)=\dtob(0,v)\le k.
\]
Thus, $v=c_j-f+g$ with $(K_j\oplus\mathbf0)[g]\le k$, as claimed.
\end{proof}

\subsection{Range Oracle}\label{sec:sweep_oracle}

We now sweep over the block indices and maintain their powered
products $K_j$ (\cref{def:powered_ckw_interval_products}); see
\cref{fig:overview_sweep}.  Along the way, we build a range oracle for
the powered products $K_{j,\fragmentco{\ell}{e}}$ of intervals of
factor pairs, which \cref{sec:block_geometry} also uses.  A query
specifies a block index $j$, an interval $\fragmentco{\ell}{e}$, and a
vector $z$, and it returns $K_{j,\fragmentco{\ell}{e}}\oplus z$, correct
up to the threshold $W$; the interval $\fragmentcc{0}{t}$ selects the
full product $K_j$.  The sweep maintains the sequence of matrices whose
product is $K_j$.

\begin{definition}[Tokens and powered words]
\label{def:tokens}
A \emph{token} is a pair $(G,X)$ consisting of a nonempty integer interval $G$ and
a matrix $X$.  For $j\in J_{\mathrm{pad}}$, every plain run $G$ of $j$
with $|G|\ge\chi$ gives the \emph{run token} $(G,\overline A)$, and
every index $i\in\fragmentcc{0}{t}$ that lies in no such run gives the
\emph{singleton token} $(\fragmentcc{i}{i},I_{j,i})$.  The
\emph{powered word} of $j$ is the sequence of these tokens, ordered by
their intervals.
\end{definition}

The intervals of the tokens partition $\fragmentcc{0}{t}$, and the
product of their matrices, in this order, is~$K_j$.  We call the
indices $i\in\fragmentco{0}{t}$ with $i\in\mathcal X_P$ or
$j+i\in\mathcal X_T$ the \emph{exceptional positions} of $j$.  When $j$
increases by one, the exceptional positions from $\mathcal X_P$ stay
fixed, while those from $\mathcal X_T$ move one step to the left, so a
long plain run may change its length but keeps its matrix
$\overline A$.  The next lemma separates the changes of matrices and
token types from this predictable movement of the intervals.  It
adapts the crossing and short-run events of \cite[Claim~4.37]{CKW22}
and \cite[Claim~7.39]{gk24}; here, a constant cutoff and a fixed
long-run matrix replace trimming every run to a common length bound.

\begin{lemma}[Events of the sweep]
\label{lem:sweep_events}
Given the preprocessing record $\mathfrak P$
(\cref{def:preprocessing_record}), one can partition $J_{\mathrm{pad}}$
in $\Oh(d^2\lg m)$ time into $\Oh(d^2)$ intervals such that, on each
interval, the powered word of $j$ consists of $\Oh(d)$ tokens whose
types and matrices do not depend on $j$.  The sequences of these
matrices on consecutive intervals differ by $\Oh(d^2)$ insertions,
deletions, and substitutions in total, which are listed in
$\Oh(d^2\lg m)$ time.
\end{lemma}
\begin{proof}
The indices in $\fragmentcc{0}{t}$ outside the plain runs of $j$ are
$t$ and the $\Oh(d)$ exceptional positions of $j$, each giving one
singleton token (one in total if a position from $\mathcal X_P$ and one
from $\mathcal X_T$ coincide).  The plain runs are the gaps between
these indices.  A gap with at least $\chi$ elements gives one run
token, and a shorter gap gives fewer than $\chi$ singleton tokens with
the matrix $A$ (\cref{lem:plain_outside_exceptional}).  There are at
most $|\mathcal X_P|+|\mathcal X_T|$ exceptional positions and at most
$|\mathcal X_P|+|\mathcal X_T|+1$ gaps, each giving fewer than $\chi$
tokens.  Hence, the powered word of $j$ consists of at most
$\chi(|\mathcal X_P|+|\mathcal X_T|+1)=\Oh(d)$ tokens
(\cref{lem:exceptional_list_construction}).

We \emph{isolate}, as singleton intervals, the $\Oh(d^2)$ indices
$j\in J_{\mathrm{pad}}$ of the form $j=u-i+\delta$ with
$u\in\mathcal X_T$, $i\in\mathcal X_P\cup\{-1,t\}$, and
$\delta\in\fragmentcc{-\chi}{\chi}$, where the fixed positions $-1$ and
$t$ mark the two ends of $\fragmentco{0}{t}$.  At every other $j$, each moving position $u-j$
with $u\in\mathcal X_T$ differs by more than $\chi$ from each fixed
position.  Hence, on each interval between isolated indices, the order
of all these positions is fixed, and so is the set of those in
$\fragmentco{0}{t}$; moreover, $j+t\notin\mathcal X_T$ because no
moving position equals $t$.  A gap whose length changes with $j$ lies
between a fixed and a moving position, so it has at least $\chi$
elements and gives a run token with the matrix $\overline A$; all other
gaps have lengths independent of $j$ and hence give the same number
of tokens, of the same types, for all $j$ in the interval.  By \cref{lem:plain_outside_exceptional}, a singleton at a
position from $\mathcal X_P$ pairs a fixed pattern factor with
$Q_T^\circ$, one at a position from $\mathcal X_T$ pairs $Q_P^\circ$
with a fixed text factor, and the endpoint pair of $j$ does not depend
on $j$ because $j+t\notin\mathcal X_T$.  Hence, on each interval, the
types and matrices of the tokens do not depend on $j$.

To compute the partition, we sort the $\Oh(d^2)$ triples
$(u,i,\delta)$ by their isolated indices $u-i+\delta$ in
$\Oh(d^2\lg m)$ time.  We then sweep over the intervals in increasing
order.  For the current block index~$j$, we keep the fixed positions
and the moving positions $u-j$ in a balanced search tree, ordered by
position; we initialize it by merging $\mathcal X_P$ with
$\mathcal X_T-j$ for $j\coloneqq\min J_{\mathrm{pad}}$.  Between
isolated indices, the order does not change, and the tokens change only
their intervals.  When the sweep enters or leaves an isolated index,
we process its triples one by one: each updates the gaps next to its moving position $u-j$, which
takes $\Oh(\lg m)$ time and changes $\Oh(\chi)=\Oh(1)$ tokens.  Each
node also stores the number of tokens at the positions in its subtree
and in the gaps after these positions, so each change comes with the rank and
the interval of its token.  Hence, the changes are listed in
$\Oh(d^2\lg m)$ time.
\end{proof}

\begin{example}[Events of the sweep]\label{ex:sweep_events}
We continue \cref{ex:powered_word} with its illustrative lists
$\mathcal X_P=\{41\}$ and $\mathcal X_T=\{226\}$ and $t=210$, so that the exceptional position of $j$
from $\mathcal X_T$ is $226-j$.  Write $\mathsf S_P$ and
$\mathsf S_T$ for the singleton tokens at the exceptional positions
from $\mathcal X_P$ and $\mathcal X_T$.  For all
$j\in\fragmentcc{59}{142}$, the tokens of the internal factor
pairs have the same types and matrices,
\[
    \underbrace{A\ \cdots\ A}_{41}\ \ \mathsf S_P\ \ \overline A\ \ \mathsf S_T\ \ \overline A,
\]
although the intervals of $\mathsf S_T$ and its two neighbors change
with $j$.  For $j\in\fragmentcc{17}{58}$, the run after $\mathsf S_T$
has $j-17<\chi=42$ elements, each a singleton token, so the tokens
change at every step; at $j=59$, this run becomes a run token.  The
tokens also change at every step for $j\in\fragmentcc{143}{227}$: the
run between $\mathsf S_P$ and $\mathsf S_T$ has $184-j<\chi$ elements
until $j=184$, $\mathsf S_T$ and $\mathsf S_P$ coincide at $j=185$
(a crossing), and then $\mathsf S_T$ moves through the prefix
$\fragmentco{0}{41}$.  For
$j\le16$ and $j\ge227$, the position $226-j$ lies outside
$\fragmentco{0}{t}$, and the tokens are $41$ copies of $A$,
$\mathsf S_P$, and $\overline A$.
In the proof of \cref{lem:sweep_events}, each pair of $\mathsf S_T$
with $\mathsf S_P$ or an end of $\fragmentco{0}{t}$ yields at most
$2\chi+1=85$ isolated indices, such as $\fragmentcc{143}{227}$ for
$\mathsf S_T$ and $\mathsf S_P$.
\end{example}

To build the range oracle, we sweep over the intervals of
\cref{lem:sweep_events}, apply the listed changes to a persistent
balanced product tree over the matrices of the current powered word, and
keep the version of this tree for every interval.  A query combines
the CMOs of $\Oh(\lg m)$ subtrees.

\begin{lemma}[Range oracle]
\label{lem:range_oracle}
Given the preprocessing record $\mathfrak P$
(\cref{def:preprocessing_record}), one can construct in
$\Oh(d^3\lg^2 m)$ time a data structure that answers the following
queries in $\Oh(d\lg^2 m)$ time each.  Given $j\in J_{\mathrm{pad}}$,
a nonempty integer interval
$\fragmentco{\ell}{e}\subseteq\fragmentcc{0}{t}$, and a finite
nonnegative vector $z$ indexed by $\fragment{0}{\Delta}$, the query
$\mathsf{Range}(j,\ell,e,z)$ returns a finite nonnegative vector $y$
such that $y\meq{W}K_{j,\fragmentco{\ell}{e}}\oplus z$.
\end{lemma}
\begin{proof}
We sweep over the intervals of \cref{lem:sweep_events}, maintain the
matrices of the current powered word in a persistent balanced product
tree, and keep its version for each interval.

\emph{Matrix handles.}  We construct the fern catalogue by
\cref{lem:ckw_fern_catalogue}.  A singleton token
$(\fragmentcc{i}{i},I_{j,i})$ obtains a read-only handle of a CMO
$W$-equivalent to $I_{j,i}$ by the catalogue query $(j,i)$, in
$\Oh(\lg m)$ time, whenever it is created or its matrix changes.  All run tokens share one CMO of
$\overline A$, which we construct from its $(\Delta+1)^2$ entries,
computed naively, in $\Oh(d^3)$ time (\cref{lem:bd_monge_cmo_operations}); this needs a
bounded-difference Monge matrix, which $\overline A$ is by
\cref{lem:canonical_run_power_monotonicity}.  When the sweep first
encounters a run token, say at block index $j$ with smallest index
$i$, the query $(j,i)$ also returns a CMO of some $\widehat A\meq{W}A$
(\cref{lem:plain_outside_exceptional}), which queries use for short
parts of runs.  If no run token ever occurs, neither CMO is needed.

\emph{Product trees.}  As in \cite[Lemma~9.1]{gk24}, we store the
tokens of the current powered word, in order, at the leaves of a
balanced binary tree.  Each leaf stores the CMO handle of its token.
Each internal node stores a CMO of the min-plus product of its children
and the number of leaves in its subtree.  This number counts tokens,
not factor pairs, because the lengths of the tokens may depend on~$j$.  We build the first tree using
$\Oh(d)$ products and then process the $\Oh(d^2)$ updates of
\cref{lem:sweep_events} by balanced insertions and deletions and by
substitutions at the leaves.  Each update locates its token by rank,
using the leaf counts.  We use a functionally persistent tree,
such as an AVL tree~\cite{AVL62} with path copying~\cite{DSST89}: an
update copies the nodes that it changes and their ancestors,
$\Oh(\lg m)$ in total, computes the products of the copies, and never modifies
existing nodes or CMOs.  After the updates for each interval, we store
the current root with the interval.  A CMO product costs
$\Oh(d\lg m)$ time by \cref{lem:bd_monge_cmo_operations}, so all
updates take $\Oh(d^3\lg^2 m)$ time.  Together with the catalogue
and the CMOs of $\overline A$ and $\widehat A$, the construction takes
$\Oh(d^3\lg^2 m)$ time.

\emph{Range queries.}  To answer $\mathsf{Range}(j,\ell,e,z)$, we find
the interval of \cref{lem:sweep_events} containing $j$ by predecessor
search and use the tree of its stored root.  Merging the sorted lists $\mathcal X_P$
and $\mathcal X_T-j$ yields the exceptional positions of $j$ in
$\fragmentco{0}{t}$, which determine the intervals of the $\Oh(d)$ tokens
of $j$ (\cref{def:tokens}) in $\Oh(d)$ time.  By
\cref{lem:sweep_events}, these tokens correspond in order to the leaves
of that tree.  The tokens meeting $\fragmentco{\ell}{e}$ form a
consecutive range of complete tokens and at most two partial tokens.  A partial token must be a run, because integer cuts cannot
split a singleton; if both cuts lie in the same run, we take its
common part only once.  The complete tokens are the leaves with ranks
in some interval, which we split into $\Oh(\lg m)$ subtrees by descending
from the root using the leaf counts; we retrieve their CMOs.  For the selected part of each partial run, of
length $h$, we use the CMO of $\overline A$ if $h\ge\chi$ and $h$
copies of $\widehat A$ otherwise.  We apply these
$\Oh(\lg m+\chi)=\Oh(\lg m)$ CMOs to $z$ in reverse order, which takes
$\Oh(d\lg^2 m)$ time by \cref{lem:bd_monge_cmo_operations}, and the
resulting vector $y$ is finite and nonnegative.

\emph{Correctness.}  Every leaf CMO is $W$-equivalent to the matrix of
its token, and $\widehat A\meq{W}A$.  By min-plus congruence
(\cref{obs:threshold_congruence}), each subtree CMO is
$W$-equivalent to the product of the matrices of its leaves, so the
query applies the matrices prescribed by
\cref{def:powered_ckw_interval_products}, up to $W$-equivalence.  Since
$z$ is nonnegative, min-plus congruence gives
$y\meq{W}K_{j,\fragmentco{\ell}{e}}\oplus z$.
\end{proof}

\subsection{Batched Output on Eligible Block Indices}\label{sec:sweep_output}

We group consecutive eligible block indices with equal powered
products and evenly spaced start intervals (\cref{lem:profile_partition});
see \cref{fig:overview_sweep}.  One range query per group yields all
occurrence starts assigned to the group, as $\Oh(d)$ progressions of
difference $\tau$ (\cref{prop:eligible_output}), which
\cref{sec:central_algorithm} uses.

\begin{lemma}[Intervals with equal powered products]
\label{lem:profile_partition}
Given the preprocessing record $\mathfrak P$, one can compute in
$\Oh(d^2\lg m)$ time a partition $\Pi$ of $J_{\mathrm{elig}}$ into
$\Oh(d^2)$ integer intervals on each of which the powered product
$K_j$ and the differences $c_j-j\tau$, $r_j-j\tau$, and
$r_{j+1}-j\tau$ are constant.
\end{lemma}
\begin{proof}
We refine the partition of \cref{lem:sweep_events} so that, for every
$j$ in an interval with at least two elements, $j+1\in J_{\mathrm{pad}}$
and $j,j+1\notin\mathcal X_T$; then, \cref{lem:plain_outside_exceptional}
gives $T_j=Q_T^\circ$ and $c_j-r_j=c_{j+1}-r_{j+1}=\kappa$.

We remove the $\Oh(d^2/q)$ indices of the
sorted list $J_{\mathrm{high}}$ stored in $\mathfrak P$
(\cref{lem:ckw_load_bounds}), so that the remaining intervals partition
$J_{\mathrm{elig}}$.  We also isolate, as singleton intervals, the index
$\max J_{\mathrm{pad}}$ and the $\Oh(d)$ indices $j$ with
$j\in\mathcal X_T$ or $j+1\in\mathcal X_T$.  The resulting partition
$\Pi$ has $\Oh(d^2)$ intervals, and sorting all boundaries takes
$\Oh(d^2\lg m)$ time.

As $\Pi$ refines the partition of \cref{lem:sweep_events}, $K_j$ is
constant on each interval of $\Pi$.  Let $j$ belong to an interval of
$\Pi$ with at least two elements.  Since $j$ is not isolated, the
conditions of the first paragraph hold, and
$j\in J_{\mathrm{pad}}\subseteq\mathcal I_T$, as
\cref{lem:plain_outside_exceptional} requires.  The fragment
$T_j=T\fragmentco{c_j-f}{c_{j+1}+f}$ equals $Q_T^\circ$, of length
$\tau+2f$, so $c_{j+1}-c_j=\tau$, and $c_j-r_j=c_{j+1}-r_{j+1}$ yields
$r_{j+1}-r_j=\tau$.  Therefore,
$c_j-j\tau$ and $r_j-j\tau$ are constant on each interval of $\Pi$, and
so is $r_{j+1}-j\tau=(r_j-j\tau)+\tau$.
\end{proof}

Recall that an occurrence is assigned to $j$ if its start lies in
$\fragmentco{r_j}{r_{j+1}}$ (\cref{def:jdef}).  On an interval
$\fragmentcc{j_0}{j_1}$ of $\Pi$, increasing $j$ by one shifts this
start interval and the position $c_j-f+g$ that row $g$ of $K_j$
represents by $\tau$, while $K_j$ stays fixed.  Hence, by
\cref{lem:few_mark_ckw_uniform_instantiation}, the row minima of
$K_{j_0}$ suffice: every row $g$ that yields an occurrence assigned to
$j_0$ yields the progression of starts $c_{j_0}-f+g+\nu\tau$ for
$\nu\in\fragmentcc{0}{j_1-j_0}$; a singleton counts as a progression as
well.  \Cref{alg:eligible} summarizes the procedure.

\begin{algorithm}[t]
\caption{Reporting occurrence starts assigned to eligible block indices}
\label{alg:eligible}
\KwIn{a valid instance of \SM with $J_{\mathrm{pad}}\ne\emptyset$ and its preprocessing record $\mathfrak P$}
\KwOut{the starts in $\OccE_k(P,T)$ assigned to $J_{\mathrm{elig}}$, as progressions of difference $\tau$}
compute the partition $\Pi$\Comment*[r]{\cref{lem:profile_partition}}
construct the range oracle $\mathsf{Range}$\Comment*[r]{\cref{lem:range_oracle}}
\ForEach{interval $J'=\fragmentcc{j_0}{j_1}\in\Pi$}{
  $y\gets\mathsf{Range}(j_0,0,t+1,\mathbf 0)$\Comment*[r]{row minima, \cref{lem:range_oracle}}
  \ForEach{$g\in\fragment{0}{\Delta}$ with $y[g]\le k$ and $c_{j_0}-f+g\in\fragmentco{r_{j_0}}{r_{j_0+1}}$}{
    report $\{c_{j_0}-f+g+\nu\tau:\nu\in\fragmentcc{0}{j_1-j_0}\}$\Comment*[r]{\cref{lem:few_mark_ckw_uniform_instantiation}}
  }
}
\end{algorithm}

\begin{proposition}[Batched output on eligible block indices]
\label{prop:eligible_output}
Given the preprocessing record $\mathfrak P$
(\cref{def:preprocessing_record}), one can report in
$\Oh(d^3\lg^2 m)$ time all occurrence starts assigned to
$J_{\mathrm{elig}}$ as $\Oh(d^3)$ pairwise disjoint arithmetic
progressions of difference $\tau$.
\end{proposition}
\begin{proof}
We run \cref{alg:eligible}, which uses one vector $y$ for all indices
of each interval of $\Pi$.

\emph{Correctness.}  Fix an interval $J'=\fragmentcc{j_0}{j_1}\in\Pi$
and $j=j_0+\nu\in J'$.  By \cref{lem:profile_partition}, $K_j=K_{j_0}$,
the position $c_j-f+g$ equals $c_{j_0}-f+g+\nu\tau$ for every row $g$,
and $\fragmentco{r_j}{r_{j+1}}$ is $\fragmentco{r_{j_0}}{r_{j_0+1}}$
shifted by $\nu\tau$.  By \cref{lem:range_oracle},
$y\meq{W}K_{j_0}\oplus\mathbf0=K_j\oplus\mathbf0$, and $k\le d<W$, so
$y[g]\le k$ if and only if $(K_j\oplus\mathbf0)[g]\le k$.  Hence, the
rows selected for $J'$ are exactly those with
$(K_j\oplus\mathbf0)[g]\le k$ and $c_j-f+g\in\fragmentco{r_j}{r_{j+1}}$,
and, by \cref{lem:few_mark_ckw_uniform_instantiation}, their terms
$c_j-f+g$ with index $\nu$ are exactly the starts in $\OccE_k(P,T)$
assigned to $j$.  In particular, the term with index $\nu$ of every
progression reported for $J'$ lies in the start interval of $j_0+\nu$.

\emph{Disjointness.}  \Cref{alg:eligible} reports one progression for
each pair $(J',g)$ of an interval of $\Pi$ and a selected row.  The start
intervals are pairwise disjoint (\cref{lem:working_budget_ckw_slicing}),
so a common term $v=c_j-f+g$ of two progressions determines $j$, then
$J'$ and $\nu$, and finally $g$; hence, both come from the same pair
$(J',g)$.

\emph{Running time.}  Computing $\Pi$ and the oracle takes $\Oh(d^3\lg^2 m)$ time.  For each of the
$\Oh(d^2)$ intervals of $\Pi$, binary searches yield $r_{j_0}$ and
$r_{j_0+1}$ from $\mathfrak P$ and $c_{j_0}$ from $\brp{\A_T}$ in
$\Oh(\lg m)$ time, the range query takes
$\Oh(d\lg^2 m)$ time, and scanning the $\Delta+1$ rows takes $\Oh(d)$
time.  Each interval yields at most $\Delta+1=\Oh(d)$ progressions, so
there are $\Oh(d^3)$ of them, and the total time is $\Oh(d^3\lg^2 m)$.
\end{proof}

\section{Algorithm for High-Load Block Indices}
\label{sec:block_geometry}

This section reports, in $\Oh(d^3\lg^2 m)$ time, the occurrence starts
assigned to high-load block indices one by one (\cref{prop:high_load_output}); see
\cref{sec:overview_high}.  \Cref{sec:central_algorithm} combines them
with the output of \cref{sec:powered_ckw_sweep}.

\Cref{sec:slab_partition,sec:tall_slab_localization} prepare the
geometry.  Tight paths from
occurrence starts assigned to $j$ stay in a \emph{band} around
$\eD{b_j}$ (due to \cref{lem:relevant_path,lem:ckw_start_relevance}).
We partition it along cut rows into $\Oh(d/q)$ \emph{slabs}, each
\emph{short} (few factors) or \emph{light} (few edits of the supplied
alignments), by \cref{lem:band_block_partition}.
In a light slab that is not short, called \emph{tall}, a relevant
witness that meets only one rail has a clean middle part
(\cref{def:clean_paths,lem:tall_block_clean_middle}), handled by the
powered products of \cref{def:powered_ckw_interval_products}, and one
that meets several rails stays in one of few narrow state intervals
(\cref{lem:tall_block_witness_localization}).
\Cref{fig:overview_band} illustrates the partition and this dichotomy.

\Cref{sec:high_load_transfers,sec:high_load_output}
turn this geometry into an algorithm for each high-load block index
$j$.  We propagate estimates of the suffix
distances $\dtob$ upward through the cut rows.  The estimates
are exact at the relevant states, which include every state on an
optimal witness for an occurrence assigned to~$j$.
A \emph{transfer} computes a slab's entrance vector from its exit
vector (\cref{lem:short_block_profile,lem:tall_light_block_profile}).  Short slabs use exact
factor products; tall light slabs combine candidates for these two
kinds of witnesses.  Finally,
\cref{prop:high_load_output} reads the occurrence starts
from the top vector and bounds the total running time.

Throughout this section, we fix a valid instance of \SM with $J_{\mathrm{pad}}\ne\emptyset$
and use the parameters of \cref{def:common_factorization_parameters}.
High-load indices exist only if $\sqrt d\le q\le\kappa$
(\cref{def:ckw_load_partition}), so we also assume $\sqrt d\le q\le5d$.
With \cref{rem:parameter_bounds}, this gives $d\le q^2$ and
$d<\tau\le\kappa+q\le6q^2$.

\subsection{Slab Partition}
\label{sec:slab_partition}

Recall that we use the threshold $W=3\kappa$ for relevant vertices
(\cref{def:relevant_vertices}) and that $w=\floor{W/q}+1$ is the least
number of periods with $wq>W$.  The band
around a rail is bounded by proper rails, so that both of its sides
extend from the top row to the bottom row.

\begin{definition}[Band]\label{def:band}
Fix an integer $b$ with $\fragmentcc{b-w}{b+w}\subseteq\prD$.
The \emph{band centered at $b$} is the union of the closed strips of
$\eAG(P,T)$ (\cref{def:rail_strips}) that lie between $\eD{b-w}$ and
$\eD{b+w}$.
\end{definition}

Every vertex relevant to $b$ (that is, every vertex of
$\tClose_{W,b}$) has its reference offset within $W<wq$ of $bq$, so it
lies in the band; by \cref{lem:relevant_path}, so does every tight path
starting at such a vertex.

A rail may meet a lifted cut row in several vertices, which form
an interval of that row.  As for the pieces of other paths
(\cref{def:path_pieces}), we distinguish the latest of them.

\begin{definition}[Latest rail contacts]
\label{def:last_rail_contacts}
\let\qed\relax %
For every $c\in\prD$ and $i\in\fragmentcc{0}{t+1}$, the
\emph{latest rail contact} of $\eD{c}$ on the $i$th lifted cut row
(\cref{def:cut_rows}) is its vertex on this row with the largest
second coordinate:
\[
    \widetilde u_{c,i}\coloneqq
      \max\left(\eD{c}\cap
        \bigl(\{\widetilde a_i\}\times\fragment{0}{\ell_T}\bigr)\right).
        \tag*{\lipicsEnd}
\]
\end{definition}

Since $\eD{c}$ is proper, it crosses every lifted cut row, so the
latest contact is well-defined.  Every vertex $x$ on the $i$th lifted
cut row satisfies $\phP(x)=\widehat a_i$, and every vertex of $\eD{c}$
has reference offset $cq$, so
$\phPT(\widetilde u_{c,i})=(\widehat a_i,\widehat a_i+cq)$.  We process
a band in slabs, each delimited by two lifted cut rows.

\begin{definition}[Slabs]\label{def:band_block}
Fix a band centered at $b$ and a nonempty integer interval
$\fragmentco{\ell}{e}\subseteq\fragmentcc{0}{t}$.
The \emph{slab}
$B_{b,\fragmentco{\ell}{e}}$ is the subgraph of the band induced by its
vertices in the rows $\fragmentcc{\widetilde a_{\ell}}{\widetilde a_{e}}$.
Its vertices on the lifted cut rows $\ell$ and $e$ form its
\emph{entrance frontier} and \emph{exit frontier}, respectively.
\end{definition}

The height and the canonical cost of a slab determine how it is
processed.

\begin{definition}[Tall, short, and light slabs]\label{def:slab_classes}
The \emph{height} and \emph{canonical cost} of a slab
$B=B_{b,\fragmentco{\ell}{e}}$ are
\[
    H_B\coloneqq\widehat a_{e}-\widehat a_{\ell},
    \qquad
    \Lambda_B\coloneqq
       \cancost(x,y),
\]
where $x$ is the leftmost vertex of the entrance frontier and $y$ is
the rightmost vertex of the exit frontier.  The slab is \emph{tall} if $H_B>18q^2$ and
\emph{short} otherwise; it is \emph{light} if $\Lambda_B<q/6$.
\end{definition}

The height $H_B$ is the length of the slab on the pattern-reference
axis.  The canonical cost $\Lambda_B$ counts the edits of the supplied
alignments between the leftmost entrance vertex and the rightmost exit
vertex, so it bounds the canonical cost of every path inside the
slab.  A slab can be both short and light.

\begin{observation}[Canonical costs within a slab]
\label{obs:block_canonical_bound}
For every slab $B$ and vertices $u,u'\in B$ with $u'$ reachable
from $u$, we have $\cancost(u,u')\le\Lambda_B$.
\end{observation}
\begin{proof}
Since the bounding rails are monotone, both coordinates of $u$ and
$u'$ lie between the corresponding coordinates of the leftmost
entrance vertex and the rightmost exit vertex.
Hence, the restrictions of the supplied alignments counted by
$\cancost(u,u')$ are parts of those counted by $\Lambda_B$.
\end{proof}

In particular, the observation bounds the canonical cost of every
rail segment from a vertex of $B$ to the latest contact of its rail on
the exit frontier.  The next lemma cuts a band greedily into slabs
that are short or light.  Away from the top and bottom rows, the union
of two consecutive slabs has canonical cost at least $q/6$, so there
are only $\Oh(d/q)$ slabs.

\begin{lemma}[Partition into short or light slabs]
\label{lem:band_block_partition}
Given an integer $b$ with $\fragmentcc{b-w}{b+w}\subseteq\prD$ and the
breakpoint representations of $\A_P,\A_T$, one can partition the band
centered at $b$ into $\Oh(d/q)$ consecutive slabs in
$\Oh(d\lg m+(d/q)\lg^2 m)$ time.  Every resulting slab
$B_{b,\fragmentco{\ell}{e}}$ is short or is light with
$\fragmentco{\ell}{e}\subseteq\fragmentco{1}{t}$.
\end{lemma}
\begin{proof}
In this proof, we call a slab of the form $B_{b,\fragmentco{i}{i+1}}$ a \emph{tile}.
The tiles for $i\in\fragmentcc{0}{t}$ partition the band, and their
heights are positive because $\tau>0$ and $y_P-t\tau\ge\kappa>0$.
We merge tiles greedily into maximal runs that are short or light, and
the maximality bounds the number of slabs.

\emph{Tiles are short.}  Every tile except the last has height $\tau$,
and, by the choice of $t$ (\cref{def:common_factorization_parameters}),
the last has height $y_P-t\tau<\kappa+\tau<2\tau$.  Since
$\tau\le6q^2$, every tile has height less than $12q^2<18q^2$.

\emph{Greedy partition.}  The two boundary tiles, with $i\in\{0,t\}$,
form slabs on their own; we call the other tiles \emph{interior}.
Starting at the first unassigned interior tile, we take the longest run of interior tiles starting there whose
union is short or light; a single tile qualifies because it is short.
We repeat this until all tiles are assigned.  The boundary tiles are
short, and every other slab, which we also call \emph{interior}, is short or
light and satisfies $\fragmentco{\ell}{e}\subseteq\fragmentco{1}{t}$
because it consists of interior tiles.

\emph{Number of slabs.}  We extend every interior slab except the last
by the next tile.  By the maximality of the greedy choice, each
extended slab $B'$ is neither short nor light, so
$\Lambda_{B'}\ge q/6$.  The interior slabs are disjoint and have
height at least $\tau$, so every tile lies in at most two extended
slabs, and the edits of $\A_P$ counted by the values $\Lambda_{B'}$ are
counted at most twice.  The edits of $\A_T$ counted by $\Lambda_{B'}$
lie in the pattern-reference interval of $B'$, shifted by $bq$ and
enlarged by at most $wq$ on either side.  Since
$wq\le W+q=3\kappa+q<4\tau$, every reference coordinate lies in
$\Oh(1)$ of these enlarged intervals.  Consequently,
$\sum_{B'}\Lambda_{B'}=\Oh(d_P+d_T)=\Oh(d)$, so there are $\Oh(d/q)$
extended slabs.  Adding the last interior slab and the two boundary
tiles, we obtain $\Oh(d/q+1)=\Oh(d/q)$ slabs because $q\le5d$.

\emph{Running time.}  We precompute prefix costs and predecessor
structures for the two breakpoint representations in
$\Oh((d_P+d_T)\lg m)=\Oh(d\lg m)$ time.  Then, the canonical cost of a union
of consecutive tiles can be computed with a constant number of
predecessor and prefix-sum queries, in $\Oh(\lg m)$ time.  The height
and this cost do not decrease when the union grows, so, at each greedy
step, we binary-search for the last tile for which the union is short
or light.  This takes
$\Oh(\lg^2 m)$ time per slab, and the claimed bound follows from the
bound on the number of slabs.
\end{proof}

\subsection{Localizing Tall-Slab Witnesses}
\label{sec:tall_slab_localization}

\begin{figure}[t]
\centering
\begin{tikzpicture}[x=0.4cm,y=-0.4cm]
  \fill[color5!16] (-1,0) rectangle (21,2);
  \fill[color5!16] (-1,8) rectangle (21,10);
  \fill[color3!14,opacity=0.75] (3.6,0) -- (10.4,0) -- (20.4,10) -- (13.6,10) -- cycle;
  \foreach \y in {1,3,4,...,7,9} {\draw[black!25] (-1,\y) -- (21,\y);}
  \foreach \y/\l in {0/\ell,2/\ell',8/e',10/e} {
    \draw[black!45] (-1,\y) -- (21,\y);
    \node[font=\scriptsize,anchor=mid west] at (21.1,\y) {cut $\l$};
  }
  \foreach \k in {0,...,5} {\draw[color2,line width=0.8pt] (2*\k,0) -- (2*\k+10,10);}
  \foreach \k/\l in {2/c-1,3/c,4/c+1,5/c+2} {
    \node[font=\scriptsize,color2,anchor=base] at (2*\k,-0.5) {$\eD{\l}$};
  }
  \draw[color2,line width=1.7pt,line cap=round,line join=round]
     (-0.8,0) -- (1,1) -- (4.4,4.4) -- (4.4,5.0) -- (5.2,5.8) -- (5.8,5.8) -- (9,9) -- (9.3,10);
  \draw[color4,line width=1.7pt,line cap=round,line join=round]
     (6.8,0) -- (6.8,0.8) -- (10,4) -- (12.8,4.8) -- (18,10);
  \fill[color4] (10,4) circle (2.4pt);
  \fill[color4] (12.8,4.8) circle (2.4pt);
  \node[font=\scriptsize,color4!80!black] at (9.4,4.5) {$v_c$};
  \node[font=\scriptsize,color4!80!black] at (13.5,4.45) {$v_{c+1}$};
  \fill (16,10) circle (2.4pt);
  \fill (18,10) circle (2.4pt);
  \node[font=\scriptsize] at (17,10) {$\times$};
  \node[font=\scriptsize,anchor=base] at (16,10.95) {$\widetilde u_{c,e}$};
  \node[font=\scriptsize,anchor=base] at (18,10.95) {$\widetilde u_{c+1,e}$};
  \node[font=\scriptsize,color3!80!black,anchor=mid west] at (13.6,2.5) {$V_c$};
\end{tikzpicture}
\caption{Witnesses in a tall light slab
(\cref{lem:tall_block_clean_middle,lem:tall_block_witness_localization}),
schematically in reference coordinates.  Every witness meets a rail in
each guard (yellow).  One meeting only one rail (purple) is clean
between the cuts $\ell'$ and $e'$.  The orange witness meets the rails
$c$ and $c+1$, so $(c,c+1)$ is imbalanced (${\times}$); this pair
forms a run, so the witness stays between rails $c-1$ and
$c+2$, with cut-row states in $V_c$ (magenta).}
\label{fig:tall_witnesses}
\end{figure}
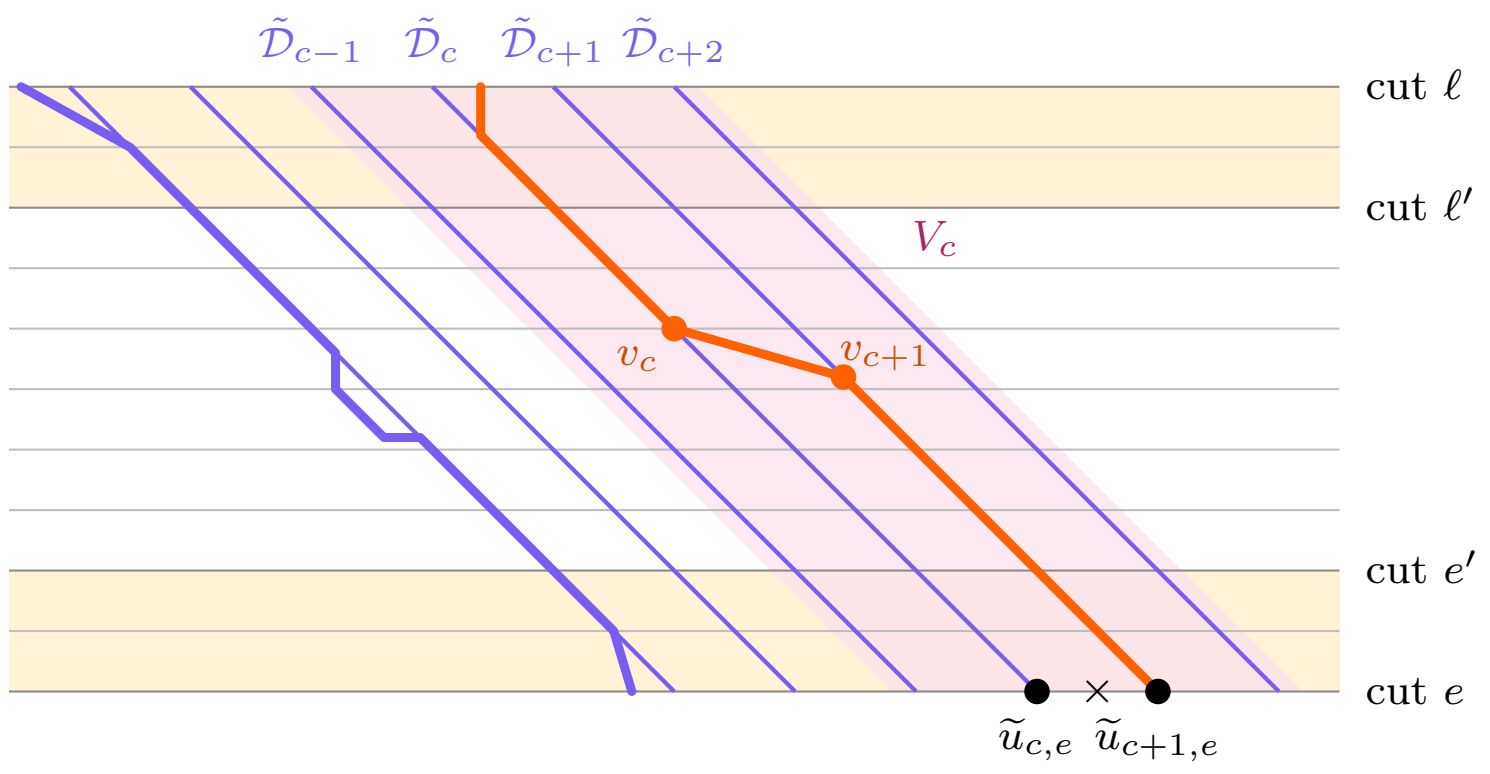

Short slabs are processed directly using catalogue matrices.  A tall
light slab, however, may consist of many factors.  Its witnesses are of
two kinds: those that meet only one rail and those that meet several
rails.  \Cref{fig:tall_witnesses} shows both kinds and the claims behind
the next two lemmas.

We set aside a top and a bottom \emph{guard}, consisting of the factors
between the cuts $\ell$ and $\ell'$ and between $e'$ and $e$,
respectively.  The next lemma shows that a witness that meets only one
rail is clean between the guards, so that the range oracle of
\cref{lem:range_oracle} handles it.

\begin{lemma}[Clean middles in a tall light slab]
\label{lem:tall_block_clean_middle}
Fix $j\in J_{\mathrm{pad}}$ and a tall light slab
$B=B_{b_j,\fragmentco{\ell}{e}}$ with
$\fragmentco{\ell}{e}\subseteq\fragmentco{1}{t}$, and put
$\ell'\coloneqq\ell+\floor{9q^2/\tau}$ and
$e'\coloneqq e-\floor{9q^2/\tau}$.  Then, $\ell<\ell'<e'<e$, and every
tight path from a vertex of $\tClose_{W,b_j}$ on the entrance frontier
to the exit frontier that meets at most one rail has a clean portion
(\cref{def:clean_paths}) between the cuts $\ell'$ and $e'$.
\end{lemma}
\begin{proof}
By \cref{lem:relevant_path}, every tight path in the statement stays
relevant and hence inside $B$.  We first show that a long shortest path
in a light slab meets a rail, so that such a path meets a rail in each
guard.  Then, we show that, if it meets only one rail, all its
detours in between lie in the covers.

\begin{claim}
\label{clm:tall_block_rail_contact}
Every shortest path from $u$ to $u'$ contained in $B$ with
$|\phP(u,u')|\ge3q^2$ meets a rail.
\end{claim}
\begin{claimproof}
Suppose that such a path meets no rail.  By \cref{lem:canonical_rails},
it stays strictly between two consecutive rails $c$ and $c+1$, so the
reference offsets of its endpoints lie in $\fragmentcc{cq}{(c+1)q}$.

\emph{Upper bound.}  After shifting the text-reference coordinates by
$-cq$, the two reference projections of the path become fragments of
$Q^\infty$ whose endpoints differ by at most $q$.  Matching their
common part and deleting or inserting the rest shows that their edit
distance is at most $2q$.  Since the path is a shortest path,
\cref{obs:expanded_projection}, the triangle inequality, and
\cref{obs:block_canonical_bound} bound its cost by $2q+\Lambda_B$.

\emph{Lower bound.}  The path shares no reference-diagonal edge with a
rail, so \cref{lem:noblue} bounds its cost from below by
$\floor{3q^2/q}-\Lambda_B=3q-\Lambda_B$.  This exceeds $2q+\Lambda_B$
because $\Lambda_B<q/6$, a contradiction.
\end{claimproof}

\emph{The cuts.}  Put $L\coloneqq\tau\floor{9q^2/\tau}$, so that
$\ell'=\ell+L/\tau$ and $e'=e-L/\tau$.  Since $\tau\le6q^2$, the guard
height $L$ satisfies $3q^2<L\le9q^2$.  Since $e\le t$, the slab has
height $H_B=(e-\ell)\tau$, and tallness gives $H_B>18q^2\ge2L$.  Hence,
$\ell<\ell'<e'<e$.

\emph{Paths that meet at most one rail.}
Consider a tight path $\eA$ as in the statement.  Its portions between
the cuts $\ell$ and $\ell'$ and between the cuts $e'$ and $e$ are tight,
hence shortest, paths in $B$ whose pattern-reference projections have
length $L>3q^2$.  By \cref{clm:tall_block_rail_contact}, $\eA$ meets a
rail in each of these portions, and these two contacts lie on the same
rail $b'$ because $\eA$ meets at most one rail.
Consider a detour $\mathcal Y$ of the subpath of $\eA$ between the
two contacts (\cref{def:detours}).  Its endpoints $x,x'$ lie on rails,
hence on rail~$b'$.  Since
$\mathcal Y$ is tight, \cref{lem:rail_stability} gives
\[
    \cost(\mathcal Y)=\tdtob(x)-\tdtob(x')\le\cancost(x,x').
\]
Moreover, $\mathcal Y$ shares no reference-diagonal edge with any
rail.  Hence, by \cref{cor:nobluesame}, its pattern-reference
projection lies in a fragment of $\Es_P$ or its text-reference
projection lies in a fragment of $\Es_T$.  Every edge of the portion
of $\eA$ between the cuts $\ell'$ and $e'$ that does not belong to a
rail lies in such a detour $\mathcal Y$, so its reference projections
satisfy the same containment.  Hence, this portion is clean.
\end{proof}

A witness that meets several rails pays about $q$ for each change of
rail.  As the slab is light, following a rail to the exit saves less
than $q/6$, so the exit distances of the two rails differ by more than
$q/2$.  By \cref{thm:combi}, few such pairs exist, so few narrow state
intervals contain all such witnesses.

\begin{lemma}[State intervals for paths that meet several rails]
\label{lem:tall_block_witness_localization}
Fix $j\in J_{\mathrm{pad}}$ and a light slab
${B=B_{b_j,\fragmentco{\ell}{e}}}$ with
$\fragmentco{\ell}{e}\subseteq\fragmentco{1}{t}$.  There is a family
$\mathcal V$ of integer intervals in $\fragment{0}{\Delta}$, each of
size at least $q$ and of total size $\Oh(M_j)$, such that every tight
path from a vertex of $\tClose_{W,b_j}$ on the entrance frontier to the
exit frontier that meets at least two rails has all its states at the
cuts $\ell,\ldots,e$ (\cref{def:path_pieces}) in one interval
$V\in\mathcal V$.

Given the breakpoint representations of $\A_P,\A_T$ and the values
$\tdtob$ at the latest rail contacts on the exit frontier that are
relevant to $b_j$, one can compute such a family in $\Oh(d\lg m)$ time.
\end{lemma}
\begin{proof}
By \cref{lem:relevant_path}, every tight path in the
statement stays relevant and hence inside $B$.  A path that moves to a
neighboring rail forces the exit distances of the two rails apart.  We
cover each maximal run of such pairs of rails by one state interval and
bound their number using \cref{thm:combi}.

\emph{Imbalanced pairs.}  We call a pair $(c,c+1)$ of rails with
$c\in\fragmentco{b_j-w}{b_j+w}$ \emph{imbalanced} if both latest contacts
$\widetilde u_{c,e}$ and $\widetilde u_{c+1,e}$ on the exit frontier are
relevant and
\[
    \bigl|\tdtob(\widetilde u_{c,e})
          -\tdtob(\widetilde u_{c+1,e})\bigr|>q/2.
\]
The given data determine all imbalanced pairs.

\begin{claim}[Crossing rails forces imbalance]
\label{clm:tall_block_rail_crossing}
If a tight path from a relevant vertex of the entrance frontier to the
exit frontier meets rails of indices $b'<b''$, every pair $(c,c+1)$ with
$c\in\fragmentco{b'}{b''}$ is imbalanced.
\end{claim}
\begin{claimproof}
We first show that if a relevant vertex $x$ of $B$ lies on a rail, then
the latest contact $x'$ of this rail on the exit frontier is relevant as
well.  Indeed, \cref{lem:rail_stability,obs:cancost_factorizes} give
$\tdtob(x')+\cancost(x')\le\tdtob(x)+\cancost(x)$, and the reference
offset in the penalty is the same at $x$ and $x'$.

By \cref{lem:canonical_rails}, the path meets every rail between $b'$
and $b''$.  For $c\in\fragmentco{b'}{b''}$, choose contacts $v_c$ and
$v_{c+1}$ of the path with the rails $c$ and $c+1$; they are relevant
by \cref{lem:relevant_path}.  Their reference
offsets differ by $q$, so, by
\cref{lem:offset_change,obs:block_canonical_bound}, the part of the
path between them costs at least $q-\Lambda_B$.  This part is tight, so
$|\tdtob(v_c)-\tdtob(v_{c+1})|\ge q-\Lambda_B$.  The canonical cost
from either contact to the latest contact of its rail on the exit
frontier is at most $\Lambda_B$, so \cref{lem:rail_stability} gives
\[
    \bigl|\tdtob(\widetilde u_{c,e})
          -\tdtob(\widetilde u_{c+1,e})\bigr|
      \ge |\tdtob(v_c)-\tdtob(v_{c+1})|-2\Lambda_B
      \ge q-3\Lambda_B>q/2,
\]
where the last step uses $\Lambda_B<q/6$.  Both exit contacts are
relevant by the first paragraph.
\end{claimproof}

\emph{The intervals.}
Call a maximal integer interval $\fragmentco{c}{c'}$ such that the pair
$(c'',c''+1)$ is imbalanced for every $c''\in\fragmentco{c}{c'}$ a
\emph{run} of imbalanced pairs.  For every such run, put
\[
    V_c\coloneqq\fragment{0}{\Delta}\cap
       \fragmentcc{f+(c-1-b_j)q-\Lambda_B}{f+(c'+1-b_j)q+\Lambda_B},
\]
and let $\mathcal V$ consist of these sets.  The exit contacts of the
rails $c$ and $c'$ are relevant, so $|c-b_j|q\le W$ and
$|c'-b_j|q\le W$.  Since $W<f$, the set $V_c$ contains the integers
from $f+(c-b_j)q$ to $f+(c'-b_j)q$, so $|V_c|\ge q$.

\emph{Paths that meet several rails.}
Let $\eA$ be a tight path as in the statement, and let $b'<b''$ be the
smallest and the largest index of a rail that it meets.  By
\cref{clm:tall_block_rail_crossing}, all pairs $(c'',c''+1)$ with
$c''\in\fragmentco{b'}{b''}$ are imbalanced, so $\fragmentco{b'}{b''}$
is contained in a run $\fragmentco{c}{c'}$, that is, $c\le b'$ and
$b''\le c'$.  The contacts of $\eA$ with the rails $b'$ and $b''$ are
relevant, so $|b'q-b_jq|\le W<wq$ and
$|b''q-b_jq|\le W<wq$.  Hence, the rails $b'-1$ and $b''+1$ belong to
the band.  The path avoids them, so \cref{lem:canonical_rails} confines
it to the region between them, where every vertex $x$ satisfies
$(b'-1)q\le\mathrm{off}(x)\le(b''+1)q$.

Now, consider a vertex $x$ of $\eA$ on the $i$th lifted cut row.  The
alignment $\A_T$ pairs $z_{j,i}+f=c_{j+i}$ with $\widehat a_i+b_jq$
(\cref{def:internal_cuts}) and $\pT(x)$ with
$\widehat a_i+\mathrm{off}(x)$.  Hence, the state
$\widetilde\rho_{j,i}(x)=\pT(x)-z_{j,i}$ differs from
$f+\mathrm{off}(x)-b_jq$ by at most the number of edits of $\A_T$
between these two reference coordinates.  Both coordinates lie between
the text-reference coordinates of the leftmost entrance vertex and the
rightmost exit vertex, so this number is at most $\Lambda_B$
(\cref{def:slab_classes}).  By
\cref{obs:cut_states_near_rail}, the states of $\eA$ also lie in
$\fragment{0}{\Delta}$, so all of them lie in
\[
    \fragment{0}{\Delta}\cap
    \fragmentcc{f+(b'-1-b_j)q-\Lambda_B}{f+(b''+1-b_j)q+\Lambda_B}
       \subseteq V_c.
\]

\emph{Total size and construction time.}
Let $n_{\mathrm{imb}}$ denote the number of imbalanced pairs.  We
apply \cref{thm:combi} to the rail interval
$\fragmentcc{b_j-w}{b_j+w}$ and the latest contacts on the exit frontier.
Since $j$ is padded and $R_*\ge w+\eta^\star$, the enlargement of this
interval (\cref{def:rail_interval_enlargement}) is proper, and its
total mark load is at most $M_j$.  The canonical cost between the two
extreme exit contacts is at most $\Lambda_B<q/6$, so
\[
    \tfrac q2\, n_{\mathrm{imb}}
      \le\Lambda_B+4M_j<\tfrac q6+4M_j.
\]
If $n_{\mathrm{imb}}=0$, then $\mathcal V$ is empty.  Otherwise,
$q/6\le qn_{\mathrm{imb}}/6$, and rearranging gives
$n_{\mathrm{imb}}<12M_j/q$.  The set $V_c$ of a run
$\fragmentco{c}{c'}$ has at most $(c'-c+2)q+2\Lambda_B+1=\Oh((c'-c)q)$
elements, and the runs partition the indices of the imbalanced pairs,
so
\[
    \sum_{V\in\mathcal V}|V|=\Oh(qn_{\mathrm{imb}})=\Oh(M_j).
\]
Using prefix counts of edits and predecessor queries on the breakpoint
representations of $\A_P$ and $\A_T$ (\cref{fct:alignments}), we
compute $\Lambda_B$ and the $\Oh(w)=\Oh(d/q)$ latest rail contacts on
the exit frontier.  Scanning their given distances yields the runs of
imbalanced pairs and their sets $V_c$, in $\Oh(d\lg m)$ total time.
\end{proof}

\begin{remark}[Guards and state intervals]
\label{rem:guards_and_state_intervals}
Guards and state intervals restrict witnesses in different directions.
A guard extends along the slab, over $\Theta(q^2)$ pattern-reference
positions, so that every witness meets a rail inside it
(\cref{clm:tall_block_rail_contact}).  An interval of $\mathcal V$
extends across the band, over $\Oh(q)$ states per imbalanced pair, and
holds the states at every cut of the slab, the guards included.  Hence,
the two do not constrain each other: the guards serve the witnesses
that meet one rail, and the intervals serve those that meet several
rails.  For example, if $d=100$, $\kappa=300$, and $q=40$, then each
guard consists of $45$ factor pairs, a cut row has $4201$ states, and,
if $M_j=20$, the intervals of $\mathcal V$ have total size at most
$665$.
\end{remark}

\subsection{Distance Estimates and Transfers}
\label{sec:high_load_transfers}

We now define what the algorithm maintains at cut rows and how to
carry it through a slab.  Fix $j\in J_{\mathrm{pad}}$.  At every cut $i$,
the algorithm stores one value per state $g\in\fragment{0}{\Delta}$,
that is, per vertex $(a_i,z_{j,i}+g)$ of the $i$th cut row
(\cref{lem:centered_ckw_factor_containment}).  The value is the exact suffix
distance of this vertex if one of its lifts is relevant to $b_j$, and it
never falls below this distance unless both exceed $W$.

\begin{definition}[Relevant states and distance estimates]
\label{def:frontier_state_data}
\let\qed\relax %
Fix $j\in J_{\mathrm{pad}}$.  For
$i\in\fragmentcc{0}{t+1}$, define the vector $h_i$ on
$\fragment{0}{\Delta}$ by $h_i[g]\coloneqq\dtob(a_i,z_{j,i}+g)$.  A state $g$ is
\emph{relevant at cut $i$} if some vertex of $\tClose_{W,b_j}$ on the
$i$th lifted cut row has state $g$ (\cref{def:cut_row_states}).  A
vector $y$ on $\fragment{0}{\Delta}$ is a
\emph{distance estimate at cut~$i$} if
\begin{enumerate}
    \item $y[g]\ge h_i[g]$ for every state $g$ with $y[g]\le W$, and
    \item $y[g]=h_i[g]$ for every state $g$ that is relevant at cut $i$.\lipicsEnd
\end{enumerate}
\end{definition}

We omit the fixed index $j$ from this notation.  The states
$\fragment{0}{\Delta}$ cover all relevant vertices: by
\cref{obs:cut_states_near_rail}, every vertex of $\tClose_{W,b_j}$ on a
lifted cut row has its state in $\fragment{0}{\Delta}$.  Moreover, all
lifts of a state $g$ have suffix distance $h_i[g]$
(\cref{lem:semiglobal_expanded_transfer}), so a distance estimate gives
the exact suffix distance of every relevant vertex on the cut row,
including the vertices of witnesses
(\cref{lem:relevant_path,lem:ckw_start_relevance}).

A \emph{transfer} turns a distance estimate at the exit cut of a slab
into one at its entrance cut.  It applies a matrix that
dominates the exact factor product to the exit vector.  The next lemma
shows that the result is a distance estimate as long as the matrix
charges no tight path from a relevant entrance vertex more than its
cost.  Our transfers for short and tall light slabs only need to check
this condition.

\begin{lemma}[Transfers between cut rows]
\label{lem:exact_ckw_relevant_transfer}
Fix $j\in J_{\mathrm{pad}}$, a nonempty integer interval
$\fragmentco{\ell}{e}\subseteq\fragmentcc{0}{t}$, and a matrix
$F\ge I_{j,\fragmentco{\ell}{e}}$ such that
$F[\widetilde\rho_{j,\ell}(x_\ell),\widetilde\rho_{j,e}(x_e)]\le\cost(\mathcal A)$
for every tight path $\mathcal A$ from a vertex $x_\ell\in\tClose_{W,b_j}$ on
the $\ell$th lifted cut row to a vertex $x_e$ on the $e$th lifted cut
row.
If $y_e$ is a distance estimate at cut $e$, then every
vector $y_\ell\meq{W}F\oplus y_e$ is a distance estimate at
cut $\ell$.  Moreover, $F=I_{j,\fragmentco{\ell}{e}}$ satisfies this
assumption.
\end{lemma}
\begin{proof}
We check the two conditions of \cref{def:frontier_state_data} for
$y_\ell$.  The first follows from \cref{lem:factor_product_suffix_bound},
and the second from the assumption on $F$, applied to a prefix of a
shortest path.  Finally, we show that $F=I_{j,\fragmentco{\ell}{e}}$
satisfies the assumption by projecting each piece of a path to its
factor pair.

\emph{The first condition.}  Let $y_\ell[g]\le W$.  By $W$-equivalence,
$y_\ell[g]=(F\oplus y_e)[g]=F[g,g']+y_e[g']$ for some state $g'$.  Since
$F\ge I_{j,\fragmentco{\ell}{e}}$ is nonnegative, $y_e[g']\le W$, so the
first condition for $y_e$ gives $y_e[g']\ge h_e[g']$.  By
$F\ge I_{j,\fragmentco{\ell}{e}}$ and
\cref{lem:factor_product_suffix_bound},
$y_\ell[g]\ge I_{j,\fragmentco{\ell}{e}}[g,g']+h_e[g']\ge h_\ell[g]$.

\emph{The second condition.}  Let $g$ be relevant at cut $\ell$.  We take a vertex $x_\ell\in\tClose_{W,b_j}$ with state $g$ on the
$\ell$th lifted cut row and a shortest path from $x_\ell$ to the bottom row,
and we let $\mathcal A$ be the prefix of this path that ends at its
latest vertex $x_e$ on the $e$th lifted cut row.  This prefix is tight,
and, by \cref{lem:relevant_path}, the state
$g'\coloneqq\widetilde\rho_{j,e}(x_e)$ is relevant.
Hence, $\cost(\mathcal A)=h_\ell[g]-h_e[g']$
(\cref{lem:semiglobal_expanded_transfer}), and the second condition for
$y_e$ gives $y_e[g']=h_e[g']$.  By the assumption on $F$, and since
$h_\ell[g]\le W$ for the relevant state $g$,
\[
    (F\oplus y_e)[g]\le F[g,g']+y_e[g']\le\cost(\mathcal A)+h_e[g']
       =h_\ell[g]\le W.
\]
By $W$-equivalence, $y_\ell[g]=(F\oplus y_e)[g]\le h_\ell[g]\le W$, and
the first condition gives $y_\ell[g]\ge h_\ell[g]$.

\emph{The assumption for $F=I_{j,\fragmentco{\ell}{e}}$.}  Let
$\mathcal A$ be a tight path from $x_\ell$ to $x_e$ as in the
statement.  By \cref{lem:relevant_path}, all vertices of
$\mathcal A$ are relevant,
so its states $g_\ell,\ldots,g_e$ at the cuts $\ell,\ldots,e$
(\cref{def:path_pieces}) belong to $\fragment{0}{\Delta}$ by
\cref{obs:cut_states_near_rail}; here, $g_\ell=\widetilde\rho_{j,\ell}(x_\ell)$
and $g_e=\widetilde\rho_{j,e}(x_e)$.  By \cref{obs:expanded_projection},
the $i$th piece of $\mathcal A$ projects to a path of the same cost from
$(a_i,z_{j,i}+g_i)$ to $(a_{i+1},z_{j,i+1}+g_{i+1})$, which is a path in
the alignment graph of the factor pair of $j$ with index $i$.  Hence,
its cost is at least $I_{j,i}[g_i,g_{i+1}]$, and summing over
$i\in\fragmentco{\ell}{e}$ gives
$I_{j,\fragmentco{\ell}{e}}[\widetilde\rho_{j,\ell}(x_\ell),\widetilde\rho_{j,e}(x_e)]\le\cost(\mathcal A)$.
\end{proof}

By \cref{lem:band_block_partition}, every slab is short or tall and
light, and the two kinds need different transfer algorithms.  A short slab
consists of only $\Oh(q^2/d)$ factor pairs, so we apply their catalogue
matrices to the exit vector one by one; the endpoint pair, if the slab
contains it, is handled like the internal ones.

\begin{lemma}[Transfer through a short slab]
\label{lem:short_block_profile}
Fix $j\in J_{\mathrm{pad}}$ and a short slab
$B=B_{b_j,\fragmentco{\ell}{e}}$.  Given the fern catalogue of
\cref{lem:ckw_fern_catalogue} and a distance estimate $y_e$ at the exit
cut $e$, one can compute a distance estimate $y_\ell$ at the entrance
cut $\ell$ in $\Oh(q^2\lg m)$ time.
\end{lemma}
\begin{proof}
We retrieve from the catalogue the CMOs of matrices $W$-equivalent to
the factor matrices $I_{j,i}$ with $i\in\fragmentco{\ell}{e}$ and apply
them to $y_e$ in reverse order (\cref{lem:bd_monge_cmo_operations}).  By min-plus
congruence, the resulting vector
$y_\ell$ satisfies $y_\ell\meq{W}I_{j,\fragmentco{\ell}{e}}\oplus y_e$, so
it is a distance estimate at cut $\ell$ by
\cref{lem:exact_ckw_relevant_transfer}.

\emph{Running time.}  Every internal factor has pattern-reference
height $\tau$, and the slab contains at most one endpoint factor, so it
consists of at most $H_B/\tau+1=\Oh(q^2/d)$ factors, because $d\le q^2$.
Each application costs $\Oh(d\lg m)$ time by
\cref{lem:bd_monge_cmo_operations}, so the total time is
$\Oh(q^2\lg m)$.
\end{proof}

For a tall light slab,
\cref{lem:tall_block_clean_middle,lem:tall_block_witness_localization}
provide the cuts $\ell'$ and $e'$ and a family $\mathcal V$ of state
intervals.  We compute one
\emph{common candidate}, with exact products in the two guards and the
powered product (\cref{def:powered_ckw_interval_products}) between them, and one \emph{interval candidate} for
each $V\in\mathcal V$, with exact products restricted to the states in
$V$; see \cref{fig:tall_transfer}.  No candidate entry of at most $W$
is too small.  At a relevant entrance state, the common candidate is
exact if a shortest path from it has a clean middle, and an interval
candidate is exact otherwise, so the entrywise minimum is a distance
estimate at cut~$\ell$.

\begin{lemma}[Transfer through a tall light slab]
\label{lem:tall_light_block_profile}
Fix $j\in J_{\mathrm{pad}}$ and a tall light slab
$B=B_{b_j,\fragmentco{\ell}{e}}$ with
$\fragmentco{\ell}{e}\subseteq\fragmentco{1}{t}$.  Given the breakpoint
representations of $\A_P,\A_T$, the fern catalogue of
\cref{lem:ckw_fern_catalogue}, the range oracle of
\cref{lem:range_oracle}, and a distance estimate $y_e$ at the exit cut
$e$, one can compute a distance estimate $y_\ell$ at the entrance cut
$\ell$ in $\Oh((q^2+qM_j)\lg^2 m)$ time.
\end{lemma}
\begin{proof}
We find the relevant exit contacts, which \cref{lem:tall_block_witness_localization}
needs, compute the common candidate and the interval candidates, and
check the assumption of \cref{lem:exact_ckw_relevant_transfer} for their
entrywise minimum.

\emph{Relevant exit contacts.}  By \cref{obs:cut_states_near_rail}, a
latest rail contact $\widetilde u_{c,e}$ on the exit frontier can be
relevant only if its state lies in $\fragment{0}{\Delta}$.  In that
case, it is relevant if and only if
\[
    y_e[\widetilde\rho_{j,e}(\widetilde u_{c,e})]
        +\tpena_{b_j}(\widetilde u_{c,e})\le W.
\]
If the inequality holds, then the tested entry is at most
$W$, so the first condition of \cref{def:frontier_state_data} and
\cref{lem:semiglobal_expanded_transfer} make the contact relevant.  Its state is then
relevant, so the entry is exact.  Conversely, a relevant contact has a
relevant state, whose entry is exact.  With prefix
costs stored, computing the penalty of a contact takes one predecessor
query on the breakpoint representations (\cref{fct:alignments}).  We extract the
relevant contacts and their distances from $y_e$, apply
\cref{lem:tall_block_witness_localization} to obtain the family
$\mathcal V$, and take the cuts $\ell'$ and $e'$ of
\cref{lem:tall_block_clean_middle}.  These steps take
$\Oh(d\lg m)$ time.

\begin{figure}[t]
\centering
\begin{tikzpicture}[x=0.25cm,y=-0.25cm]
  \foreach \i in {0,1,10,11} {
    \fill[black!7] (\i,\i) -- (\i+10,\i) -- (\i+11,\i+1) -- (\i+1,\i+1) -- cycle;
    \draw[black!35] (\i,\i) -- (\i+10,\i) -- (\i+11,\i+1) -- (\i+1,\i+1) -- cycle;
  }
  \fill[color2!12] (2,2) -- (12,2) -- (20,10) -- (10,10) -- cycle;
  \foreach \y in {3,...,9} {\draw[color2!25] (\y,\y) -- (\y+10,\y);}
  \draw[color2!60] (2,2) -- (12,2) -- (20,10) -- (10,10) -- cycle;
  \node[font=\scriptsize,color2!80!black,fill=color2!12,inner sep=1.5pt] at (11,6) {$K_{j,\fragmentco{\ell'}{e'}}$};
  \foreach \y/\l in {0/\ell,2/\ell',10/e',12/e} {\node[font=\scriptsize,anchor=mid east] at (\y-0.4,\y) {cut $\l$};}
  \draw[color4,line width=2.6pt,opacity=0.55] (0,0) -- (10,0);
  \draw[color4,line width=2.6pt,opacity=0.55] (12,12) -- (22,12);
  \node[font=\scriptsize,anchor=mid west] at (10.4,0) {$y_\star$};
  \node[font=\scriptsize,anchor=mid west] at (22.4,12) {$y_e$};
  \node[font=\small,anchor=base] at (11,16.6) {(a) common candidate};
  \begin{scope}[xshift=30*0.25cm]
  \draw[black!25] (0,0) -- (10,0) -- (22,12) -- (12,12) -- cycle;
  \fill[color3!14] (3.5,0) -- (7.5,0) -- (19.5,12) -- (15.5,12) -- cycle;
  \foreach \y in {2,5,6,9,10} {\draw[black!20] (\y+3.5,\y) -- (\y+7.5,\y);}
  \foreach \a/\b in {1/3,4/7,8/11} {\draw[black!35] (\a+3.5,\a) -- (\a+7.5,\a) -- (\b+7.5,\b) -- (\b+3.5,\b) -- cycle;}
  \foreach \i in {0,3,7,11} {
    \fill[color4!22] (\i+3.5,\i) -- (\i+7.5,\i) -- (\i+8.5,\i+1) -- (\i+4.5,\i+1) -- cycle;
    \draw[color4] (\i+3.5,\i) -- (\i+7.5,\i) -- (\i+8.5,\i+1) -- (\i+4.5,\i+1) -- cycle;
  }
  \draw[color4,line width=2.6pt,opacity=0.55] (3.5,0) -- (7.5,0);
  \draw[color4,line width=2.6pt,opacity=0.55] (15.5,12) -- (19.5,12);
  \node[font=\scriptsize,anchor=mid west] at (10.4,0) {$y_V$};
  \node[font=\scriptsize,anchor=mid west] at (22.4,12) {$y_e[V]$};
  \foreach \a/\b/\p in {1/3/2,4/7/3,8/11/3} {
    \node[font=\tiny,anchor=mid west] at ({(\a+\b)/2+10.4},{(\a+\b)/2}) {$\bigl(A[V,V]\bigr)^{\oplus\p}$};
  }
  \foreach \i/\s in {0/1,3/0.5,7/0.5,11/0} {\node[font=\tiny,color4!80!black,anchor=mid west] at (\i+10.4+\s,\i+\s) {irregular};}
  \node[font=\scriptsize,color3!80!black,anchor=base] at (17.5,13.3) {$V$};
  \node[font=\small,anchor=base] at (11,16.6) {(b) interval candidate};
  \end{scope}
\end{tikzpicture}
\caption{Transfer through a tall light slab
(\cref{lem:tall_light_block_profile}), schematic; thick orange segments
are the vectors at the cuts $e$ and $\ell$.  (a)~The common candidate
$y_\star$ applies the guard pairs (gray) one by one and the middle
(purple) by one range query.  (b)~An interval candidate $y_V$ keeps the
states in $V$ (magenta), applying irregular pairs (orange) separately and
runs of $p$ regular pairs via $\bigl(A[V,V]\bigr)^{\oplus p}$.}
\label{fig:tall_transfer}
\end{figure}
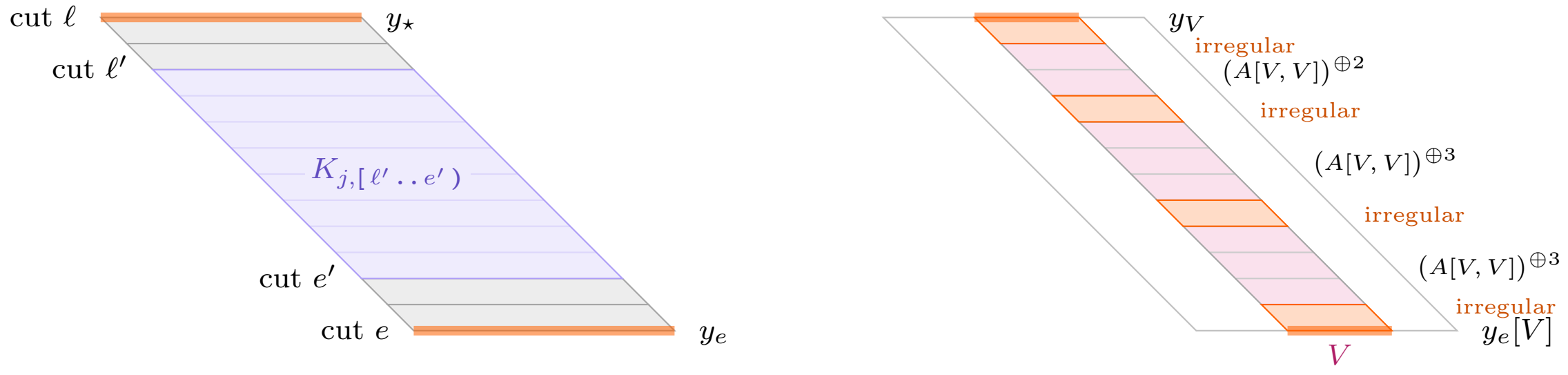

\emph{The common candidate.}  Define
${F_B^\star\coloneqq I_{j,\fragmentco{\ell}{\ell'}}\oplus K_{j,\fragmentco{\ell'}{e'}}\oplus I_{j,\fragmentco{e'}{e}}}$.
By \cref{obs:powered_product_dominance},
\begin{equation}\label{eq:tall_common_product_dominance}
    F_B^\star\ge I_{j,\fragmentco{\ell}{e}}.
\end{equation}

\begin{claim}
\label{clm:tall_common_candidate}
One can compute a vector $y_\star$ with
$y_\star\meq{W}F_B^\star\oplus y_e$ in $\Oh(q^2\lg^2m)$ time.
\end{claim}
\begin{claimproof}
Starting from $y_e$, we apply the CMOs of the factors of the bottom
guard in reverse order (\cref{lem:bd_monge_cmo_operations}), pass the
result $z$ to the query $\mathsf{Range}(j,\ell',e',z)$, and apply the
CMOs of the factors of the top guard in reverse order to its answer.  By
min-plus congruence (\cref{obs:threshold_congruence}) and the guarantee
of \cref{lem:range_oracle}, the
resulting vector $y_\star$ satisfies
${y_\star\meq{W}F_B^\star\oplus y_e}$, as claimed.

Retrieving and applying the CMO of a factor pair takes
${\Oh(\Delta\lg m)=\Oh(d\lg m)}$ time (\cref{lem:bd_monge_cmo_operations}).
The guards consist of $\Oh(q^2/\tau)$ pairs, so they take
$\Oh(q^2\lg m)$ time overall as $d<\tau$.  With the range query, which
takes $\Oh(d\lg^2m)$ time, the common candidate costs
$\Oh(q^2\lg^2m)$ time because $d\le q^2$.
\end{claimproof}

\emph{The interval candidates.}  For each $V\in\mathcal V$, we restrict
every factor matrix to the states in $V$ and define
\[
    I^V_{j,\fragmentco{\ell}{e}}\coloneqq\bigoplus_{i=\ell}^{e-1}I_{j,i}[V,V].
\]
Restricting the intermediate states can only increase the product, so
\begin{equation}\label{eq:interval_restricted_dominance}
    I^V_{j,\fragmentco{\ell}{e}}\ge I_{j,\fragmentco{\ell}{e}}[V,V].
\end{equation}

\begin{claim}
\label{clm:tall_interval_candidates}
One can compute, for all $V\in\mathcal V$, vectors $y_V$ on $V$ with
$y_V\meq{W}I^V_{j,\fragmentco{\ell}{e}}\oplus y_e[V]$ in
$\Oh(qM_j\lg^2m)$ total time.
\end{claim}
\begin{claimproof}
To apply $I^V_{j,\fragmentco{\ell}{e}}$ to the restricted exit vector
$y_e[V]$, we exploit that most factor pairs of the slab are plain.  The
factor pair of $j$ with index $i\in\fragmentco{\ell}{e}$ has the pattern
window ${\fragmentco{i\tau}{(i+1)\tau}}$ and the text window
${\fragmentco{(j+i)\tau-f}{(j+i+1)\tau+f}}$.  We call this pair
\emph{irregular} if
\begin{enumerate}[label=(\roman*),ref=(\roman*)]
    \item\label{it:irregular_window} its text window is not contained in
    ${\fragmentco{(j+\ell)\tau-wq}{(j+e)\tau+wq}}$,
    \item\label{it:irregular_pattern_edit} a reference support of an
    edit of $\A_P$ weakly overlaps its pattern window, or
    \item\label{it:irregular_text_edit} a reference support of an edit
    of $\A_T$ weakly overlaps its text window.
\end{enumerate}
Otherwise, it is \emph{regular}.  By \cref{lem:periodic_factor_charging}, both factors of
a regular pair equal their prototypes, so its matrix is $A$.  Each
maximal run of regular pairs and each irregular pair forms a
\emph{group}.

We apply the groups to $y_e[V]$ in reverse order.  For an irregular
pair, we retrieve from the catalogue the CMO of a matrix $W$-equivalent
to its matrix and apply it to the current vector with rows and columns
restricted to $V$ (\cref{lem:bd_monge_cmo_operations}).  For a run of
$p$ regular pairs, we apply a CMO of a matrix $W$-equivalent to
$\bigl(A[V,V]\bigr)^{\oplus p}$, which we obtain by repeated squaring
from the catalogue CMO of any regular pair, cropped once per interval
$V$ to $V\times V$.  By min-plus
congruence, the resulting vector $y_V$ on $V$ satisfies
${y_V\meq{W}I^V_{j,\fragmentco{\ell}{e}}\oplus y_e[V]}$, as claimed.

The groups are the
same for all intervals.  Condition~\ref{it:irregular_window} applies only to $\Oh(1)$ pairs
at each end of the slab, because $f=\Oh(\tau)$.  Since $\kappa,q\le5d$, we have
${f=21d>3\kappa+q=W+q\ge wq}$, so the condition applies to the pairs with
indices $\ell$ and $e-1$.  Among the other pairs, at most two have a
text window that shares an endpoint with
${\fragmentco{(j+\ell)\tau-wq}{(j+e)\tau+wq}}$, and the windows of the
remaining pairs lie strictly inside the reference ranges of the slab.
Hence, an edit that makes a remaining pair irregular by
condition~\ref{it:irregular_pattern_edit} or~\ref{it:irregular_text_edit} is counted by $\Lambda_B<q/6$,
and it makes $\Oh(1+\Delta/\tau)=\Oh(1)$ pairs irregular.
Thus, there are $\Oh(q)$ groups, and they can be enumerated from the breakpoint
representations in $\Oh(q\lg m)$ time.  For $V\in\mathcal V$, the
cropped CMO of $A$ is constructed in $\Oh(d\lg m)$ time.  Applying an irregular
pair costs $\Oh(|V|\lg m)$ time.  For a run of $p\le m$ pairs, repeated
squaring uses $\Oh(\lg m)$ products of $\Oh(|V|\lg m)$ time each, so
constructing and applying the power costs $\Oh(|V|\lg^2m)$ time.  With $\Oh(q)$ groups, the interval
candidate for $V$ costs
${\Oh(d\lg m+q|V|\lg^2m)=\Oh(q|V|\lg^2m)}$ time because
$d\le q^2\le q|V|$.  By \cref{lem:tall_block_witness_localization},
$\sum_{V\in\mathcal V}|V|=\Oh(M_j)$, so all interval candidates cost
$\Oh(qM_j\lg^2m)$ time.
\end{claimproof}

\emph{Combining the candidates.}  For every entrance state
$g\in\fragment{0}{\Delta}$, we set
\[
    y_\ell[g]\coloneqq
       \min\bigl(\{y_\star[g]\}\cup
                  \{y_V[g]:V\in\mathcal V,\ g\in V\}\bigr).
\]
This takes $\Oh(\Delta+\sum_{V\in\mathcal V}|V|)=\Oh(d+M_j)$ time.
Extend every matrix $I^V_{j,\fragmentco{\ell}{e}}$ by $\infty$ outside
$V\times V$, and let $F$ be the entrywise minimum of $F_B^\star$ and
these extended matrices.  By
\cref{eq:tall_common_product_dominance,eq:interval_restricted_dominance},
$F\ge I_{j,\fragmentco{\ell}{e}}$, and, by
\cref{clm:tall_common_candidate,clm:tall_interval_candidates} and
min-plus congruence,
$y_\ell\meq{W}F\oplus y_e$.  Below, we verify the assumption of
\cref{lem:exact_ckw_relevant_transfer}, separately for paths with and
without a clean middle; the lemma then shows that $y_\ell$ is a
distance estimate at cut $\ell$.

\emph{The assumption of \cref{lem:exact_ckw_relevant_transfer}.}
Let $\eA$ be a tight path from a vertex $x_\ell\in\tClose_{W,b_j}$ on the
$\ell$th lifted cut row to a vertex $x_e$ on the $e$th lifted cut row,
and let $g_i$ be its state at cut $i\in\fragment{\ell}{e}$
(\cref{def:path_pieces}).  We show that
$F[g_\ell,g_e]\le\cost(\eA)$.  By \cref{lem:relevant_path}, all
vertices of $\eA$ are relevant, so every portion of $\eA$ between cuts
$i_1<i_2$ is a tight path from a vertex of $\tClose_{W,b_j}$.  Hence,
the second part of \cref{lem:exact_ckw_relevant_transfer} bounds
$I_{j,\fragmentco{i_1}{i_2}}[g_{i_1},g_{i_2}]$ by the cost of this
portion.  If the portion between the cuts $\ell'$ and $e'$ is clean,
then \cref{lem:ckw_clean_powered_exactness} bounds
$K_{j,\fragmentco{\ell'}{e'}}[g_{\ell'},g_{e'}]$ by its cost, because
relevant vertices have reference offsets within $W$ of $b_jq$.  Adding
the costs of this portion and of the two guard portions, we obtain
\[
    F[g_\ell,g_e]\le F_B^\star[g_\ell,g_e]
      \le I_{j,\fragmentco{\ell}{\ell'}}[g_\ell,g_{\ell'}]
         +K_{j,\fragmentco{\ell'}{e'}}[g_{\ell'},g_{e'}]
         +I_{j,\fragmentco{e'}{e}}[g_{e'},g_e]
      \le\cost(\eA).
\]
If the portion between the cuts $\ell'$ and $e'$ is not clean, then
$\eA$ meets at least two rails by \cref{lem:tall_block_clean_middle},
so \cref{lem:tall_block_witness_localization} gives $V\in\mathcal V$
that contains $g_\ell,\ldots,g_e$.  Adding the costs of the pieces of
$\eA$, we obtain
\[
    F[g_\ell,g_e]\le I^V_{j,\fragmentco{\ell}{e}}[g_\ell,g_e]
      \le\sum_{i=\ell}^{e-1}I_{j,i}[g_i,g_{i+1}]
      \le\cost(\eA).
\]

\emph{Overall running time.}
In total, the relevant exit contacts, the candidates
(\cref{clm:tall_common_candidate,clm:tall_interval_candidates}), and
their minimum take
$\Oh(d\lg m+q^2\lg^2m+qM_j\lg^2m)=\Oh((q^2+qM_j)\lg^2m)$ time.
\end{proof}

\subsection{High-Load Output}
\label{sec:high_load_output}

\Cref{alg:high_load} combines the transfers.  For each high-load index
$j$, it transfers the zero vector at the bottom cut $t+1$ upward
through the slabs of the band centered at $b_j$.  The resulting
distance estimate at the top cut determines the occurrence starts in
the start interval $\fragmentco{r_j}{r_{j+1}}$, so the reports for
different indices are disjoint.  The total load bounds the running
time, and $|J_{\mathrm{high}}|$ bounds the output size.

\begin{proposition}[Output for all high-load start intervals]
\label{prop:high_load_output}
Given the preprocessing record $\mathfrak P$
(\cref{def:preprocessing_record}), one can report in
$\Oh(d^3\lg^2 m)$ time all $\Oh(d|J_{\mathrm{high}}|)$ occurrence
starts assigned to $J_{\mathrm{high}}$.
\end{proposition}
\begin{algorithm}[t]
\caption{Reporting occurrence starts assigned to high-load block indices}
\label{alg:high_load}
\KwIn{a valid instance of \SM with $J_{\mathrm{pad}}\ne\emptyset$ and its preprocessing record $\mathfrak P$}
\KwOut{the starts in $\OccE_k(P,T)$ assigned to $J_{\mathrm{high}}$}
build the fern catalogue and the range oracle\Comment*[r]{\cref{lem:ckw_fern_catalogue,lem:range_oracle}}
\ForEach{$j\in J_{\mathrm{high}}$}{
  partition the band centered at $b_j$ into slabs\Comment*[r]{\cref{lem:band_block_partition}}
  $y\gets\mathbf 0$ on $\fragment{0}{\Delta}$\Comment*[r]{distance estimate at cut $t+1$}
  \ForEach{slab $B=B_{b_j,\fragmentco{\ell}{e}}$, from the bottom to the top}{
    transfer $y$ from cut $e$ to cut $\ell$\Comment*[r]{\cref{lem:short_block_profile,lem:tall_light_block_profile}}
  }
  report every $v=z_{j,0}+g\in\fragmentco{r_j}{r_{j+1}}$ with $y[g]\le k$\Comment*[r]{\cref{lem:semiglobal_expanded_transfer,lem:ckw_start_relevance}}
}
\end{algorithm}
\begin{proof}
\emph{Algorithm.}  If $J_{\mathrm{high}}$ is empty, there is nothing
to report.  Otherwise, $\sqrt d\le q\le\kappa$ by
\cref{def:ckw_load_partition} and $\kappa\le5d$ by
\cref{rem:parameter_bounds}, so the standing assumption of this section
holds, and we run \cref{alg:high_load}.  For each $j\in J_{\mathrm{high}}$,
it keeps a distance estimate $y$ (\cref{def:frontier_state_data}) and
transfers it through short slabs by \cref{lem:short_block_profile} and
through the other (tall and light) slabs by
\cref{lem:tall_light_block_profile}.

\emph{Correctness.}  Fix $j\in J_{\mathrm{high}}$.  Since $h_{t+1}=\mathbf 0$,
the zero vector is a distance estimate at cut $t+1$.  Since $j$ is
padded and $R_*\ge w$, \cref{lem:band_block_partition} applies to
$b_j$, and each slab $B_{b_j,\fragmentco{\ell}{e}}$ is short or is tall
and light with $\fragmentco{\ell}{e}\subseteq\fragmentco{1}{t}$, so
\cref{lem:short_block_profile,lem:tall_light_block_profile} turn a
distance estimate at its exit cut into one at its entrance cut.  By induction, the final vector $y_0$ is a distance
estimate at the top cut.

If $y_0[g]\le k$, then $h_0[g]\le y_0[g]\le k$ because $k<W$, so, by
\cref{lem:semiglobal_expanded_transfer}, every reported position is an
occurrence start.  Conversely, by \cref{lem:ckw_start_relevance},
every occurrence start assigned to $j$ has a relevant lift on the top
cut, so its state $g$ is relevant at cut $0$.  By
\cref{obs:cut_states_near_rail}, $g\in\fragment{0}{\Delta}$, so
${y_0[g]=h_0[g]\le k}$, and the algorithm reports the start.  Thus, the
report for $j$ is exactly ${\OccE_k(P,T)\cap\fragmentco{r_j}{r_{j+1}}}$.
The start intervals are disjoint (\cref{lem:working_budget_ckw_slicing}),
so no start is reported twice.

\emph{Running time.}  The catalogue and the range oracle are
constructed in $\Oh(d^3\lg^2 m)$ time
(\cref{lem:ckw_fern_catalogue,lem:range_oracle}).  For each
$j\in J_{\mathrm{high}}$, the band has $\Oh(d/q)$ slabs, and each
transfer costs $\Oh((q^2+qM_j)\lg^2m)$ time by
\cref{lem:short_block_profile,lem:tall_light_block_profile}.  Since
$M_j\ge q/10$, processing the band takes
${\Oh((d/q)(q^2+qM_j)\lg^2m)=\Oh(dM_j\lg^2m)}$ time.  This also covers the construction of the partition, in
$\Oh(d\lg m+(d/q)\lg^2m)$ time, and the $\Oh(d\lg m)$
time for the initial vector and the final scan.  Summing over all
high-load indices and using $\sum_{j\in J_{\mathrm{high}}}M_j=\Oh(d^2)$
(\cref{lem:ckw_load_bounds}) gives $\Oh(d^3\lg^2m)$ time.
Finally, each top cut has $\Delta+1=\Oh(d)$ states, so at most
$\Oh(d|J_{\mathrm{high}}|)$ positions are reported.
\end{proof}

\section{Assembly and Main Theorem}
\label{sec:central_algorithm}

This section combines the procedures for the three kinds of block
indices; see \cref{sec:overview_high}.  Extreme block
indices use direct verification (\cref{fct:verifyRj}); eligible and
high-load indices use \cref{sec:powered_ckw_sweep,sec:block_geometry},
respectively.  Disjoint start intervals cover all occurrence starts, so
the three outputs form $\OccE_k(P,T)$.
\Cref{sec:progression_normalization} converts the progressions from
difference $\tau$ to difference~$q$, and \cref{sec:central_periodic_algorithm} solves \SM
(\cref{lem:sthold}).  Finally, \cref{sec:main_theorem_proof} proves
\cref{thm:edalgII} by combining \cref{lem:sthold} with the reduction of
Pattern Matching with Edits to \SM (\cref{fact:reduction}).

\subsection{Arithmetic-Progression Normalization}
\label{sec:progression_normalization}

The algorithm of \cref{prop:eligible_output} returns progressions of
difference $\tau$, and that of \cref{prop:high_load_output} returns
single starts, which can be interpreted as progressions of any
difference.  \SM
asks for progressions of difference $q$, which divides $\tau$.  We
compute the maximal progressions of difference $q$ and bound their
number using a structural result of~\cite{CKW20}.  A similar conversion
in~\cite[Section~6.3]{CKW22} relies on an event-driven sweep; we use
interval operations instead.

\begin{lemma}[Changing the common difference from $\tau$ to $q$]
\label{lem:tau_to_q_postprocessing}
For a valid instance of \SM, given $\OccE_k(P,T)$ as the union of
$\Oh(d^3)$ arithmetic progressions of difference $\tau$
(\cref{def:structural_ckw_parameters}), one can represent $\OccE_k(P,T)$ as
$\Oh(d^3)$ pairwise disjoint arithmetic progressions of difference
$q$ in $\Oh(d^3\lg m)$ time.
\end{lemma}
\begin{proof}
Put $S\coloneqq\OccE_k(P,T)$.  We output the progressions of
difference $q$ contained in $S$ that are maximal under inclusion, called
\emph{maximal progressions} below.  We bound their number $N_q$ using~\cite{CKW20} and find
the smallest and largest term of each maximal progression by interval
operations.

\emph{A structural bound.}  We first show that $S$ can be represented
by $\Oh(d^3)$ progressions of difference~$q$; this particular
representation is used only for counting and is not computed.  If $S=\emptyset$, there is
nothing to show.  Otherwise,  let $T'\coloneqq T\fragmentco{v_0}{e_0}$,
where $v_0$ is the leftmost start and $e_0$ is the rightmost end of a
$k$-error occurrence of $P$ in $T$.  All $k$-error occurrences lie in
$T'$, and every occurrence in $T'$ is also an occurrence in $T$, so
$S=\OccE_k(P,T')+v_0$, where ${X+a\coloneqq\{x+a:x\in X\}}$.  Some occurrence is a prefix of $T'$, and some
is a suffix of $T'$.  Moreover, since both sides are
integers, the length promise $|T|<\ceil{\threehalfs m}+k$ implies
$|T'|\le|T|<\threehalfs m+k$.  Hence, all hypotheses of
\cite[Main Theorem~7]{CKW20} hold for $P$ and $T'$ with threshold $k$
and structural parameter $d$, and translating the resulting
decomposition by $v_0$ gives the claimed representation of~$S$.  Each
maximal progression contains at least one of these progressions because
their union is $S$, and maximal progressions are disjoint, so
$N_q=\Oh(d^3)$.

\emph{Smallest and largest terms.}  An element $v\in S$ is the smallest term of
its maximal progression if and only if ${v\in S^-\coloneqq S\setminus(S+q)}$, and
the largest term if and only if ${v\in S^+\coloneqq S\setminus(S-q)}$, so
$|S^-|=|S^+|=N_q$.  The maximal progressions are disjoint, so their smallest and largest
terms alternate within each residue class modulo $q$.  Hence, sorting
$S^-$ and $S^+$ by the residue modulo $q$ and then by value, and
pairing their elements in order within each residue, yields the
maximal progressions.

\emph{Computing $S^-$ and $S^+$.}  Let $N_\tau$ be the number of input
progressions.  The sets $S$, $S+q$, and $S-q$ are unions of $N_\tau$
progressions of difference $\tau$ each.  We group these
progressions by the residue of their terms modulo $\tau$ and map each
term $v$ to $\floor{v/\tau}$, so that every progression becomes an
integer interval.  Within each residue modulo $\tau$, we sort and merge the intervals
of each union and compute the two differences by linear scans, in
$\Oh(N_\tau\lg N_\tau)$ time in total.  Every endpoint of a
resulting interval differs by at most $1$ from an endpoint of an input
interval, so $\Oh(N_\tau)$ intervals represent $S^-$ and $S^+$.  We
enumerate the $2N_q$ elements of $S^-$ and $S^+$, and we sort and pair
them as above in
$\Oh(N_q\lg N_q)$ time.  The total running time is
$\Oh((N_\tau+N_q)\lg(N_\tau+N_q))=\Oh(d^3\lg m)$ because $d\le m$
and $N_\tau,N_q=\Oh(d^3)$.
\end{proof}

\subsection{Algorithm for \texorpdfstring{\SM}{AlignedPeriodicMatches}}
\label{sec:central_periodic_algorithm}

\begin{algorithm}[!htb]
\caption{Solving a valid instance of \SM}
\label{alg:central}
\KwIn{a valid instance $(P,T,k,d,Q,\brp{\A_P},\brp{\A_T})$ of \SM}
\KwOut{$\OccE_k(P,T)$ as $\Oh(d^3)$ pairwise disjoint progressions of difference $q$}
compute $J$, the progressions representing $(r_j)_{j\in J}$, and $J_{\mathrm{pad}}$\Comment*[r]{\cref{fct:rj,lem:padded_ckw_shift_interval}}
verify the $\Oh(1)$ block indices of $J\setminus J_{\mathrm{pad}}$ directly\Comment*[r]{\cref{fct:verifyRj}}
\If{$J_{\mathrm{pad}}\ne\emptyset$}{
  build the preprocessing record $\mathfrak P$\Comment*[r]{\cref{lem:preprocessing_record}}
  report the occurrence starts assigned to $J_{\mathrm{elig}}$ by \cref{alg:eligible}\Comment*[r]{\cref{prop:eligible_output}}
  report the occurrence starts assigned to $J_{\mathrm{high}}$ by \cref{alg:high_load}\Comment*[r]{\cref{prop:high_load_output}}
}
\Return the union of the reports as progressions of difference $q$\Comment*[r]{\cref{lem:tau_to_q_postprocessing}}
\end{algorithm}

Our algorithm for \SM, given as \cref{alg:central}, verifies the
extreme block indices directly, runs \cref{alg:eligible,alg:high_load}
on the eligible and high-load block indices, and converts the union of
their outputs with \cref{lem:tau_to_q_postprocessing}.  We now show
that this algorithm meets the bounds of \cref{lem:sthold}.

\newalg*
\begin{proof}
We analyze the three parts of \cref{alg:central} in order: the extreme
block indices, the padded block indices, and the combination of their
outputs.

\emph{Extreme block indices.}  \Cref{fct:rj} builds the representation
of $(r_j)_{j\in J}$ in $\Oh(d)$ time, and
\cref{lem:padded_ckw_shift_interval} yields the endpoints of
$J_{\mathrm{pad}}$ in constant time; $J\setminus J_{\mathrm{pad}}$
consists of at most $78$ indices at the two ends of $J$.  Their direct
verification by \cref{fct:verifyRj} takes $\Oh(d^2)$ time and returns
$\Oh(d)$ occurrence starts, treated as singleton progressions.  If
$J_{\mathrm{pad}}=\emptyset$, these are all occurrence starts by
\cref{lem:working_budget_ckw_slicing}, because $k\le d$.

\emph{Padded block indices.}  If $J_{\mathrm{pad}}\ne\emptyset$, as
\cref{sec:ckw_common,sec:powered_ckw_sweep,sec:block_geometry} require,
then building the preprocessing record $\mathfrak P$ takes
$\Oh(d^2\lg m)$ time (\cref{lem:preprocessing_record}).  The record
determines the partition
$J_{\mathrm{pad}}=J_{\mathrm{elig}}\mathbin{\dot\cup}J_{\mathrm{high}}$.
The algorithms of \cref{prop:eligible_output,prop:high_load_output}
report exactly the occurrence starts assigned to $J_{\mathrm{elig}}$
and to $J_{\mathrm{high}}$, respectively, each in $\Oh(d^3\lg^2 m)$
time.  The former returns $\Oh(d^3)$ progressions of difference
$\tau$.  The latter returns $\Oh(d|J_{\mathrm{high}}|)=\Oh(d^3/q)$
starts by the bound $|J_{\mathrm{high}}|=\Oh(d^2/q)$ of
\cref{lem:ckw_load_bounds}, and we treat them as singleton
progressions.

\emph{Combining the outputs.}  By
\cref{lem:working_budget_ckw_slicing}, the start intervals are
pairwise disjoint and contain all occurrence starts.  Hence, the
outputs are disjoint, and their union is $\OccE_k(P,T)$,
represented by $\Oh(d^3)$ progressions of difference $\tau$
computed in $\Oh(d^3\lg^2 m)$ time.
\Cref{lem:tau_to_q_postprocessing} converts this representation into
$\Oh(d^3)$ pairwise disjoint progressions of difference $q$
in $\Oh(d^3\lg m)$ time.
\end{proof}

\subsection{Proof of the Main Theorem}
\label{sec:main_theorem_proof}

Finally, we prove \cref{thm:edalgII} by combining the reduction of
Pattern Matching with Edits to \SM (\cref{fact:reduction}) with our
algorithm for \SM (\cref{lem:sthold}).

\edalgII
\begin{proof}
If $k\ge m$, then every position $v\in\fragment{0}{n}$ is an
occurrence start, because $P$ can be aligned with the empty fragment
$T\fragmentco{v}{v}$ using $m\le k$ deletions.  In this case, the
answer is a single arithmetic progression of difference $1$, which
we return in constant time.  Hence, we may assume
$k\in\fragmentco{1}{m}$.

Then, \cref{fact:reduction} applies: it computes $\OccE_k(P,T)$ as
$\Oh(k^3\cdot\ceil{n/m})$ pairwise disjoint arithmetic progressions
with a common difference in $\Oh(k^3\cdot\ceil{n/m})$ time plus the
time needed to solve valid instances of \SM.
For completeness, we derive the bound on $\sum_i(d^{(i)})^3$ stated in
\cref{fact:reduction}: the proof of \cite[Fact 1.1]{CKW22} produces
patterns $P^{(i)}$ with $m/(8k)\le|P^{(i)}|\le m$ and
$\sum_i|P^{(i)}|=\Oh(n)$, and structural budgets
$d^{(i)}=\ceil{8k|P^{(i)}|/m}$.
The lower bound on $|P^{(i)}|$ gives $8k|P^{(i)}|/m\ge1$, so
$d^{(i)}\le16k|P^{(i)}|/m$.  Using $|P^{(i)}|\le m$ and $\sum_i|P^{(i)}|=\Oh(n)$, we obtain
\[
    \sum_i (d^{(i)})^3
     \le \tfrac{16^3k^3}{m^3}\sum_i |P^{(i)}|^3
     \le \tfrac{16^3k^3}{m}\sum_i |P^{(i)}|
     =\Oh\left(k^3\cdot\ceil{\tfrac nm}\right).
\]
By \cref{lem:sthold} and $|P^{(i)}|\le m$, we solve all instances in
total time
\[
    \Oh\left(\sum_i (d^{(i)})^3\lg^2 m\right)
    =\Oh\left(k^3\lg^2 m\cdot\ceil{\tfrac nm}\right),
\]
and their outputs consist of $\Oh(k^3\cdot\ceil{n/m})$ arithmetic
progressions in total.
\end{proof}

\section*{Acknowledgements}
I thank Karl Bringmann, Alejandro Cassis, Nick Fischer, and Philip
Wellnitz for many useful discussions throughout this project.  I am
also grateful to Philip for his contributions to an early draft of this
paper, including a proof of an earlier version of
\cref{lem:planar_unit_increase}.

\paragraph*{Use of AI tools.}
The main ideas of this paper, including the planar variation theorem
(\cref{thm:combi}) and how to prove and use it, were conceived in
December 2022, and discussed with all the original collaborators.  The initial approach was extremely complicated and
lacked many non-trivial details.  Until May 2026, while occasionally discussing it with the collaborators mentioned above,
I gradually
simplified it and developed the language needed to formalize it,
including, among others, expanded alignment graphs, tight paths, rails
and the strips between them, relevant vertices, dense fragments, dense
covers, detours, mark loads of rails, the common factorization, and
slabs.  Along the way, I reworked the proof of \cref{thm:combi}
several times, each time with a substantially simpler approach.

By then, the manuscript had about 45 pages, with essentially complete
proofs of the combinatorial sections.  I started using GPT models
(initially \mbox{GPT-5.5}) for discussions and drafting low-level
arguments.  In July 2026, I asked \mbox{GPT-5.6}~Sol to convert the notes into
a complete paper, which it did with a lot of guidance.  On the positive
side, this confirmed that the remaining details required no conceptual
contribution beyond the existing ones.  On the negative side, the resulting 150-page
manuscript was overcomplicated and lacked structure and narration.
Extensive discussions with \mbox{GPT-5.6}~Sol eliminated many
redundancies, reduced the length to 80 pages, and shaped the paper's
current structure and the statements of the main results.  I then used
\mbox{GPT-6}~Astra and \mbox{GPT-6.1}~Sol to further streamline the terminology and
the structure, especially in
\cref{sec:powered_ckw_sweep,sec:block_geometry}.

In October 2026, I had Claude Opus~5.5 autonomously rewrite the entire paper,
adding narration and figures but keeping the proofs' structure.  Only then
did I process the technical sections one by one.  This included
autonomous simplifications by \mbox{GPT-6}~Pro and \mbox{GPT-6}~Astra,
exposition improvements by Claude Opus~5.5, my own deep pass (at
30--45 minutes per page, in which I asked Claude Opus~5.5 for further
restructuring and numerous local improvements of individual paragraphs and
sentences), and a correctness audit by \mbox{GPT-6}~Astra.  During my
pass, I also asked for the notion of selected edits
(\cref{def:selected_edits}) to be extracted to simplify the predecessor
of \cref{lem:rail_path_savings}, and I asked \mbox{GPT-6}~Astra whether
this notion helps in \cref{sec:mark_distance_variation}.  A short
conversation led to a much simpler proof of \cref{thm:combi}, which
increases the weights along plane cuts in the original alignment graph
instead of individual edges of a carefully constructed auxiliary
graph.  This is the only major conceptual insight
from language models.  It
shortened \cref{sec:mark_distance_variation} by a few pages and gave
cleaner accounting with smaller constants. I take full responsibility for the entire manuscript.

\bibliographystyle{alphaurl}
\bibliography{main}

\appendix
\crefalias{section}{appendix}
\crefalias{subsection}{appendix}

\section{Notation}\label{sec:notation}

This appendix lists the notation that is used beyond a single proof,
in the order in which the technical sections introduce it, with a pointer to each definition.  The first rows recall
the problem parameters.

{\small
\begin{longtable}{@{}p{0.22\linewidth}p{0.55\linewidth}p{0.178\linewidth}@{}}
\toprule
Symbol or term & Meaning & Defined in\\
\midrule
\endfirsthead
\toprule
Symbol or term & Meaning & Defined in\\
\midrule
\endhead
\bottomrule
\endlastfoot
\multicolumn{3}{@{}l}{\textbf{Problem (\cref{sec:introduction,sec:technical_overview})}}\\[2pt]
$P,T$; $m,n$ & pattern and text; their lengths & Sec.~\ref{sec:introduction}\\
$k$; $\OccE_k(P,T)$ & threshold; starting positions of $k$-error occurrences & Sec.~\ref{sec:introduction}\\
$Q,q$; $d$ & primitive period and its length; structural budget, with $d\ge2k$ and $q\le m/(8d)$ & Sec.~\ref{sec:overview_reduction}\\
$\A_P,\A_T$; $d_P,d_T$ & supplied alignments $P\onto Q^\infty\fragmentco{0}{y_P}$ and $T\onto Q^\infty\fragmentco{x_T}{y_T}$; their costs, with $d_P\le d$ and $d_T\le3d$ & Sec.~\ref{sec:overview_reduction}\\
\addlinespace
\multicolumn{3}{@{}l}{\textbf{Preliminaries (\cref{sec:preliminaries})}}\\[2pt]
$\lg z$ & $\log_2(\max(2,z))$, for $z\in\mathbb R$ & Sec.~\ref{sec:prelims_notation}\\
$Q^\infty$ & bi-infinite repetition of $Q$ & Def.~\ref{def:bi_infinite_repetition}\\
$\AG(X,Y)$ & alignment graph & Def.~\ref{def:alignment_graph}\\
$\ed^{\A}(F,G)$ & cost of the restriction of $\A$ that aligns $F$ with $G$ & Def.~\ref{def:alignment_restriction_cost}\\
$\brp{\A}$ & breakpoint representation: the tails of the edits of $\A$ with the characters involved, between two dummy entries & Def.~\ref{def:breakpoint_representation}\\
\addlinespace
\multicolumn{3}{@{}l}{\textbf{Block indices (\cref{sec:ckw_windows})}}\\[2pt]
reference coordinates & positions in $Q^\infty$ & Def.~\ref{def:structural_ckw_parameters}\\
$\kappa$; $\tau$ & combined edit budget $d+d_P+d_T\le5d$; block length, the least multiple of $q$ exceeding $\kappa$ & Def.~\ref{def:structural_ckw_parameters}\\
$\hP,\hT$ & reference fragments $Q^\infty\fragmentco{0}{y_P}$ and $Q^\infty\fragmentco{x_T}{y_T}$ & Def.~\ref{def:periodic_reference_fragments}\\
period shift $b$ & integer satisfying the endpoint bound for aligned fragments & Def.~\ref{def:phase_shift}\\
$J$; $b_j$ & block indices; origin $j\tau/q$ & Def.~\ref{def:jdef}\\
$r_j$; $\fragmentco{r_j}{r_{j+1}}$ & leftmost text position aligned with a reference coordinate $\ge j\tau-\kappa$; start interval of $j$ & Def.~\ref{def:jdef}\\
\addlinespace
\multicolumn{3}{@{}l}{\textbf{Expanded graph and rails (\cref{sec:expanded_canonical_geometry})}}\\[2pt]
$\eAG(P,T)$; $\ell_P,\ell_T$ & expanded alignment graph on pairs of vertices of $\A_P$ and $\A_T$; numbers of edges of $\A_P$ and $\A_T$ & Def.~\ref{def:expanded_alignment_graph}\\
$\pP,\phP,\pT,\phT$ & positions in $P$, $\hP$, $T$, $\hT$ of an expanded vertex & Def.~\ref{def:expanded_alignment_graph}\\
$\pPT$; $\phPT$; $\mathrm{off}(u)$ & input projection; reference projection; reference offset $\phT(u)-\phP(u)$ & Def.~\ref{def:expanded_projections}\\
reference-diagonal edge & edge whose reference projection is a diagonal edge & Def.~\ref{def:reference_diagonal_edge}\\
$\dtob$; $\tdtob$ & distance to the bottom row; its expanded version & Def.~\ref{def:semiglobal_suffix_distances}\\
tight path & path whose cost equals the decrease of the suffix distance & Def.~\ref{def:tight_paths}\\
$\cancost(u,v)$; $\cancost(u)$ & canonical cost: edits of $\A_P$ and $\A_T$ between $u$ and $v$; the same up to the end & Def.~\ref{def:canonical_cost}\\
$\canD$; $\prD$ & rail indices; indices of proper rails & Defs.~\ref{def:blue} and~\ref{def:proper_rails}\\
$\hD{b}$; $\eD{b}$ & reference rail $\hat t-\hat p=bq$; rail, its lift to $\eAG(P,T)$ & Def.~\ref{def:blue}\\
rail contact; strip & vertex shared with a rail; region between two consecutive proper rails & Defs.~\ref{def:blue} and~\ref{def:rail_strips}\\
left/right of $\eD{c}$ & $\mathrm{off}(u)<cq$; $\mathrm{off}(u)>cq$, or $\mathrm{off}(u)=cq$ and $u\notin\eD{c}$ & Def.~\ref{def:proper_rails}\\
$\tpena_b(u)$; $\tClose_{W,b}$ & penalty $|\mathrm{off}(u)-bq|+\cancost(u)$; relevant vertices, with $\tdtob+\tpena_b\le W$ & Def.~\ref{def:relevant_vertices}\\
\addlinespace
\multicolumn{3}{@{}l}{\textbf{Dense covers (\cref{sec:dense_covers_marks})}}\\[2pt]
(weakly) overlap; (strictly) disjoint & share a position (weakly: the closed position intervals intersect); the negations & Def.~\ref{def:fragment_overlap}\\
support; $\ed^{\A}(F)$ & fragment $S\fragmentco{x}{x'}$ of an edge $(x,y)\to(x',y')$; fragment alignment cost: number of edits with support in $F$ & Def.~\ref{def:one_argument_alignment_cost}\\
$(\rho_*,\rho_+)$-dense & $|F|<\rho_*(\ed^{\A}(F)+\rho_+)$ for prefixes and suffixes, $|F|<\rho_*\ed^{\A}(F)$ otherwise & Def.~\ref{def:dense}\\
$\Es_{\A}(\rho_*,\rho_+)$ & dense cover: hulls of the components of dense fragments & Def.~\ref{def:dense_cover}\\
$\Es_P,\Es_T$ & reference covers, with parameters $(5q,2d)$ and $(5q,0)$ & Def.~\ref{def:canonical_dense_covers}\\
reference support & for an alignment onto a fragment of $Q^\infty$: support of an edit in the inverse alignment & Def.~\ref{def:reference_supports}\\
detour & part of a path between its reference-diagonal rail edges & Def.~\ref{def:detours}\\
\addlinespace
\multicolumn{3}{@{}l}{\textbf{Mark loads and path savings (\cref{sec:mark_savings})}}\\[2pt]
strip $b$; weakly meet & region $bq\le\hat t-\hat p\le(b+1)q$; cover fragments containing positions $\hat p,\hat t$ with $bq\le\hat t-\hat p\le(b+1)q$ & Def.~\ref{def:reference_strips}\\
$w_P,w_T$; $\mu(b)$ & marking weights; mark load of strip $b$: sum of the pair marks $\min(w_P(\E_P),w_T(\E_T))$ over the pairs that weakly meet it & Def.~\ref{def:marking}\\
$S_b$ & edits selected for strip $b$: for each pair weakly meeting strip $b$, all edits on its side of smaller marking weight; $|S_b|\le\mu(b)$ & Def.~\ref{def:selected_edits}\\
$z_b$ & latest vertex of the proper rail $\eD{b}$, on the bottom row & Def.~\ref{def:rail_bottom_endpoints}\\
$\cost^\circ(\mathcal A)$ & cost of a path with free moves along the bottom row & Def.~\ref{def:rail_bottom_endpoints}\\
\addlinespace
\multicolumn{3}{@{}l}{\textbf{Planar variation theorem (\cref{sec:mark_distance_variation})}}\\[2pt]
$\eta^\star$; $C^+$ & $\floor{8d/q}+1$; enlargement of a rail interval $C$ by $\eta^\star$ & Def.~\ref{def:rail_interval_enlargement}\\
$G_{I,\ep}$; $R_b$ & band graph; rails restricted to the suffix rectangle & Def.~\ref{def:comparison_graph}\\
$L_e$; $K_e$ & row or column line of an edit $e\in S_b$; edges of the closed strip between $R_b$ and $R_{b+1}$ that pass $e$, including both rails & Sec.~\ref{sec:eliminate_hops}\\
\addlinespace
\multicolumn{3}{@{}l}{\textbf{Matrices and ferns (\cref{sec:fern_toolbox})}}\\[2pt]
$A\oplus B$; $A^{\oplus u}$ & min-plus product; $u$-fold min-plus power & Def.~\ref{def:min_plus_product}\\
Monge; CMO & Monge inequalities; core-based matrix oracle & Sec.~\ref{sec:monge_toolbox}\\
$\oAG(F,G)$ & alignment graph with reverse edges of cost $2$ & Def.~\ref{def:augmented_ag}\\
$\udmat{F}{G}{\Delta}$ & restricted distance matrix, from the leftmost $\Delta+1$ top vertices to the rightmost $\Delta+1$ bottom vertices & Def.~\ref{def:restricted_distance_matrix}\\
$a\meq{k}b$ & $k$-equivalence: $a=b$ or both exceed $k$ & Def.~\ref{def:threshold_equivalence}\\
$(\Delta,k)$-fern & matrix that is $k$-equivalent to $\udmat{F}{G}{\Delta}$ & Def.~\ref{def:fern}\\
covered seam & cut that every path of cost at most $k$ crosses within the shared states & Def.~\ref{def:covered_seam}\\
string sequence; drift & sequence of strings with distinct consecutive values; sum of consecutive edit distances & Def.~\ref{def:string_sequence}\\
\addlinespace
\multicolumn{3}{@{}l}{\textbf{Common factorization and preprocessing (\cref{sec:ckw_common})}}\\[2pt]
$f$; $\Delta$ & text margin $21d$; overlap $2f$ of consecutive text factors & Def.~\ref{def:common_factorization_parameters}\\
$W$; $w$; $R_*$ & threshold $3\kappa$ of the ferns; $\floor{W/q}+1$; $w+\floor{f/q}+1$ & Def.~\ref{def:common_factorization_parameters}\\
$t$ & number $\floor{(y_P-\kappa)/\tau}$ of internal cuts & Def.~\ref{def:common_factorization_parameters}\\
$J_{\mathrm{pad}}$ & padded block indices: all rails within $R_*$ of $b_j$ are proper & Def.~\ref{def:common_factorization_parameters}\\
$a_i$; $c_u$ & pattern cut at $i\tau$; text cut at $u\tau$ & Def.~\ref{def:internal_cuts}\\
$P_i,T_u$; $Q_P^\circ,Q_T^\circ,Q_R^\circ$ & internal factors; prototypes of $P_i$, $T_u$, and $T_{j,R}$ (an internal factor equal to its prototype is \emph{plain}) & Defs.~\ref{def:internal_cuts} and~\ref{def:prototypes}\\
$c_j^+$; $C_j$; $P_R,T_{j,R}$ & text position paired with $j\tau+y_P$; text fragment of $j$; endpoint factors & Def.~\ref{def:centered_carrier}\\
$I_{j,i}$; $I_{j,\fragmentco{\ell}{e}}$; $I_j$ & factor matrices; their product over the indices in $\fragmentco{\ell}{e}$; full product & Def.~\ref{def:natural_ckw_factor_matrices}\\
$H_i^P,H_i^T$; $\mathcal X_P,\mathcal X_T$ & reference windows; exceptional lists & Def.~\ref{def:exceptional_lists}\\
fern catalogue & $\Oh(\lg m)$-time access to ferns of the factor matrices $I_{j,i}$ & Lem.~\ref{lem:ckw_fern_catalogue}\\
$M_j$; $J_{\mathrm{high}}$, $J_{\mathrm{elig}}$ & load of the block index $j$; high-load and eligible block indices & Def.~\ref{def:ckw_load_partition}\\
$\mathfrak P$ & preprocessing record & Def.~\ref{def:preprocessing_record}\\
\addlinespace
\multicolumn{3}{@{}l}{\textbf{Eligible block indices (\cref{sec:powered_ckw_sweep})}}\\[2pt]
$A$ & matrix of the prototypes & Def.~\ref{def:canonical_run_powers}\\
in-phase state; $S$ & state in $f+q\mathbb Z$; the set of these states & Def.~\ref{def:canonical_run_powers}\\
$\overline A$ & long-run matrix, $\overline A[g,h]=\min_{c\in S}(|g-c|+|h-c|)$ & Def.~\ref{def:canonical_run_powers}\\
$\chi$ & cutoff $42$ for long plain runs & Def.~\ref{def:canonical_run_powers}\\
plain run & maximal interval $G\subseteq\fragmentco{0}{t}$ with $i\notin\mathcal X_P$ and $j+i\notin\mathcal X_T$ for all $i\in G$ & Def.~\ref{def:powered_ckw_interval_products}\\
$K_{j,\fragmentco{\ell}{e}}$, $K_j$ & powered products: plain runs with at least $\chi$ pairs in $\fragmentco{\ell}{e}$ replaced by $\overline A$ & Def.~\ref{def:powered_ckw_interval_products}\\
$\widehat a_i,\widetilde a_i$ & cut row in reference and expanded coordinates & Def.~\ref{def:cut_rows}\\
$z_{j,i}$; $\widetilde\rho_{j,i}$ & text anchor; state of a vertex at cut $i$ & Def.~\ref{def:cut_row_states}\\
piece & subpath between the vertices of a path at consecutive cuts & Def.~\ref{def:path_pieces}\\
clean path & path whose edges off the rails lie in the covers & Def.~\ref{def:clean_paths}\\
witness for $v$ & shortest path from a top-row lift of $(0,v)$ to the bottom row & Def.~\ref{def:witnesses}\\
token; powered word & pair of an interval of factor indices and a matrix; the tokens of $j$ in order & Def.~\ref{def:tokens}\\
$\mathsf{Range}(j,\ell,e,z)$ & range oracle: applies $K_{j,\fragmentco{\ell}{e}}$ to the vector $z$ & Lem.~\ref{lem:range_oracle}\\
$\Pi$ & partition of $J_{\mathrm{elig}}$ into intervals with equal powered products & Lem.~\ref{lem:profile_partition}\\
\addlinespace
\multicolumn{3}{@{}l}{\textbf{High-load block indices (\cref{sec:block_geometry})}}\\[2pt]
band & region between the rails $b-w$ and $b+w$ & Def.~\ref{def:band}\\
$\widetilde u_{c,i}$ & latest contact of the rail $c$ with the lifted cut $i$ & Def.~\ref{def:last_rail_contacts}\\
slab $B_{b,\fragmentco{\ell}{e}}$ & part of a band between two cuts & Def.~\ref{def:band_block}\\
$H_B$; $\Lambda_B$ & height of a slab; its canonical cost; \emph{tall} if $H_B>18q^2$, \emph{light} if $\Lambda_B<q/6$ & Def.~\ref{def:slab_classes}\\
$h_i$ & vector of the suffix distances of the states at cut $i$ & Def.~\ref{def:frontier_state_data}\\
relevant state; distance estimate & state of a vertex of $\tClose_{W,b_j}$ on the cut; vector of upper bounds on $h_i$ (up to $W$) that is exact at relevant states & Def.~\ref{def:frontier_state_data}\\
\end{longtable}
}

\end{document}